\documentclass{CSL}%%%%where CSL is the template name
\usepackage[authoryear]{natbib}

\jname{International Journal of Astrobiology}%

\newcommand{\tick}{\textcolor{green}{\ding{52}}}
\newcommand{\cross}{\textcolor{red}{\ding{56}}}

\usepackage[version=4]{mhchem}

\definecolor{dgreen}{rgb}{0.05,0.5,0.1}
\definecolor{burgundy}{rgb}{0.50,0.00,0.0}%13}

\usepackage{pifont}

\usepackage{aas_macros}

\begin{document}

\supertitle{Research Paper}

\title[Habitability constraints by nutrient availability]{Habitability constraints by nutrient availability in atmospheres of rocky exoplanets}

\author[Herbort et al.]{Oliver Herbort$^{1}$, Peter Woitke$^{2}$, Christiane Helling$^{2,3}$, Aubrey L. Zerkle$^{4}$}

\address{\add{1}{Department for Astrophysics, University of Vienna, T\"urkenschanzstrasse 17, A-1180 Vienna, AT} \\ \add{2}{Space Research Institute, Austrian Academy of Sciences, Schmiedlstrasse 6, A-8042 Graz, AT} \\ \add{3}{TU Graz, Fakult\"at f\"ur Mathematik, Physik und Geod\"asie, Petersgasse 16, A-8010 Graz, AT} \\ \add{4}{Blue Marble Space Institute of Science, Seattle, WA, 98104, USA}}

\corres{\name{Oliver Herbort} \email{oliver.herbort@univie.ac.at}}

\begin{abstract}
Life as we know it requires the presence of liquid water and the availability of nutrients, which are mainly based on the elements C, H, N, O, P, and S (CHNOPS) and trace metal micronutrients.
We aim to understand the presence of these nutrients within atmospheres that show the presence of water cloud condensates, potentially allowing the existence of aerial biospheres.\\
In this paper we introduce a framework of nutrient availability levels based on the presence of water condensates and the chemical state of the CHNOPS elements.
These  nutrient availability levels are applied to a set of atmospheric models based on different planetary surface compositions resulting in a range of atmospheric compositions. 
The atmospheric model is a bottom-to-top equilibrium chemistry atmospheric model which includes the atmosphere-crust interaction and the element depletion due to the formation of clouds.\\
While the reduced forms of CNS are present at the water cloud base for most atmospheric compositions, P and metals are lacking.
This indicates the potential bio-availability of CNS, while P and metals are limiting factors for aerial biospheres. 

\end{abstract}


\selfcitation{Herbort O., Woitke P., Helling Ch., Zerkle A. L.  (2024). Habitability constraints by nutrient availability in atmospheres of rocky exoplanets https://doi.org/xx.xxxx/xxxxx}

\received{17 Aug 2023}

\revised{31 Jan 2024}

\accepted{11 Mar 2024}

\maketitle

\section{Introduction}\label{sec:IntroARE3}
Some of the key questions for humankind are what the conditions favourable for the creation, evolution, and persistence of life on Earth and other planets are. 
In recent times, many exoplanets have been detected and the question of the potential for biological and chemical processes to exist on these planets arose.
These show a vast diversity, some without analogue in our own solar system.
Especially for the rocky planets, a large diversity of different atmospheric compositions is expected \citep[e.g.][]{Leconte2015, Grenfell2020}.
While the extent of these atmospheres can range from a large envelope of primordial gas to almost no atmosphere at all, the composition can be very diverse, covering reducing and oxidizing atmospheres. 
For highly irradiated planets, the atmosphere can be composed of evaporated minerals due to the high temperature, which can melt the planetary surface.
Recent observations with JWST for example constrain the presence of atmospheres on Trappist-1b \citep{Ih2023} and hint to a water-rich atmosphere of GJ 486b \citep{Moran2023}.
The upcoming next generation ground and space based telescopes (e.g. ELT \citep{Gilmozzi2007}, ARIEL \citep{Tinetti2018}) will allow to further investigate the composition of atmospheres of rocky exoplanets \citep[see e.g.][]{Lopez-Morales2019, Wunderlich2020, Ito2021}.
The detailed understanding of the planets itself becomes important for interpreting observations, especially for the detection of biosignatures \citep{Catling2018, Meadows2018, Lisse2020, Lyons2020}.
These signatures of life are mostly confined to water based life, as it is the only known form of life.
However, other solvents like \ce{NH3} or \ce{CH4} could potentially replace the liquid water for the formation of life \citep[e.g.][]{Ballesteros2019}.

In order to understand and constrain the existence of liquid water on the surface of a planet, the habitable zone has been defined \citep[e.g.][]{Huang1959, Kasting1993}.
This concept has been further adjusted by including energy budgets and constraints on other molecules such as \ce{CO2} for further needs for the evolution of life \citep[e.g.][]{Hoehler2007, Ramirez2020} and can explain the water on Earth and on the young Mars \citep{2018A&ARv..26....2L}.
However, water can also be found outside of this habitable zone, for example in sub-glacial oceans on Europa and Enceladus \citep{Tjoa2020} as well as in clouds of rocky planets, sub-Neptunes and gas giants \citep{Benneke2019, Charnay2021, Madhusudhan2021, Mang2022}.
This shows that a large diversity of water environments is to be expected for exoplanets.
Whereas the liquid water in oceans and below frozen surfaces provide the potential for subsurface biospheres on rocky planets and moons, the presence of water clouds on any planet could allow for some aerial biospheres.
\Fpagebreak

The potential of an aerial biospheres for Venus, or as a general scenario, has been discussed previously \citep[e.g.][]{Morowitz1967, Woese1979, Dartnell2015}, and recently gained further interest \citep{Greaves2020} \citep{Kotsyurbenko2021, Rimmer2021, Patel2022} following from the proposed detection of phosphine (\ce{PH3}) in Venus' atmosphere \citep{Greaves2020}.

While \citep{Bains2021, Bains2022} discussed that there is no known process available to abiotically produce a \ce{PH3} feature in the Venus atmosphere, \citet{Omran2021} investigate the general possibility of phosphine production on rocky exoplanets.
Assuming the production by some kind of living organism, this would introduce the necessity of a biosphere which is only present in a cloud layer.
Individual organisms would need to remain afloat for long enough to metabolize and reproduce, before eventually falling towards the surface of Venus \citep{Seager2021a, Patel2022, Seager2021}.
This concept of aerial biospheres enlarges the possibilities of potential habitability from the presence of liquid water on the surface to all planets with liquid water clouds.
See \citet{Seager2021} for a discussion on the potential of aerial biospheres on mini-Neptunes.
The water clouds of such planets could provide the chemical potential for the formation of life \citep{Hallsworth2021}.
This motivates a further investigation of the chemical environment of such aerial biospheres.

For the formation of life, the presence of liquid water is not the only hurdle to overcome, but also further elements and conditions need to be biologically available \citep[see e.g.][]{Deamer2022}.
All life as we know it is is made up of carbon, hydrogen, nitrogen, oxygen, phosphorus, and sulphur, in the following referred to as CHNOPS elements \citep{Horneck2016}. 
If these elements are available, chemical reactions triggered by stellar radiation, lightning, high energetic particles, and cosmic rays can lead to the production of molecules, which allow the formation of amino acids and other pre-biotic molecules from which life can eventually form \citep[see e.g.][]{Bailey2014, Ferus2017, Ranjan2021arXiv, Barth2021XUV}.
For these reactions, the red-ox state in which the elements are present is of particular importance, because elemental speciation dictates bioavailability.

The importance of the CHNOPS elements for  life can be illustrated by the Redfield ratio. 
\citet{Redfield1934} described, that organic matter in all of Earth's oceans comprises of a fairly constant ratio of CNP. 
This has been further expended to further elements \citep[e.g.][]{Anderson1994, Ho2003}, revealing a stoichiometric ratio of C$_{124}$ N$_{16}$ P$_{1}$ S$_{1.3}$, with further traces of metals. 
This universal incorporation of these nutrient elements underlines their importance to life as we know it and shows that the P and S are limiting factors for life.

In order to investigate the pre-biotic atmospheres from which life might evolve, we investigate and characterise the chemical state of the elements present in atmospheres of rocky exoplanets, with liquid water thermally stable  at some point in the atmosphere.
We use a chemical equilibrium model for the lower parts of atmospheres of rocky exoplanets with various surface compositions, which allow the study of all elements present in the gas phase.
We especially focus on the different CHNOPS elements which are necessary for the formation of pre-biotic molecules in exoplanetary atmospheres.

%==============================================
\section{Defining nutrient availability levels}\label{sec:HabLevDef}
%==============================================
\begin{table}[t!]
\caption{Definition of the nutrient availability levels.}
\label{tab:Levels}
		\vspace{-0mm}
 		\begin{center}
		\begin{tabular}{lccccc} 
		\hline \hline
Nutrient availability~ &\ce{H2O}[l]	&C&N&S&P	\\ \hline
Not habitable	&\cross	&-   		&-			&-			&-		\\
Level 0		&\tick	&\cross	&\cross	&\cross	&\cross		\\
Level 1C	&\tick	&\tick	&\cross	&\cross	&-			\\
Level 1N	&\tick	&\cross	&\tick	&\cross	&-			\\
Level 1S	&\tick	&\cross	&\cross	&\tick	&-			\\
Level 2CN	&\tick	&\tick	&\tick	&\cross	&-			\\
Level 2CS	&\tick	&\tick	&\cross	&\tick	&-			\\
Level 2SN	&\tick	&\cross	&\tick	&\tick	&-			\\
Level 3		&\tick	&\tick	&\tick	&\tick	&-			\\
Level 3red	&\tick	&~red~	&~red~	&~red~	&-			\\
Level 3ox	&\tick	& ox	& ox	& ox	&-			\\
P			&\tick	&-		&-		&-		&\tick		\\ \hline
   		\end{tabular}  		
   		\end{center}
\textbf{Notes:} 
The tick marks and crosses indicate the presence and absence of the element in the header of the table, respectively. 
An element is considered present, if at least one molecule incorporating the element in question is present in the gas phase at concentrations above $10^{-9}$.
The state of the elements indicated with {`-'} do not matter for the definition of the respective nutrient availability level.\\
`red': indicates that the reduced form of the respective element (\ce{CH4}, \ce{NH3}, \ce{H2S}) is present at concentrations higher than $10^{-9}$.\\
`ox': indicates the presence of the oxidised state (\ce{CO2}, \ce{NO_x}, \ce{SO2}) at concentrations higher than $10^{-9}$.\\
While these six molecules are the most prominent CNS bearing molecules, further molecules contributing to the presence of these elements are \ce{CO}, \ce{COS}, \ce{HNO3}, \ce{S2O}, \ce{H2SO4}, and \ce{S_x}.
\end{table}

The characterisation of the habitability potential of different atmospheric compositions should take not only the presence of liquid water, but also the availability of essential elements into account.
Therefore we introduce a concept of nutrient availability levels.

The fundamental substance for any discussion of biology as we know it is liquid water.
Therefore the presence of liquid water is used as a prerequisite for every nutrient availability level.
We considered any atmosphere without water condensates as uninhabitable. We defined atmospheres with only liquid water, but no nutrient elements, as having level~0 nutrient availability.

We define further nutrient availability levels as having the stability of water as a precondition and are based on the presence of the elements C, N, and S, as H and O are by definition already present in the form of \ce{H2O}. 
We define an element $i$ as present, if the concentration of an $i$ bearing molecule is greater than one part per billion ($10^{-9}$, ppb).
As the current understanding of the formation of life does not allow to preciously define a lower threshold for the needed concentrations of any molecule, we set this threshold to 1\,ppb in the gas phase as this allows for the inclusion of atmospheric trace species. In principle, also atmospheric species with lower concentrations could accumulate locally by for example dissolution in cloud droplets. However, such processes are not included in this work and therefore the trace species concentration of 1\,ppb is applied to define the presence of a molecule.
If one or two of C, N, and S are present, the atmosphere is classified as level~1 and level~2, respectively.
In order to distinguish the different kinds of level~1 and 2, the present elements are added to the notation.
For example, if C is present, but not N and S, the atmospheric composition classifies as level~1C.
Following on this, the nutrient availability level~3 requires the presence of all three elements.
 
For the nutrient availability level~3 we further distinguish between the chemical redox state of the CNS elements.
If all three elements are present in a reduced form (\ce{CH4} for C, \ce{NH3} for N, and \ce{H2S} for S), the nutrient availability level is classified as level~3red.
The need for this additional distinction is motivated as under such reducing conditions processes such as lightning can result in the formation of pre-biotic molecules, such as \ce{HCN}, amino acids, and nucleobasis \citep[e.g.][]{Miller1953, Miller1959, Toupance1975, Ferris1978, Parker2011, Ferus2017, Pearce2022}.
Similarly to the reduced case, we define the oxidised regime as level~3ox with the presence of \ce{CO2}, \ce{NO2}, and \ce{SO2}.
It is also possible, that different redox states of an element coexist in an atmosphere. 
If for carbon, nitrogen, and sulphur the oxidised and reduced from coexist, with all species present at concentrations more than $10^{-9}$, the atmosphere is categorised as nutrient availability level~3redox.
In \citet{Woitke2020a} we have shown that the coexistence of \ce{CO2} and \ce{CH4} is a fundamental result of chemical equilibrium for specific element abundances.

Molecular nitrogen shows a very strong triple bond, which makes it much less accessible for life than other nitrogen sources such and \ce{NH3} or \ce{NO2}.
Therefore we distinguish between the presence of nitrogen in a more easily accessible form and \ce{N2}.
We note that processes multiple processes exist, which transform \ce{N2} to the more easily accessible forms of nitrogen \citep[see e.g.][]{Burris1945, Stuecken2016}.
Examples for these can be of natural origin (e.g. lightning or UV and cosmic ray radiation) as well biotic by microbes.

In addition to the nutrient availability levels based on CNS, the presence of phosphorus is a further constraint for the formation of life.
As it is a limiting factor for the biosphere on Earth \citep{Syverson2021}, we use it as a category on its own.
All nutrient availability levels are summarised in Table~\ref{tab:Levels}.

These nutrient availability levels are defined to investigate whether or not atmospheres with liquid water clouds can provide basic necessities for possible aerial biosphere. In order to principally allow the presence of an aerial biosphere, all nutrients need to be available.
However, even in the case of their presence, further questions of whether such an aerial biosphere can exist need to be investigated. This includes but is not limited to processes that form more complex chemicals from the nutrients, but also whether life can remain afloat for long enough periods of time.

%==============================================
\section{Atmospheric model}\label{sec:methods}
%==============================================
In order to investigate the implications of the concept of the different nutrient availability levels defined in the previous section on atmospheres of rocky planets, we create atmospheric models with different element abundances to which we apply the concept of the nutrient availability levels.
The atmospheric model is presented in \citet{Herbort2022} and uses the chemical equilibrium solver {\textsc{GGchem}} \citep{Woitke2018}. 
The bottom-to-top atmosphere model describes the lower atmosphere with a polytropic hydrostatic equilibrium atmosphere which takes element depletion by cloud formation into account.
The base of the atmosphere is in chemical phase equilibrium with the crust, as described in \citet{Herbort2020}.
In every atmospheric layer, the chemical equilibrium is solved and all thermally stable condensates are removed while the gas phase is the total element abundance of the atmospheric layer above. 
This depletes the element abundances of all elements affected by cloud formation.
Each bottom-to-top atmosphere is fully defined by the surface pressure $(p_\mathrm{surf})$, surface temperature $(T_\mathrm{surf})$, set of total element abundances $(\epsilon_\text{tot})$, and the polytropic index $(\gamma)$.

As in \citet{Herbort2022}, we investigate $(p, T)$ profiles with a constant polytropic index $\gamma=1.25$.
The surface conditions are defined such that the atmosphere reaches the reference temperature of $T_\mathrm{ref}=300\,$K in the atmosphere at $p_\mathrm{ref}=1\,$bar.
This is guaranteed for a  $(p_\mathrm{gas},T_\mathrm{gas})$ profile defined by 
\begin{align}
T_\mathrm{gas} = \frac{T_\mathrm{ref}}{ p_\mathrm{ref}^{1-1/\gamma}} · p_\mathrm{gas}^{1-1/\gamma}.\label{eq:Tofp}
\end{align}
We investigate a surface temperature range from 300\,K to 1000\,K.
The surface pressures for the different surface temperatures are given by Eq.~\ref{eq:Tofp} to ensure that all atmospheric profiles reach the reference point.

In our calculations, the included 18 elements (H, C, N, O, F, Na, Mg, Al, Si, P, S, Cl, K, Ca, Ti, Cr, Mn, and Fe) can form 471 gas species and 208 condensates.
The total element abundances $\epsilon_\text{tot}$ for the different compositions used in this paper are listed in Table~\ref{tab:abundances} and briefly introduced in the following.
The compositions of Bulk Silicate Earth \citep[BSE, ][]{Schaefer2012}, Continental Crust \citep[CC, ][]{Schaefer2012}, Mid Oceanic Ridge Basalt \citep[MORB, ][]{Arevalo2010}, and CI chondrite \citep[CI, ][]{Lodders2009} are used to represent a variety of different rock compositions.
In order to investigate different levels of hydration, we use sets of element abundances based on BSE with increased abundances of H and O (BSE12, BSE15).
The enrichment of respectively 12 and 15 weight percent H and O, results in a crust composition rich in hydrated minerals \citep[for further details see][]{Herbort2020}.
The Earth set of total element abundances is the result of a fit of our atmosphere-crust chemical and phase equilibrium model.
At $p = 1.013\,$bar and $T = 288.15\,$K this set of element abundances results in an modern Earth like gas phase, except for the \ce{CH4}, which is in disequilibrium \citep[Further details see Appendix A][]{Herbort2022}.
The Archean model represents a strongly reducing atmosphere (presence of \ce{CH4} and \ce{NH3}) and stable \ce{H2O}[l].
In order to also include non-solar-system based elemental ratios, three models based on element abundances from polluted white dwarfs \citep[PWD,][]{Melis2016} are included. As not all elements are included, the missing abundances (especially H) are completed from BSE, MORB, and CI.
In order to investigate hydrogen rich atmospheres, a model with solar abundances \citep{Asplund2009} is taken into account.
For further discussions and effects of these elemental compositions for different temperature regimes, see \citet{Herbort2020, Herbort2022}.

%==============================================
\section{Results} \label{sec:resultsARE3}
%==============================================

\begin{figure*}
\centering
\includegraphics[width=0.49\textwidth]{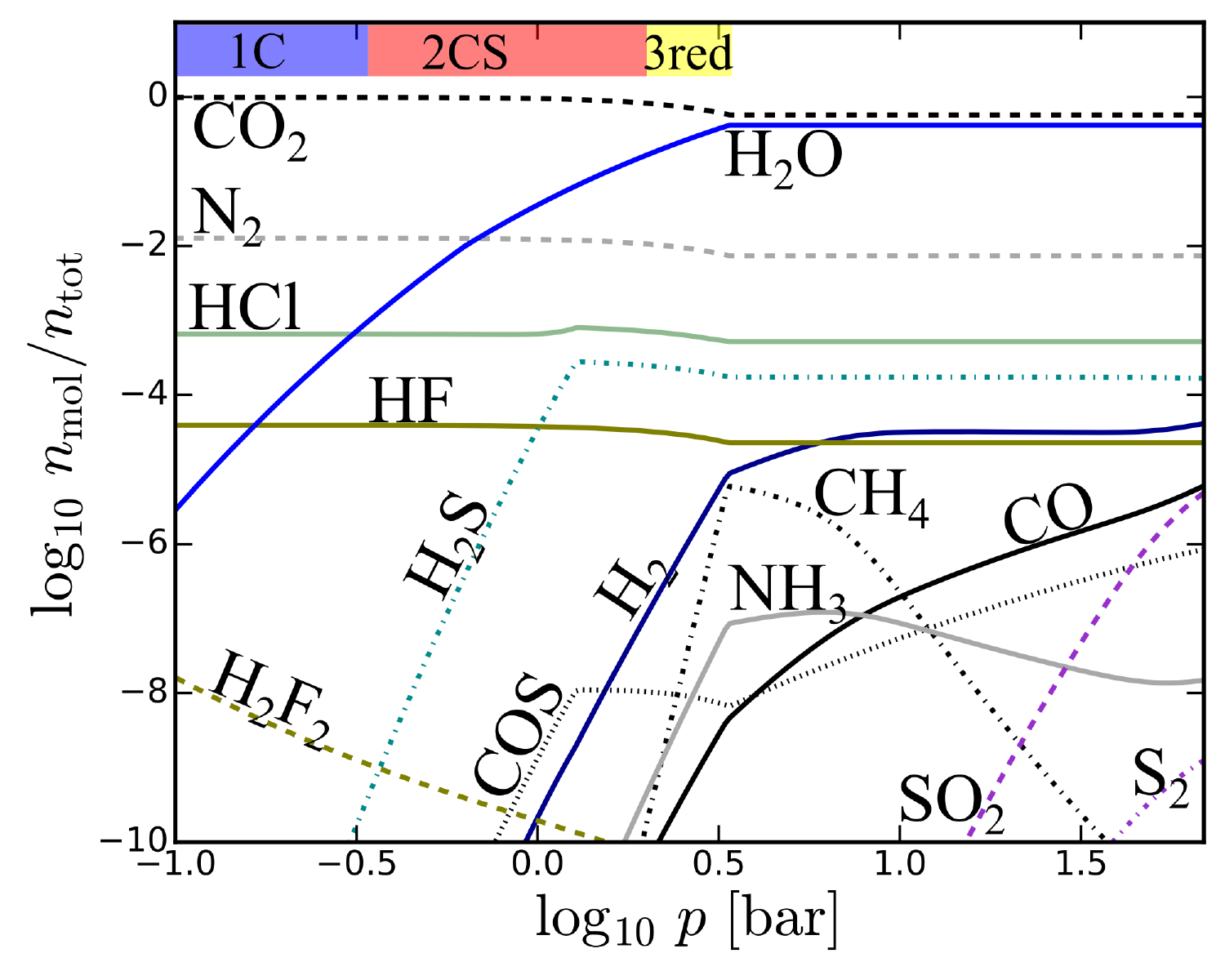}
\includegraphics[width=0.49\textwidth]{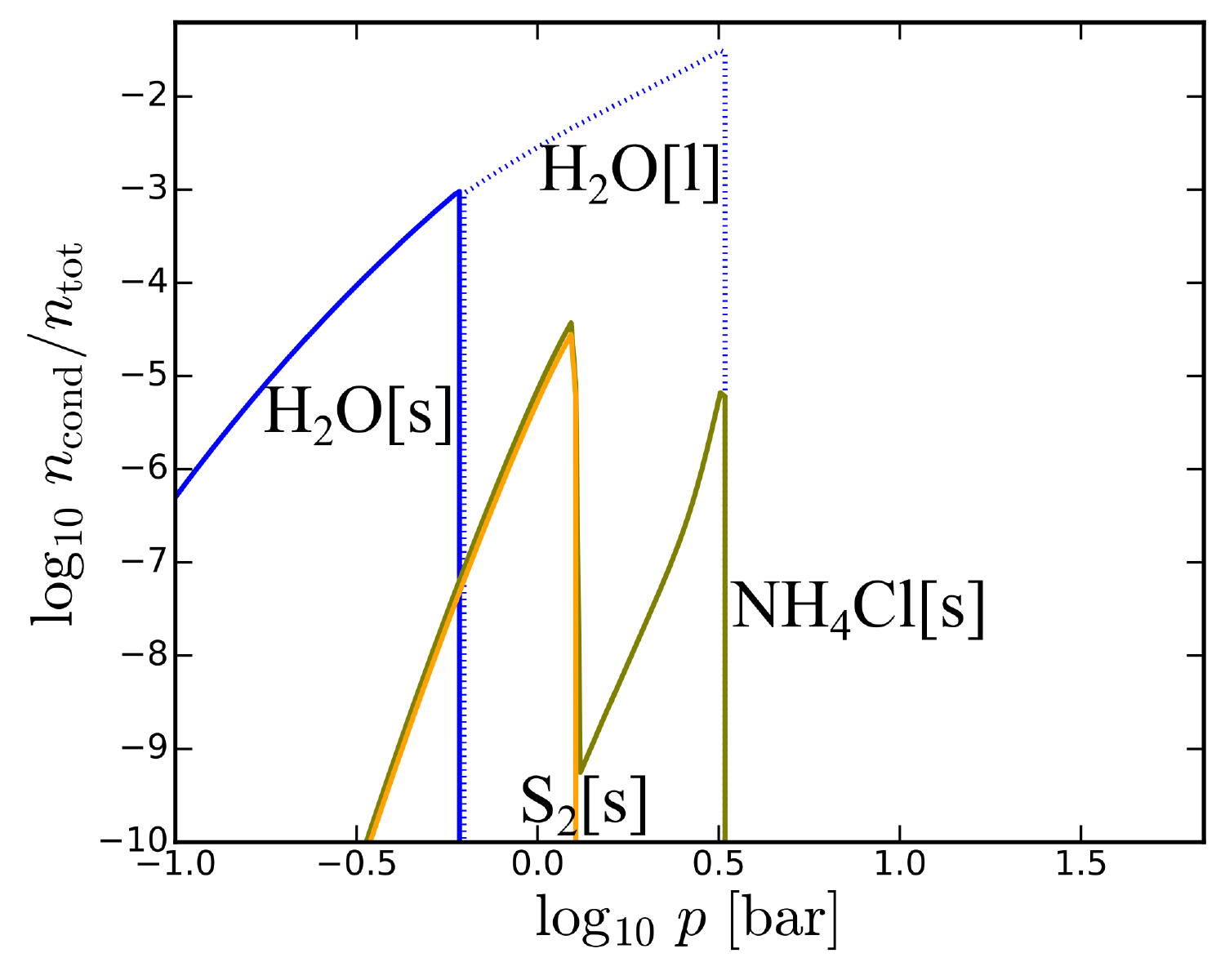}
\caption{Results of the 1D atmospheric model for CC total element abundances and surface conditions of {$T_\mathrm{surf}=700\,$K} and {$p_\mathrm{surf}=70\,$bar}.
\textbf{Left panel:} Gas-phase composition of the model atmosphere with the indicated nutrient availability levels at the top.
\textbf{Right panel:} Thermally stable cloud condensates in the model atmosphere.}
\label{fig:HabLev1D}
\end{figure*}

\begin{figure*}
\centering
\includegraphics[width = .8\linewidth]{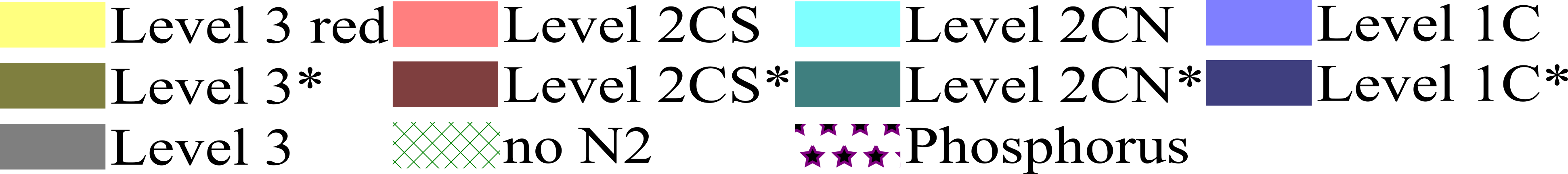}\\
\includegraphics[width = .32\linewidth, page=1]{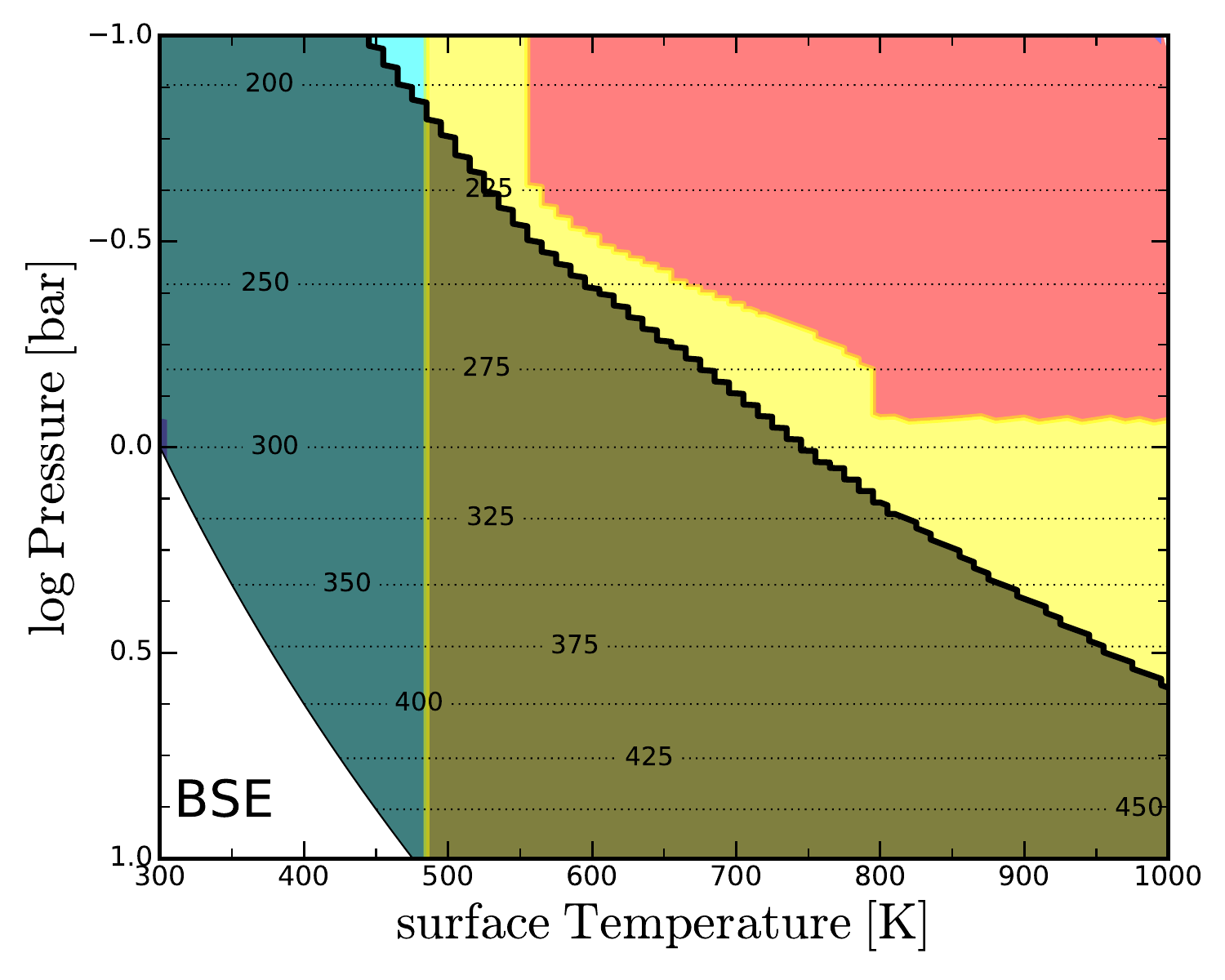}
\includegraphics[width = .32\linewidth, page=1]{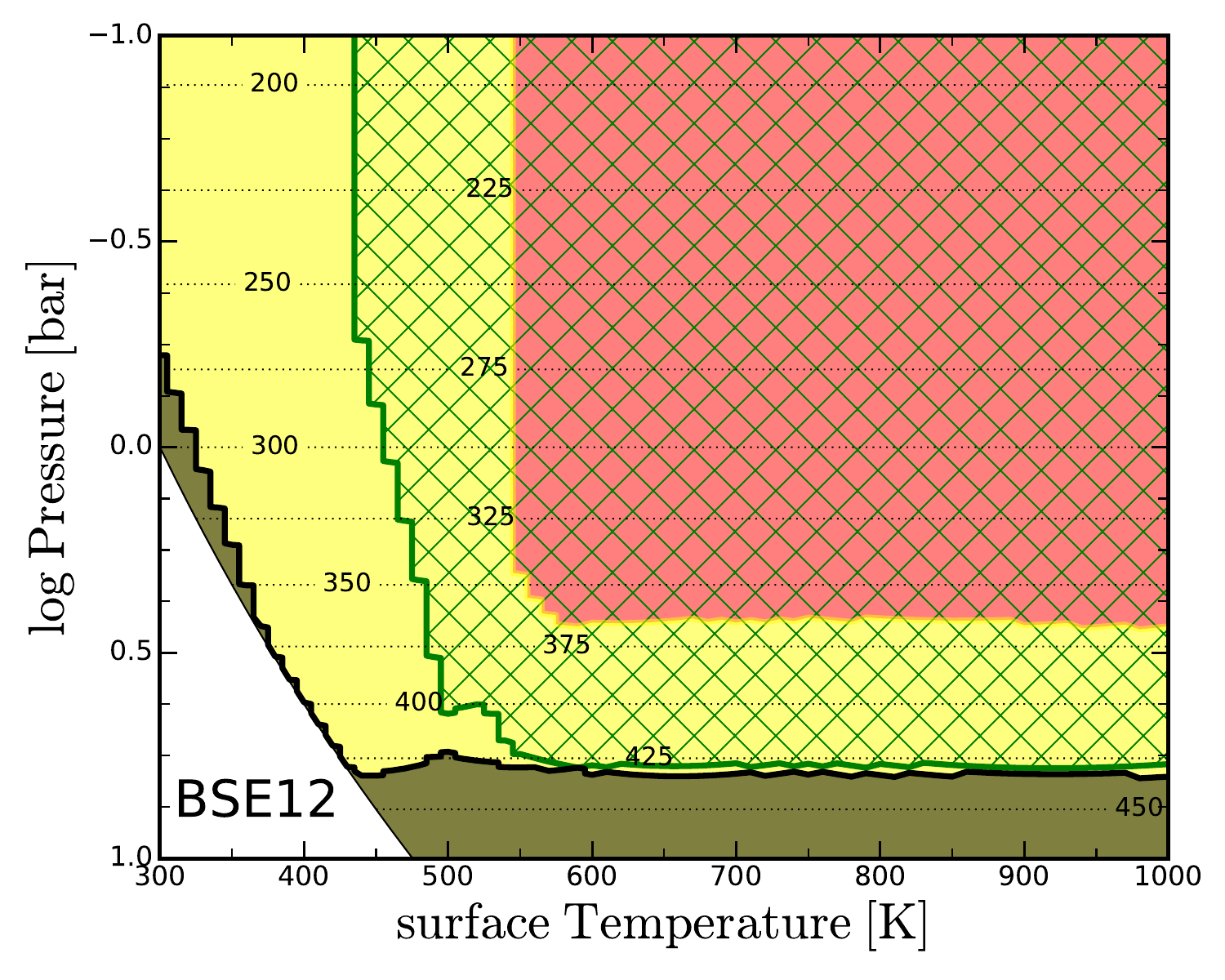}
\includegraphics[width = .32\linewidth, page=1]{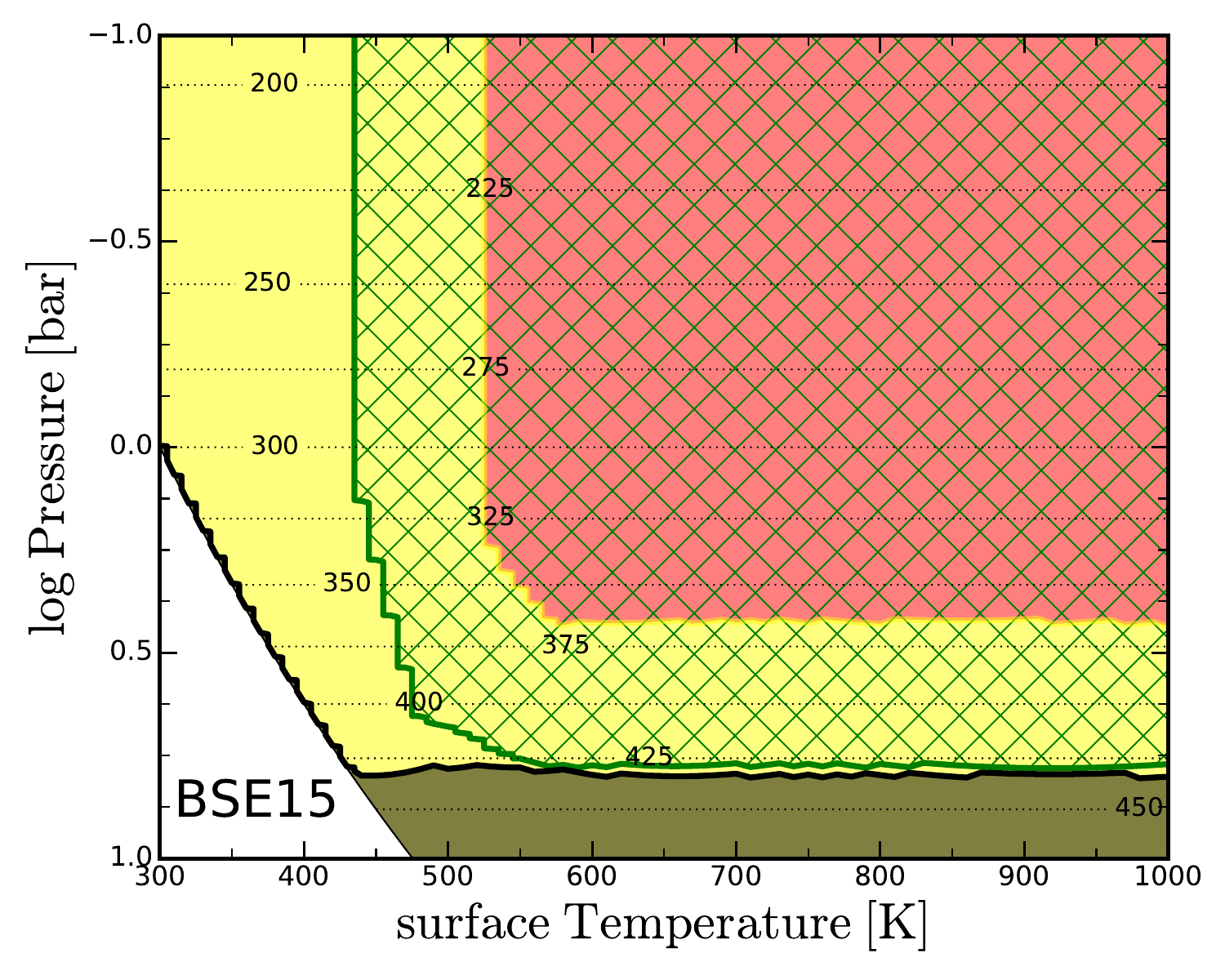}\\
\includegraphics[width = .32\linewidth, page=1]{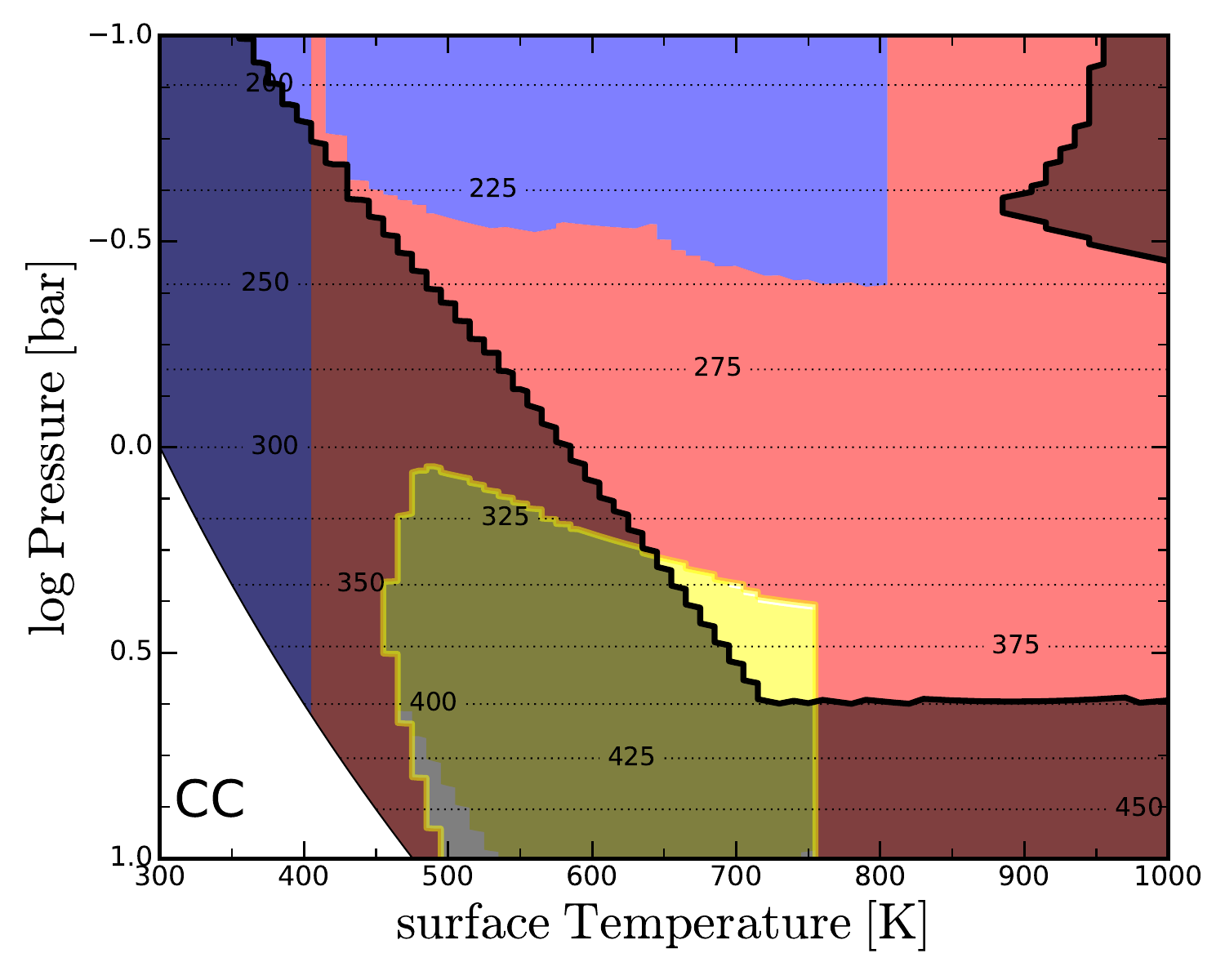}
\includegraphics[width = .32\linewidth, page=1]{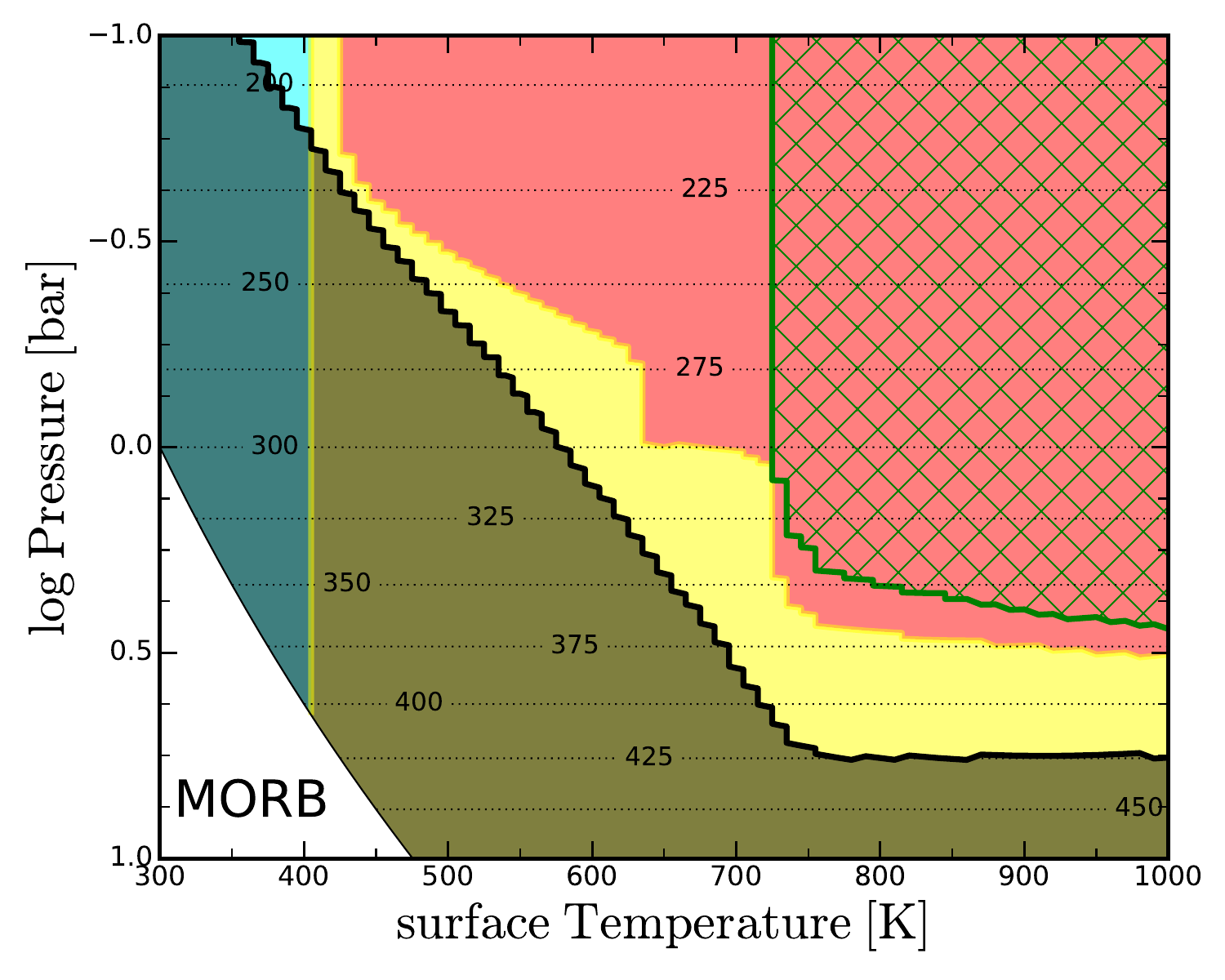}
\includegraphics[width = .32\linewidth, page=1]{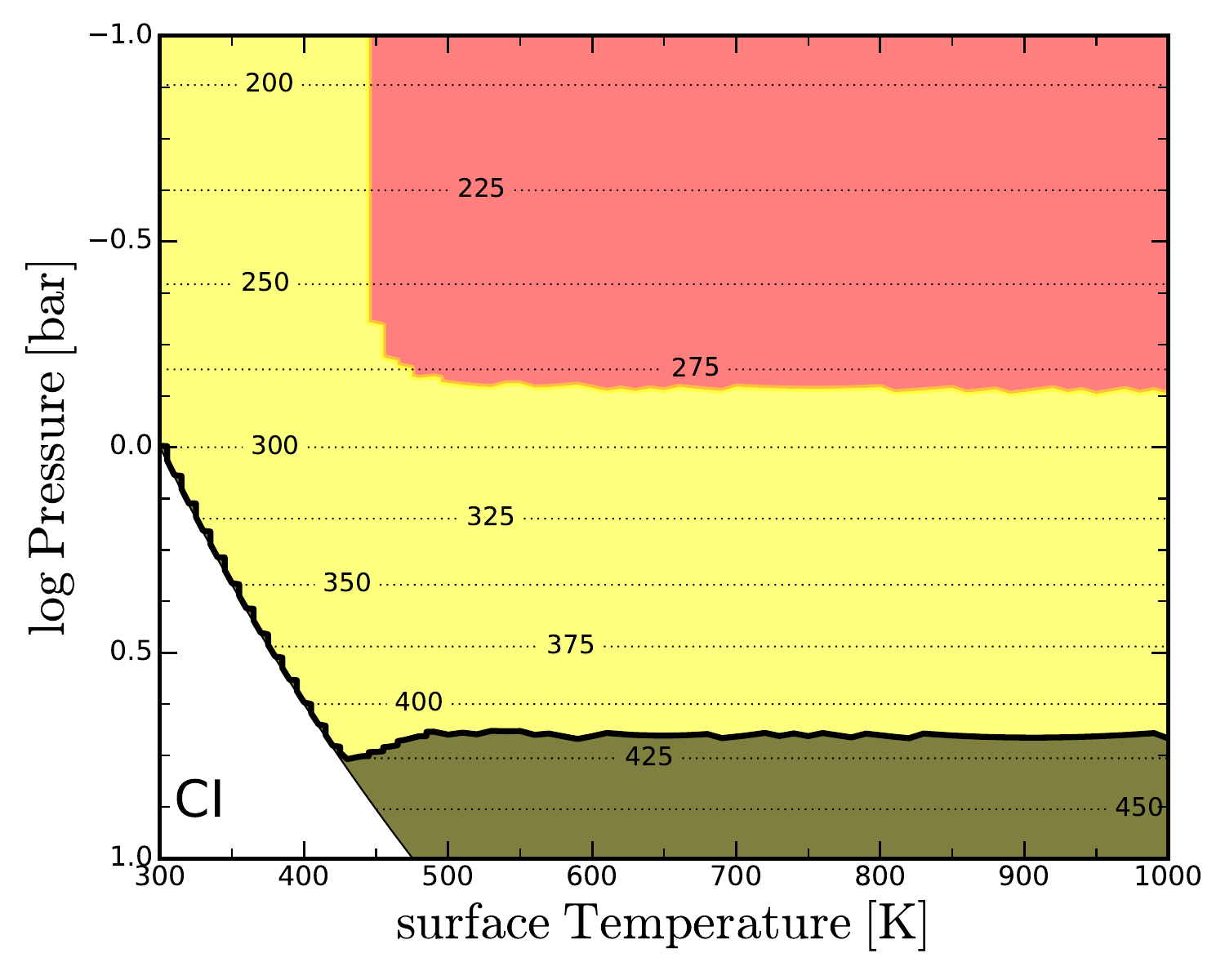}\\
\includegraphics[width = .32\linewidth, page=1]{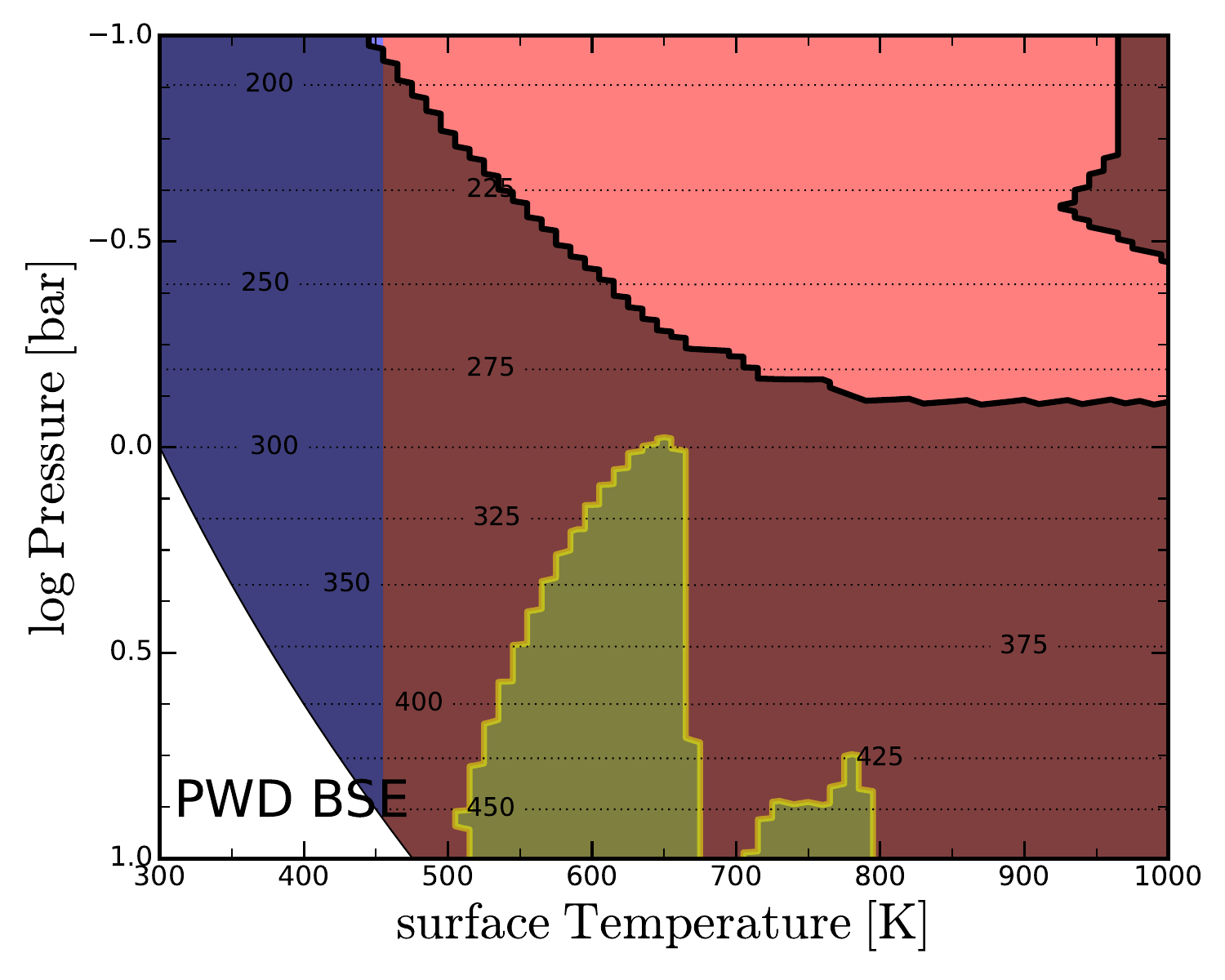}
\includegraphics[width = .32\linewidth, page=1]{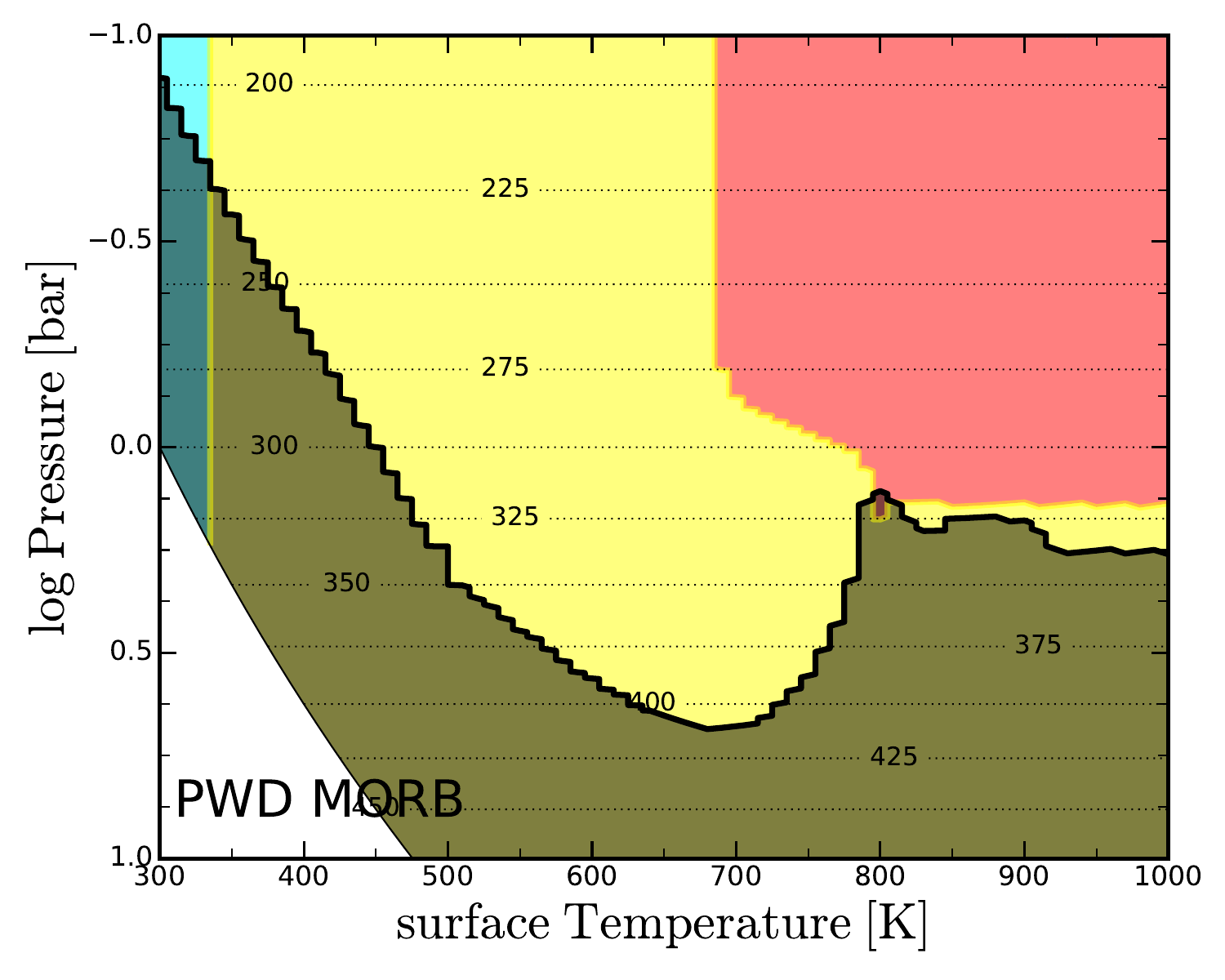}
\includegraphics[width = .32\linewidth, page=1]{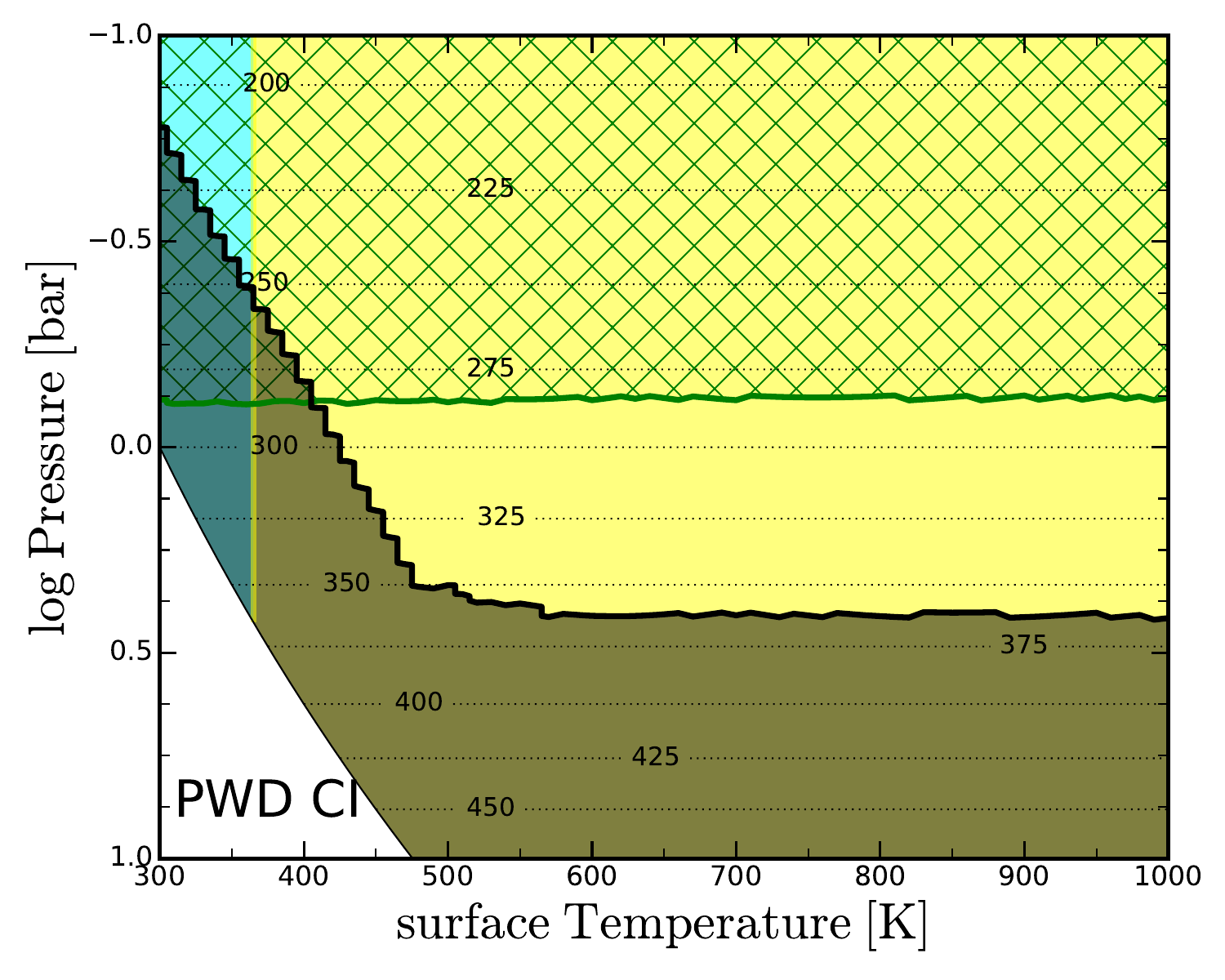}\\
\includegraphics[width = .32\linewidth, page=1]{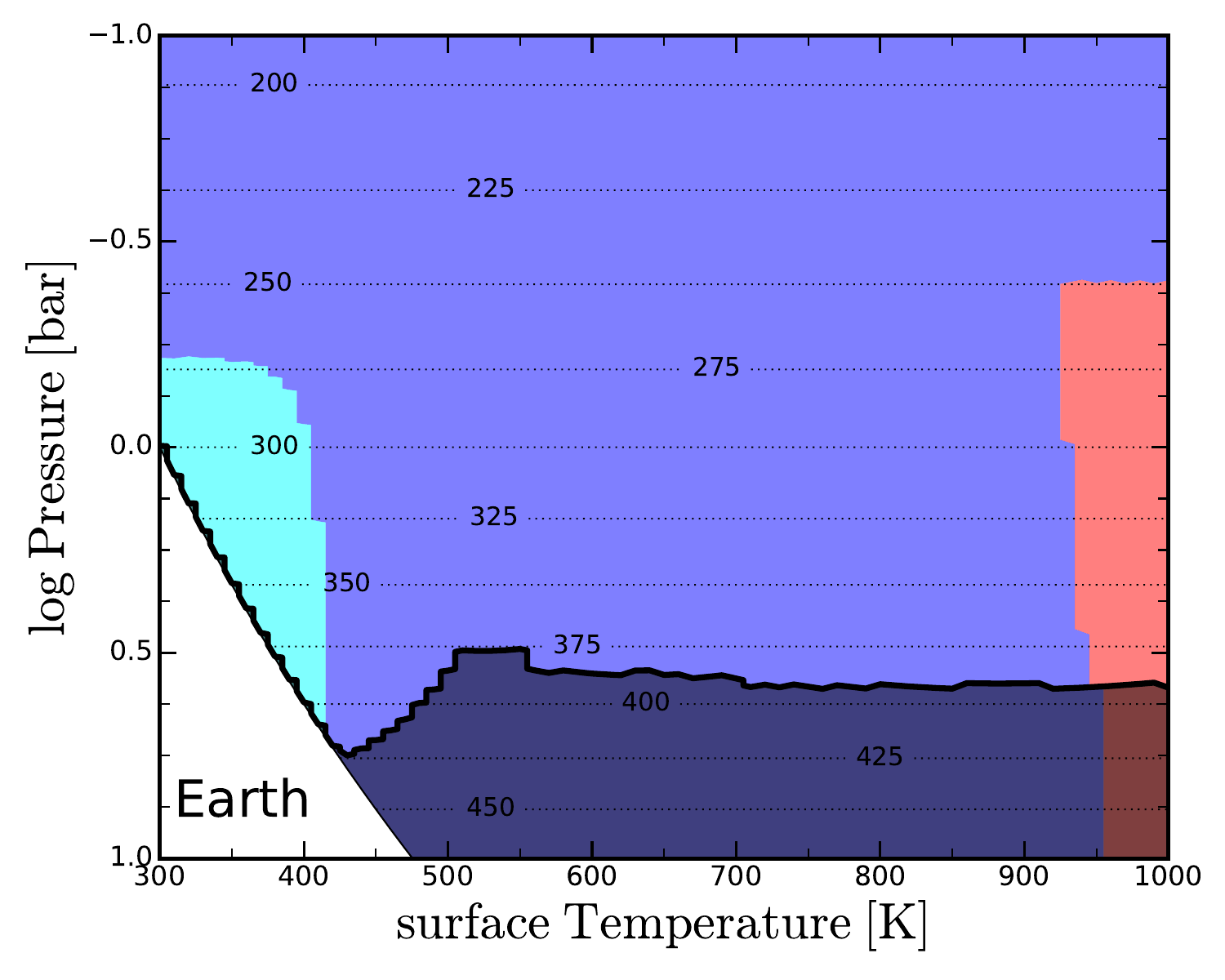}
\includegraphics[width = .32\linewidth, page=1]{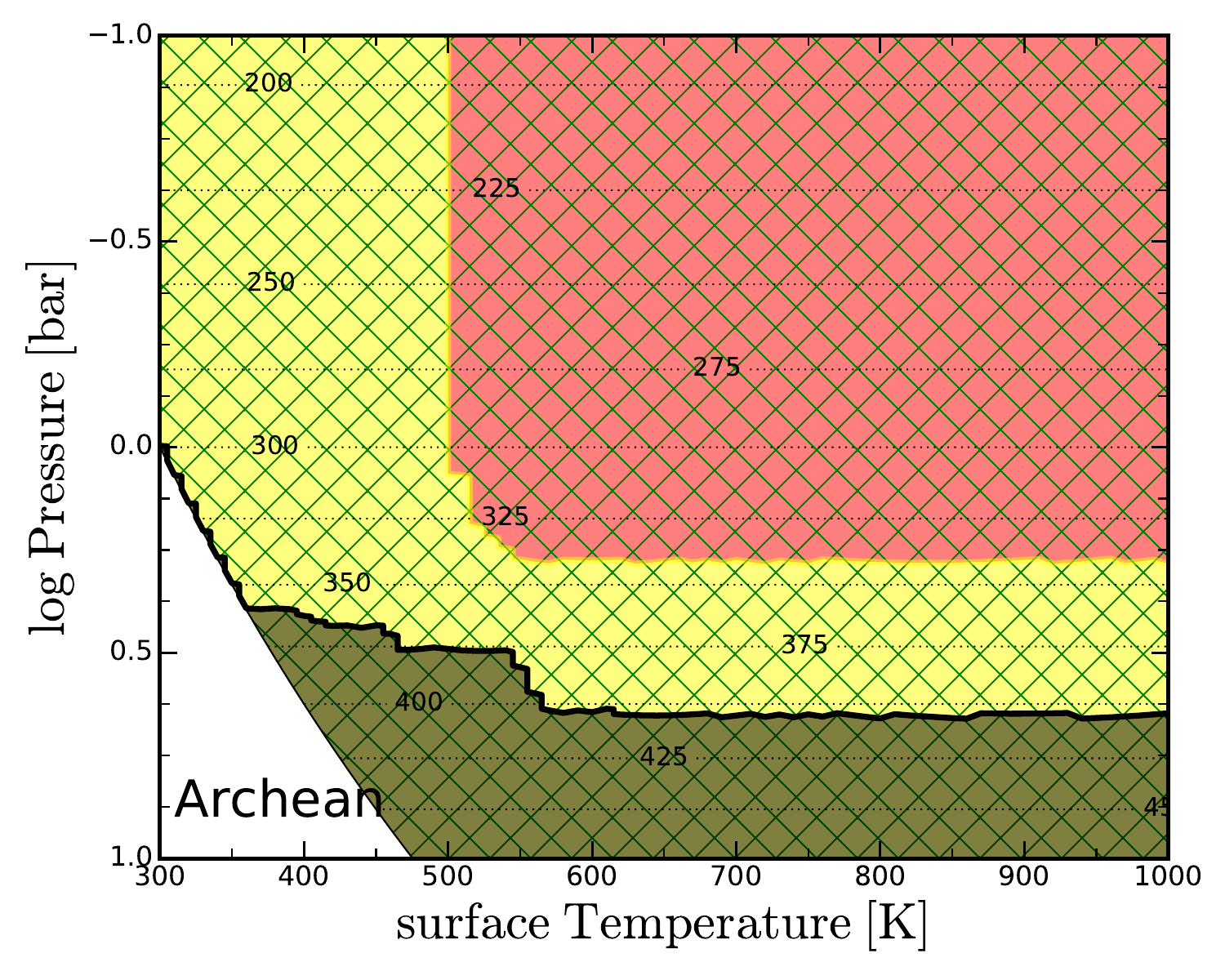}
\includegraphics[width = .32\linewidth, page=1]{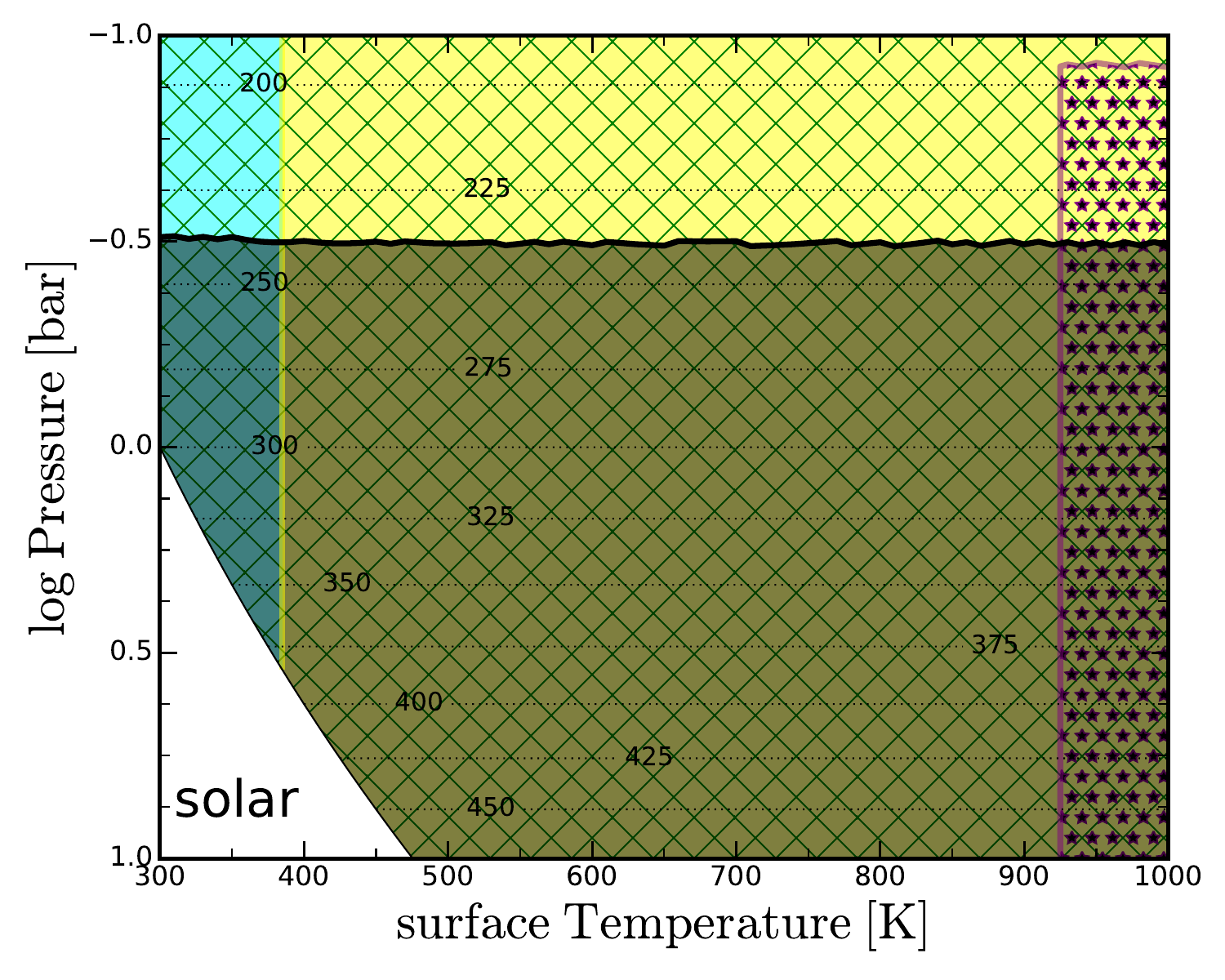}
\caption{Nutrient availability levels applied to a range of atmospheric models for different sets of total element abundances. 
The colouring of all present nutrient availability levels are given in the legend above.
The solid black line provides the pressure level of the \ce{H2O}[l,s] cloud base in the respective models.
The darker shade of the nutrient availability levels (indicated with * in the legend), are showing the corresponding nutrient availability level of gas-phase composition although the primary assumption of the presence of \ce{H2O}[l,s] is not fulfilled.
The green cross hatched region indicates $\ce{N2}/n_\mathrm{tot} < 10^{-9}$.
The horizontal dotted lines refer to the gas phase temperature $T_\mathrm{gas}$.}
\label{fig:HabLev}
\end{figure*}

\begin{table*}[t!]
\caption{Gas-phase molecules according to their highest concentration in the models.}
\label{tab:molecules}
%		\vspace{-5mm}
 		\begin{center}
% 		\resizebox{1.0\textwidth}{!}{
\addtolength{\tabcolsep}{5pt}
		\begin{tabular}{c|cccc} 
		\hline \hline
Element\hspace{1mm}&>10\%		&>0.1\% &>$10^{-6}$&>$10^{-9}$  \\ \hline
C&\ce{CO2}, \ce{CH4}&\ce{}&\ce{CO}, \ce{COS}&\ce{}\\ 
N&\ce{N2} &\ce{NH3}&\ce{}&\ce{HNO3}, \ce{NO2} (Earth $T_\mathrm{surf}<400\,$K)\\ 
P&\ce{}&\ce{}&\ce{}&\ce{P4O6}(solar $T_\mathrm{surf}>920\,$K)\\ 
S&\ce{}&\ce{H2S}, \ce{SO2}&\ce{COS}, \ce[S]$_x$&\ce{S2O}(PWD BSE), \ce{H2SO4}(Earth $T_\mathrm{surf}>920\,$K)\\
Cl&\ce{}&\ce{HCl}, \ce{Cl2}(Earth)&\ce{}&\ce{OHCl}(Earth $T_\text{gas}>425$\,K)\\ 
F&\ce{}&\ce{HF}&\ce{}&\ce{SiF4}($T_\mathrm{surf}>900\,$K)\\

others&\ce{O2}, \ce{H2}, \ce{H2O}&\ce{}&\ce{}&\ce{}\\ \hline
   		\end{tabular}
   		\addtolength{\tabcolsep}{-5pt}  		
%   		}
   		\end{center}
   		\textbf{Notes:}  Some of the present molecules are only present in few atmospheric models. For these, the defining character of these models is added in brackets. Examples are a specific temperature range or a given total  element abundance.
\end{table*}

In this section we investigate nutrient availability levels in our model atmospheres with different sets of total element abundances and subsequently discuss the effects on the different elements individually.

First we illustrate the nutrient availability levels for a single model atmosphere for CC total element abundances and surface conditions of $T_\mathrm{surf}=700\,$K and $p_\mathrm{surf}=70\,$bar shown in Fig. ~\ref{fig:HabLev1D}.
The nutrient availability levels are indicated at the top of the left panel.
At the surface, the atmosphere is dominated by \ce{CO2} and \ce{H2O}, with 10 further molecules being present at concentrations higher than $10^{-9}$. This includes \ce{H2S}, \ce{CO}, \ce{SO2}, \ce{COS}, and \ce{S2} as CHNOS bearing molecules. 
With decreasing pressure and temperature, \ce{CH4} becomes the second most important carbon bearing species while \ce{CO} and \ce{COS} decrease in concentration.
At $p_\mathrm{gas}\approx3.5\,$bar, \ce{H2O}[l] becomes a thermally stable condensate. 
Due to the presence of \ce{CH4}, \ce{NH3}, and \ce{H2S} this part of the atmosphere is of nutrient availability level~3red.
It is remarkable that C, N, and S are all abundant in their reduced form, despite the most abundant species in the atmosphere being oxidised.
The removal of \ce{H2O}[l] and \ce{NH4Cl}[s] condensates causes the depletion of \ce{H2O}, \ce{H2}, \ce{CH4}, and \ce{NH3}. 
The drop of \ce{NH3} below a concentration of $10^{-9}$ causes the nutrient availability level to change to level~2CS. 
The removal of the thermally stable \ce{S2}[s] condensate causes the decrease in \ce{H2S} concentration, which drops below $10^{-9}$ leaving behind a level~1C atmosphere.
The only molecules, which reach concentrations above $10^{-9}$ for $T_\mathrm{gas}<450\,$K and include further elements besides CHNOPS include fluorine or chlorine.
The respective concentrations of \ce{HCl} and \ce{HF} remain roughly constant throughout the atmosphere.
Similarly, \ce{N2} does not show large changes in concentration.

\subsection{Applying nutrient availability levels}\label{ssec:HabLev}
The nutrient availability levels for the twelve different sets of total element abundances with varying surface conditions are shown in Fig.~\ref{fig:HabLev}.
Although these nutrient availability levels all require the presence of liquid water, the regions without water condensates are indicated similarly to the nutrient availability levels, but shaded in black.
Of the 18 elements included in this model, only 9 elements (C, H, N, O, P, S, Cl, F, and Si) are present in the investigated atmospheric ($p_\mathrm{gas},T_\mathrm{gas}$) space in form of a gas-phase molecule with concentrations higher than $n_i/n_\mathrm{tot} > 10^{-9}$, where $n_i$ is the number density of the the molecule $i$ and $n_\mathrm{tot}$ is the total number density.
All 23 molecules, that reach concentrations above $10^{-9}$ in the investigated part of the atmosphere are listed in Table~\ref{tab:molecules} and sorted according to their peak concentration in any of the models.

Most atmospheres for the different sets of total element abundances are nutrient availability level~3red at the water cloud base.
However, for $T_\mathrm{surf} \lesssim 400\,$K, the concentrations of the sulphur bearing molecules are lower than $10^{-9}$ and the nutrient availability level drops to level~2CN or level~1C with additional \ce{N2} being present.
For these lower temperatures, the S abundance in the atmosphere is significantly lower, as more sulphur is incorporated into the crust (see Appendix~\ref{app:supl}).

The most dominant transition in the nutrient availability levels is the change from level~3red to level~2CS, which occurs in most of the models.
This depletion in available nitrogen is a result of the condensation of \ce{NH4Cl}[l], which reduces the N element abundances. 
Especially for the BSE12, BSE15, and MORB sets of total element abundances, the condensation of \ce{NH4Cl}[s] depletes the N abundances such that also the \ce{N2} concentration drops below $10^{-9}$.

Besides \ce{H2O}[l,s] and \ce{NH4Cl}[s] a total of  further 16 thermally stable condensates with normalised number densities of $n_\mathrm{cond}/n_\mathrm{tot} > 10^{-10}$ are present \citep[see also][]{Herbort2022}.
These thermally stable cloud condensates can be seen in Fig.~\ref{fig:Clouds} in Appendix~\ref{app:supl}. 
For all sets of total element abundances \ce{NaCl}[s] and \ce{KCl}[s] are stable condensates for $T_\mathrm{gas}>700\,$K.
Furthermore, every set of element abundances shows a $(p_\mathrm{gas},T_\mathrm{gas})$ region, where \ce{H2O}[s] condenses. 
Only for the solar abundances there is no regime with \ce{H2O}[l] as a condensate.
Another cloud condensate that is thermally stable in atmosphere of all element abundances, except solar, is the N bearing condensate \ce{NH4Cl}[s].

A more strict definition of the nutrient availability levels would include a concentration threshold of $10^{-6}$ instead of $10^{-9}$ for the relevant elements. 
The resulting nutrient availability levels for this are shown in Fig.~\ref{fig:HabLev-6}.
The parameter space that the level~3red nutrient availability regions are covering are significantly shrunken and mostly replaced by level~2CS.
This result shows that the presence of N can especially be a limiting factor if concentrations of the order of $10^{-8}$ for nitrogen bearing species is not sufficient.

In all models that reach nutrient availability level~3, the molecules are present in the reduced form.
This is also true for the majority of the parts of the atmosphere, where water is not a stable condensate.
The only model, where this is not the case is for the CC total element abundances with $T_\mathrm{surf}\approx500\,$K and $T_\mathrm{gas}>425\,$K.

\subsection{Hydrogen, oxygen, and water}\label{ssec:Water}
The most fundamental molecule for the discussion of habitability is water, which is not only present as a condensate, but also as a dominant gas-phase molecule in many planetary atmosphere.
Figure~\ref{fig:NutriHO} in the Appendix~\ref{app:supl} shows the gas-phase concentrations of \ce{H2O}, \ce{H2}, and \ce{O2}.
 
The BSE and PWD BSE models are the models with the lowest \ce{H2O} concentration, due to the low hydrogen mass fraction in the total element abundances.
\ce{H2O} does not reach concentrations above $10^{-6}$ for ${T_\mathrm{surf}\lesssim430\,\mathrm{K}}$ and for ${T_\mathrm{surf}\lesssim340\,\mathrm{K}}$ the \ce{H2O} concentration stays below $10^{-9}$.
The only other models, for which the \ce{H2O} concentrations drops below $10^{-9}$ are the sets of total element abundances of MORB and CC with $T_\mathrm{surf}\lesssim340\,$K.
For all other models, the \ce{H2O} concentrations are significantly higher and can become the most abundant gas-phase species, especially for high surface temperature models.

As noted above, for all sets of total element abundances, there exist atmospheric models, where \ce{H2O}[l,s] cloud condensates are stable and abundant. 
The \ce{H2O} gas-phase concentration above the cloud base is a result of the assumption of unity for the supersaturation, $S($\ce{H2O}[l,s]$)=1$.
Therefore every two models at the same $(p_\mathrm{gas}, T_\mathrm{gas})$ above the water cloud base show the same gas-phase concentration of \ce{H2O}.
This confines the maximum \ce{H2O} concentration for the atmosphere above. 

The only model which shows presence of gaseous \ce{O2} is the Earth model, as it is created to have an atmospheric composition similar to modern day Earth's atmosphere for $T_\mathrm{surf}=288\,$K at the surface \citep[Appendix A of][]{Herbort2022}.
For higher surface temperatures, the \ce{O2} concentration reduces as the \ce{CO2} concentration increases to become the most abundant gas-phase molecule, surpassing \ce{N2}.
As the Earth model is the only model with a type B atmosphere \citep[oxygen rich, see][]{Woitke2020a, Herbort2022}, it is the only model to show \ce{O2} as a gas species.
For \ce{H2} on the other hand, the picture is different, as the models for total element abundances of solar, Archean, and CI composition result in atmospheres rich in \ce{H2} for all surface temperatures \citep[type A atmospheres][]{Woitke2020a}.
For BSE12 and BSE15 total element abundances, \ce{H2} is present for all $T_\mathrm{surf}$ where water is not condensing.
For all other models \ce{H2} is only a trace gas.

\subsection{Carbon}\label{ssec:C}
Of the CNS elements, carbon is the only element, for which there are always molecules present at concentrations higher than $10^{-9}$ for the entire parameter space of all sets of total element abundances studied in this paper.
Carbon is bound in different oxidisation states, ranging from \ce{CH4} to \ce{CO2}, with some additional traces of \ce{CO} and \ce{COS} at  concentrations of $10^{-6}$ (see Fig.~\ref{fig:NutriC}).
This coexistence of \ce{CH4} and \ce{CO2} is a fundamental result of chemical equilibrium for all atmospheres of atmospheric type C \citep{Woitke2020a}.
For $T_\mathrm{gas}\lesssim350\,$K, both redox end-members of carbon are present at concentrations higher than 0.1\% in multiple models, while only  PWD CI, Earth, Archean, and Solar total element abundances show no coexistence of these redox states throughout the parameter space.
The trace forms of carbon \ce{COS} and \ce{CO} are only present at concentrations up to $10^{-6}$, if \ce{CO2} is present at more than 0.1\%.
The importance of especially \ce{CO} increases with higher gas temperature towards the lower parts of the model atmospheres. The significant presence of \ce{CO} as a major carbon species can be seen in other works, such as \citet{Fegley1997, Visscher2006, Woitke2020a} and others.

Throughout the atmospheres in this paper, the only form of carbon which is thermally stable as a condensate in the atmosphere is \ce{C}[s] (graphite).
It is only thermally stable in atmospheres where \ce{CH4} and \ce{CO2} coexist.
In these cases, the atmospheric composition in the upper parts of the atmosphere evolves towards the dominance of the more abundant one of the two carbon molecules.
The concentration of the respective other molecule decreases significantly as a result due to the ongoing depletion of carbon in the atmosphere.
As discussed in \citet{Herbort2022}, graphite is the only condensate which can be thermally stable for the entire investigated $T_\mathrm{gas}$ parameter space.
The models based on CC total element abundances are the only ones for which \ce{CH4} and \ce{CO2} coexist in the atmosphere and \ce{C}[s] does not condense.

For the carbon condensation, the different element abundances fall into three different regimes. 
For the element abundance without carbon condensation in the atmosphere and a hydrogen-rich atmosphere, also no carbon bearing condensate is stable as part of the crust.
However, for the Earth model, where also no graphite condenses in the atmosphere, all non-gaseous carbon is bound in \ce{CaCO3}[s] and \ce{CaMgC2O6}[s].
The other element abundances have either one or a multitude of carbon-bearing condensates stable at the crust.

For PWD BSE ($T_\mathrm{surf}<700\,$K and PWD MORB the carbon condensates are \ce{CaCO3}[s], \ce{C}[s], and \ce{CaMgC2O6}[s].
These condensates are also stable for the model with the highest C abundance (3.5\% mfrac for the CI), but the further condensates of \ce{MgCO3}[s], \ce{NCO3}[s], \ce{FeCO3}[s], and \ce{CaFeC2O6} are also stable.
The carbon abundance in BSE, BSE12, BSE15, and MORB is more than two orders of magnitude lower than for CI or the PWD models.
This results in a much simpler carbon condensate phase, with either \ce{C}[s] or \ce{CaCO3}[s]  being stable for the corresponding models.
Similarly for the CC element abundance, carbon is bound in the crust in form of \ce{CaMgC2O6}[s] for $T_\mathrm{surf}<680\,$K.

\subsection{Nitrogen}\label{ssec:N}
The nitrogen in the gas phase can be found in forms of \ce{N2}, \ce{NH3}, \ce{HNO3}, and \ce{NO2} and their distribution is shown in Fig.~\ref{fig:NutriN}.
While there are no N bearing condensates in the crust for any model, \ce{NH4Cl}[s] and \ce{NH4SH}[s] are thermally stable condensates in the atmospheres (see Fig.~\ref{fig:Clouds}).

For many sets of element abundances, \ce{N2} is a main N carrier at concentrations over 1\textperthousand\ for the entire investigated parameter space (BSE, CC, CI, Earth) or up to a certain threshold surface temperature ($T_\mathrm{surf}<450\,$K: BSE12, BSE15; $T_\mathrm{surf}<650\,$K: MORB, PWD BSE; $T_\mathrm{surf}<775\,$K: PWD MORB).
Otherwise, for H-rich atmospheres, \ce{NH3} becomes the main N carrier for the entire parameter space of PWD CI, Archean, and Solar, but also for $T_\mathrm{surf}>450\,$K BSE12, BSE15.
For the latter two, the dominance of \ce{N2} over \ce{NH3} is only present, when \ce{H2O}[l] is a stable crust condensate.

Although N is considered as an omnipresent gas species, we find the high atmosphere can be depleted in nitrogen for $T_\mathrm{surf}\gtrsim550\,$K and therefore higher surface pressures.
The depletion of N, is caused by the condensation of \ce{NH4Cl}[s], which removes a considerable amount of N from the gas phase.
This is the case for the the sets of total element abundances of BSE12, BSE15, and Archean.
For MORB abundances, this temperature threshold is $T_\mathrm{surf}>725\,$K.
The condensation of \ce{NH4Cl}[s] is present in most models and coincides with the condensation of \ce{H2O}[l].
Therefore the \ce{NH4Cl}[s] condensates are of additional interest for the discussion of habitability.

A further N condensate is \ce{NH4SH}[s], which is only thermally stable for $T_\mathrm{gas}<250\,$K.
This further depletes the N abundance for BSE12, BSE15, PWD MORB, and PWD CI.

\subsection{Sulphur}\label{ssec:S}
The most abundant sulphur bearing gas-phase molecule is \ce{H2S}.
Only for high $T_\mathrm{gas}$ of CC and PWD BSE total element abundances \ce{SO2} becomes more significant than \ce{H2S}, which remains present at concentrations of more than $10^{-6}$.
Furthermore, traces of \ce{COS}, \ce{S}$_x$, \ce{S2O}, and \ce{H2SO4} can be present.
In general, for higher $T_\mathrm{surf}$ a higher abundance of sulphur is found in our atmospheres (see also Fig.~\ref{fig:NutriS}).
The only models where no S bearing molecule is present at concentrations greater than $10^{-9}$ for $T_\mathrm{surf}>450\,$K is the Earth model, for which only \ce{H2SO4} is present for $T_\mathrm{surf}>900\,$K.
This significantly depleted S abundance with respect to all of the other models is a result of the thermal stability of S condensates in oxidised environments.
The investigation of S in CHNO+S atmospheres are further discussed in detail in Janssen et al (subm).
For the CC model with $400\,\mathrm{K}<T_\mathrm{surf}<800\,$K, the condensation of \ce{S2}[s] is sufficient enough to reduce the S abundance for $T_\mathrm{gas}\lesssim240\,$K to below $10^{-9}$.
For all models, with $ T_\mathrm{surf}\gtrsim450\,$K \ce{H2S} is present with concentrations higher than $10^{-9}$ in the region of the \ce{H2O}[l] cloud base.

For most of the element abundances, the majority of the sulphur is kept in the crust as condensates.
This explains the higher concentration of sulphur bearing molecules with higher temperatures, as the vapour pressure of the condensates increases with increasing temperature.
Only for  BSE12, BSE15, and Archean total element abundances and $T_\mathrm{surf}>700\,$K no S condensate is stable at the crust. 
For most abundances and $T_\mathrm{surf}$ only one S condensate is thermally stable.
These stable condensates are \ce{FeS} for BSE, MORB, PWD MORB, PWD CI (all $T_\mathrm{surf}$), BSE12, BSE15 (680\,K$<T_\mathrm{surf}<$440\,K), CI ($T_\mathrm{surf}>510\,$K), Archean ($T_\mathrm{surf}<690\,$K), solar ($T_\mathrm{surf}<670\,$K);
\ce{FeS2} for BSE12, BSE15 ($T_\mathrm{surf}<430\,$K), CC ($T_\mathrm{surf}<640\,$K), CI ($T_\mathrm{surf}<500\,$K), PWD BSE ($T_\mathrm{surf}<350\,$K);
\ce{CaSO4}[s] for CC ($T_\mathrm{surf}>650\,$K), Earth (all $T_\mathrm{surf}$), and \ce{MnS} for solar ($T_\mathrm{surf}>680\,$K).
The PWD BSE total element abundance is the only model, where multiple S condensates are stable for the same surface condition.
These stable condensates are \ce{FeS}[s] and one of \ce{CaSO4}[s] and \ce{FeS2}[s] for $T_\mathrm{surf}>800\,$K and $350\,\mathrm{K}<T_\mathrm{surf}<800\,$K, repetitively. 

The sulphur-containing cloud condensates fall into three categories.
First, at the highest temperatures, pure sulphur condensates of S2[s] and S[l] dominate.
These can be present over a large range of surface conditions.
In general, the \ce{S2}[s] cloud base lies at lower pressures than the \ce{H2O}[l] cloud bases.
Second, \ce{H2SO4}[s], which is only stable for  models with high pressures at the surface and cold local temperatures of $T_\mathrm{gas}\lesssim 300\,$K.
The \ce{H2SO4}[s] and \ce{S2}[s] condensates can coexist in the one atmospheric model.
Third, the \ce{NH4SH}[s] condensate is only stable for $T_\mathrm{gas}\lesssim 250\,$K.
As \ce{NH4SH}[s] does not coexist in an atmosphere with either \ce{S2}[l], \ce{S}[l], and \ce{H2SO4}[s], these condensates could be used for the characterisation of the atmospheric conditions.

\subsection{Phosphorus}\label{ssec:P}
The rarest of the CHNOPS elements in our model atmospheres is phosphorus, which is for all non-solar total element abundances almost completely bound in the crust, in the form of hydroxyapatite and fluorapatite, \ce{Ca5P3O13H}[s] and \ce{Ca5P3O12F}[s], respectively.
Apatite has been found to be the main phosphorus carrier for example on the Chelyabinsk meteorite \citep{Walton2021}.
This shows that P will be a limiting element for the formation and evolution of life in the atmospheres of almost all rocky exoplanets.

Only models for solar element abundances and $T_\text{surf}>900\,$K show any phosphorus-bearing molecule at concentrations higher than $10^{-9}$ in the atmosphere.
At the crust-atmosphere interaction layer of these models, \ce{PH3} is the dominant P carrier, with about 3 orders of magnitude higher concentration than \ce{P4O6}.
However, as virtually no Ca is present in the atmosphere, the remaining P cannot condense in form of apatite and only condenses much higher in the atmosphere, when \ce{H3PO4}[s] becomes thermally stable at around $p=10^{-0.5}\,$bar and $T_\mathrm{gas}=250\,$K.
Despite the overall reducing atmospheric conditions in the \ce{H2} dominated atmosphere, the most important P carrier changes close to the surface at $T_\mathrm{gas}\approx 900\,$K from \ce{PH3} to \ce{P4O6}.
This change from the reduced to the oxidised form is consistent with previous models \cite[e.g.][]{Lodders2002, Visscher2006}.

\subsection{Chlorine and fluorine}\label{ssec:FCl}
The only molecules which which reach concentrations above $10^{-9}$ for $T_\mathrm{gas} < 450\,$K and include further elements besides CHNOPS include the fluorine or chlorine.
The gas phases in the relevant temperature regime are visualised in the Appendix~\ref{app:supl} in Fig.~\ref{fig:NutriCl} and Fig.~\ref{fig:NutriF}, respectively.

Most of the halides are condensed in the crust in form of \ce{NaCl}[s] and \ce{Ca5P3O12F}[s].
Additionally \ce{MgF2}[s] tends to be stable towards higher temperatures, whereas \ce{CaF2}[s] is stable towards lower temperatures.
Only for the CC total element abundances, the additional condensate of \ce{KMg2AlSi3O10F2}[s] is thermally stable at the crust for $T_\mathrm{surf}>370\,$K. 
The further condensate of \ce{CaCl2}[s] is only found stable in the model of BSE12 total element abundances for $T_\mathrm{surf}\approx590\,$K.

The halides show increasing gas-phase concentrations for high surface temperatures, where residual Cl and F atoms remain in the gas phase.
The most important species being \ce{HCl} and \ce{HF} for all atmospheres.
In the atmospheres at $T_\mathrm{gas}\gtrsim 700\,$K, the condensation of \ce{KCl}[s] and \ce{NaCl}[s] occurs for every total element abundance. 
However, this condensation is limited by the abundance of K and Na, respectively.
As a result, the concentration of \ce{HCl} and \ce{HF} reaches up to a few $10^{-3}$ in the atmospheres based on a high surface temperature. 
The presence of these hydrogen halides in coexistence with \ce{H2O}[l] can result in acidic environments, if they dissolve in the water droplets.
Whether this acidity benefits or limits the formation of pre-biotic molecules is beyond the scope of this paper and is therefore not included in the nutrient availability levels as either helpful or harmful.

\ce{SiF4} is the only CHNOPS bearing molecule present at concentrations higher than $10^{-9}$ in any model of our atmospheric models for $T_\mathrm{gas} < 450\,$K.
It is only present for the CC and Earth total element abundances and  $T_\mathrm{surf}\gtrsim900\,$K.

%=======================================
\section{Summary and discussion}\label{sec:Disc}
%=======================================
In this paper, we defined nutrient availability levels based on the presence of water condensates and availability of nutrients.
We used this concept to explore the potential for habitable atmospheres to form under equilibrium conditions in rocky exoplanets.

We find that for most atmospheres at $(p_\mathrm{gas}, T_\mathrm{gas})$ points, where liquid water is stable, CNS bearing molecules are present at concentrations above $10^{-9}$.
Carbon is generally present in every atmosphere, while the sulphur availability in the gas phase increases with increasing  $T_\mathrm{surf}$.
For lower $T_\mathrm{surf}$ nitrogen in the form of either \ce{NH3} or \ce{N2} is in present as a major contributor to the atmospheric composition for most atmospheres.
However, for models with higher $T_\mathrm{surf}$, the presence of gaseous nitrogen can be significantly depleted by the condensation of \ce{NH4Cl}[s], which leaves an upper atmosphere devoid of nitrogen.
The limiting element of the CHNOPS elements is phosphorus, which is mostly bound in the planetary crust.
This is consistent with the scarcity of phosphorus limiting the biosphere during at least some parts of Earth history \citep{Syverson2021}.

\subsection{Implications for surface biospheres}\label{sec:SurfaceHab}
Similar to previous work, our models suggest that the limiting factor for habitability at the surface of a planet is the presence of liquid water.
If water is present at the surface, CNS are available in the gas phase of the near-crust atmosphere.
The only exception is the present day Earth model, which lacks the presence of gaseous S by construction.

The available elements for the equilibration of crust and atmosphere are dependent on the overall planetary composition and therefore formation history, as well as the tectonic regimes present \citep[e.g.][]{Tosi2017, Dorn2018, Kruijver2021, Meier2021, Honing2021}.
The evolution of a planet and its start also play a crucial role, as this determines the long-term temperature of the planet \citep{Turbet2021, Seales2021}.
The formation of hydrated minerals can also play an important role in the long-term stability of water condensates \citep{Herbort2020, Beck2021}.
An example for this is the surface of Mars, which previously hosted liquid water, which is now bound in hydrated rocks \citep[see e.g.][]{Ehlmann2010, Wernicke2021}.

Even if $(p_\mathrm{surf}, T_\mathrm{surf})$ of a planet allow the formation of liquid water and water is present in the atmosphere, even in the form of clouds, our models show that no water ocean is to be implied from these \citep[compare also][]{Ding2022}.
Especially if the surface pressure is not well constrained, the presence of water clouds is not conclusive for the presence of water at the surface of the respective planet.
Therefore additional detection methods of a surface ocean are necessary to confirm the existence \citep[e.g.][]{Williams2008, LustigYaeger2018, Ryan2022}

If water is available at the surface, the elements not present in the gas phase are stored in the crust condensates.
Processes such as chemical weathering or dissolution of condensates into water can make these elements available for life.
This provides a pathway to overcome the lack of atmospheric phosphorus and metals, which are used in enzymes that drive many biological processes.
However, there can be the difficulty in accumulating the pre-biotic molecules, if the water reservoir is too large and no exposed land is available \citep{Noack2016}.
This leads to Darwin's idea of the warm little pond with some dry-wet cycles \citep{Darwin1871, Follmann2009, Pearce2017}.
This hurdle also needs to be overcome for potential life on the icy moons in our solar system \citep[see e.g.][]{Vance2007, Tjoa2020, Taubner2020}.
If indeed it can be shown that life can form in a water ocean without any exposed land, this constraint becomes weaker and the potential for the surface habitability becomes mainly a question of water stability.

\subsection{Implications for aerial biospheres}\label{ssec:AerialHab}
Many of the models show the presence of a liquid water zone in the atmospheres, which is detached from the surface. 
These regions could be of interest for the formation of life in forms of aerial biospheres.

Previous work by for example \citet{Seager2021} focussing on sub Neptunes -- which have thicker atmospheres than planets investigated in this work -- have shown that aerial biospheres can be regions of nutrient scarcity.
We show that some of the fundamental elements (CHNOS) are likely to occur in regions of most planetary atmospheres.
These elements are mostly in a reduced state near the water cloud base, which is favourable for the formation of pre-biotic molecules.
However, our models also show that phosphorus is unlikely to occur in planetary atmospheres, as it is bound in the crust in form of apatite.
A further hurdle for more complex life is the ability to use redox-sensitive transition metals as cofactors in enzymes. Our models show these metals are also absent in most atmospheres, as they are bound in the crust.
Further cloud condensates to \ce{H2O}[l,s] like \ce{NH4Cl}[s] could play this role for life in such environments \citep[see][]{Seager2021}.
Our models show that these two cloud condensates coexist for most atmospheric models.

The lacking atoms of phosphorus and metals could in principle be provided by atmospheric updrafts induced by storms on the planet.
The closer the liquid water zone is to the surface, the more likely an updraft of surface material would be. 
However, further cloud layers are expected at higher temperatures on planets with higher surface pressure (e.g. \ce{NaCl}[s], \ce{KCl}[s],\ce{FeS}[s], or \ce{FeS2}[s]).
These condensates could be delivered to the higher atmospheres by storms and accumulated by life in the higher atmosphere.

\subsection{Implications for atmospheric biosignatures}
The atmospheric composition in this work focussed on the pre-biotic atmospheric composition.
The atmospheric composition after the emergence of life on a planet can deviate drastically from the pre-biotic atmospheric composition caused by biological activity itself \citep[e.g.][]{Holland2002, Izon2017, Gregory2021}.
Our approach does not directly aim for the understanding of biosignatures and atmospheres of planets, which are inhabited, but for the conditions in which pre-biotic chemistry can occur.

The results of our approach of a large variety of different atmospheric compositions in chemical equilibrium atmospheres can provide insights to which signatures of biology could also be produced by a simple chemical equilibrium atmosphere.
We for example, show that \ce{N2} is a major atmospheric component for planets with $T_\mathrm{surf} < 400\,$K, whereas \ce{N} in general can be lacking in atmospheres based on higher $T_\mathrm{surf}$.
\citet{Lammer2019} state that \ce{N2} itself can be a sign of a biosignature or effective tectonic regimes. 
Our models of an equilibrated crust with the atmosphere can be seen as representative of active tectonic regimes as otherwise only a small portion of the crust is able to equilibrate with the atmosphere.
Following up on this, \citet{Spross2021} state that the coexistence of \ce{N2} and \ce{O2} is only a result of biology. 
We find that this coexistence - as well as the presence of \ce{O2} in general - is only possible in a model with element abundances adopted to produce this exact atmospheric feature.

Another pair of atmospheric gas species that is discussed as a biosignature is \ce{CO2} and \ce{CH4} \citep[e.g.][]{2019BAAS...51c.158K, Wogan2020, Mikal-Evans2021}.
However, as discussed in \citet{Woitke2020a}, it is possible to have these two molecules coexists in chemical equilibrium.
In our atmospheres, we find this element pair coexisting at concentrations above $10^{-6}$ for many atmospheres with $T_\mathrm{gas}>275\,$K.
For a concentration greater than 1\textperthousand\  of both, \ce{CO2} and \ce{CH4}, $T_\mathrm{gas}>350\,$K is necessary. 
Therefore, high concentrations of both molecules at cool temperatures could be explained by some out-of-equilibrium process such at biology.
In the early Earth atmosphere during the Archean, the methane has been a product of biology \citep[methanogenesis][]{Ueno2006}.
Similarly it has been discussed that the presence of \ce{CH4} in plumes of Enceladus could potentially be a result of methanogenesis \citep{Affholder2021}.

\ce{HCN} has been found for example on the surface of Titan \citep{Lellouch2017} and the comet Hale-Bopp \citep{Jewitt1997}.
Its production in planetary atmosphere is possible by lightning, especially if \ce{CH4} and \ce{N2} are present \citep{Hodosan2017, Pearce2022}.
However, we do not find the presence of \ce{HCN} in any of our models, underlining their non-equilibrium origin. 
With the presence of \ce{HCN}, \citet{Rimmer2021HC3N} found that the production of \ce{HC3N} should occur on super-Earths rich in nitrogen and an overall reduced atmospheric composition by photochemistry.

Further molecules that we do not find to be present in any of our models are more complex hydrocarbons than \ce{CH4}.
However, these have been found in various environments ranging from interstellar matter \citep[e.g.][]{Henning1998} to places in our solar system.
Examples are the \ce{C2H4} haze on Triton \citep{Ohno2021} or various different molecules on comets such at 67P \citep{Mueller2022}.

\citet{Huang2021} found that the detection of \ce{NH3} can be a biosignature on planets, where a deep atmosphere is excluded.
We however find, that \ce{NH3} can be present throughout the atmosphere of uninhabited planets with low pressure atmospheres.
However, only if no water condensate is stable at the surface, we find that the \ce{NH3} concentration actually exceeds 1\textperthousand.
The reason for this difference can be that there is no photochemistry included in our model. Species such as \ce{NH3} and \ce{CH4} can be easily photodissociated, resulting in a significantly lower concentrations and the production of hazes.

By construction of the model, the influence of stellar irradiation to the planetary atmosphere is not discussed.
A proper understanding of the influence of the host star to the atmosphere requires the upper atmosphere and is therefore beyond the scope of this paper.
This can have a major factor for the atmospheric evolution \citep{Chen2020, Turbet2021a, Chebly2021, Locci2022, Teal2022}.
The triggered photochemistry can not only destroy pre-biotic molecules, but also enhance the formation of these as many reactions need the input of some sort of energy, this can for example be provided by the stellar irradiation (in form of UV radiation) or lightning discharges. 
Whether the strong UV radiation during flares of M~dwarfs is harmful for the evolution of life has been a matter of discussion \citep{Shields2016, OMalley-James2019}.

\section{Conclusions}\label{sec:Conclusion}
The formation of life does not only require the presence of liquid water, but also nutrients, especially in the form of the the elements CNS.
This paper investigates the presence of these elements in aerial environments where water condensates are present.
Therefore, we have introduced a new framework to extend the concept of habitability from the presence of liquid water to also include the presence and availability of nutrients.
This new system of different nutrient availability levels includes the presence of water condensates and how many nutrients based on carbon, nitrogen, and sulphur are available.

The application of these nutrient availability levels to 1D equilibrium chemistry atmospheric models which include the element depletion due to cloud formation show that all three elements (CNS) are commonly present at the water cloud base.
Our models also show that phosphorus is not commonly present in the atmosphere and could therefore be a limiting nutrient for the formation of life.

\ack[Acknowledgement]{OH acknowledges the ARIEL fellowship of the University of Vienna, the PhD stipend form the University of St Andrews' Centre for Exoplanet Science, financial support from the \"Osterreichische Akademie der Wissenschaften (OeAW).}

\ack[Funding]{PW and ChH received funding from the European Union H2020-MSCA-ITN-2019 under Grant Agreement no. 860470 (CHAMELEON).}

\ack[Declaration of interests]{The authors report no conflict of interest.}

\ack[Author ORCID]{O.Herbort, \url{https://orcid.org/0000-0002-1807-4441}; P. Woitke, \url{https://orcid.org/0000-0002-8900-3667}; Ch. Helling, \url{https://orcid.org/0000-0002-8275-1371}; A.L. Zerkle; \url{https://orcid.org/0000-0003-2324-1619}.}

%\appendix

\begin{appendix}
\onecolumn
\section{Supplementary material}\label{app:supl}
\renewcommand{\thefigure}{A\arabic{figure}}
\setcounter{figure}{0}

\renewcommand{\thetable}{A.\arabic{table}}
\setcounter{table}{0}

\begin{table*}[!ht]
\caption{Comparison of different element abundances given in \% mass fractions used in this work \citep[adapted from][]{Herbort2022}.}
\label{tab:abundances}
{\centering

\begin{tabular}{c|cccccccccc}
\hline
	&	BSE	    &	BSE12	&	BSE15	&	CC	    &	MORB	&	CI	    &		PWD 	&					Earth	&	Archean	&	solar	\\
	  &1&$1\ast$&$1\ast$&2&3&4&$5\dagger$&$1\nabla$&$1\ast$&6\\ \hline
H	&	0.006	&	1.198	&	1.457	&	0.045	&	0.023	&	1.992	&		        &					0.3341	&	2.309	&	98.4	\\
C	&	0.006	&	0.0054	&	0.005	&	0.199	&	0.019	&	3.520	&		1.2		&					1.957	&	0.0052	&	0.316	\\
N	&	8.8E-05	&	7.9E-05	&	7.7E-05	&	0.006	&	5.5E-05	&	0.298	&	            &					0.02533	&	7.6E-05	&	0.092	\\
O	&	44.42	&	49.280	&	50.320	&	47.20	&	44.5	&	46.420	&		41.0	&					50.100	&	49.880	&	0.765	\\
F	&	0.002	&	0.0018	&	0.0017	&	0.053	&	0.017	&	0.0059	&		        &					0.04816	&	0.0017	&	0.000067\\
Na	&	0.29	&	0.260	&	0.253	&	2.36	&	2.012	&	0.505	&		        &					2.145	&	0.251	&	0.0039	\\
Mg	&	22.01	&	19.69	&	19.180	&	2.20	&	4.735	&	9.790	&		18.0	&					1.999	&	19.010	&	0.094	\\
Al	&	2.12	&	1.897	&	1.847	&	7.96	&	8.199	&	0.860	&		0.27	&					7.234	&	1.831	&	0.0074	\\
Si	&	21.61	&	19.33	&	18.830	&	28.80	&	23.62	&	10.820	&		18.0	&					26.17	&	18.670	&	0.089	\\
P	&	0.008	&	0.0072	&	0.0070	&	0.076	&	0.057	&	0.098	&		0.22	&					0.06906	&	0.0069	&	0.00078	\\
S	&	0.027	&	0.024	&	0.024	&	0.070	&	0.110	&	5.411	&		3.3		&					0.06361	&	0.023	&	0.041	\\
Cl	&	0.004	&	0.0036	&	0.0035	&	0.047	&	0.014	&	0.071	&		        &					0.04271	&	0.0035	&	0.0011	\\
K	&	0.02	&	0.018	&	0.017	&	2.14	&	0.152	&	0.055	&		        &					1.945	&	0.017	&	0.00041	\\
Ca	&	2.46	&	2.201	&	2.143	&	3.85	&	8.239	&	0.933	&		6.9		&					3.499	&	2.125	&	0.0086	\\
Ti	&	0.12	&	0.107	&	0.105	&	0.401	&	0.851	&	0.046	&	            &					0.3644	&	0.104	&	0.00042	\\
Cr	&	0.29	&	0.260	&	0.253	&	0.013	&	0.033	&	0.268	&		        &					0.01181	&	0.251	&	0.00222	\\
Mn	&	0.11	&	0.098	&	0.096	&	0.072	&	0.132	&	0.195	&		        &					0.06543	&	0.095	&	0.0014	\\
Fe	&	6.27	&	5.610	&	5.463	&	4.32	&	7.278	&	18.710	&		10		&					3.926	&	5.416	&	0.172	\\ \hline
sum	&	99.77	&	99.99	&	100.00	&	99.81	&	99.99	&	100.00	&		99.73	&					100.00	&	100.00	&	100.00	\\ \hline
\end{tabular}}
\\*[1mm]$\ast$: Element abundances based on BSE abundances, but altered by the addition of \ce{H2O}.\\
$\dagger$: The PWD data is completed by taking the missing element abundances from BSE, MORB, or CI as indicated.\\
$\nabla$: The element abundances are created by a fit to Earth atmospheric concentrations.\\
{(1)~Bulk Silicate Earth: \citet{Schaefer2012}; (2)~Continental Crust: \citet{Schaefer2012}; (3)~Mid Oceanic Ridge Basalt: \citet{Arevalo2010}; (4)~CI chondrite: \citet{Lodders2009}; (5)~Polluted White Dwarf: \citet{Melis2016}; (6)~solar: \citet{Asplund2009}}
\end{table*}

\begin{table}[h!]
%\vspace{-4mm}
\caption{Condensates present in the surfaces of the different models according to their grouping.}
\label{tab:CondGroups}
 		\begin{center}
\resizebox{0.8\textwidth}{!}{
\begin{tabular}{lll||lll} 
\hline\hline
Group & Condensate&Name&Group & Condensate&Name\\ \hline		
		
Silicates	& \ce{	Al2SiO5	}		&	Kyanite	&	Nitrogen	& \ce{	NH3	}		&	Amonia	\\
	& \ce{	Ca2Al2Si3O12	}		&	Prehnite	&		& \ce{	P3N5	}		&	P-nitride	\\
	& \ce{	Ca2Al2SiO7	}		&	Gehlenite	&		& \ce{	TiN	}		&	Ti-nitride	\\
	& \ce{	Ca2MgSi2O7	}		&	\AA kermanite	&	Phosphorus	& \ce{	Ca5P3O13H	}		&	Hydroxyapatite	\\
	& \ce{	Ca3Fe2Si3O12	}		&	Andradite	&		& \ce{	Ca5P3O12F	}		&	Flourapatite	\\
	& \ce{	CaMgSi2O6	}		&	Diopside	&		& \ce{	P	}		&	Phosphorus	\\
	& \ce{	CaMgSiO4	}		&	Monticellite	&		& \ce{	P	}	[l]	&	Liquid phosphorus	\\
	& \ce{	CaSiO3	}		&	Wollastonite	&	F- Mica	& \ce{	KMg3AlSi3O10F2	}		&	Flourphlogopite	\\
	& \ce{	CaTiSiO5	}		&	Sphene	&	Halides	& \ce{	NaCl	}		&	halite	\\
	& \ce{	Fe2SiO4	}		&	Fayalite	&		& \ce{	KCl	}		&	Sylvite	\\
	& \ce{	KAlSiO4	}		&	Kalsilite	&		& \ce{	MgF2	}		&	Mg-fluoride	\\
	& \ce{	Mg2SiO4	}		&	Fosterite	&		& \ce{	CaF2	}		&	Fluorite	\\
	& \ce{	MgSiO3	}		&	Enstatite	&		& \ce{	KF	}		&	K-fluoride	\\
	& \ce{	Mn3Al2Si3O12	}		&	Spessartine	&		& \ce{	AlF6Na3	}		&	Cryolite	\\
	& \ce{	Mn2SiO4	}		&	Tephroite	&	Sulphur	& \ce{	CaCl2	}		&	Ca-dichloride	\\
	& \ce{	Na2SiO3	}		&	Na-metasilicate	&		& \ce{	CaS	}		&	Ca-sulfide	\\
	& \ce{	NaCrSi2O6	}		&	Kosmochlor	&		& \ce{	CaSO4	}		&	Anhydrite	\\
	& \ce{	NaFeSi2O6	}		&	Acmite	&		& \ce{	FeS	}		&	Troilite	\\
	& \ce{	NaAlSiO4	}		&	Nepheline	&		& \ce{	FeS2	}		&	Pyrite	\\
	& \ce{	SiO2	}		&	Quarz	&		& \ce{	MgS	}		&	Mg-sulphide	\\
Feldspar	& \ce{	CaAl2Si2O8	}		&	Anorthite	&		& \ce{	MnS	}		&	Alabandite	\\
	& \ce{	KAlSi3O8	}		&	Microcline	&	Phyllosilicates	& \ce{	Al2Si2O9H4	}		&	Kaolinite	\\
	& \ce{	NaAlSi3O8	}		&	Albite	&		& \ce{	Ca2Al2Si3O12H2	}		&	Prehnite	\\
Carbon	& \ce{	C	}		&	Graphite	&		& \ce{	Ca2Al3Si3O13H	}		&	Clinozoisite	\\
	& \ce{	CaCo3	}		&	Calcite	&		& \ce{	Ca2FeAl2Si3O13H	}		&	Epidote	\\
	& \ce{	CaFeC2O6	}		&	Ankerite	&		& \ce{	Ca2FeAlSi3O12H2	}		&	Ferri prehnite	\\
	& \ce{	CaMgC2O6	}		&	Dolomite	&		& \ce{	Ca2MnAl2Si3O13H	}		&	Piemontite	\\
	& \ce{	FeCO3	}		&	Siderite	&		& \ce{	CaAl2Si2O10H4	}		&	Lawsonite	\\
	& \ce{	MgCO3	}		&	Magnesite	&		& \ce{	CaAl2Si4O16H8	}		&	Laumonite	\\
	& \ce{	MnCO3	}		&	Rhodochrosite	&		& \ce{	CaAl4Si2O12H2	}		&	Margarite	\\
	& \ce{	SiC	}		&	Silicon carbide	&		& \ce{	Fe3Si2O9H4	}		&	Greenalite	\\
Metals	& \ce{	Mn	}		&	Manganese	&		& \ce{	FeAl2SiO7H2	}		&	Fe-chloritoid	\\
	& \ce{	Na	}		&	Sodium	&		& \ce{	KFe3AlSi3O12H2	}		&	Annite	\\
	& \ce{	Na	}	[l]	&	Liquid sodium	&		& \ce{	KMg3AlSi3O12H2	}		&	Phlogopite	\\
Iron	& \ce{	Fe	}		&	Iron	&		& \ce{	Mg3Si2O9H4	}		&	Lizardite	\\
Metal oxides	& \ce{	Al2O3	}		&	Chorundum	&		& \ce{	Mg3Si4O12H2	}		&	Talc	\\
	& \ce{	CaTiO3	}		&	Perovsike	&		& \ce{	MgAl2SiO7H2	}		&	Mg-chloritoid	\\
	& \ce{	Cr2O3	}		&	Eskolaite	&		& \ce{	MnAl2SiO7H2	}		&	Mn-chloritoid	\\
	& \ce{	FeAl2O4	}		&	Hercynite	&		& \ce{	NaAl3Si3O12H2	}		&	Paragonite	\\
	& \ce{	FeTiO3	}		&	Ilmenite	&		& \ce{	NaMg3AlSi3O12H2	}		&	Sodaphlogopite	\\
	& \ce{	MgAl2O4	}		&	Spinel	&	Hydroxides	& \ce{	FeO2H	}		&	Goethide	\\
	& \ce{	MgCr2O4	}		&	Picrochromite	&		& \ce{	AlO2H	}		&	Diaspore	\\
	& \ce{	Mn2O3	}		&	Bibyxbite	&		& \ce{	MgO2H2	}		&	Brucite	\\
	& \ce{	MnTiO3	}		&	Pyrophanite	&	Water	& \ce{	H2O	}		&	Water	\\
	& \ce{	Ti4O7	}		&	Titanium oxide	&		& \ce{	H2O	}	[l]	&	Liquid water	\\
	& \ce{	TiO2	}		&	Rutile	&		& \ce{		}		&		\\
Iron oxides	& \ce{	Fe3O4	}		&	Magnetite	&		& \ce{		}		&		\\
	& \ce{	Fe2O3	}		&	Hematite	&		& \ce{		}		&		\\
	& \ce{	FeO	}		&	Ferropericlase	&		& \ce{		}		&		\\ \hline
   		\end{tabular}  	}	
   		\end{center}
\vspace{-5mm}
\end{table}

\begin{figure*}
\centering
\includegraphics[width = .99\linewidth]{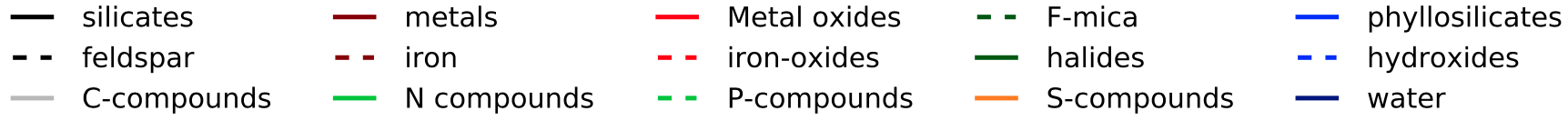}\\
\includegraphics[width = .32\linewidth, page=6]{./Figures/ggchem_Crust_BSE.pdf}
\includegraphics[width = .32\linewidth, page=6]{./Figures/ggchem_Crust_BSE12.pdf}
\includegraphics[width = .32\linewidth, page=6]{./Figures/ggchem_Crust_BSE15.pdf}\\
\includegraphics[width = .32\linewidth, page=6]{./Figures/ggchem_Crust_CC.pdf}
\includegraphics[width = .32\linewidth, page=6]{./Figures/ggchem_Crust_MORB.pdf}
\includegraphics[width = .32\linewidth, page=6]{./Figures/ggchem_Crust_CI.pdf}\\
\includegraphics[width = .32\linewidth, page=6]{./Figures/ggchem_Crust_PWD_Melis_BSE.pdf}
\includegraphics[width = .32\linewidth, page=6]{./Figures/ggchem_Crust_PWD_Melis_MORB.pdf}
\includegraphics[width = .32\linewidth, page=6]{./Figures/ggchem_Crust_PWD_Melis_CI.pdf}\\
\includegraphics[width = .32\linewidth, page=6]{./Figures/ggchem_Crust_EarthCr.pdf}
\includegraphics[width = .32\linewidth, page=6]{./Figures/ggchem_Crust_Archean.pdf}
\includegraphics[width = .32\linewidth, page=6]{./Figures/ggchem_Crust_solar.pdf}
\caption{Crust composition of the different models with the different element abundances. The condensates included in the different condensate groups are explained in Table~\ref{tab:CondGroups}.}
\label{fig:Crust}
\end{figure*}

%\section{Atmospheric composition}\label{app:Figs}
\begin{figure*}
\centering
\includegraphics[width = .8\linewidth]{./Figures/legend}\\
\includegraphics[width = .32\linewidth, page=1]{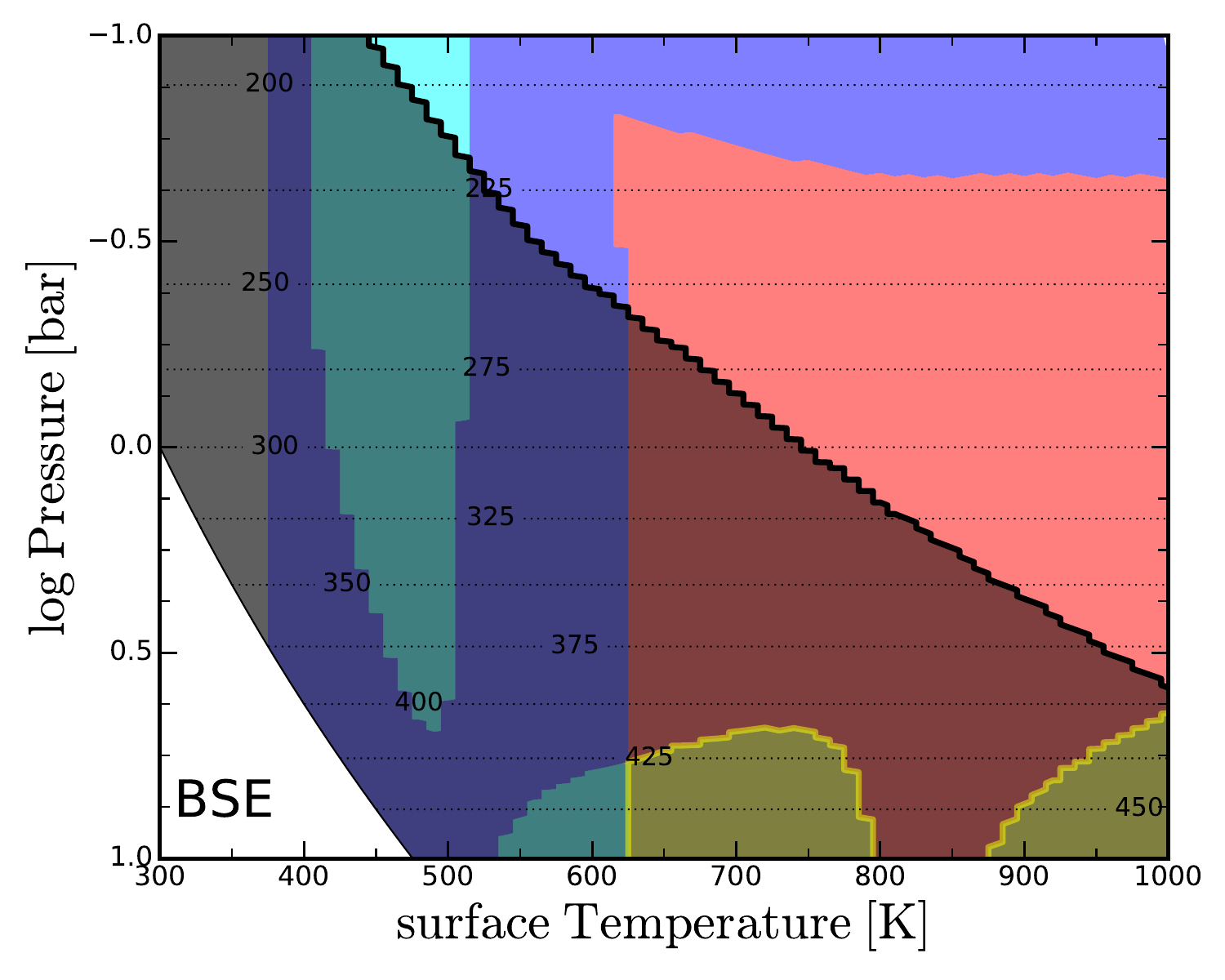}
\includegraphics[width = .32\linewidth, page=1]{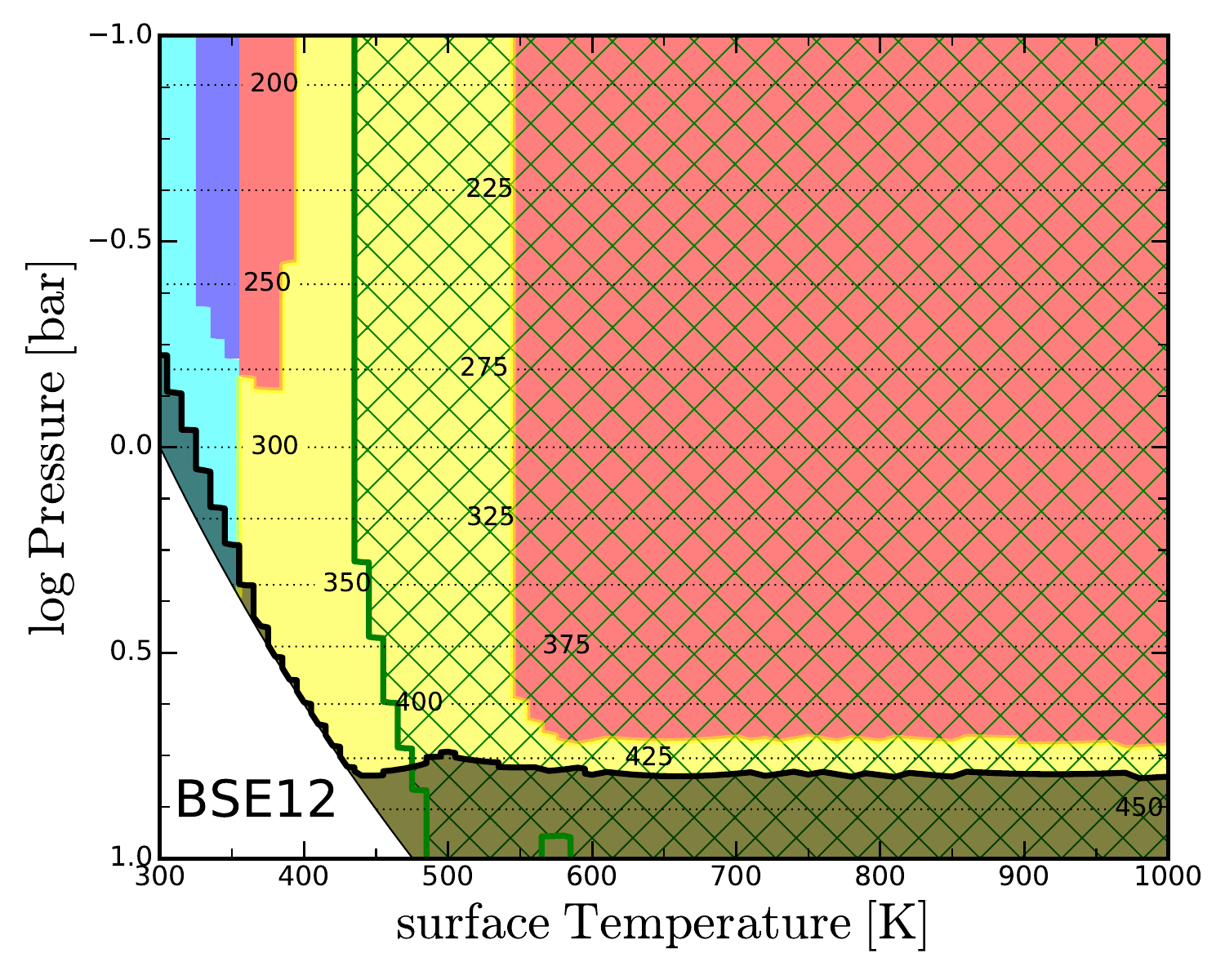}
\includegraphics[width = .32\linewidth, page=1]{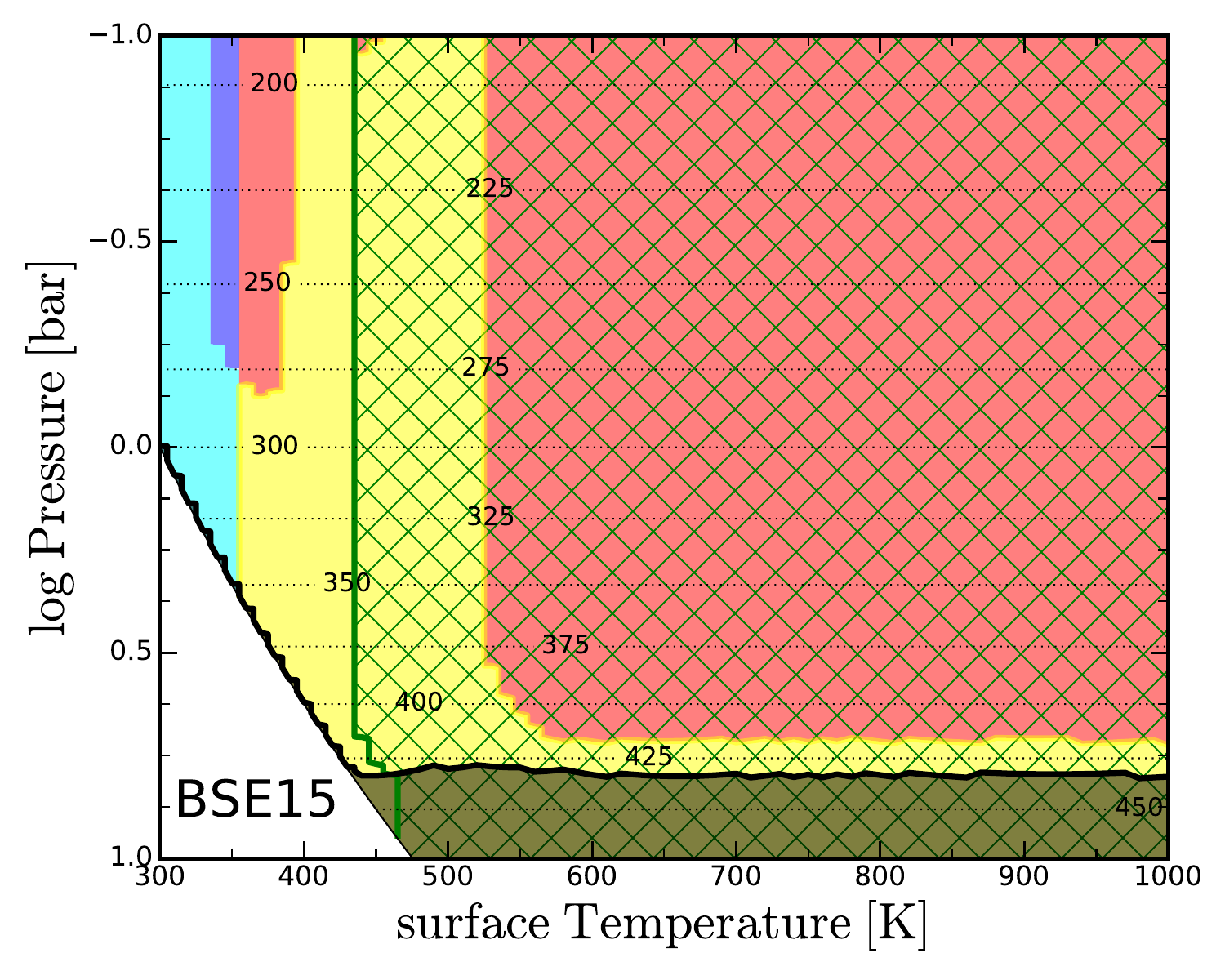}\\
\includegraphics[width = .32\linewidth, page=1]{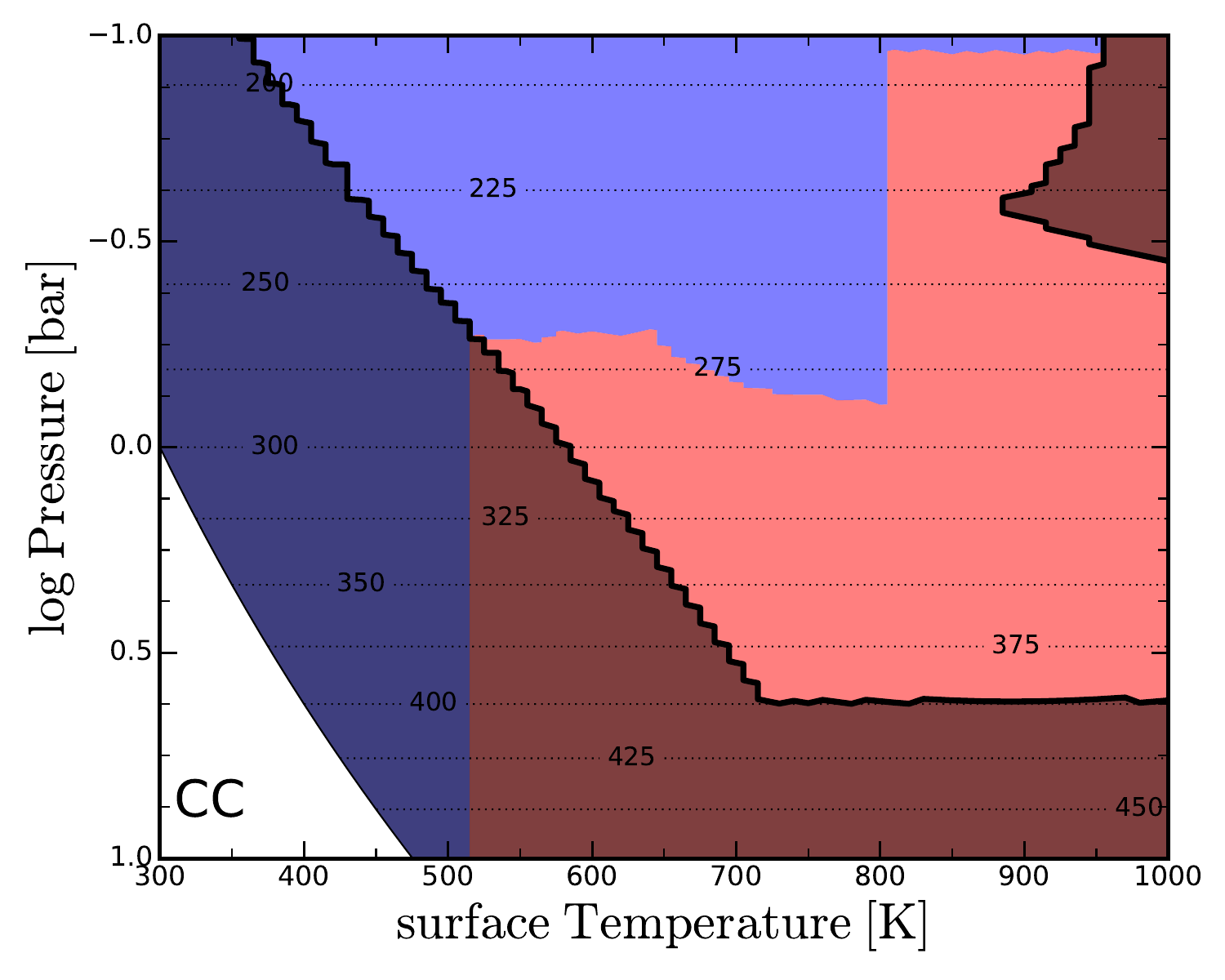}
\includegraphics[width = .32\linewidth, page=1]{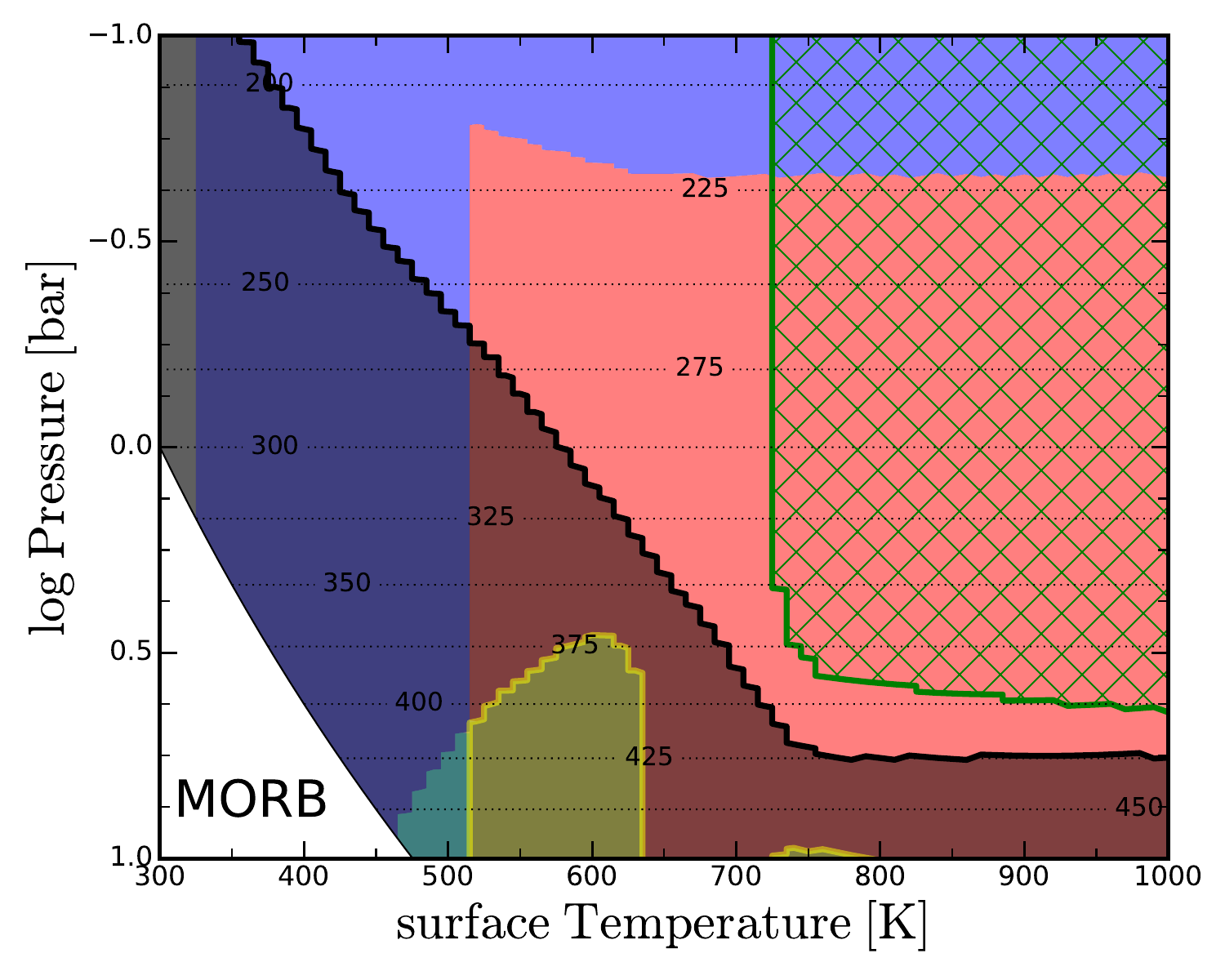}
\includegraphics[width = .32\linewidth, page=1]{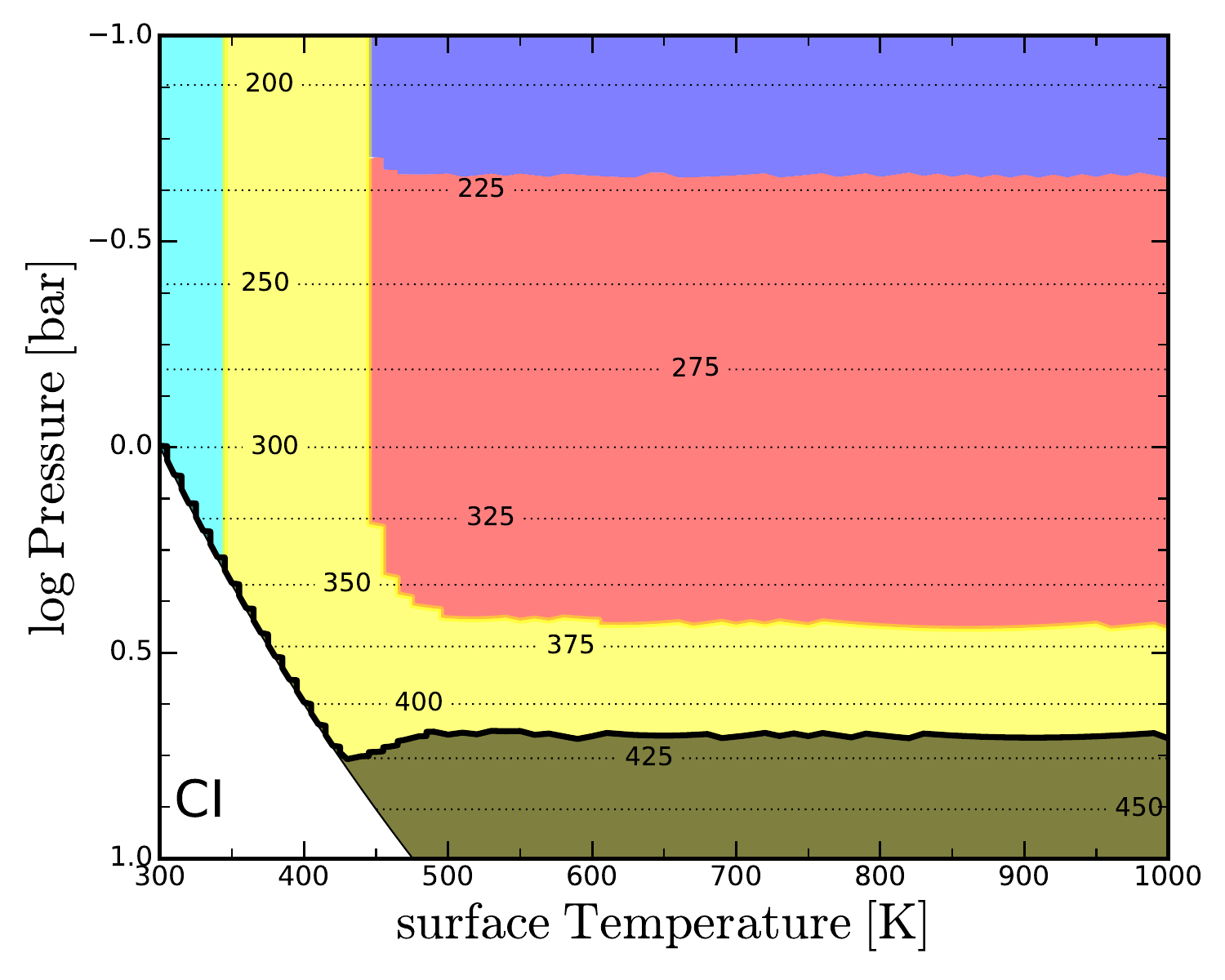}\\
\includegraphics[width = .32\linewidth, page=1]{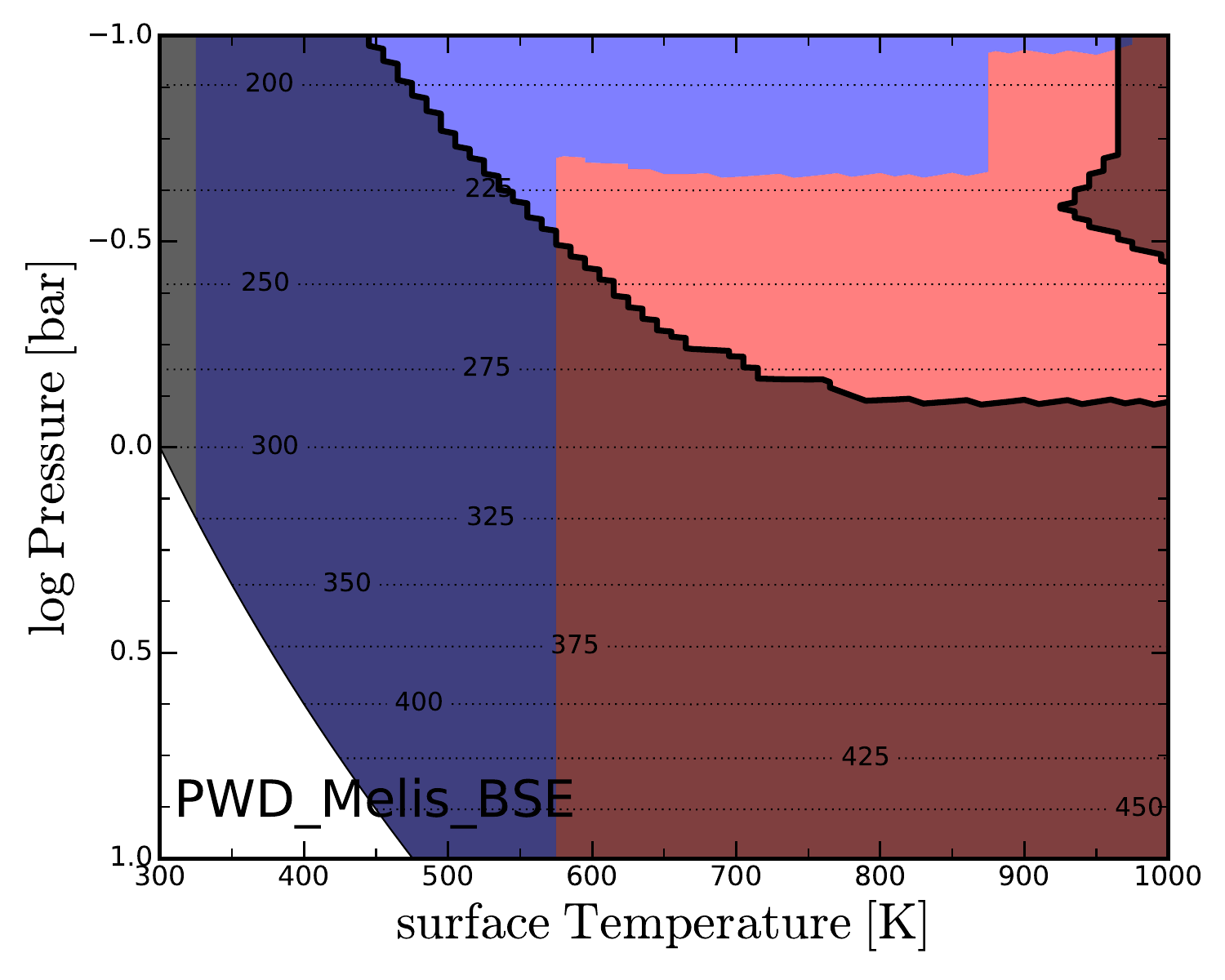}
\includegraphics[width = .32\linewidth, page=1]{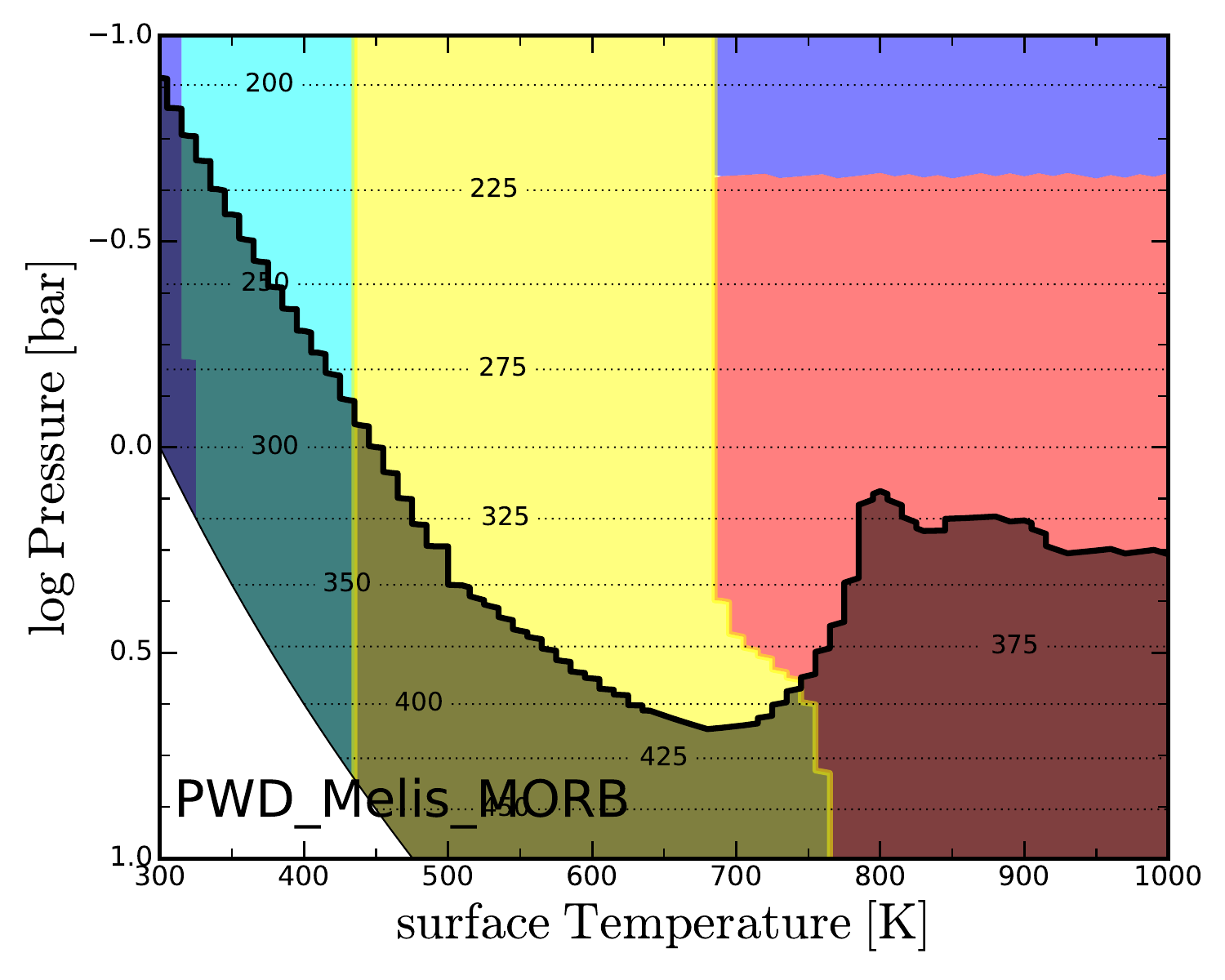}
\includegraphics[width = .32\linewidth, page=1]{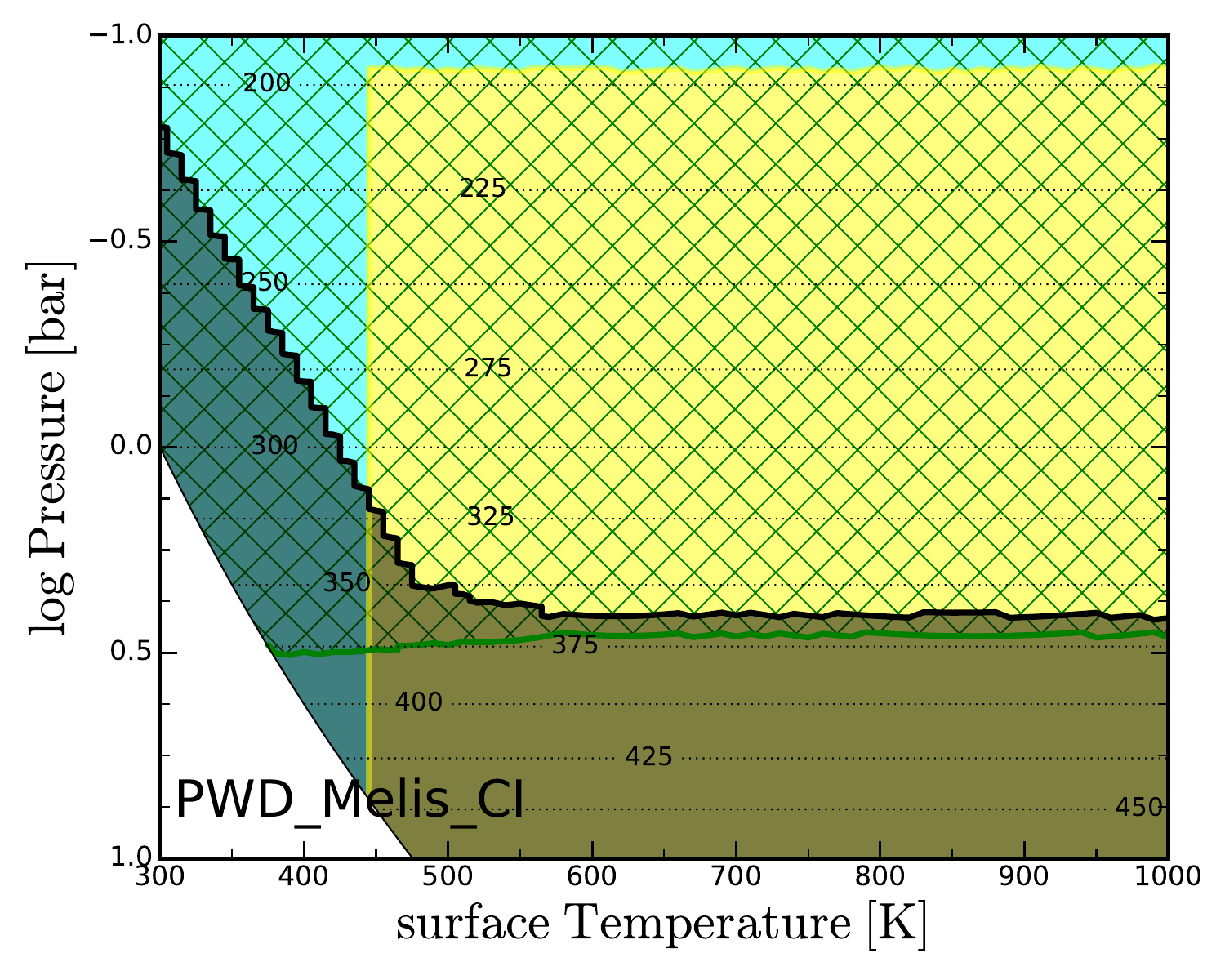}\\
\includegraphics[width = .32\linewidth, page=1]{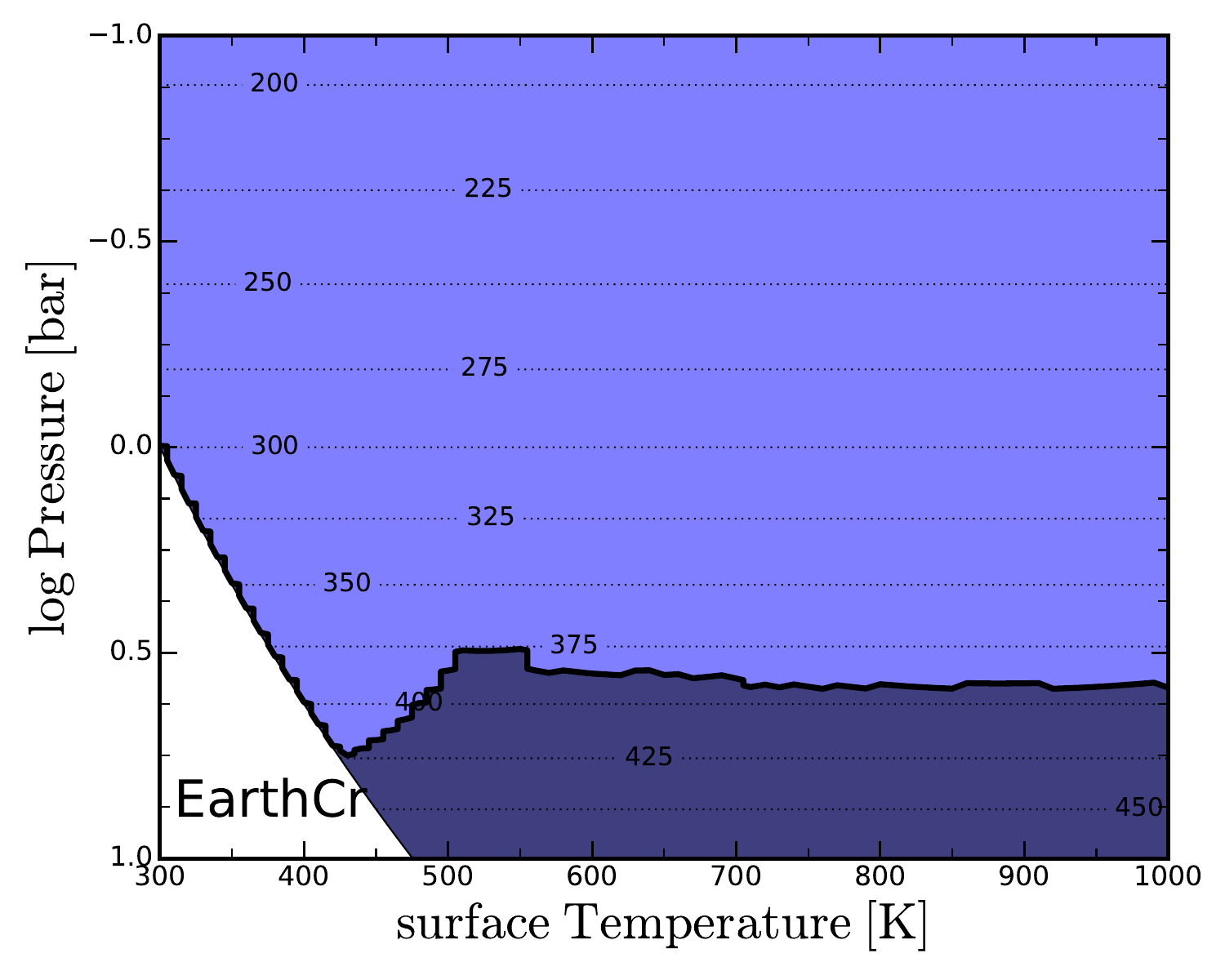}
\includegraphics[width = .32\linewidth, page=1]{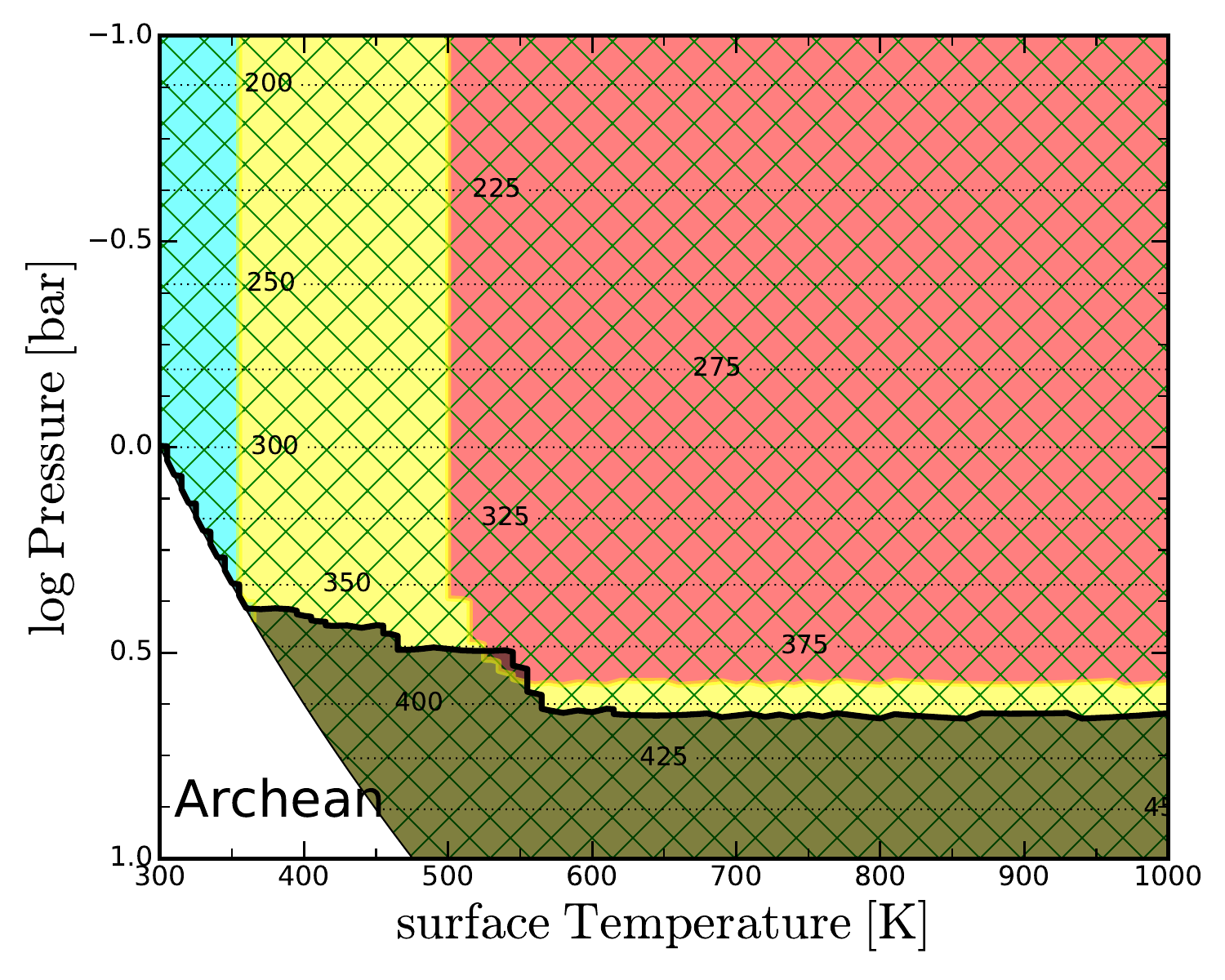}
\includegraphics[width = .32\linewidth, page=1]{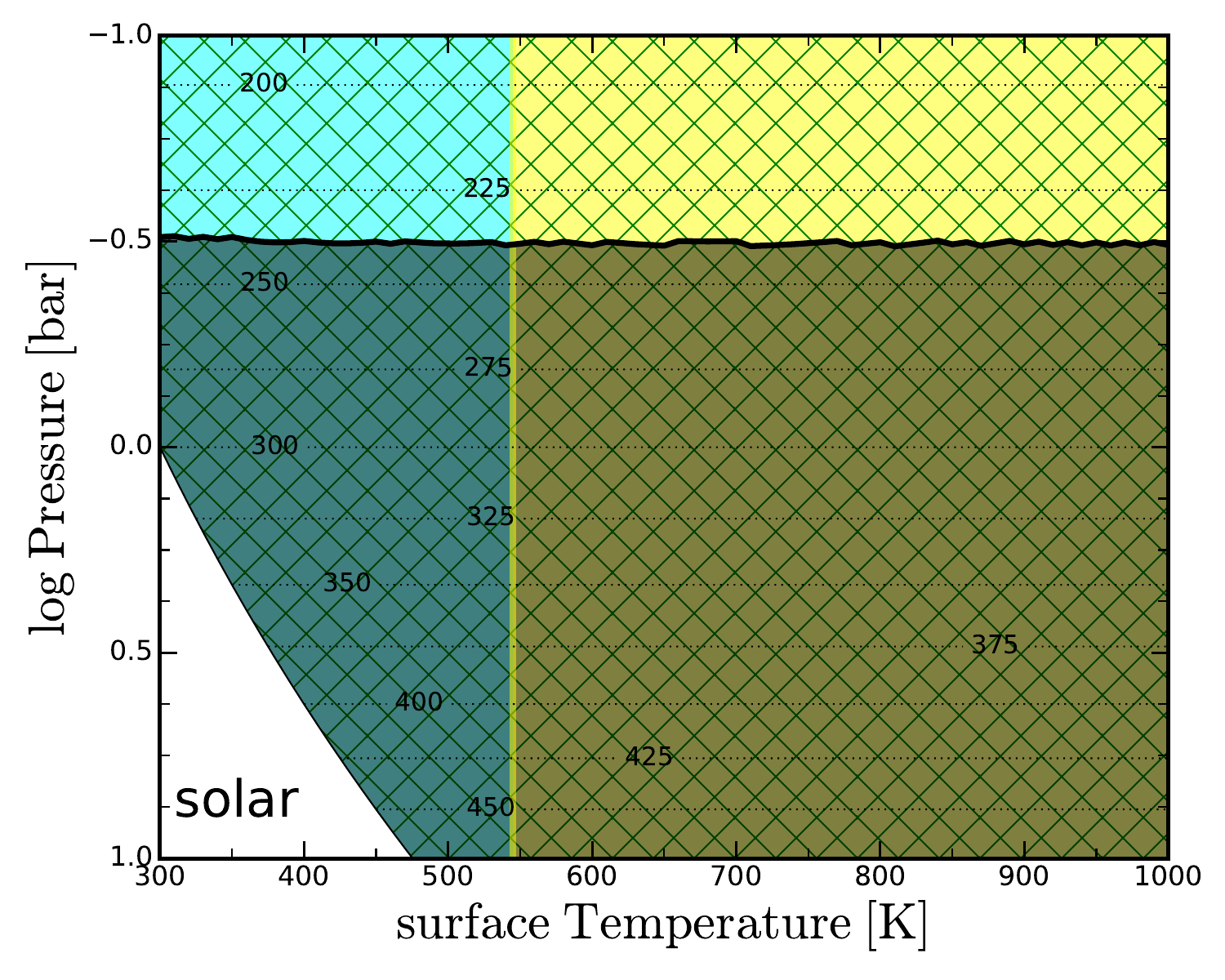}
\caption{Nutrient availability levels as defined in Table~\ref{tab:Levels} and shown in Fig.~\ref{fig:HabLev}, but with concentration thresholds of $10^{-6}$.}
\label{fig:HabLev-6}
\end{figure*}

\begin{figure*}
\centering
\includegraphics[width = .32\linewidth, page=1]{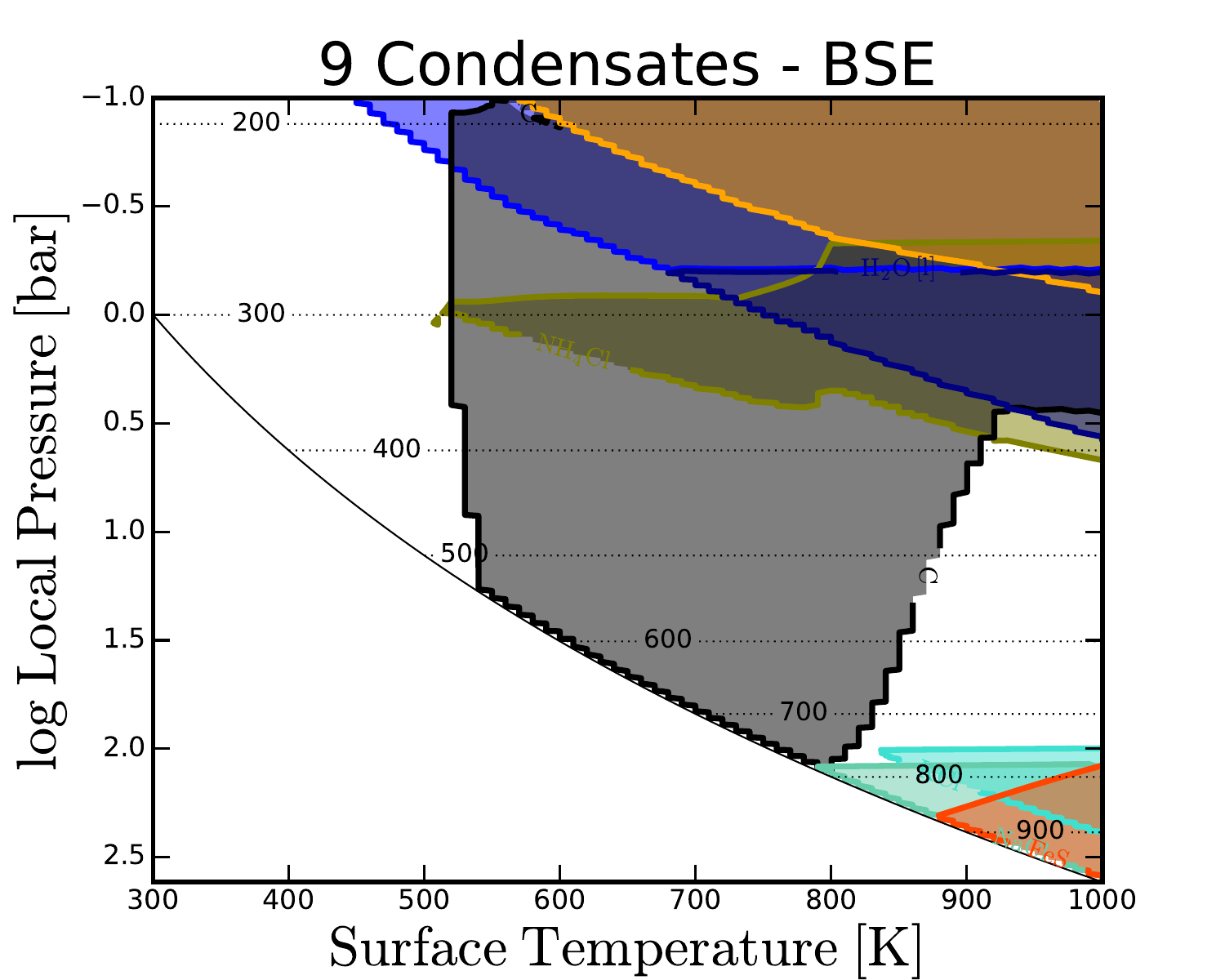}
\includegraphics[width = .32\linewidth, page=1]{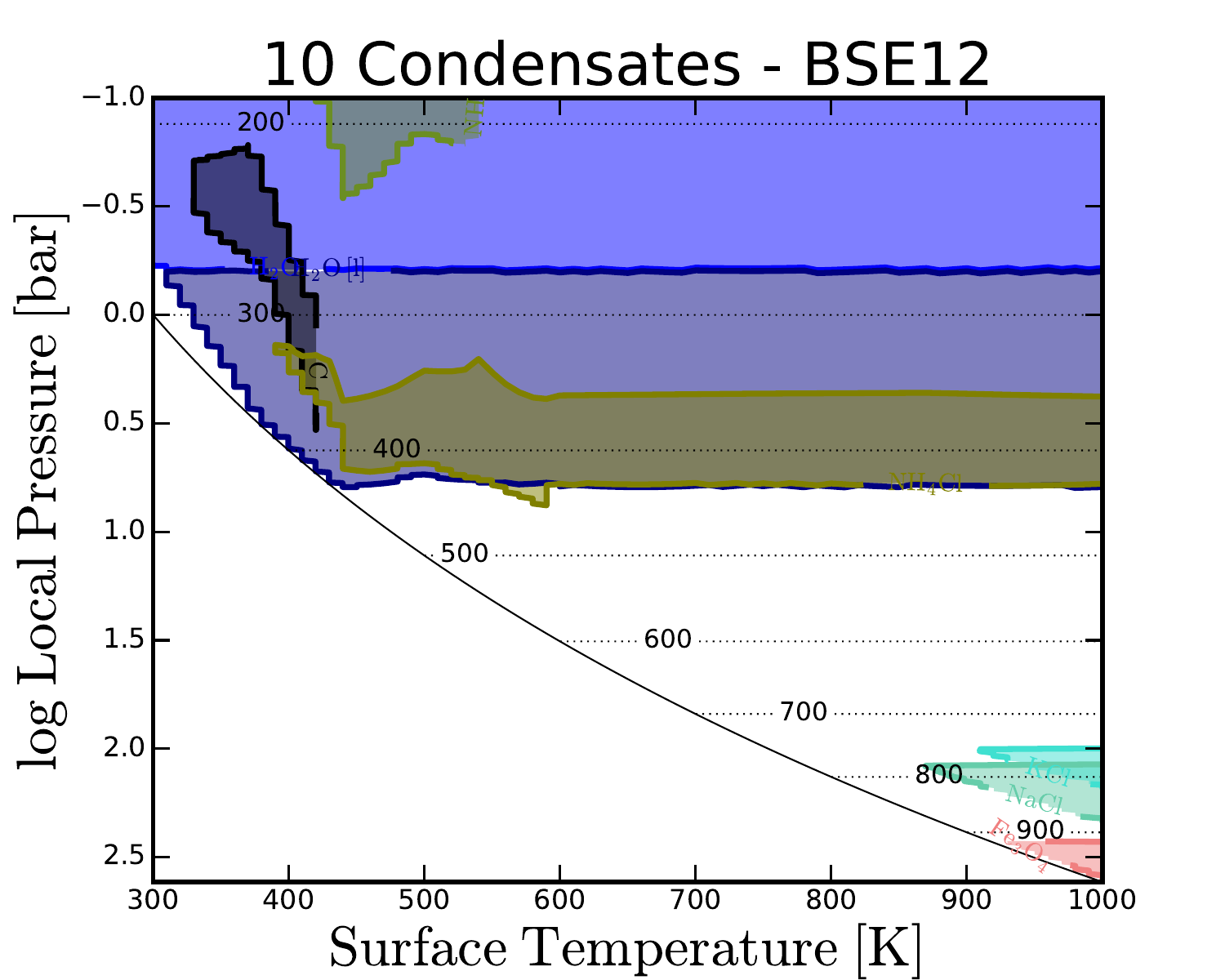}
\includegraphics[width = .32\linewidth, page=1]{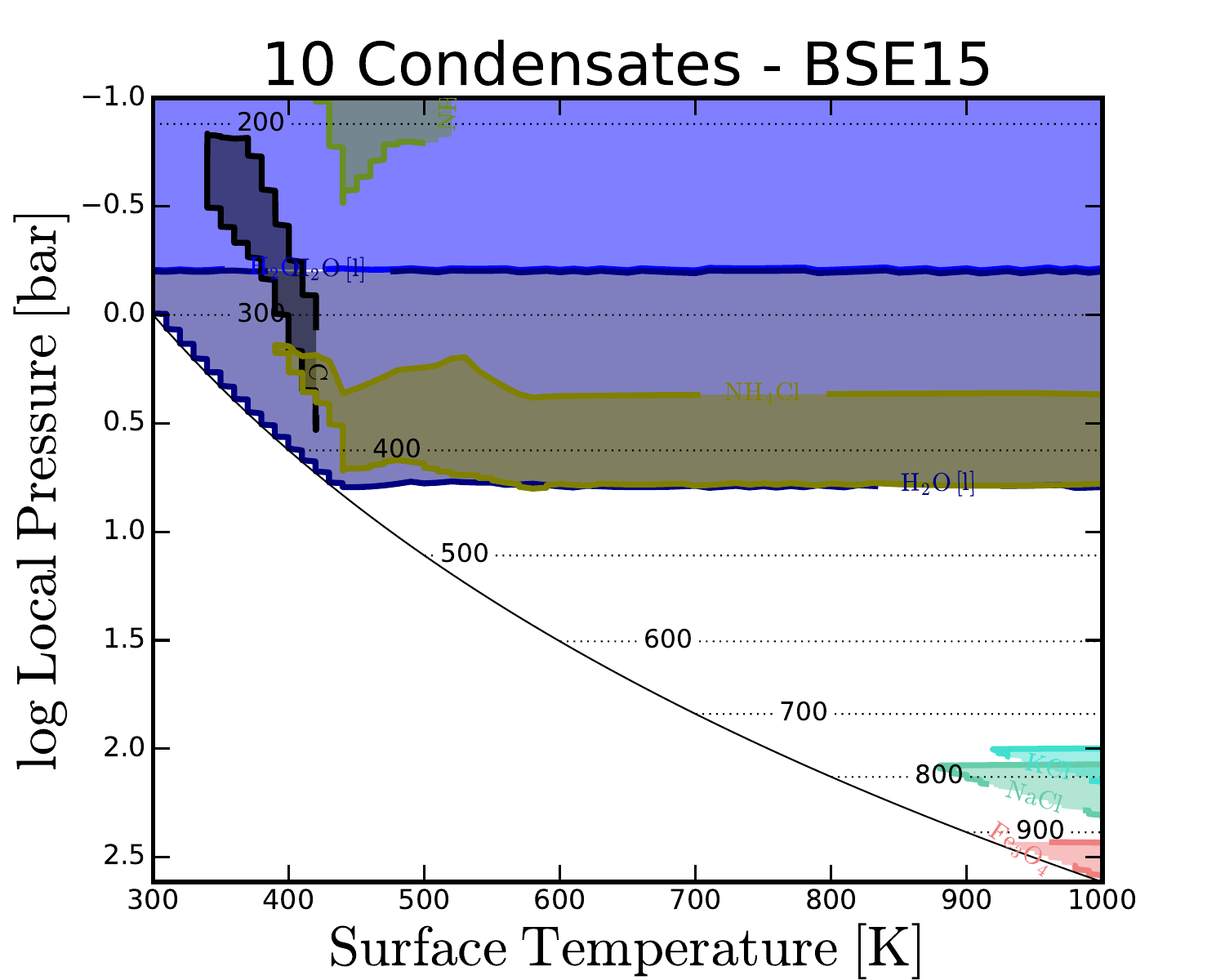}\\
\includegraphics[width = .32\linewidth, page=1]{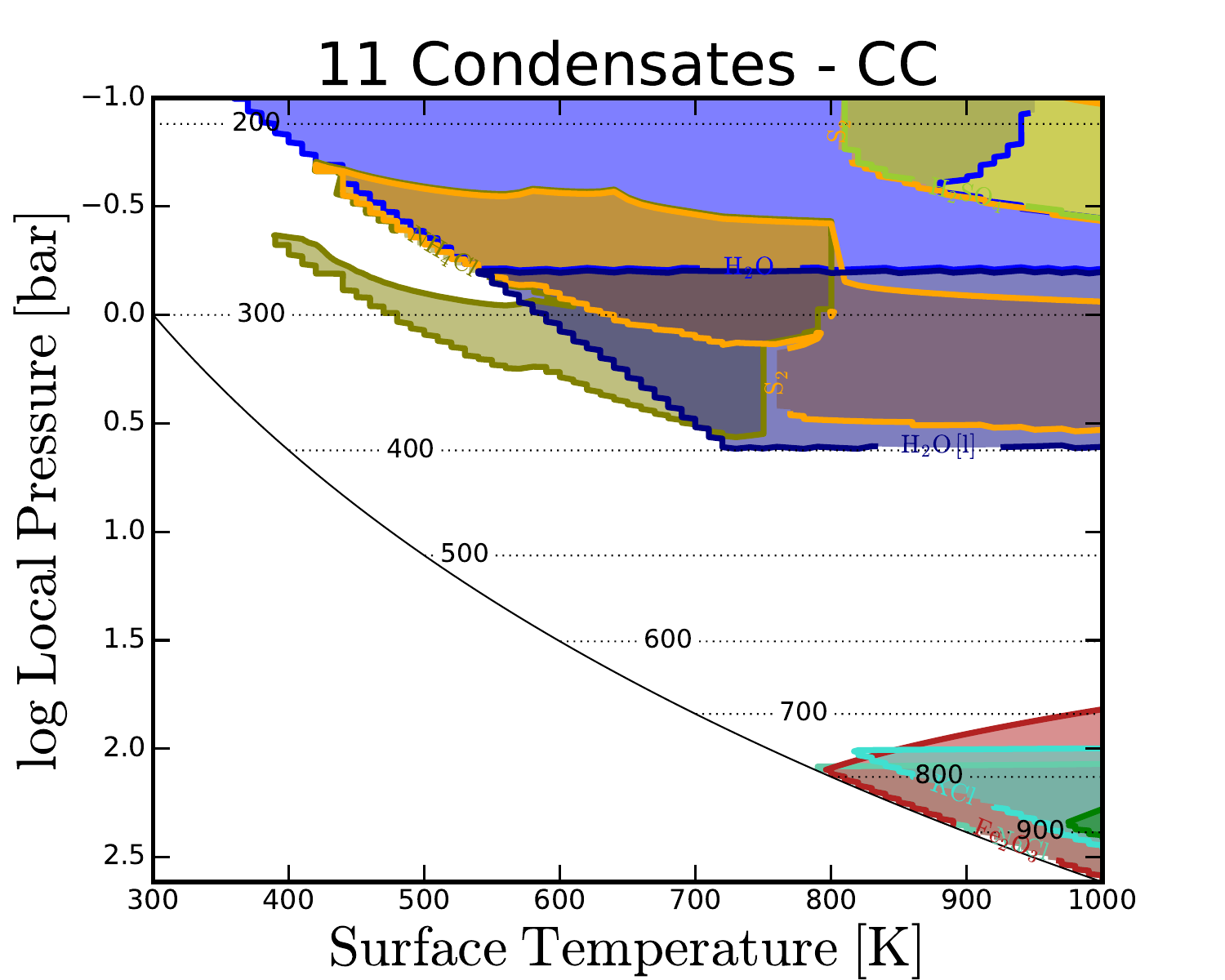}
\includegraphics[width = .32\linewidth, page=1]{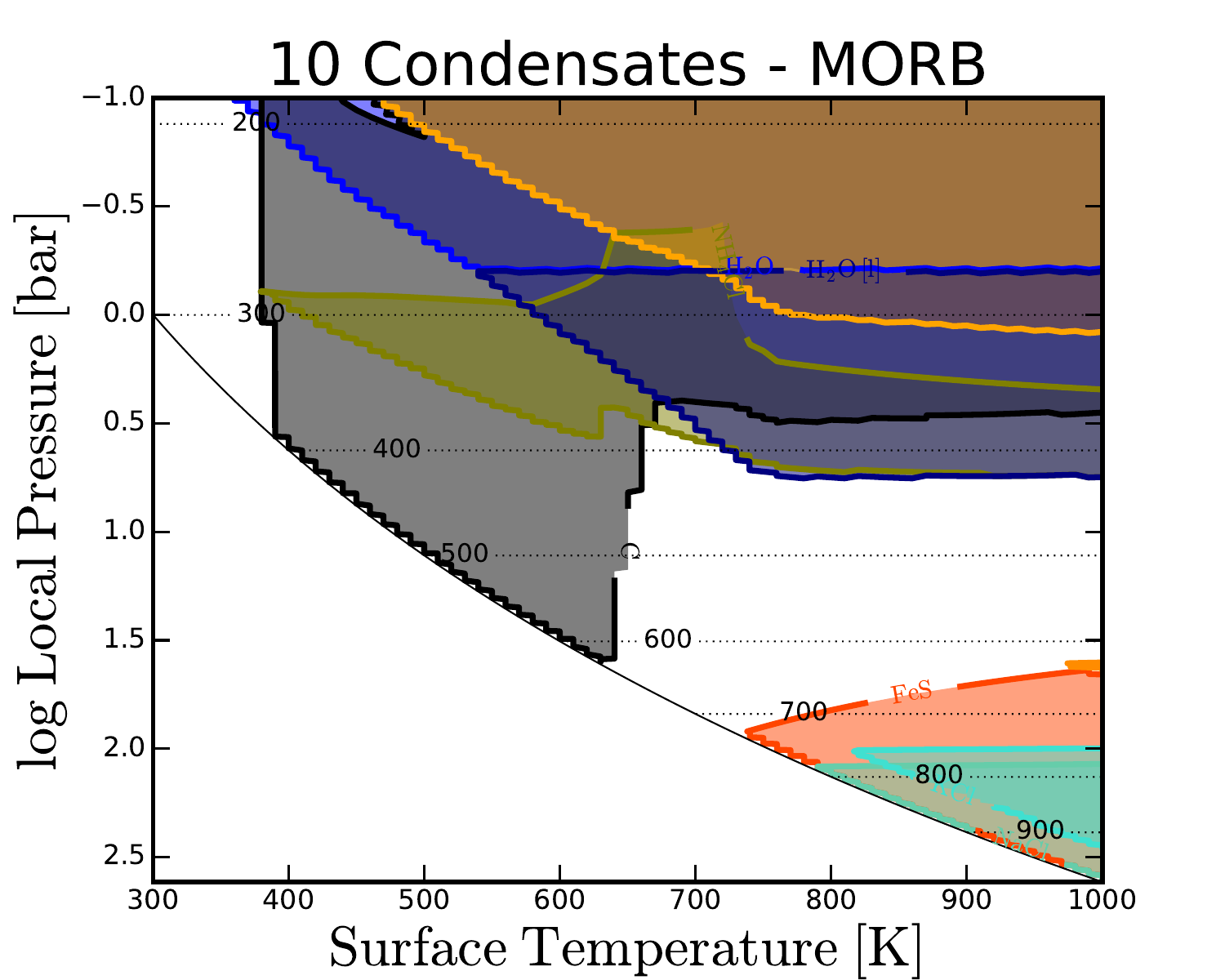}
\includegraphics[width = .32\linewidth, page=1]{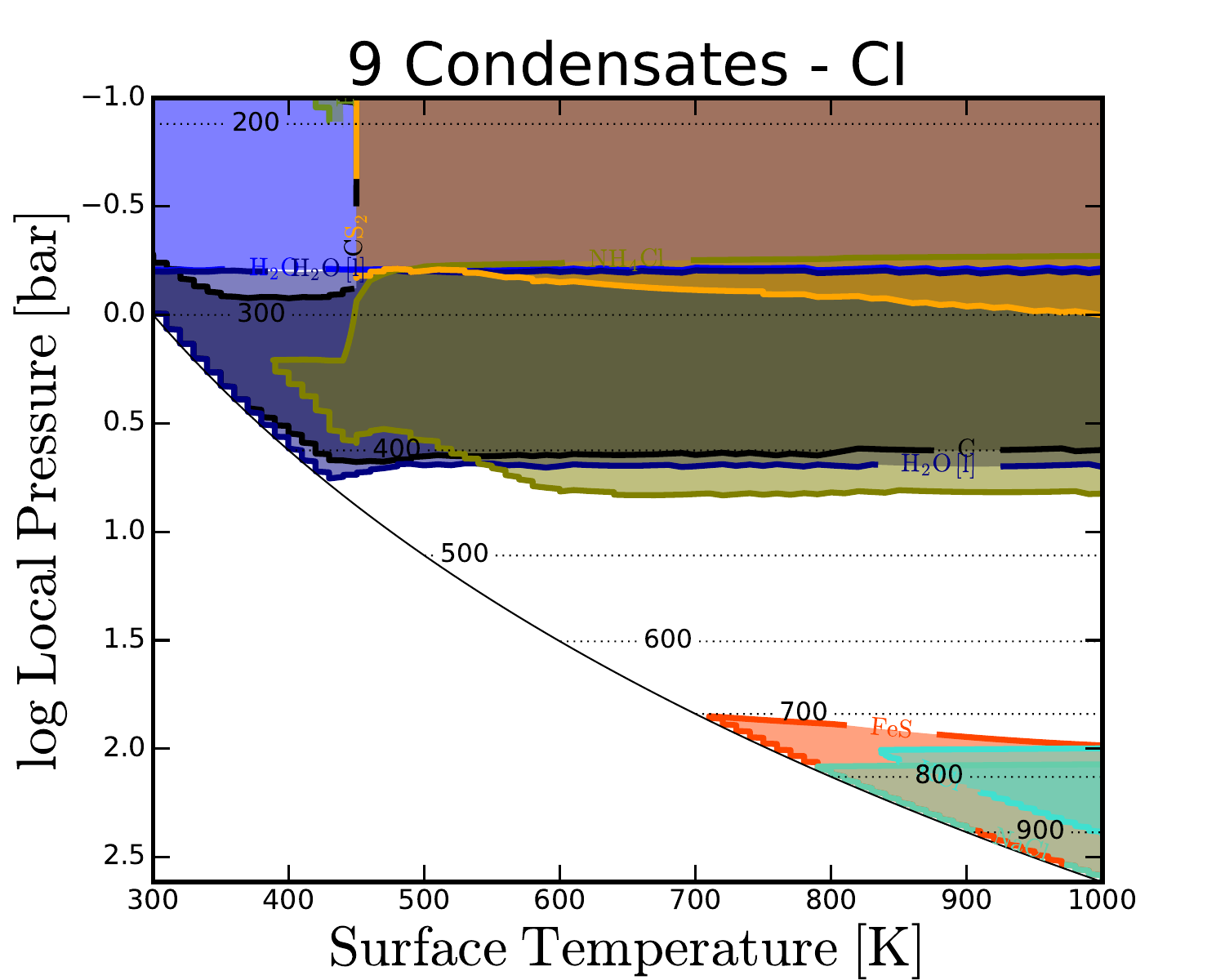}\\
\includegraphics[width = .32\linewidth, page=1]{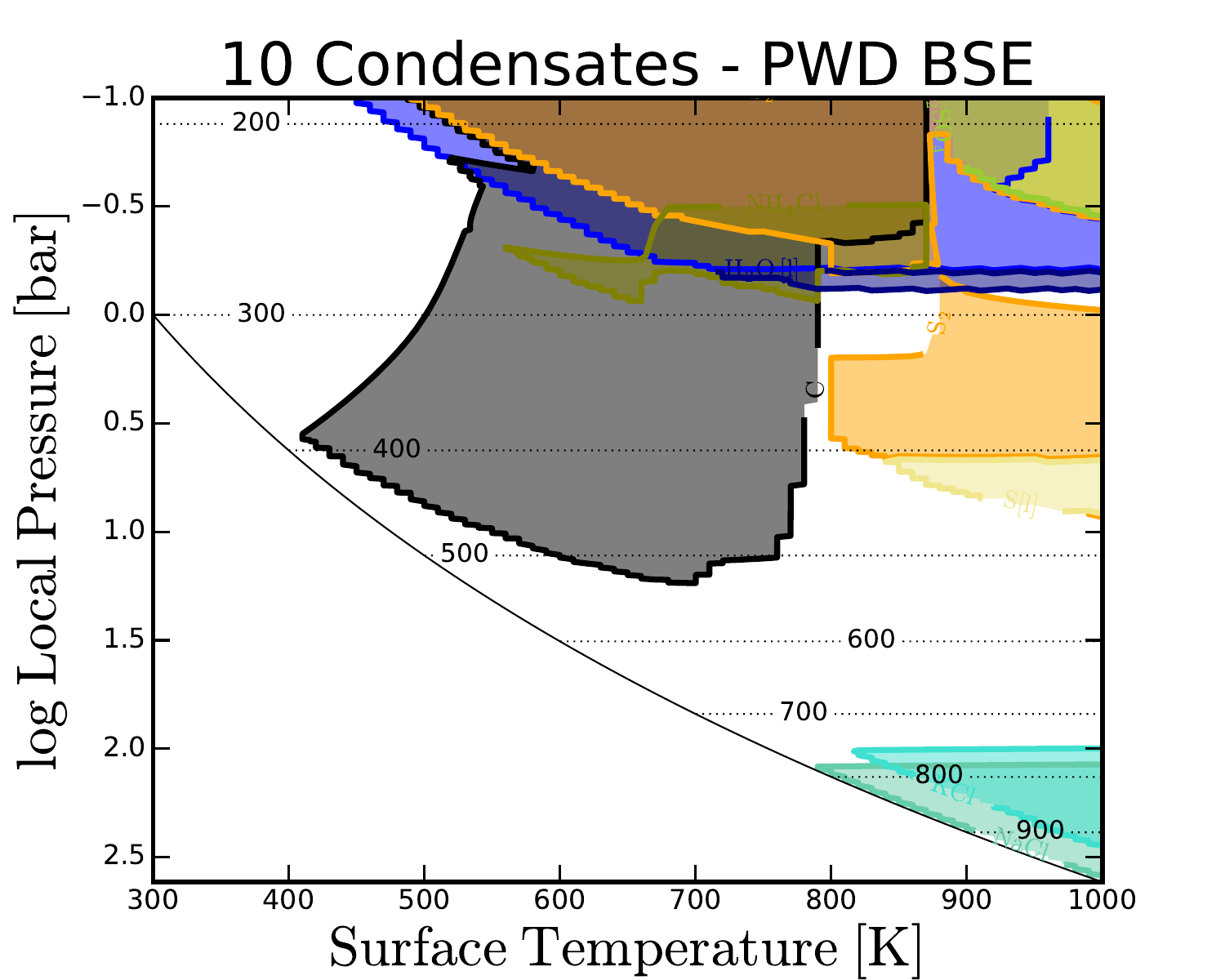}
\includegraphics[width = .32\linewidth, page=1]{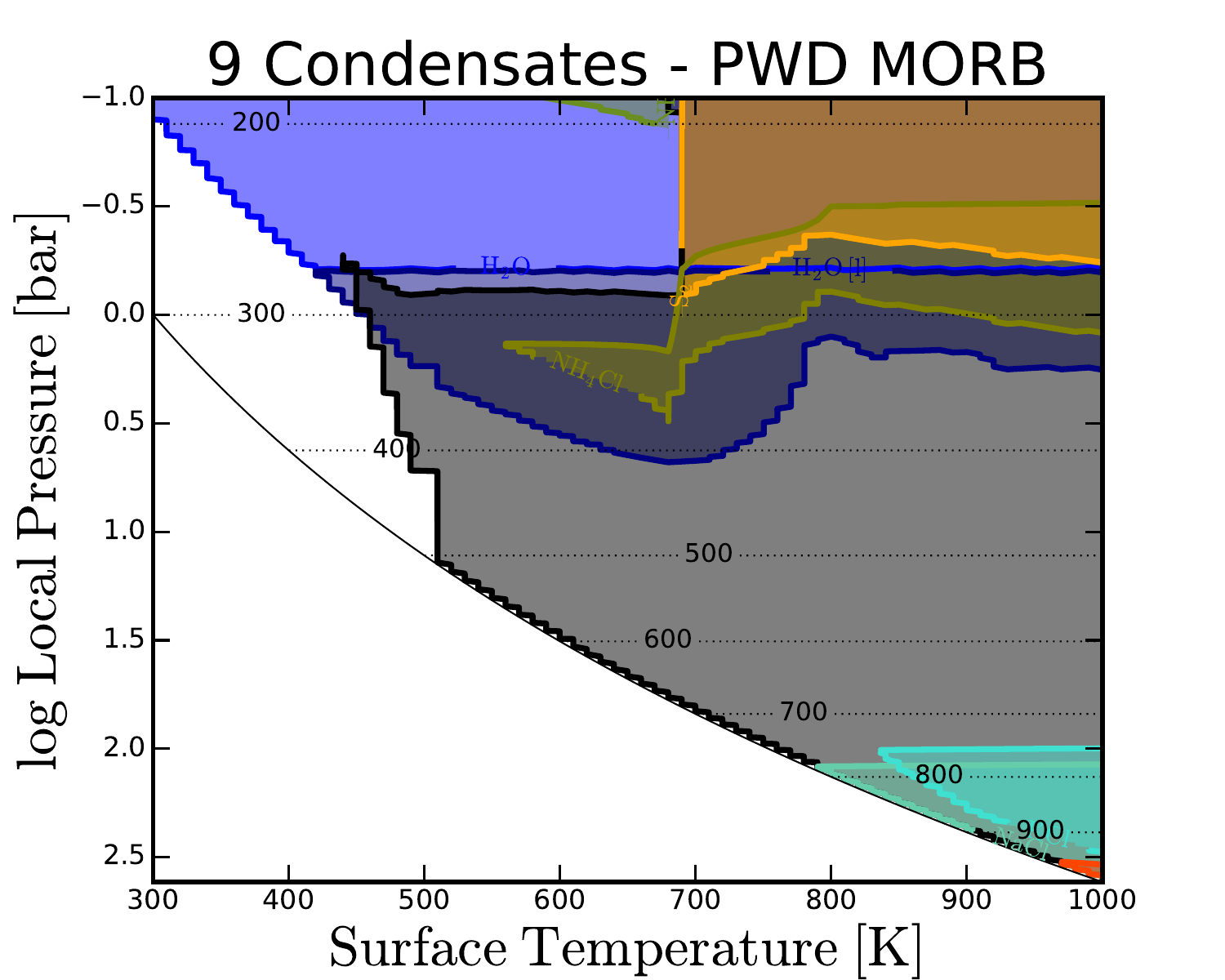}
\includegraphics[width = .32\linewidth, page=1]{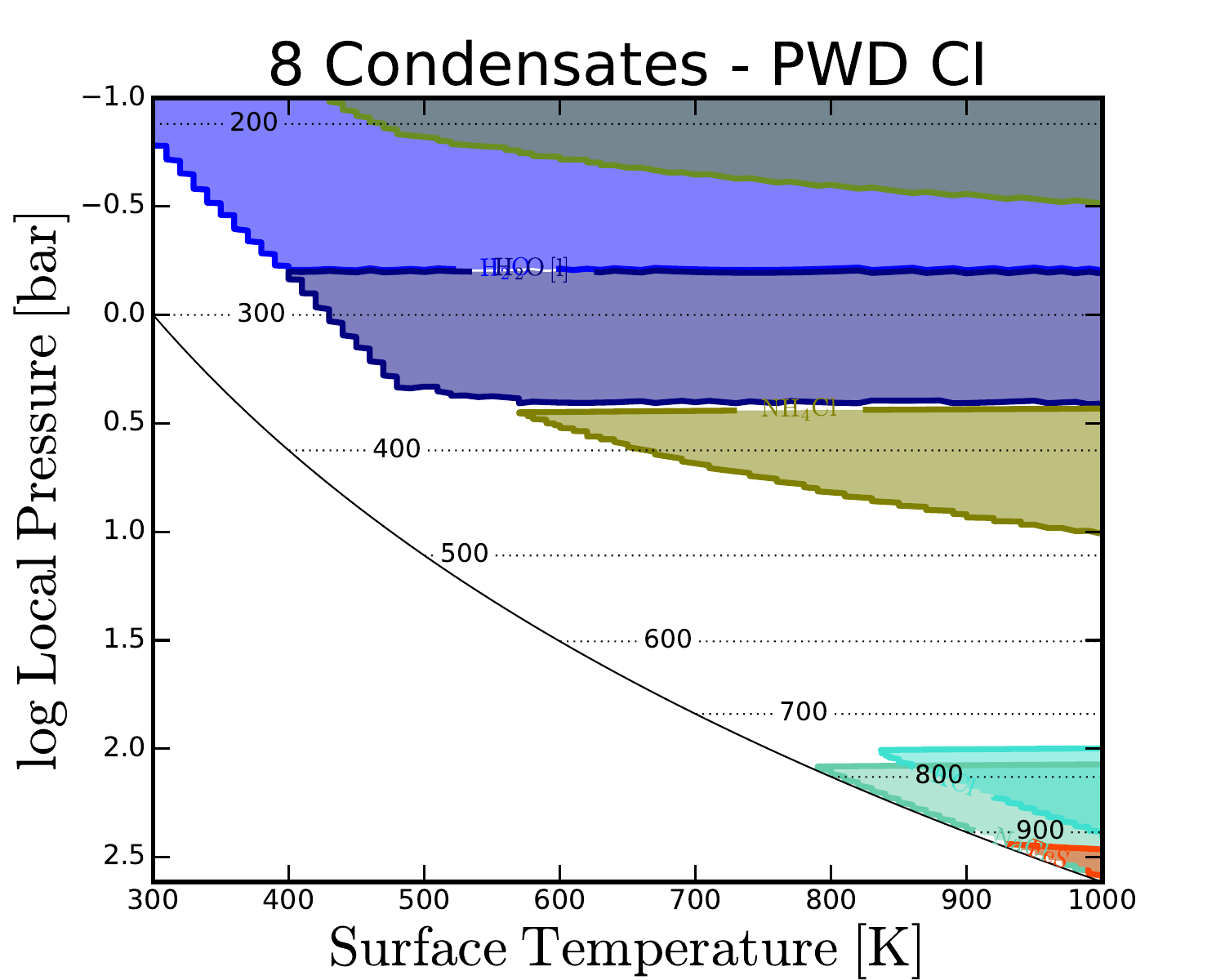}\\
\includegraphics[width = .32\linewidth, page=1]{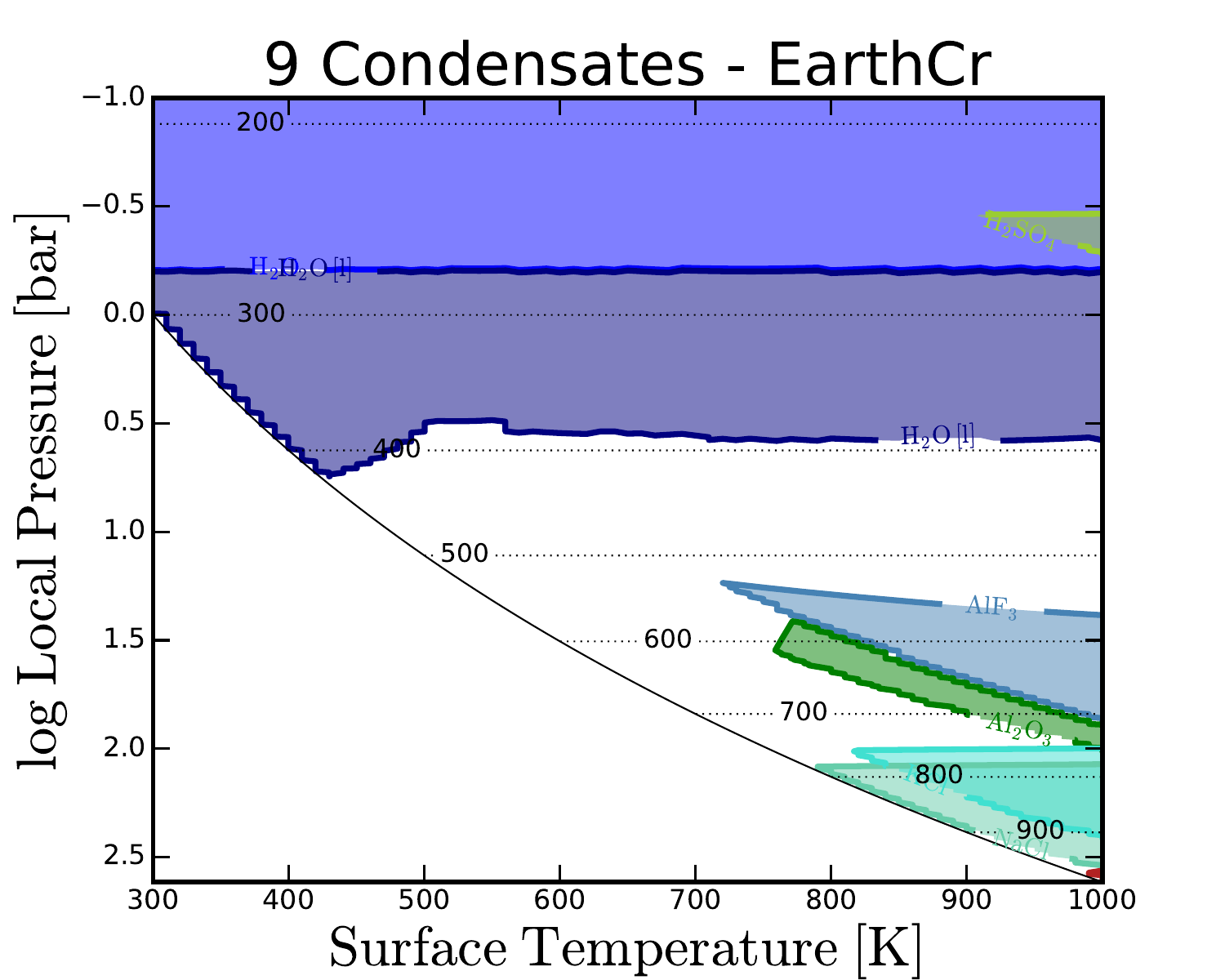}
\includegraphics[width = .32\linewidth, page=1]{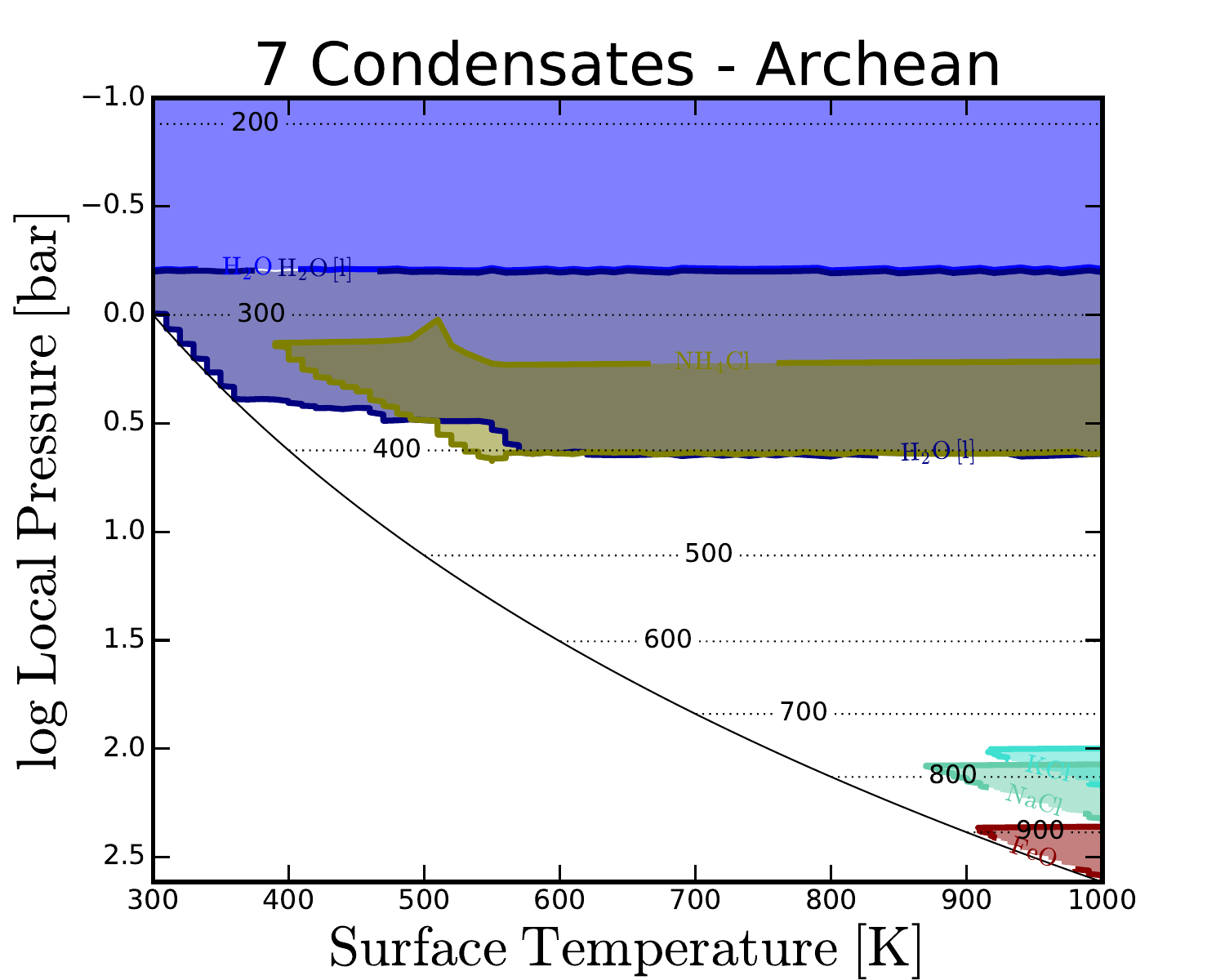}
\includegraphics[width = .32\linewidth, page=1]{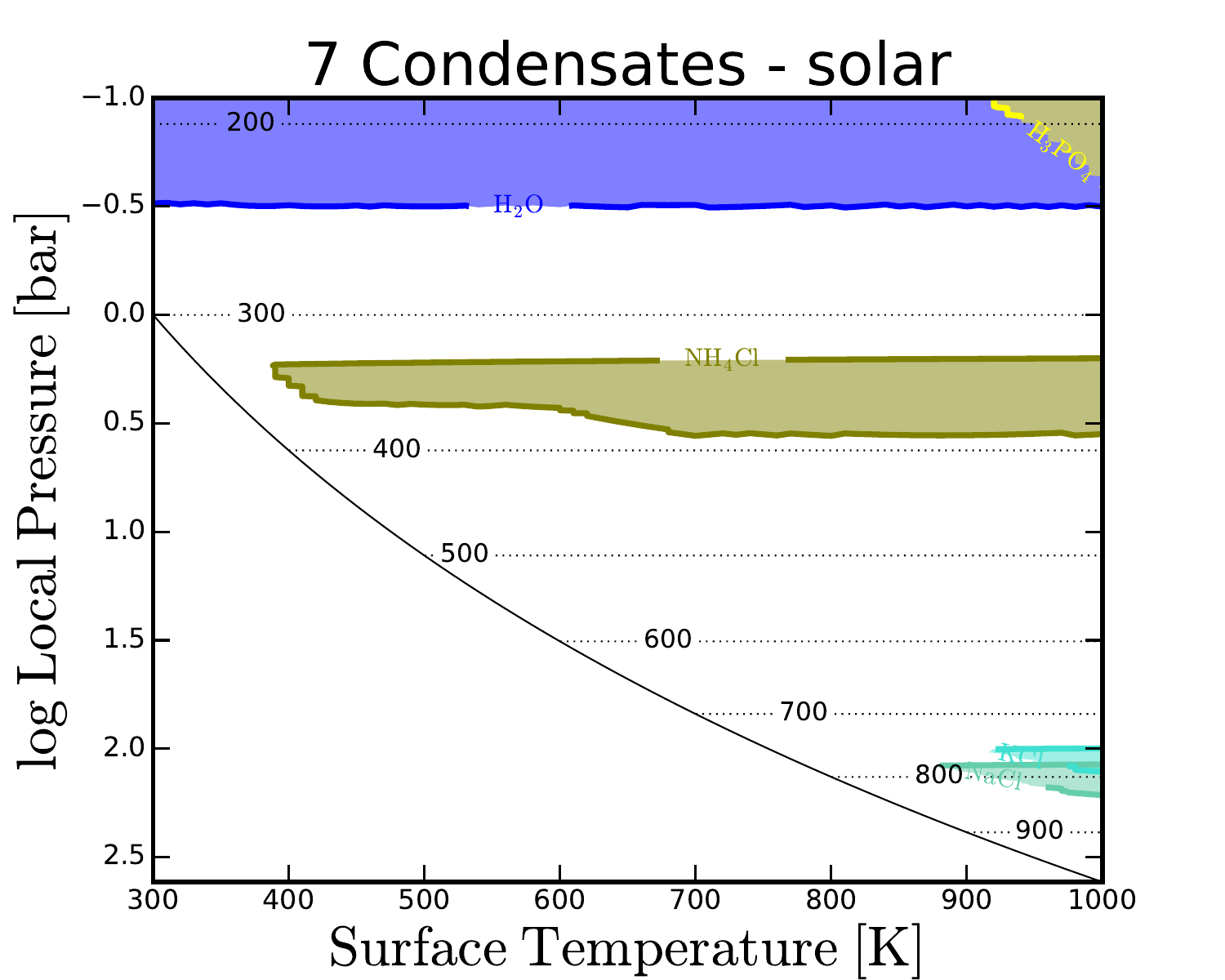}
\caption{Thermally stable cloud condensates with $\log_{10} (n_\mathrm{cond}/n_\text{cond} )> -10$ for different element abundances and surface conditions. Each panel shows one total element abundance. The thin solid black line shows the surface conditions that are used as the starting point for the respective atmospheric models. The dotted lines show the local gas phase temperature.}
\label{fig:Clouds}
\end{figure*}

\begin{figure*}
\centering
\includegraphics[width = .32\linewidth, page=4]{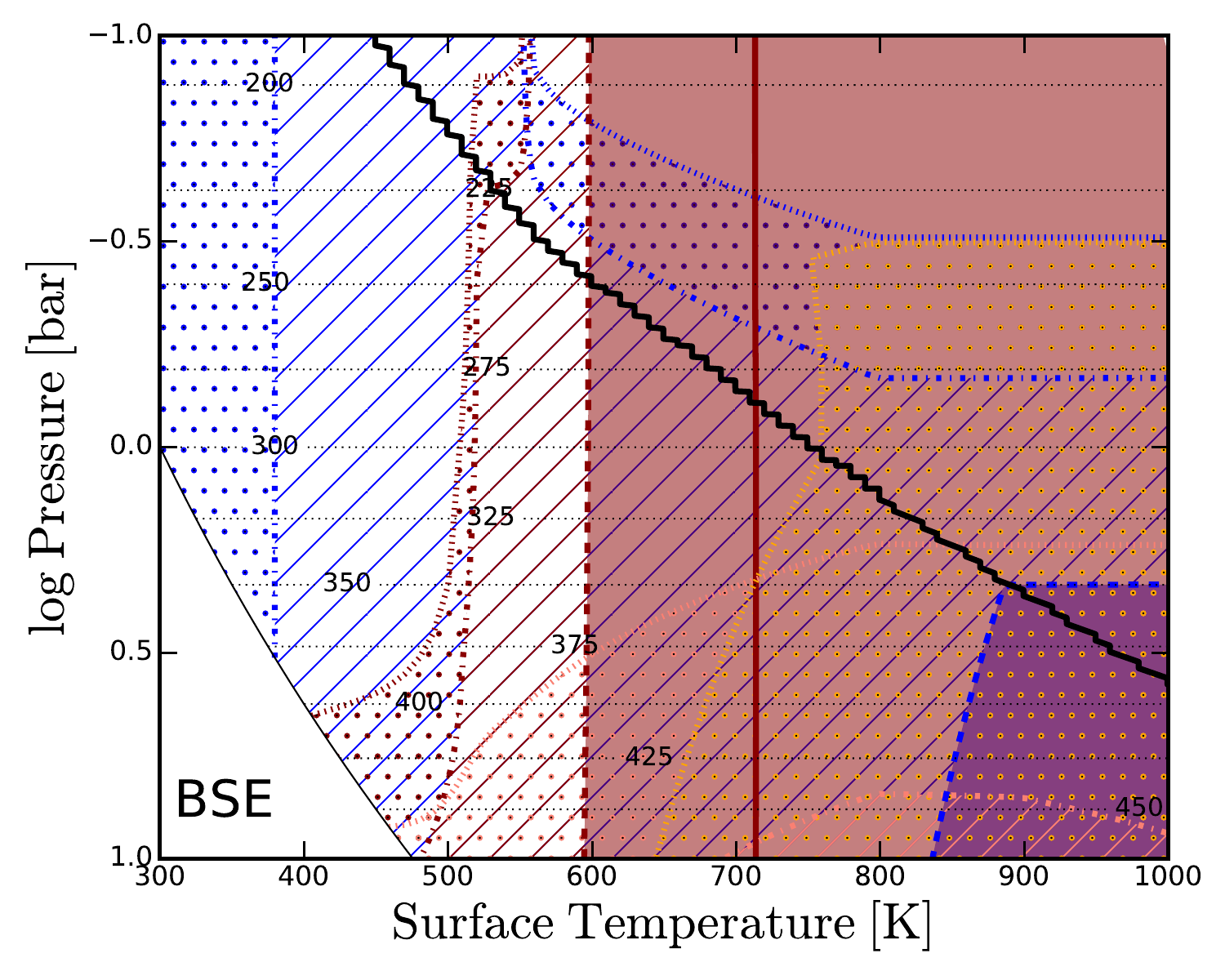}
\includegraphics[width = .32\linewidth, page=4]{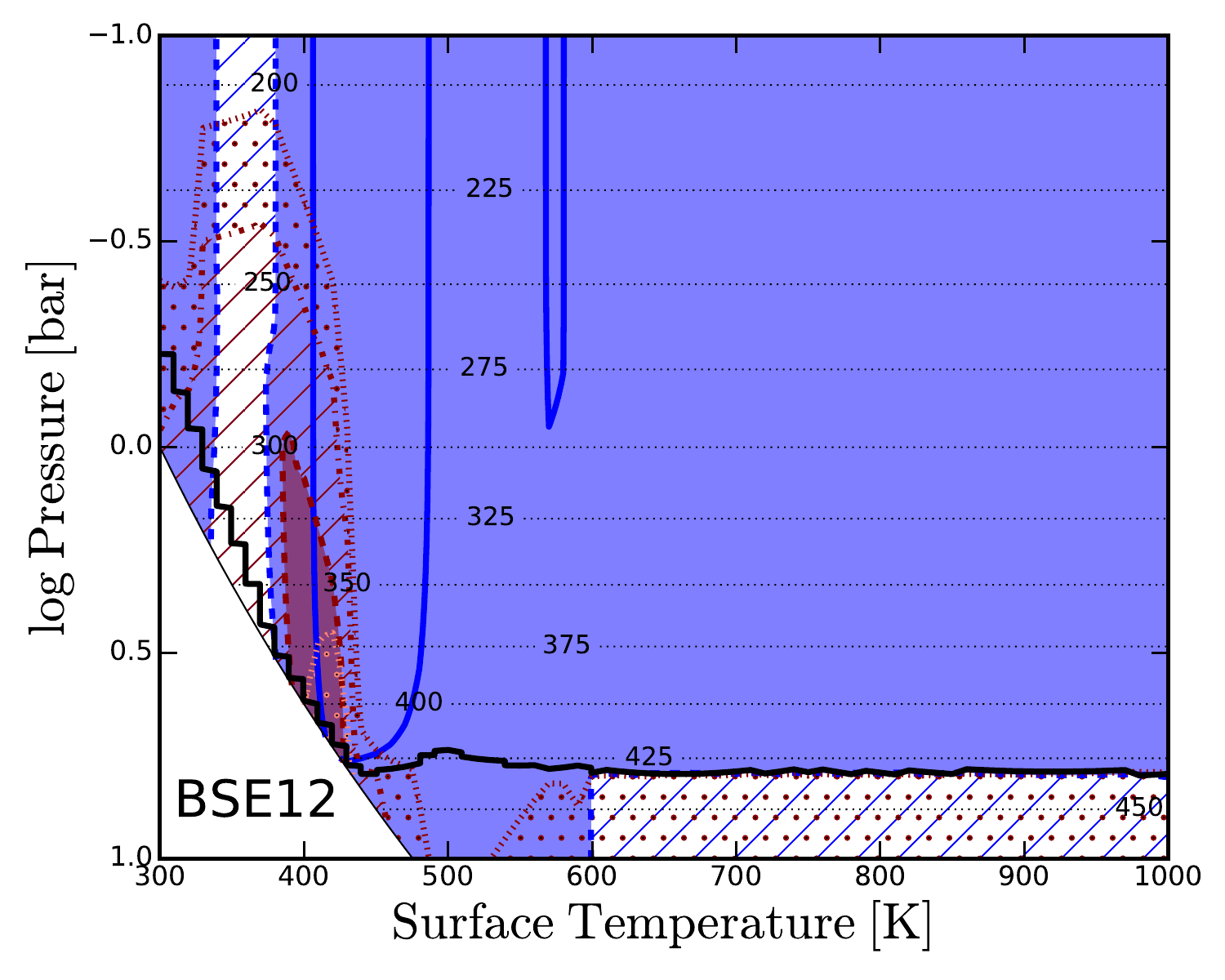}
\includegraphics[width = .32\linewidth, page=4]{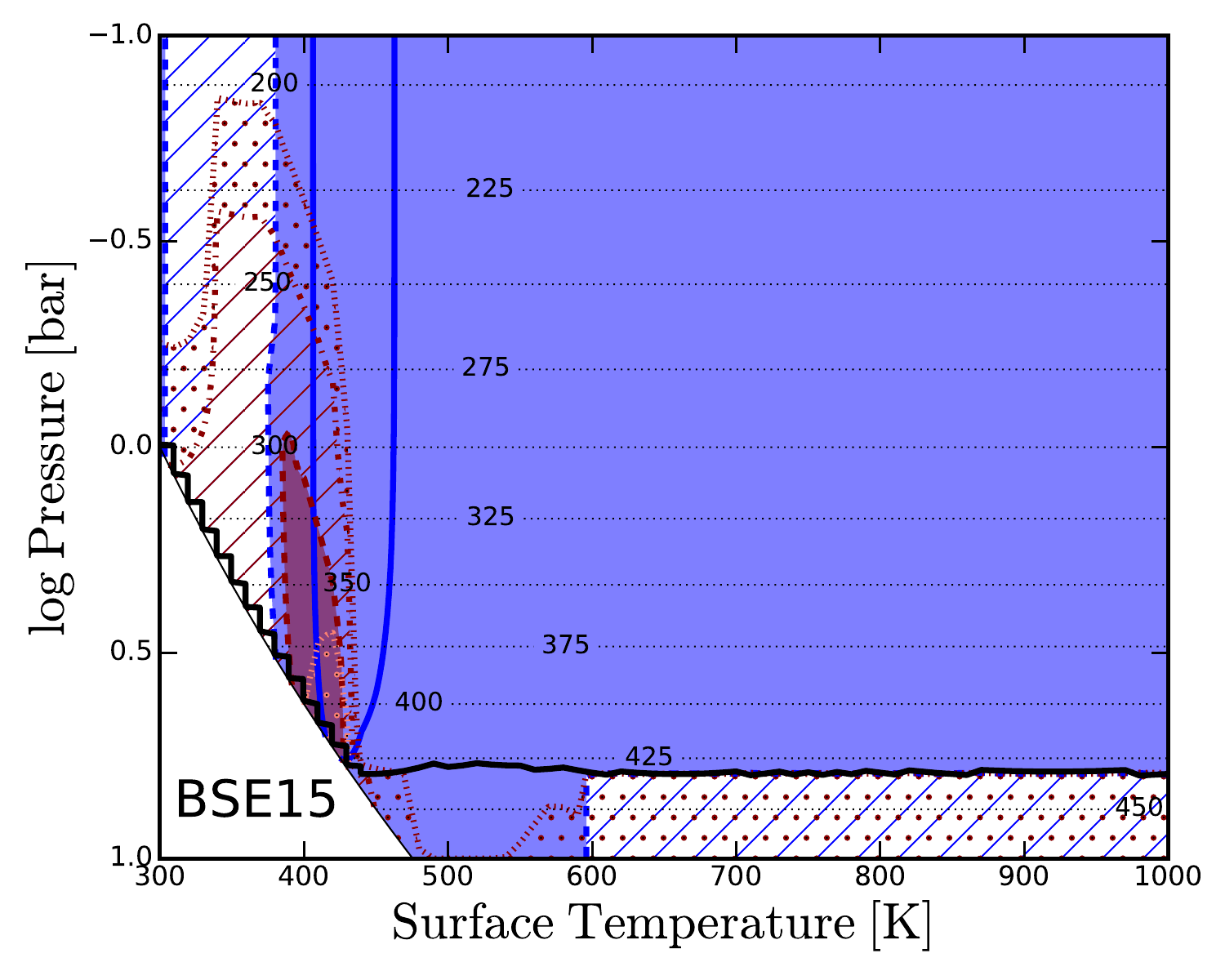}\\
\includegraphics[width = .32\linewidth, page=4]{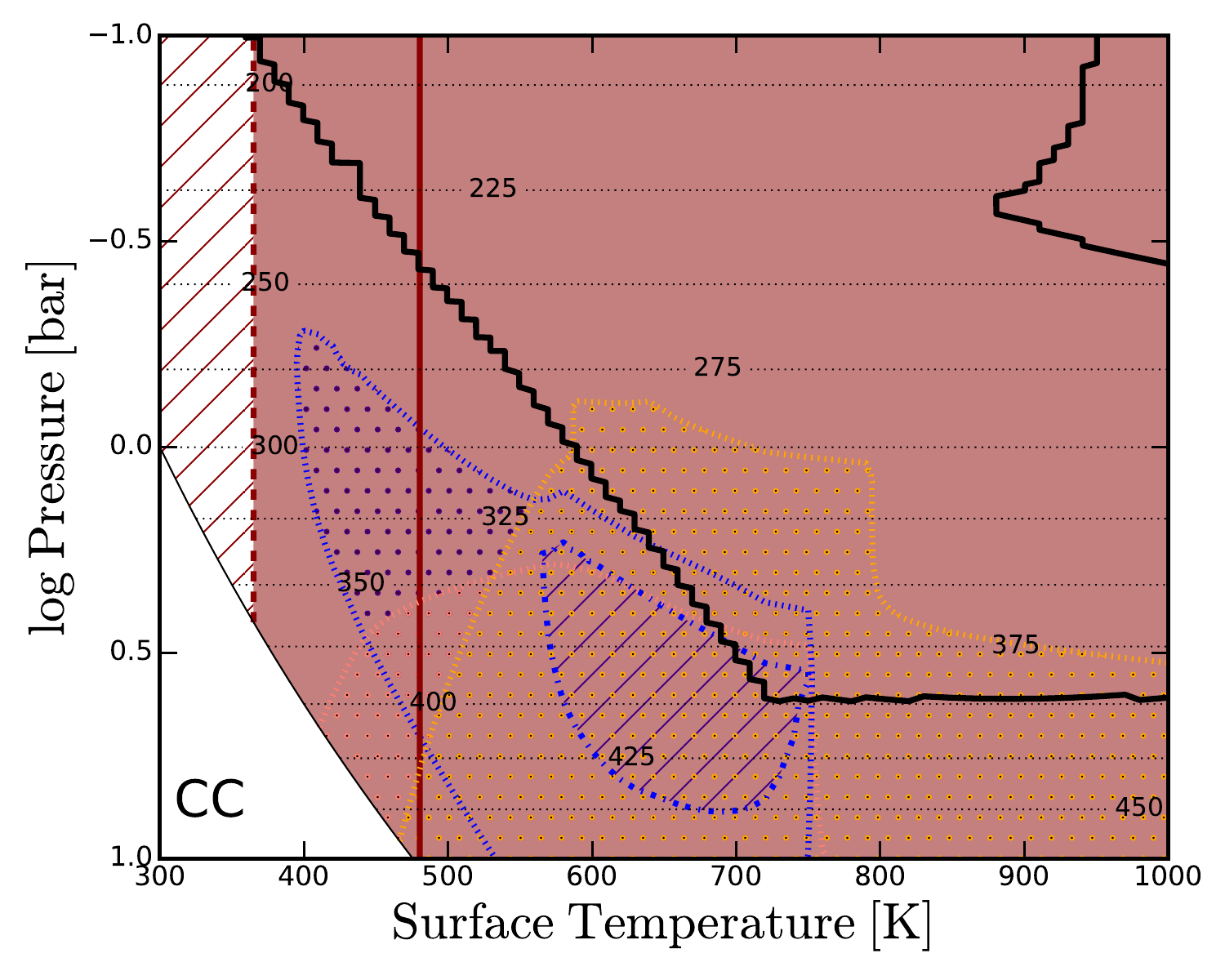}
\includegraphics[width = .32\linewidth, page=4]{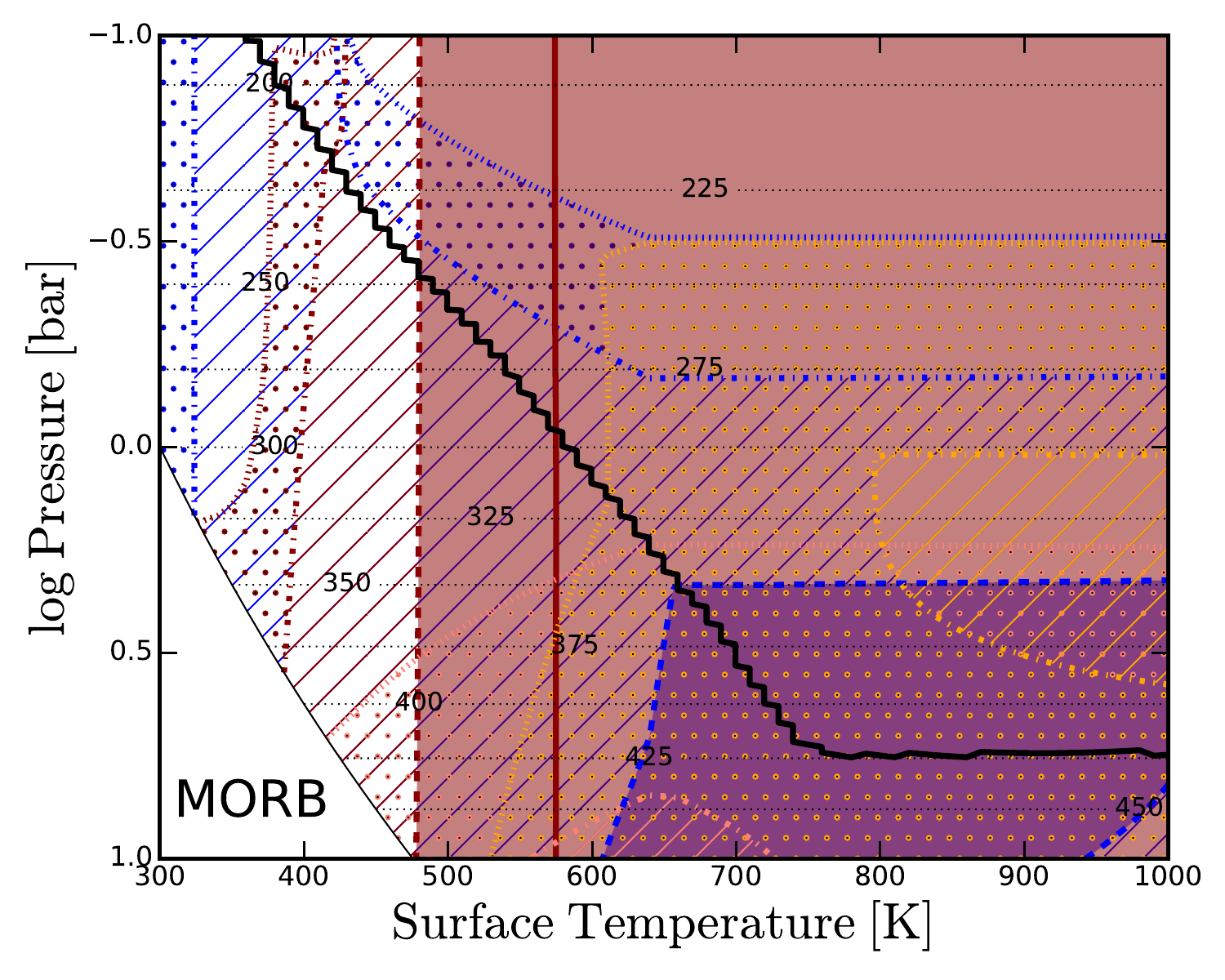}
\includegraphics[width = .32\linewidth, page=4]{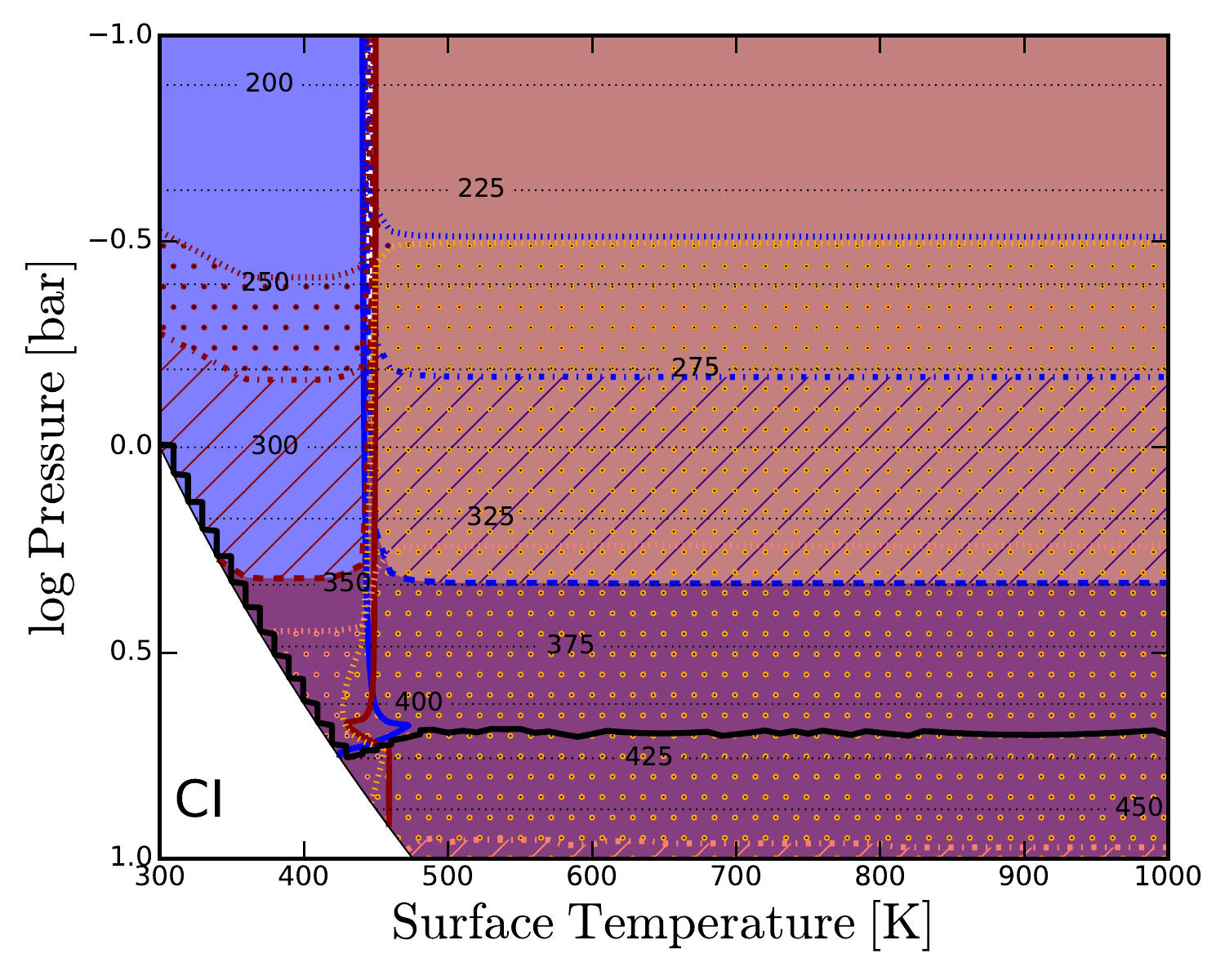}\\
\includegraphics[width = .32\linewidth, page=4]{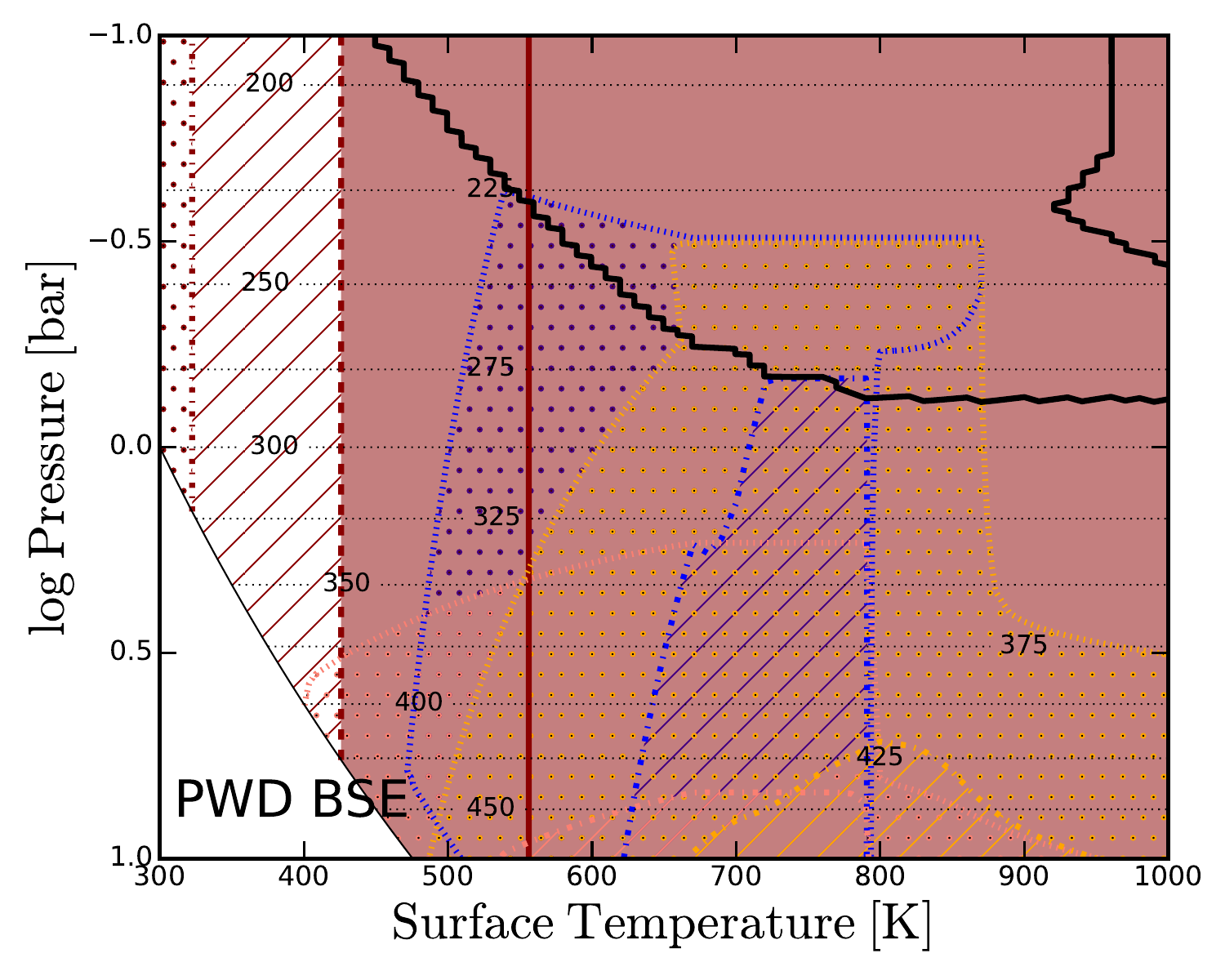}
\includegraphics[width = .32\linewidth, page=4]{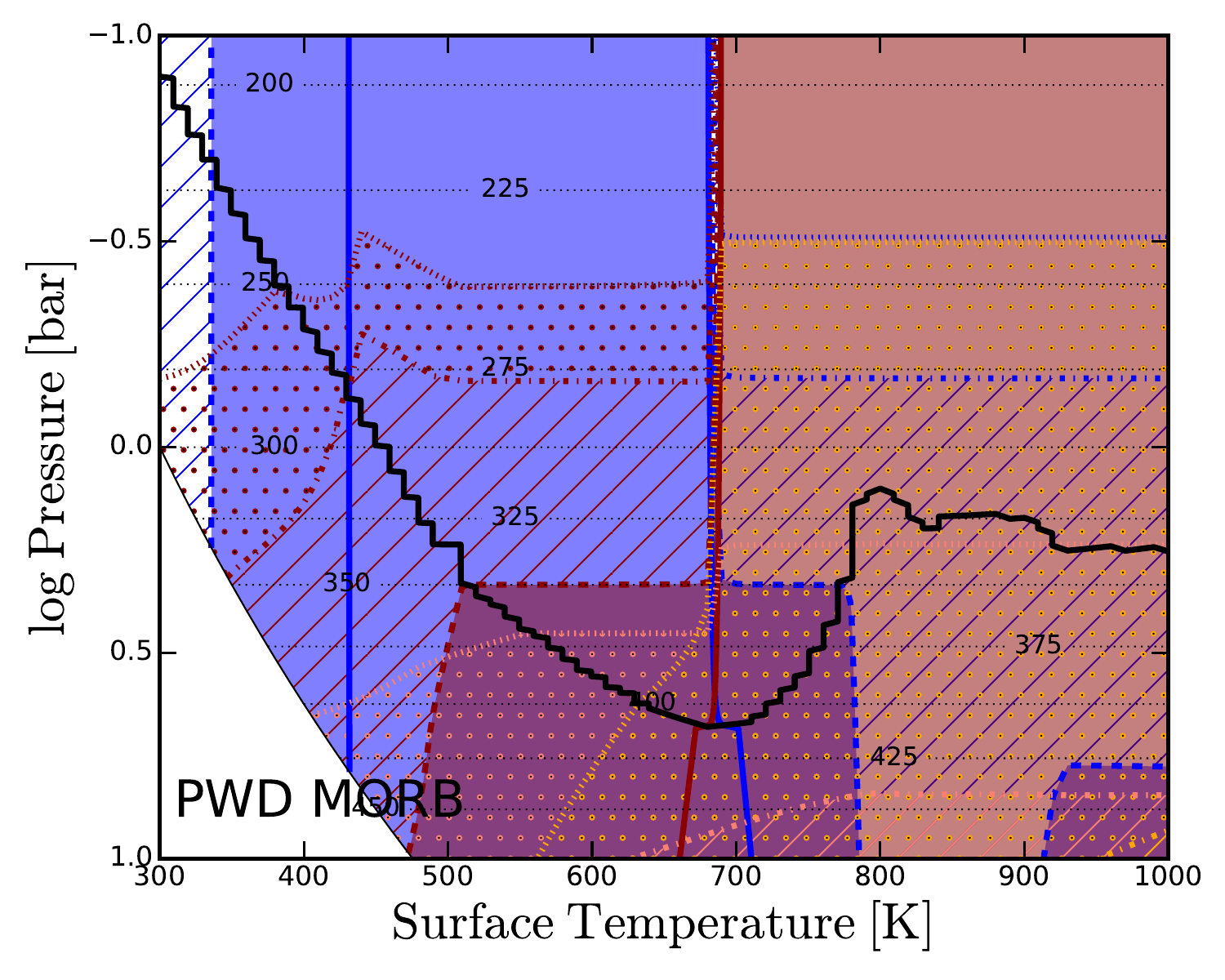}
\includegraphics[width = .32\linewidth, page=4]{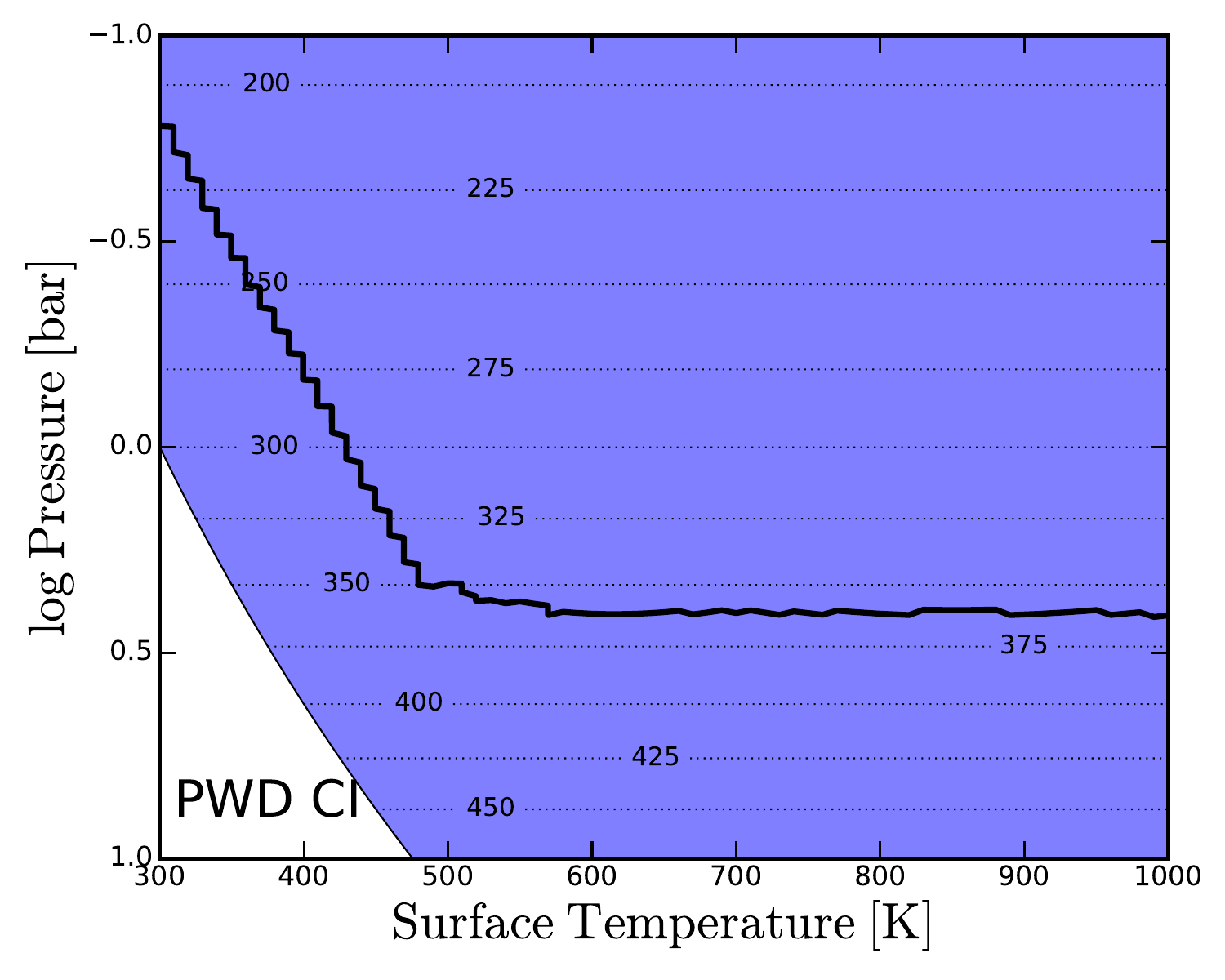}\\
\includegraphics[width = .32\linewidth, page=4]{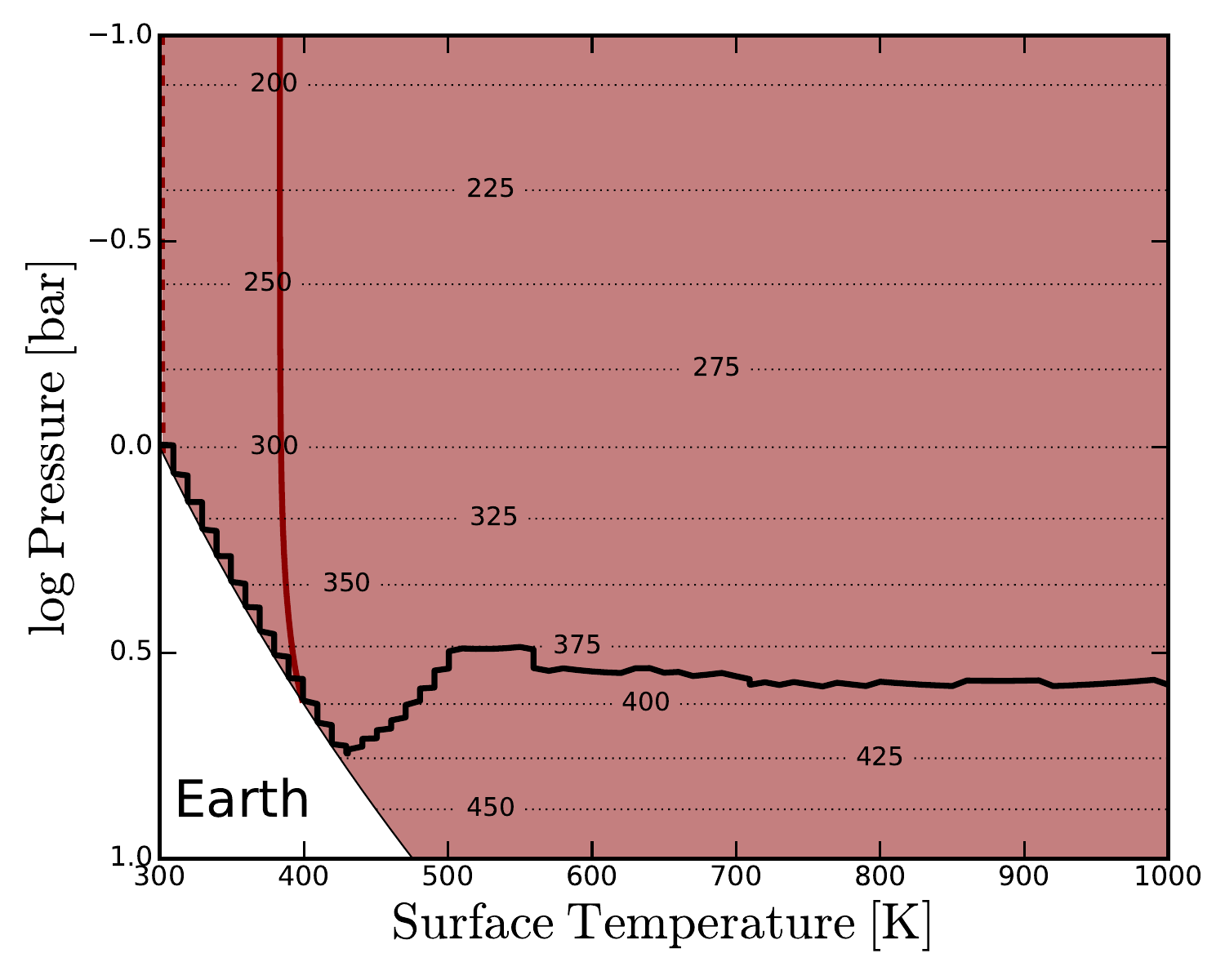}
\includegraphics[width = .32\linewidth, page=4]{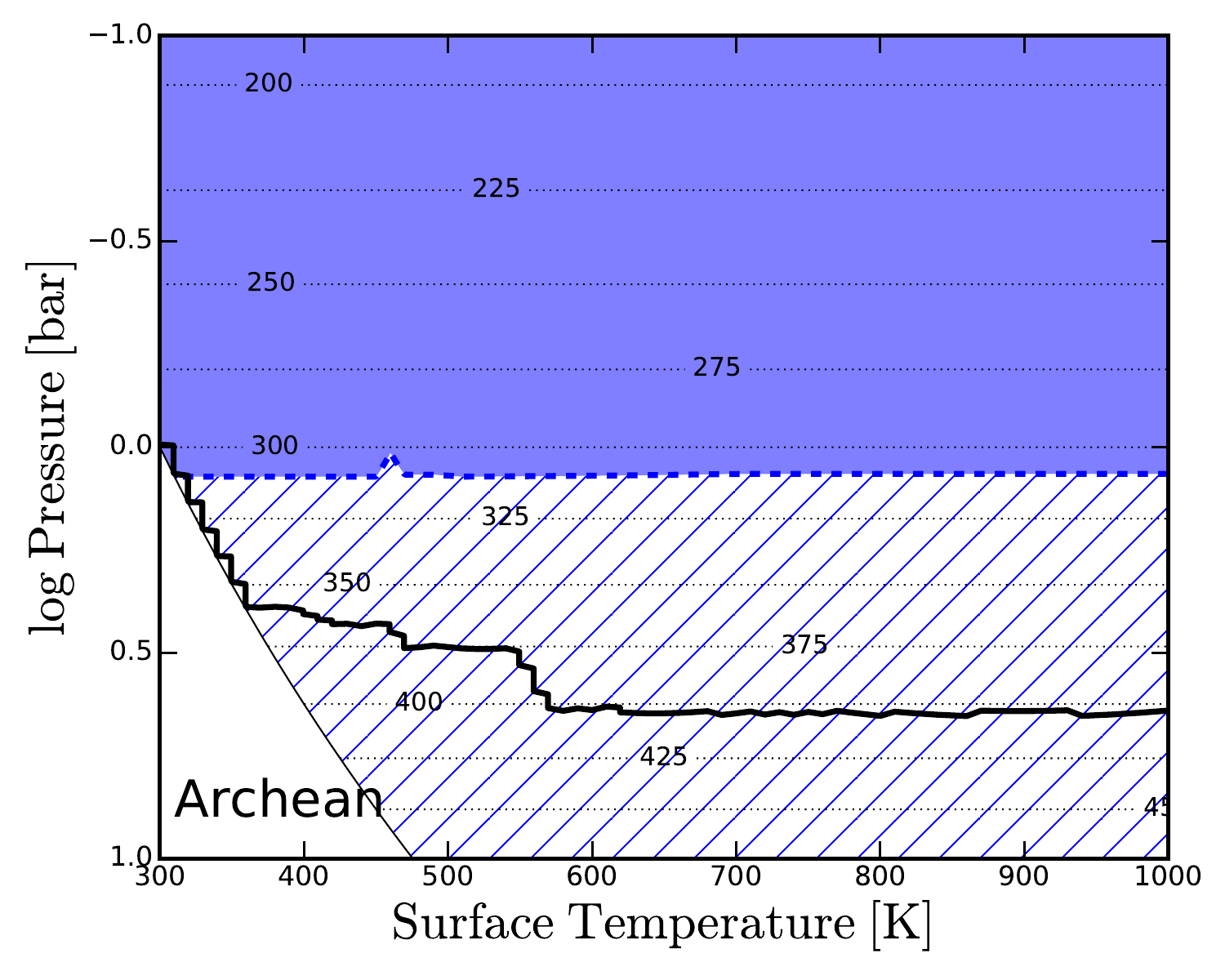}
\includegraphics[width = .32\linewidth, page=4]{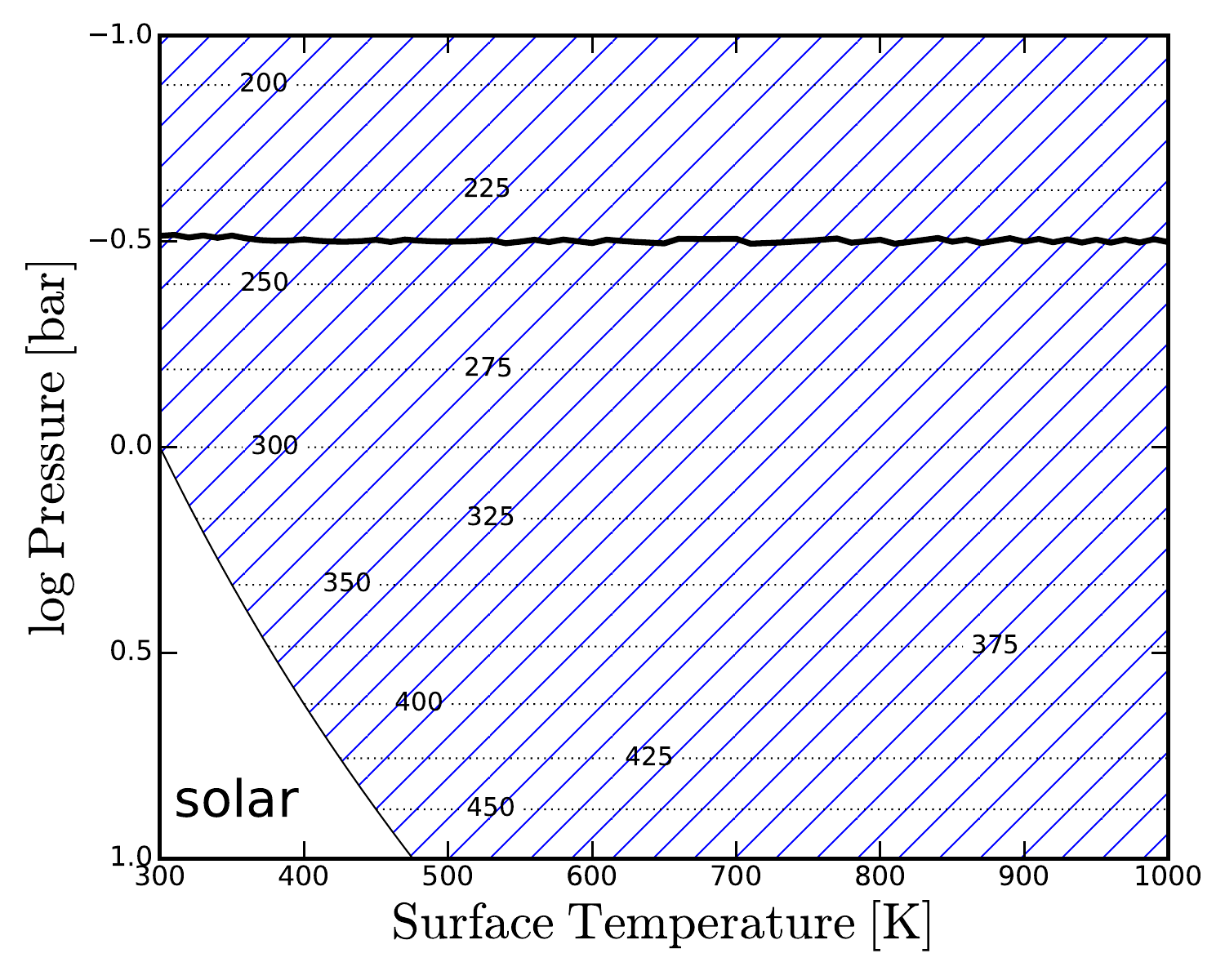}
\caption{The concentration of all carbon bearing species with higher concentration than $10^{-9}$. The different lines indicate concentration levels of 10\% (solid),  0.1\% (dashed), $10^{-6}$ (dash dotted), and $10^{-9}$ (dotted). concentrations higher than 0.1\% are filled, higher than $10^{-6}$ is hatched with lines, and concentrations above $10^{-9}$ are indicated by dots. The molecules shown are \ce{CH4} in blue, \ce{CO2} in darkred, \ce{CO} in salmon, and \ce{COS} in orange.}
\label{fig:NutriC}
\end{figure*}

\begin{figure*}
\centering
\includegraphics[width = .32\linewidth, page=5]{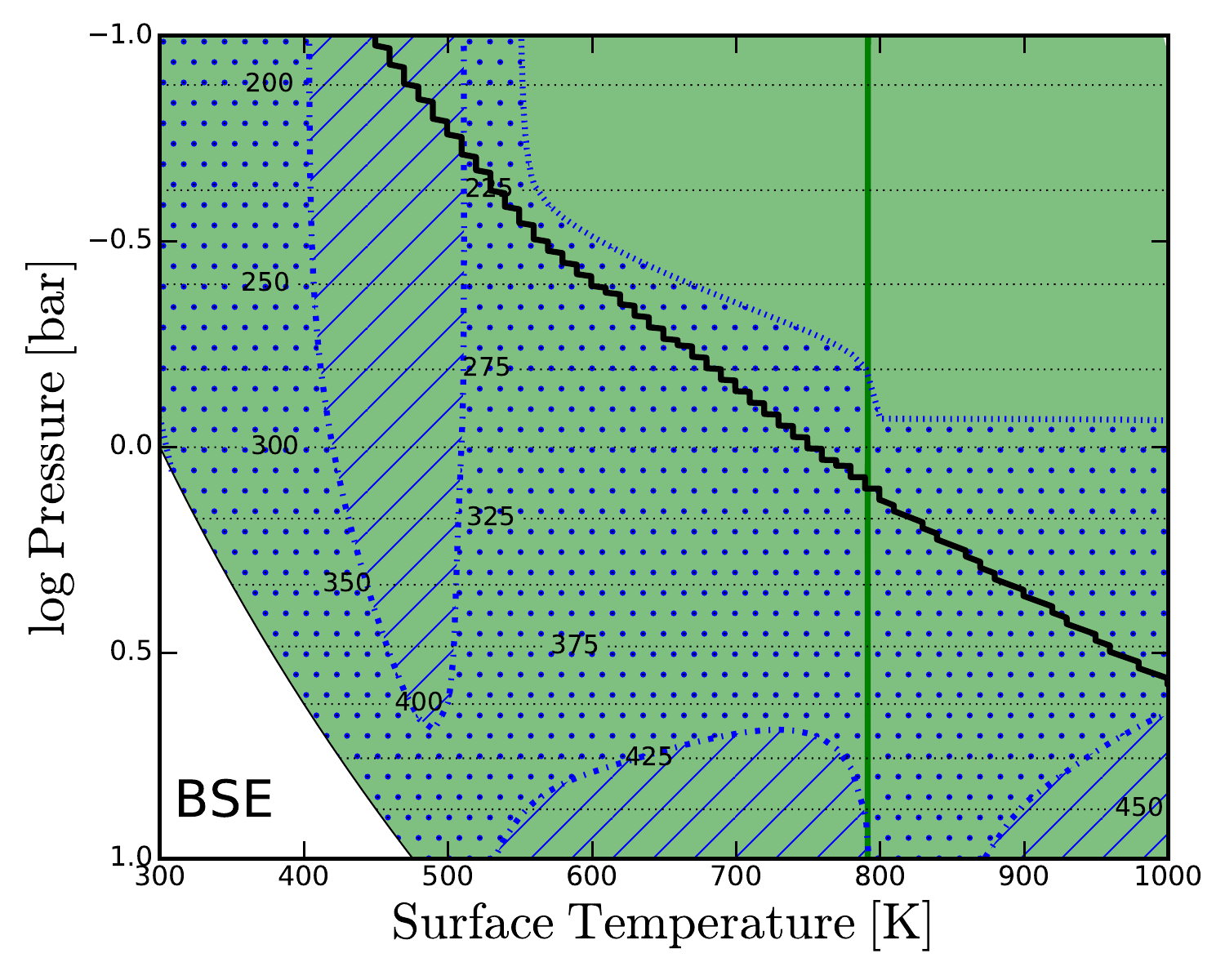}
\includegraphics[width = .32\linewidth, page=5]{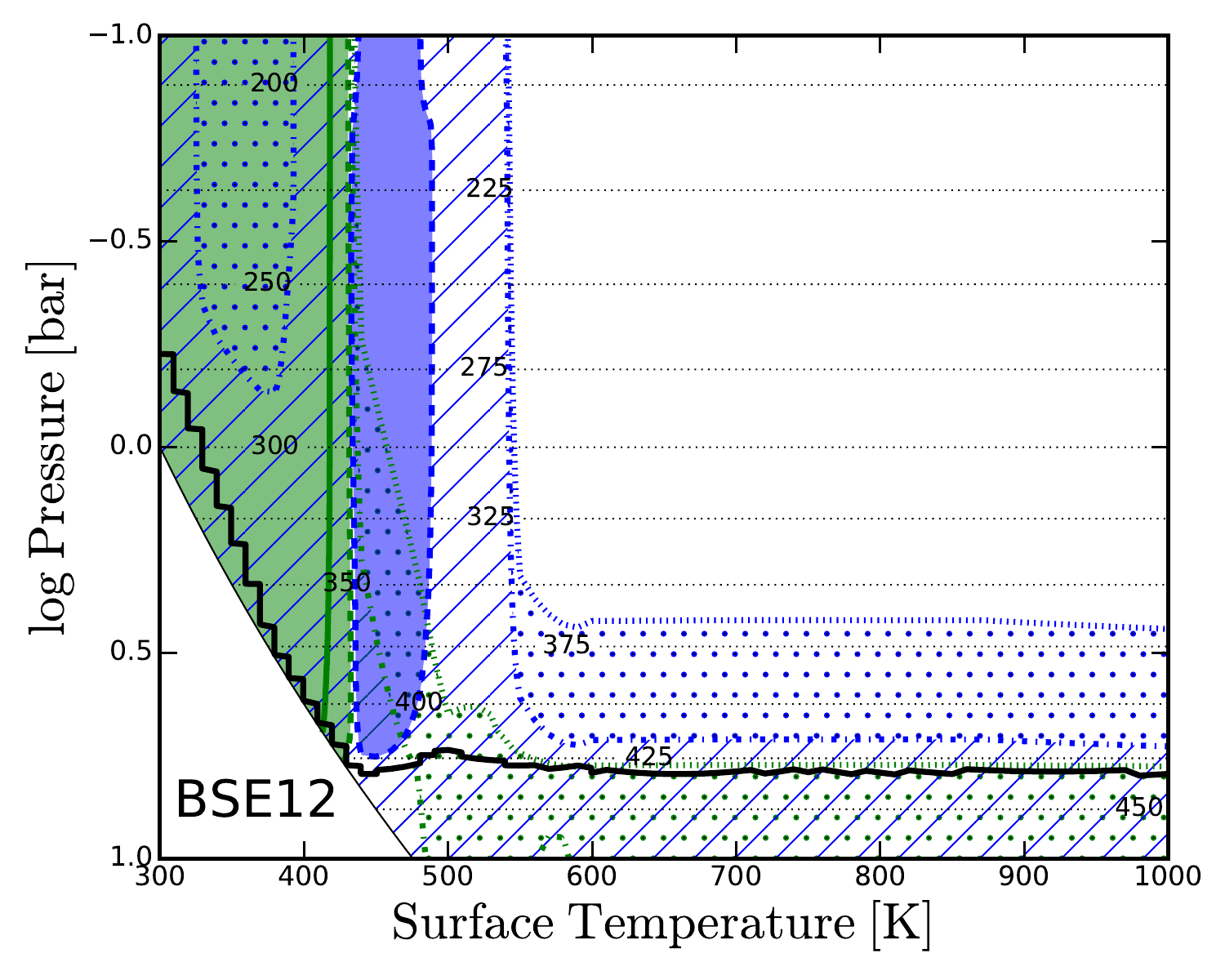}
\includegraphics[width = .32\linewidth, page=5]{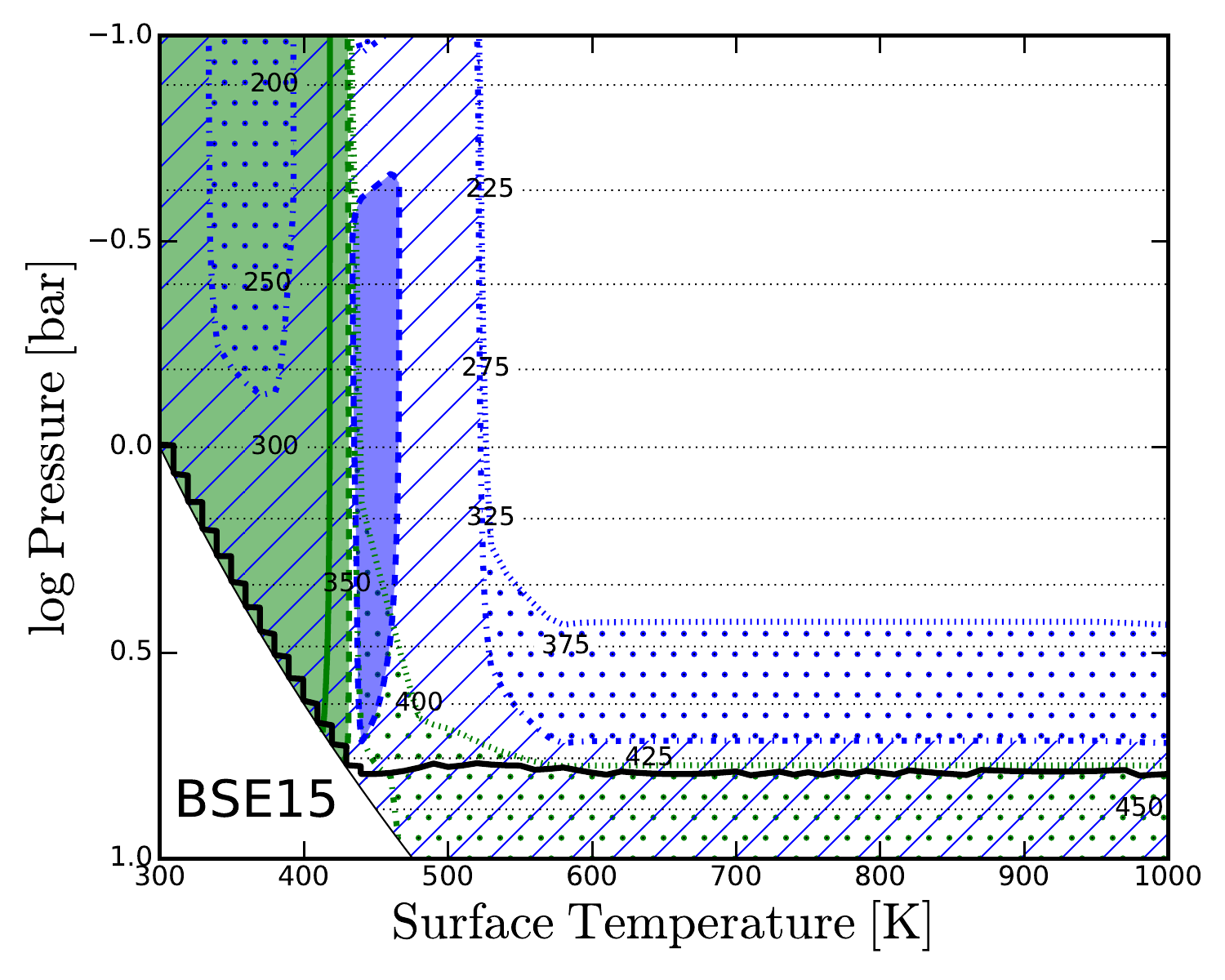}\\
\includegraphics[width = .32\linewidth, page=5]{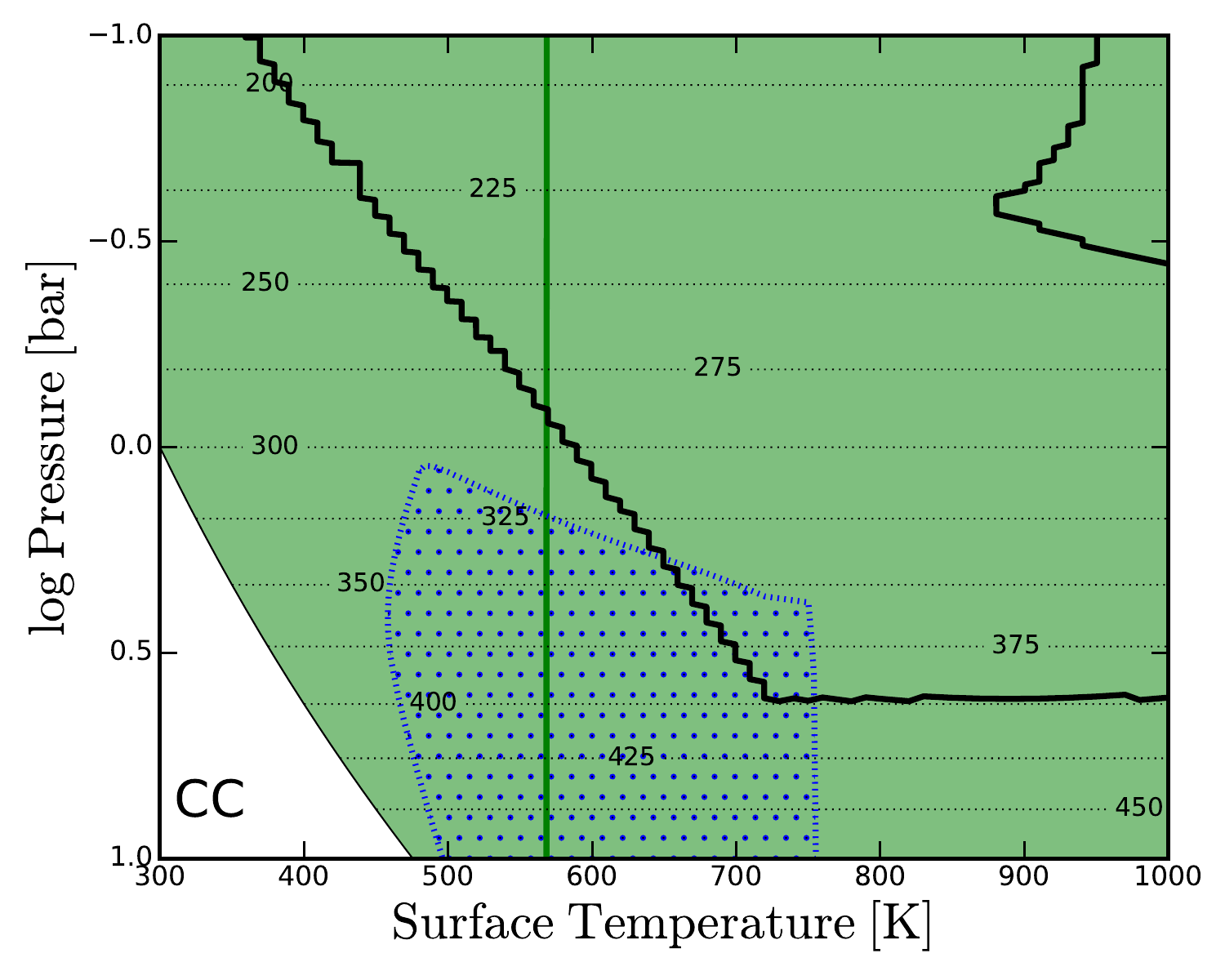}
\includegraphics[width = .32\linewidth, page=5]{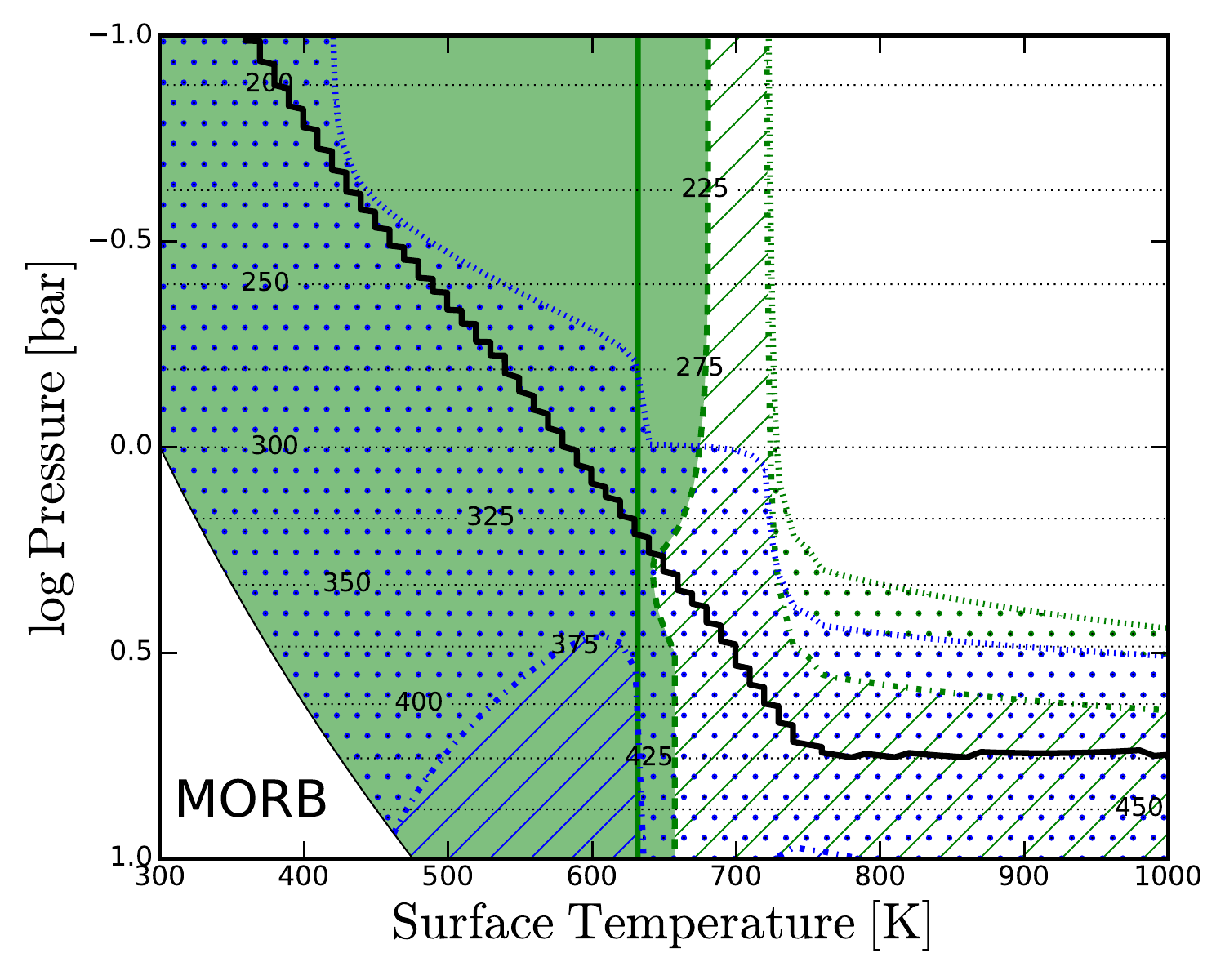}
\includegraphics[width = .32\linewidth, page=5]{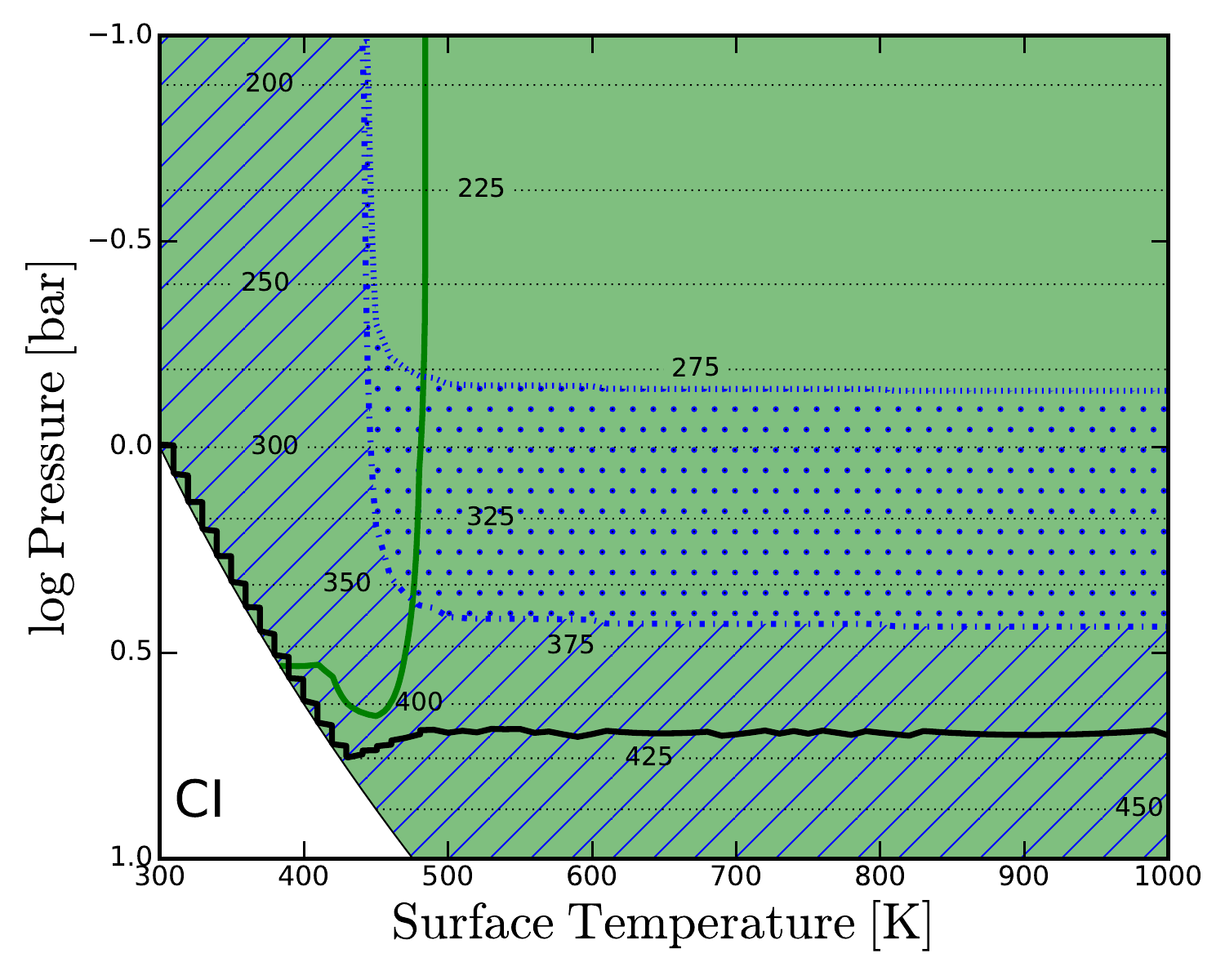}\\
\includegraphics[width = .32\linewidth, page=5]{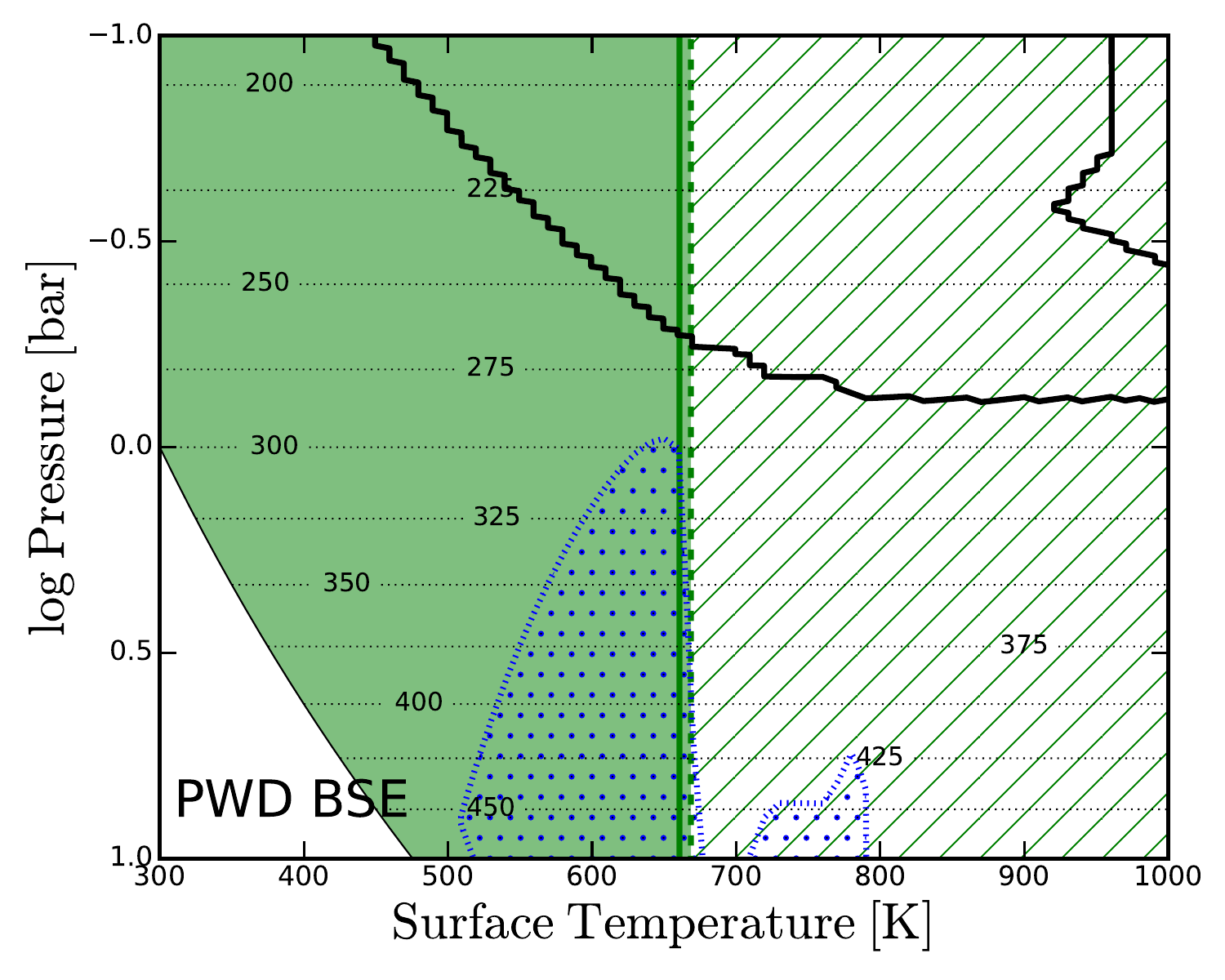}
\includegraphics[width = .32\linewidth, page=5]{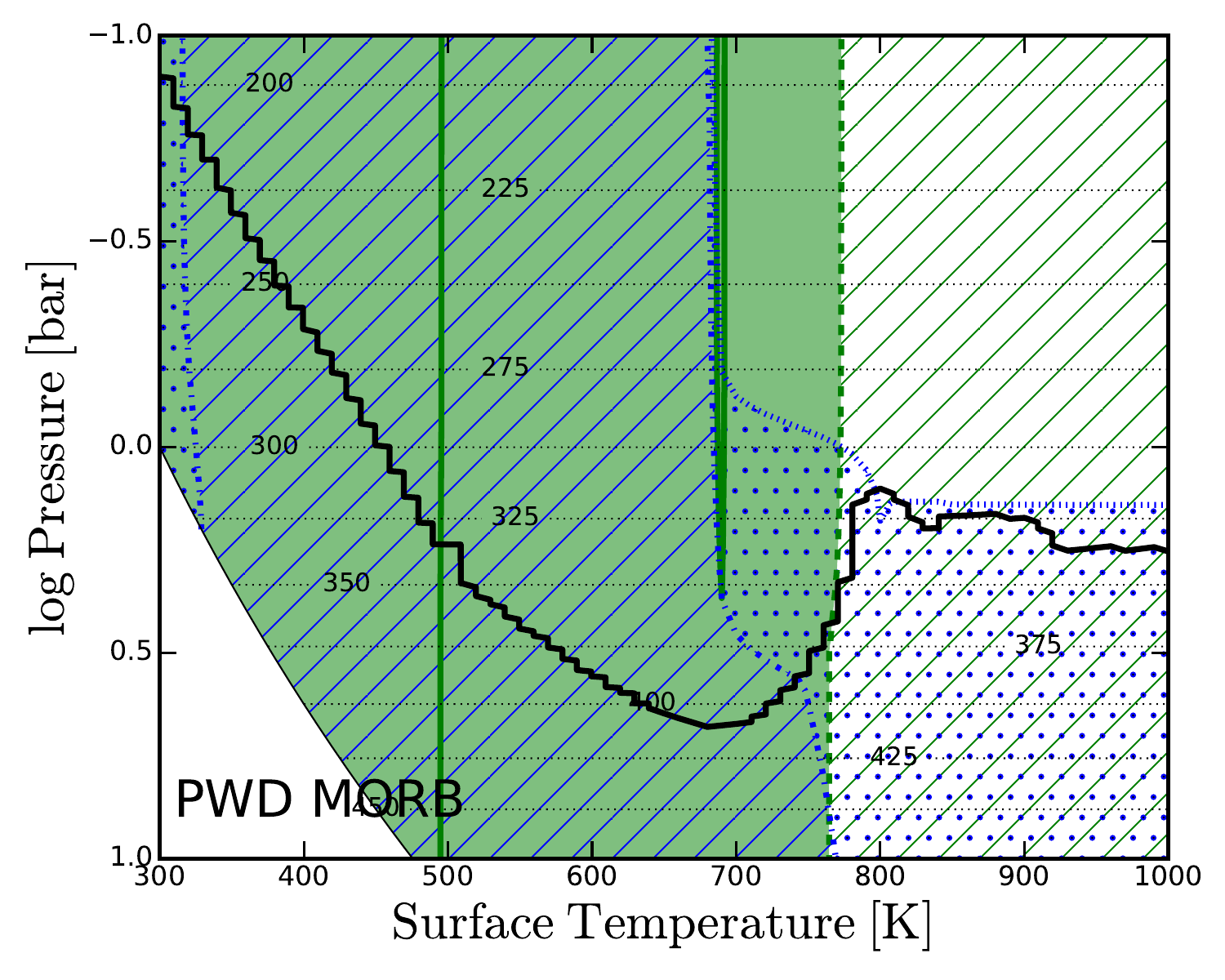}
\includegraphics[width = .32\linewidth, page=5]{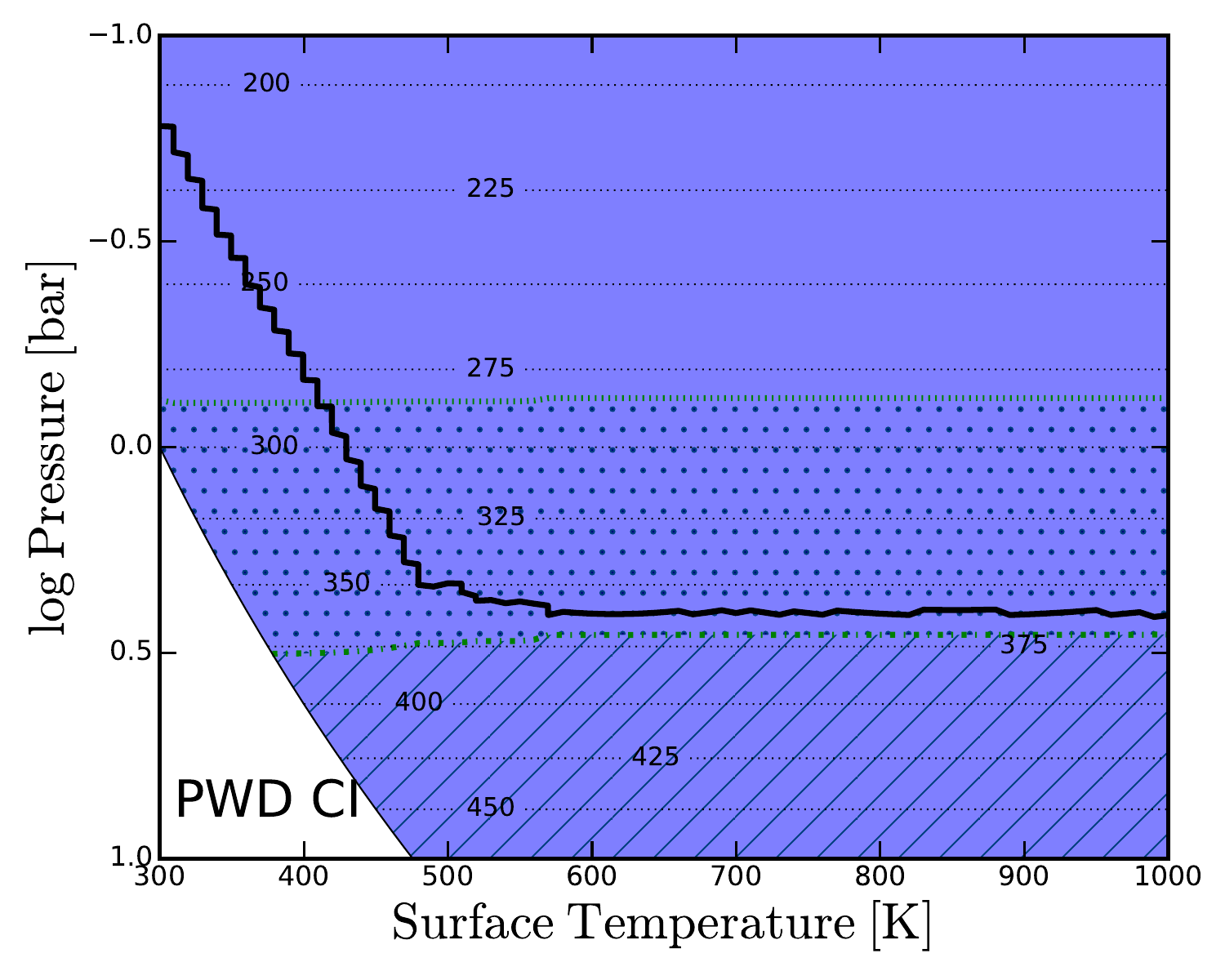}\\
\includegraphics[width = .32\linewidth, page=5]{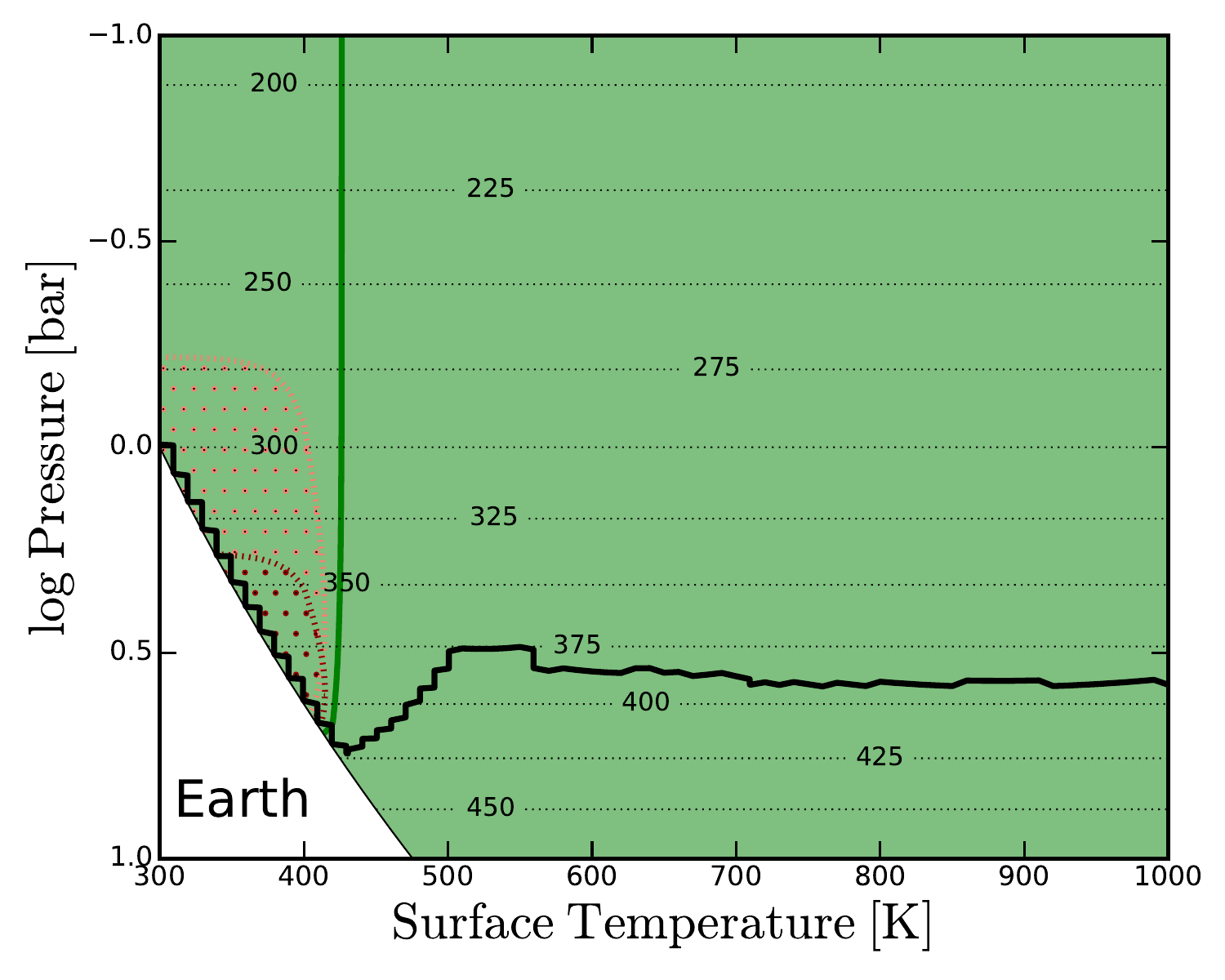}
\includegraphics[width = .32\linewidth, page=5]{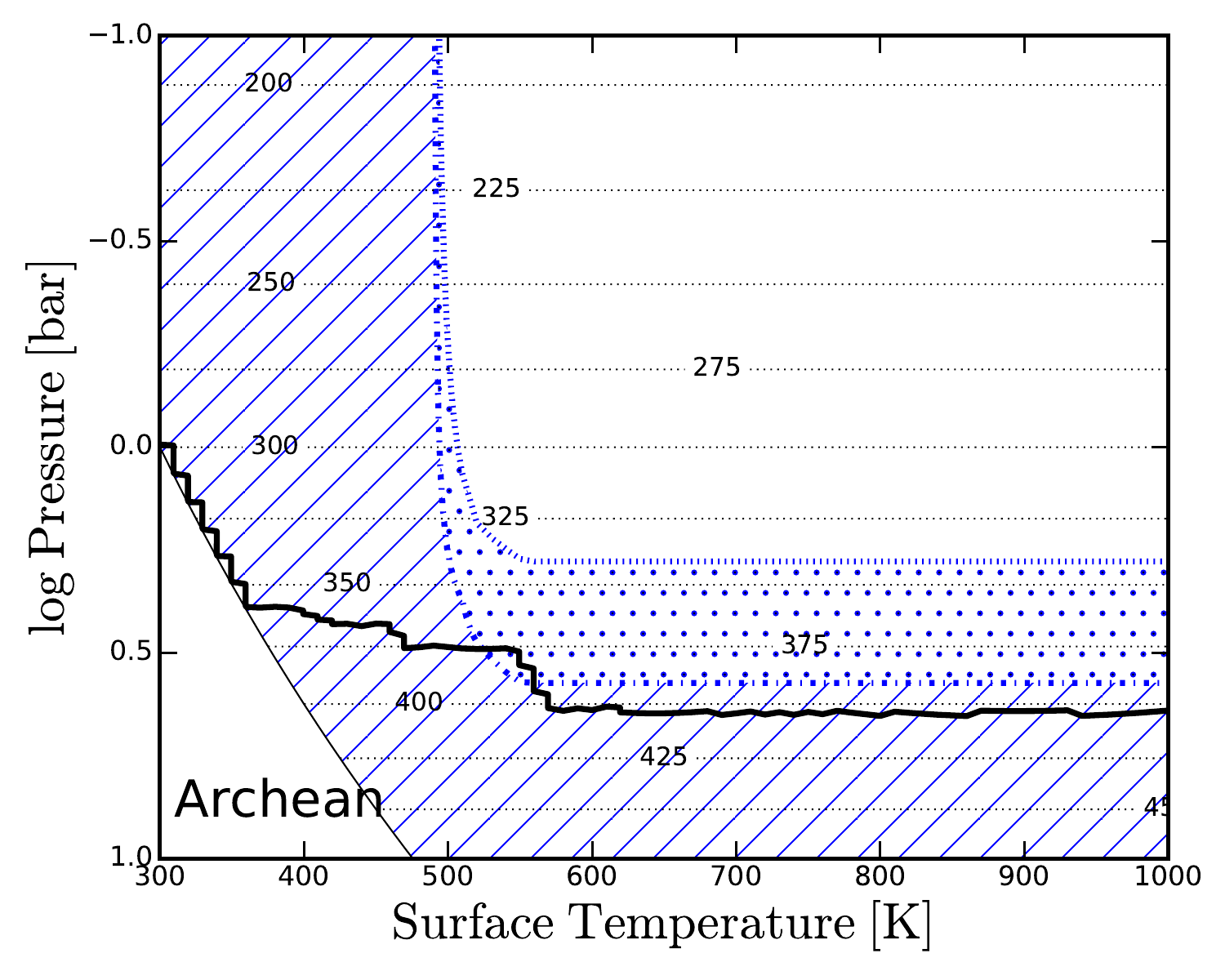}
\includegraphics[width = .32\linewidth, page=5]{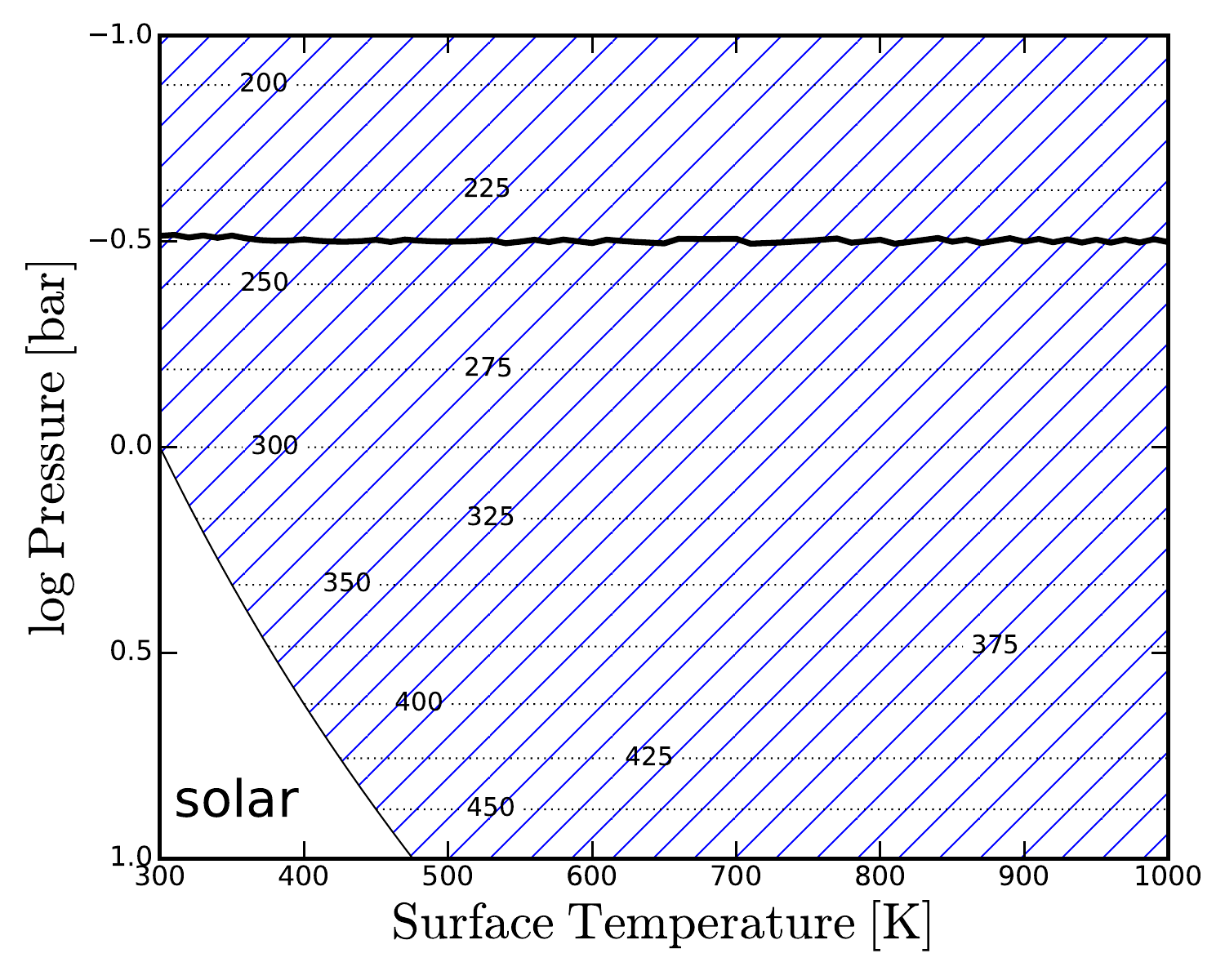}
\caption{As in Fig.~\ref{fig:NutriC}, but for all N bearing molecules. The colours refer to \ce{N2} in green, \ce{NH3} in blue, \ce{HNO3} in salmon, and \ce{NO2} in darkred.}
\label{fig:NutriN}
\end{figure*}

\begin{figure*}
\centering
\includegraphics[width = .32\linewidth, page=6]{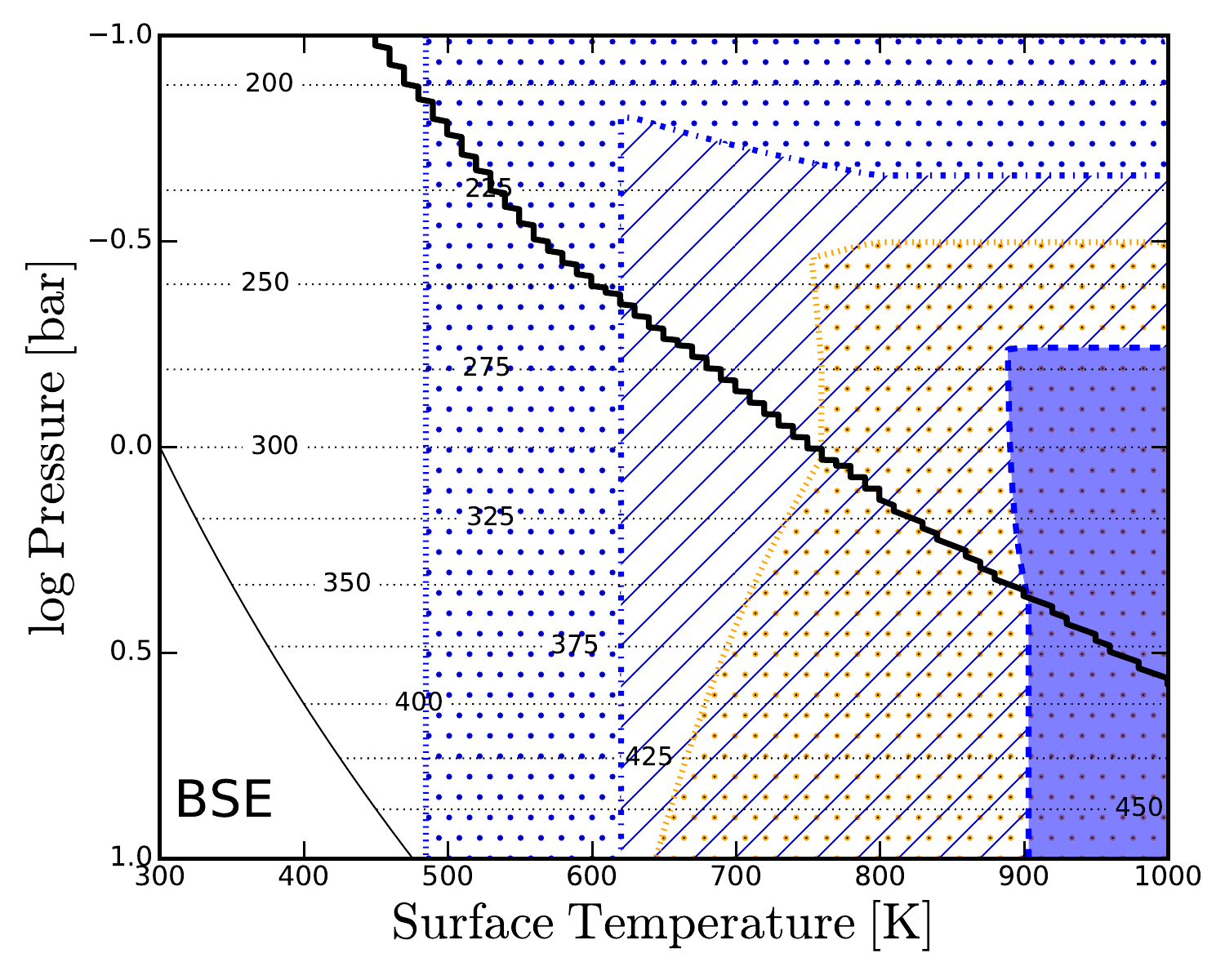}
\includegraphics[width = .32\linewidth, page=6]{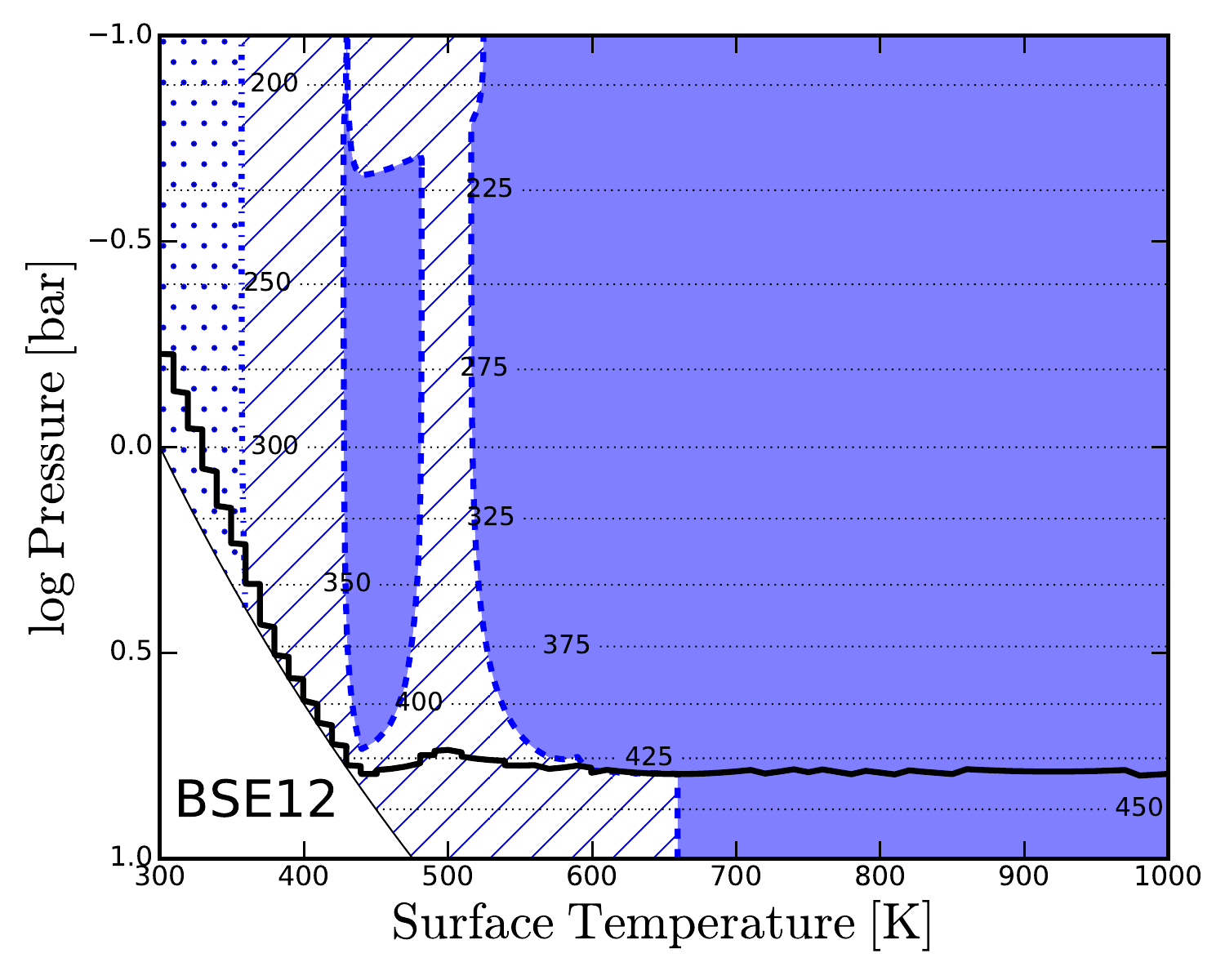}
\includegraphics[width = .32\linewidth, page=6]{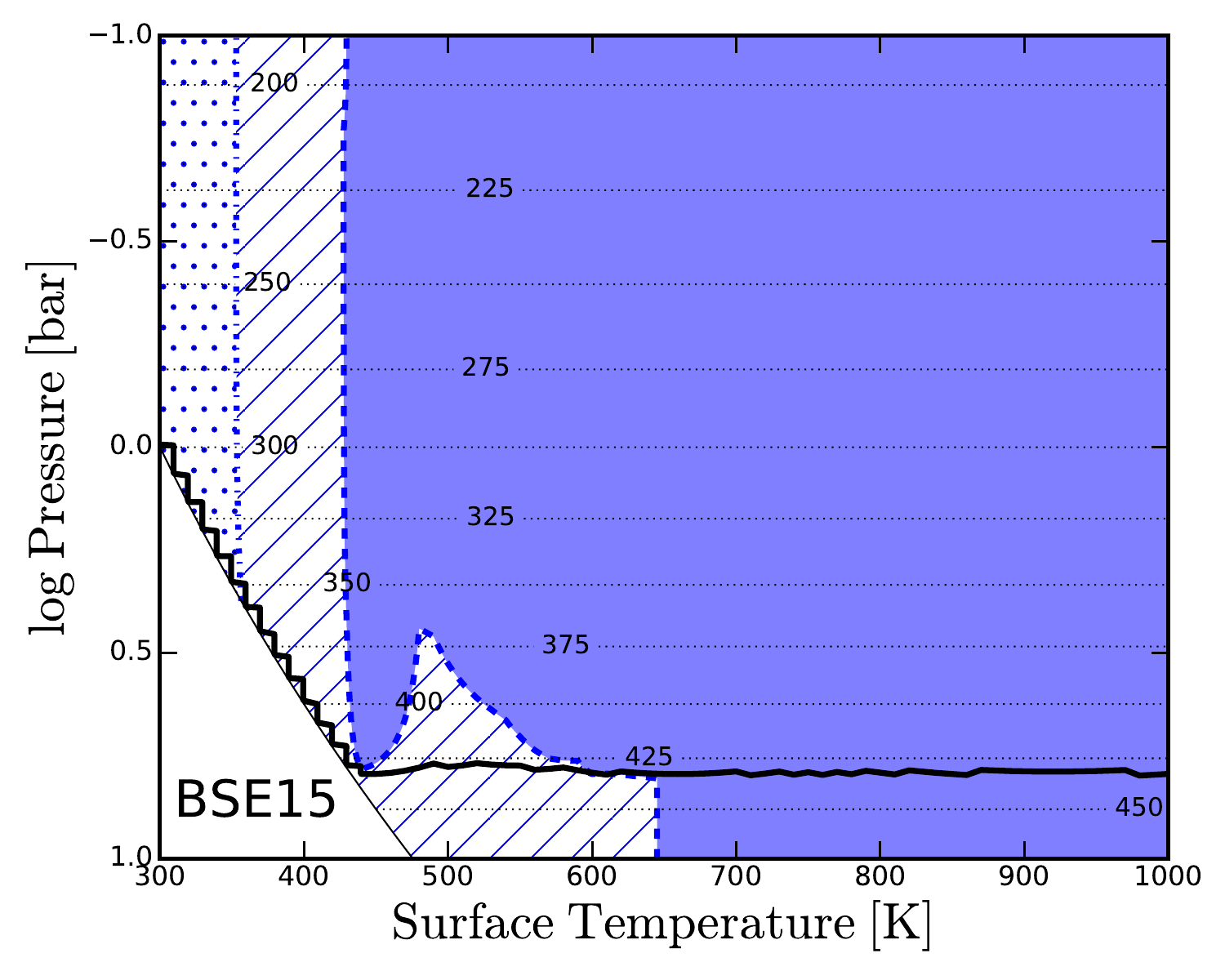}\\
\includegraphics[width = .32\linewidth, page=6]{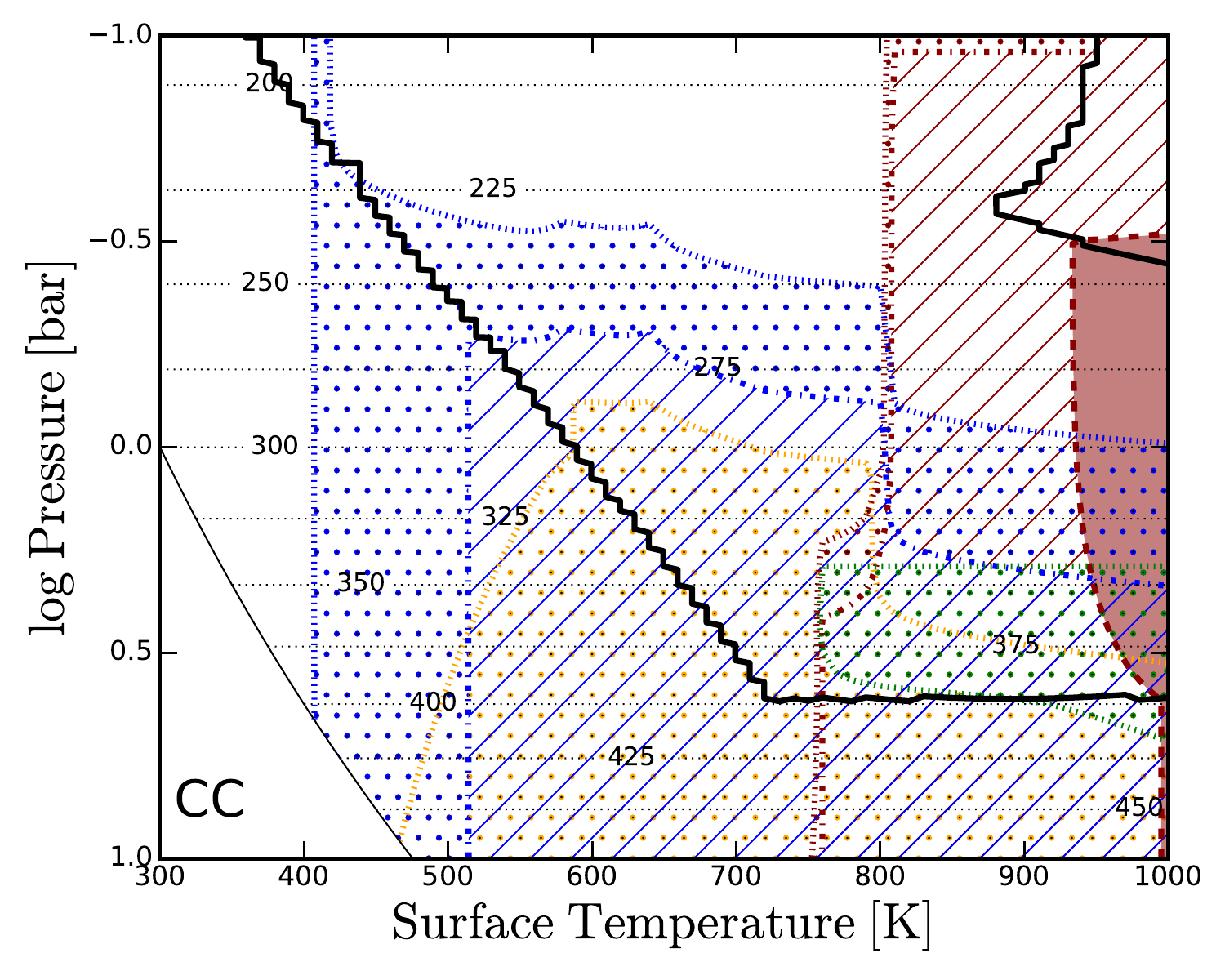}
\includegraphics[width = .32\linewidth, page=6]{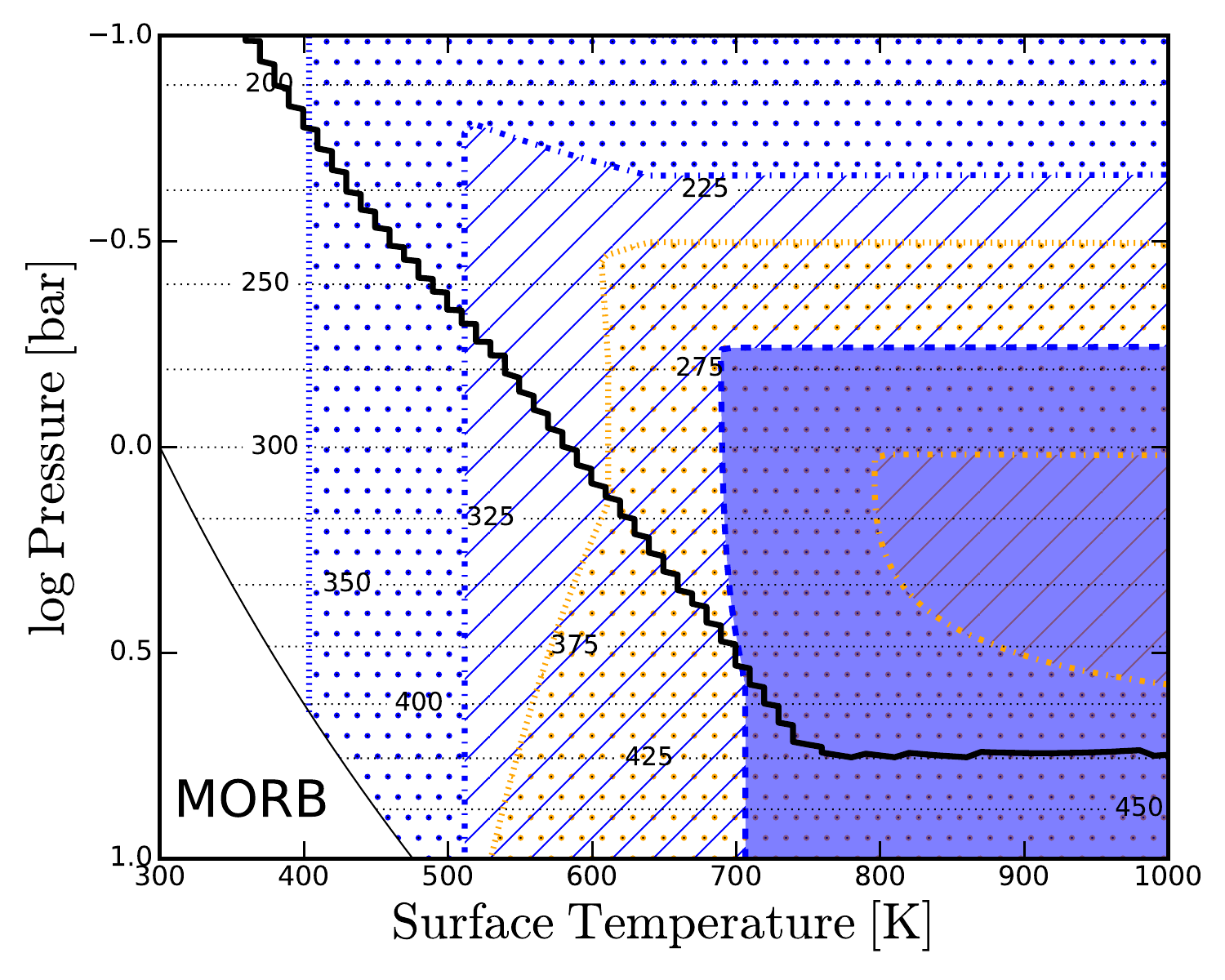}
\includegraphics[width = .32\linewidth, page=6]{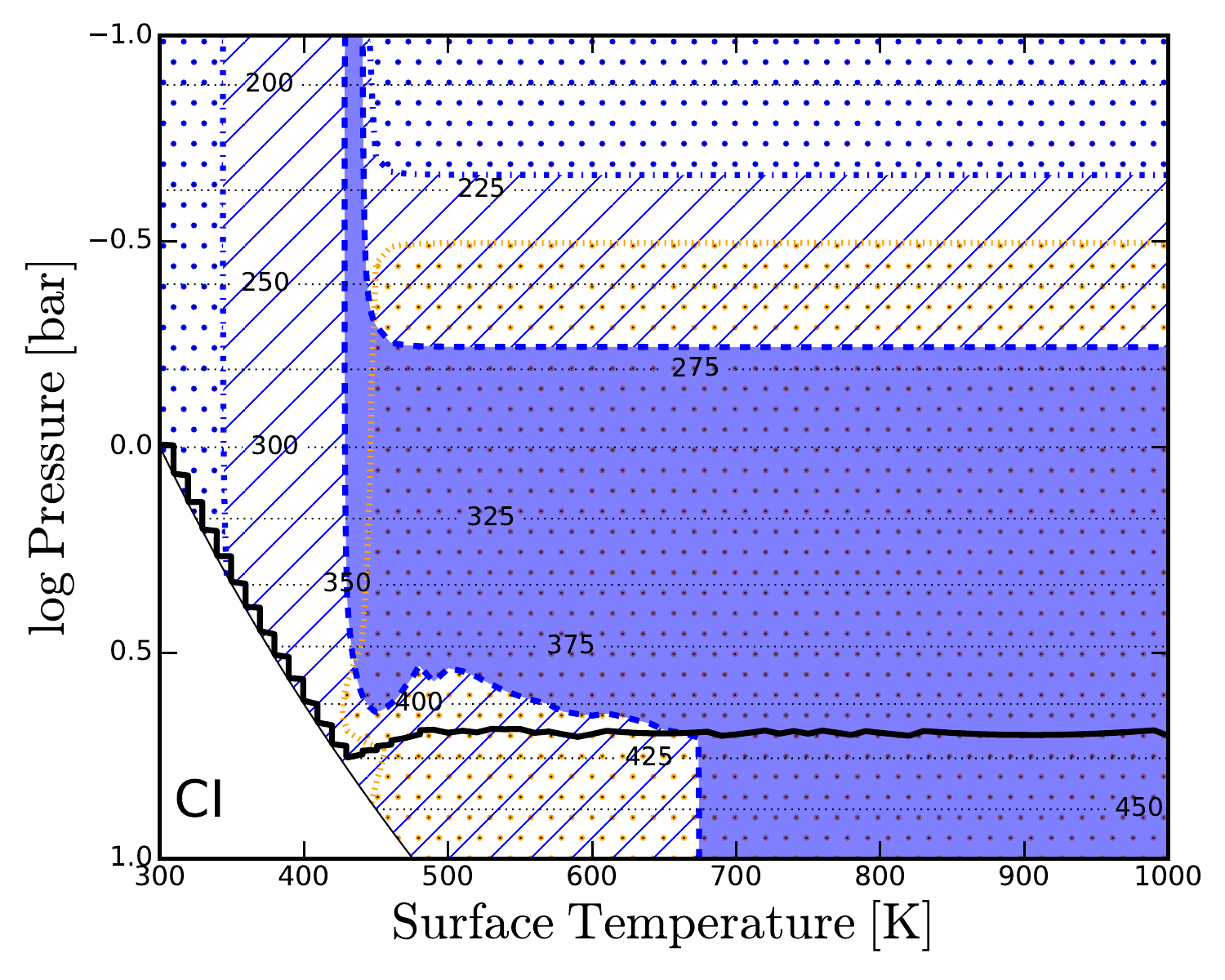}\\
\includegraphics[width = .32\linewidth, page=6]{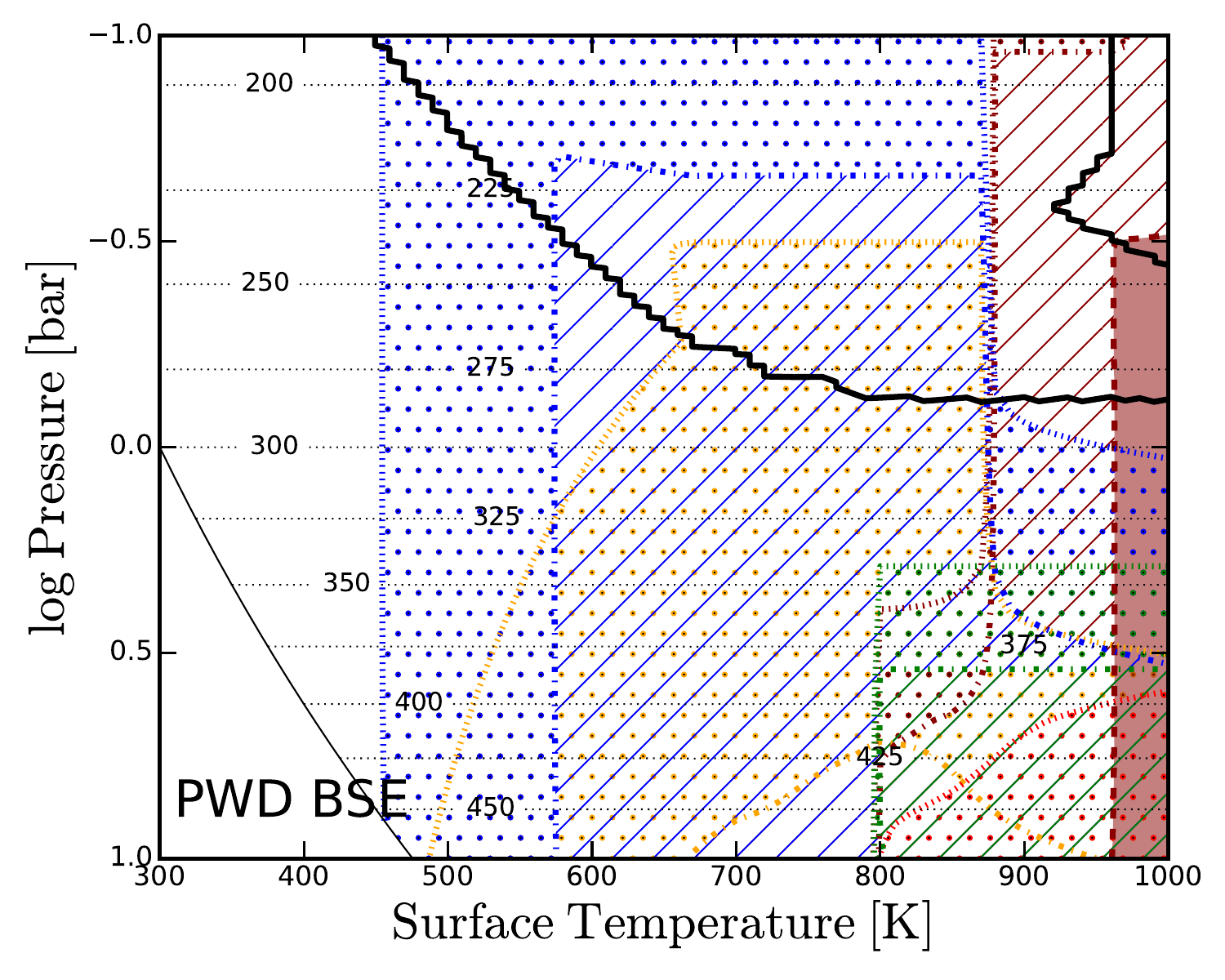}
\includegraphics[width = .32\linewidth, page=6]{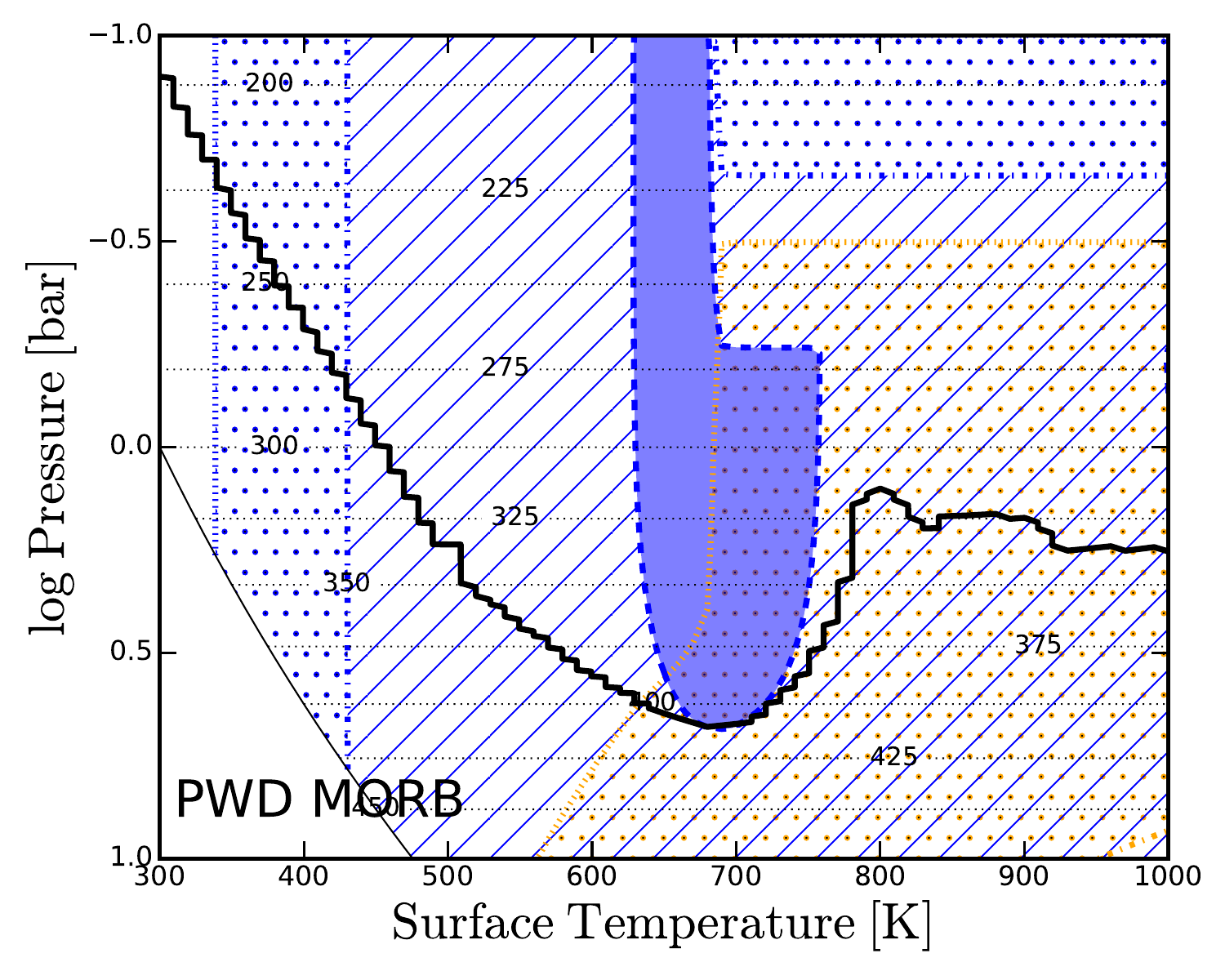}
\includegraphics[width = .32\linewidth, page=6]{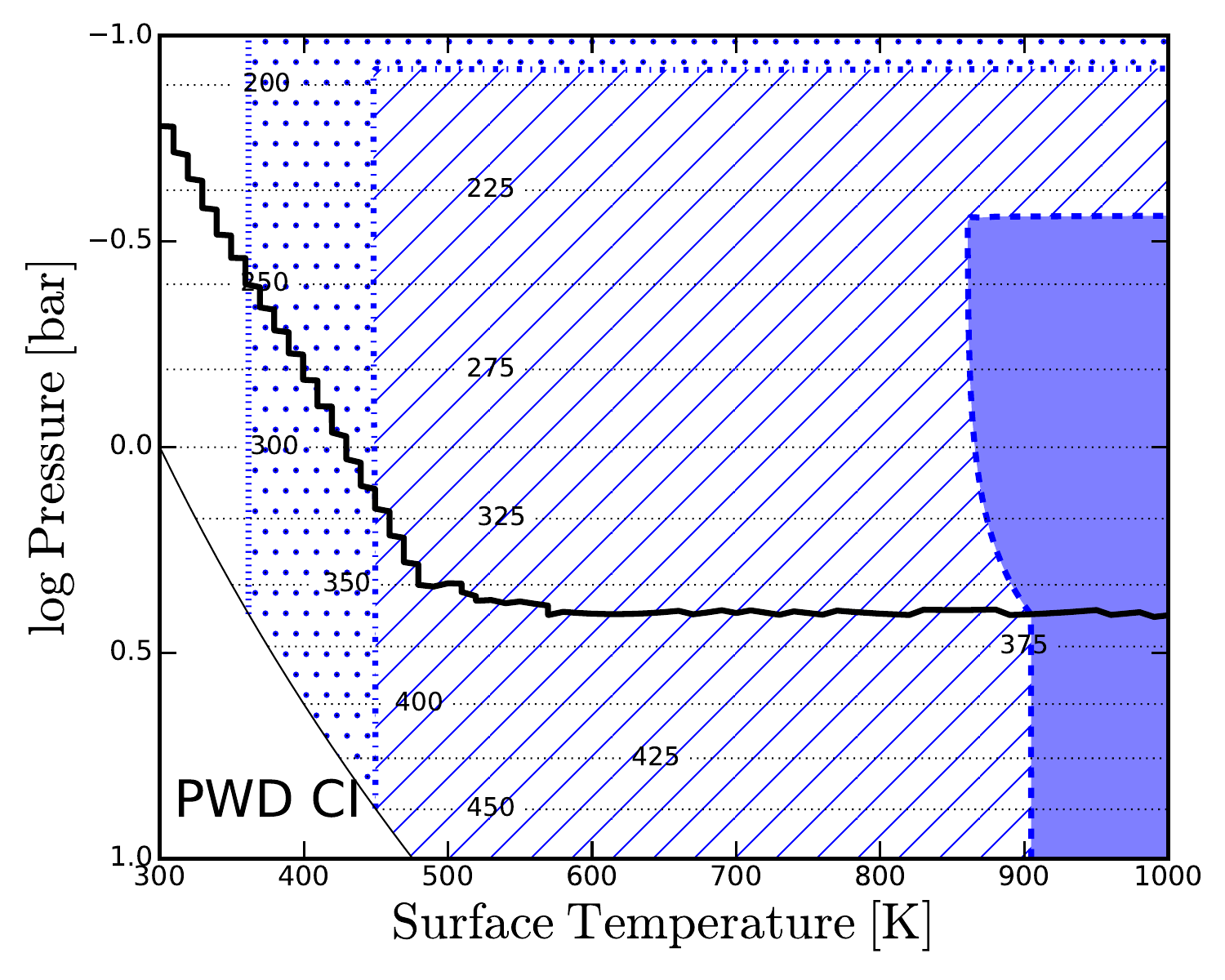}\\
\includegraphics[width = .32\linewidth, page=6]{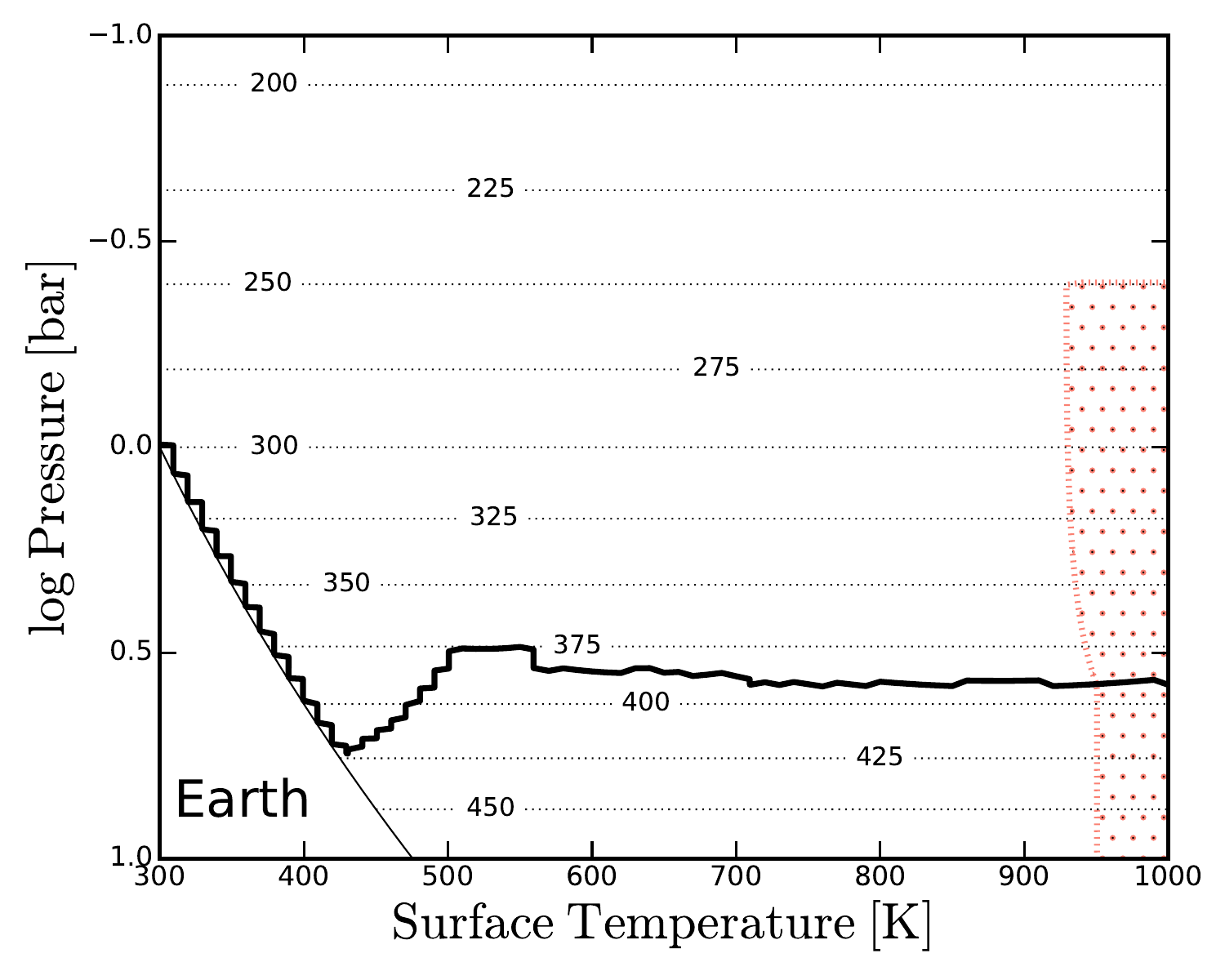}
\includegraphics[width = .32\linewidth, page=6]{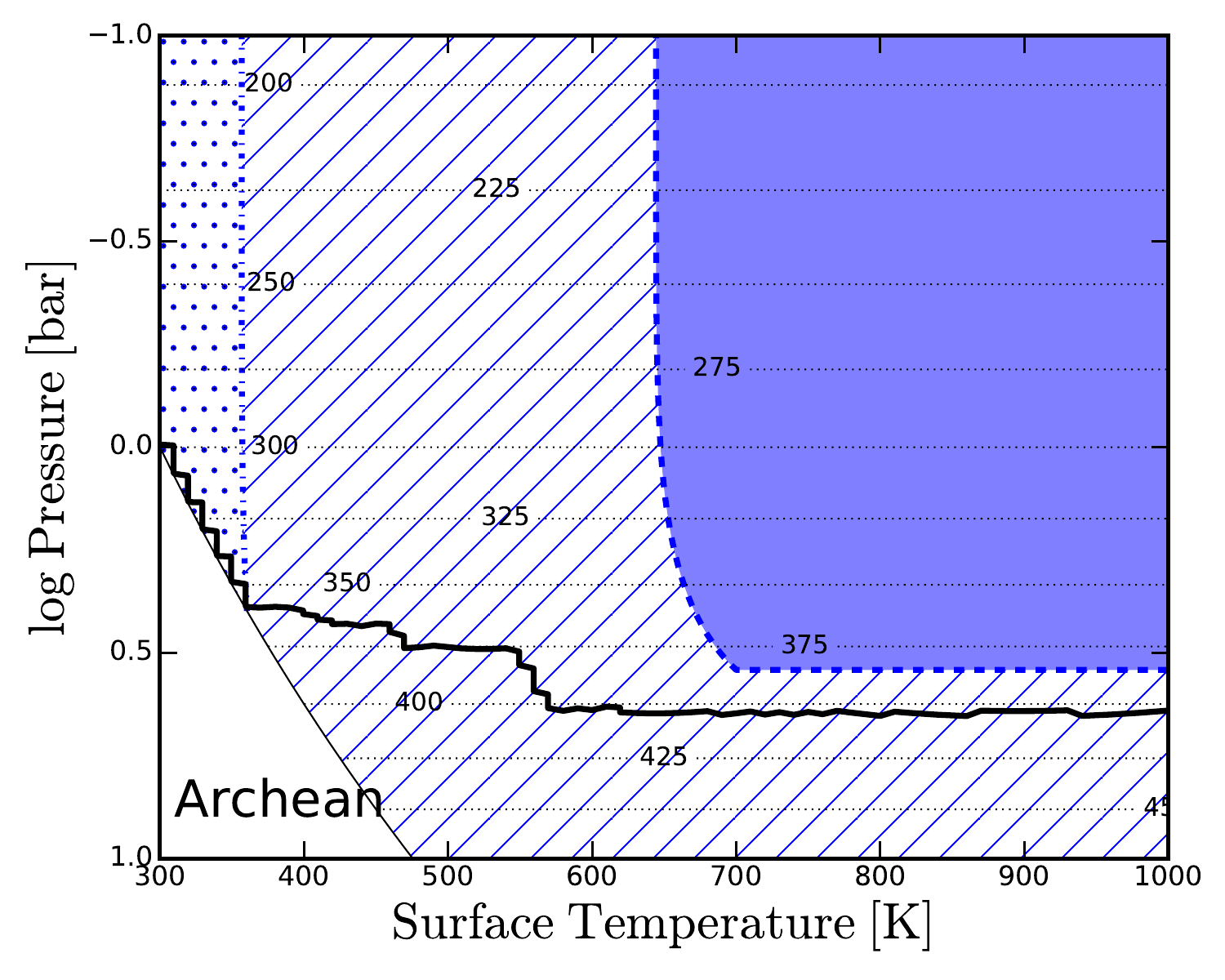}
\includegraphics[width = .32\linewidth, page=6]{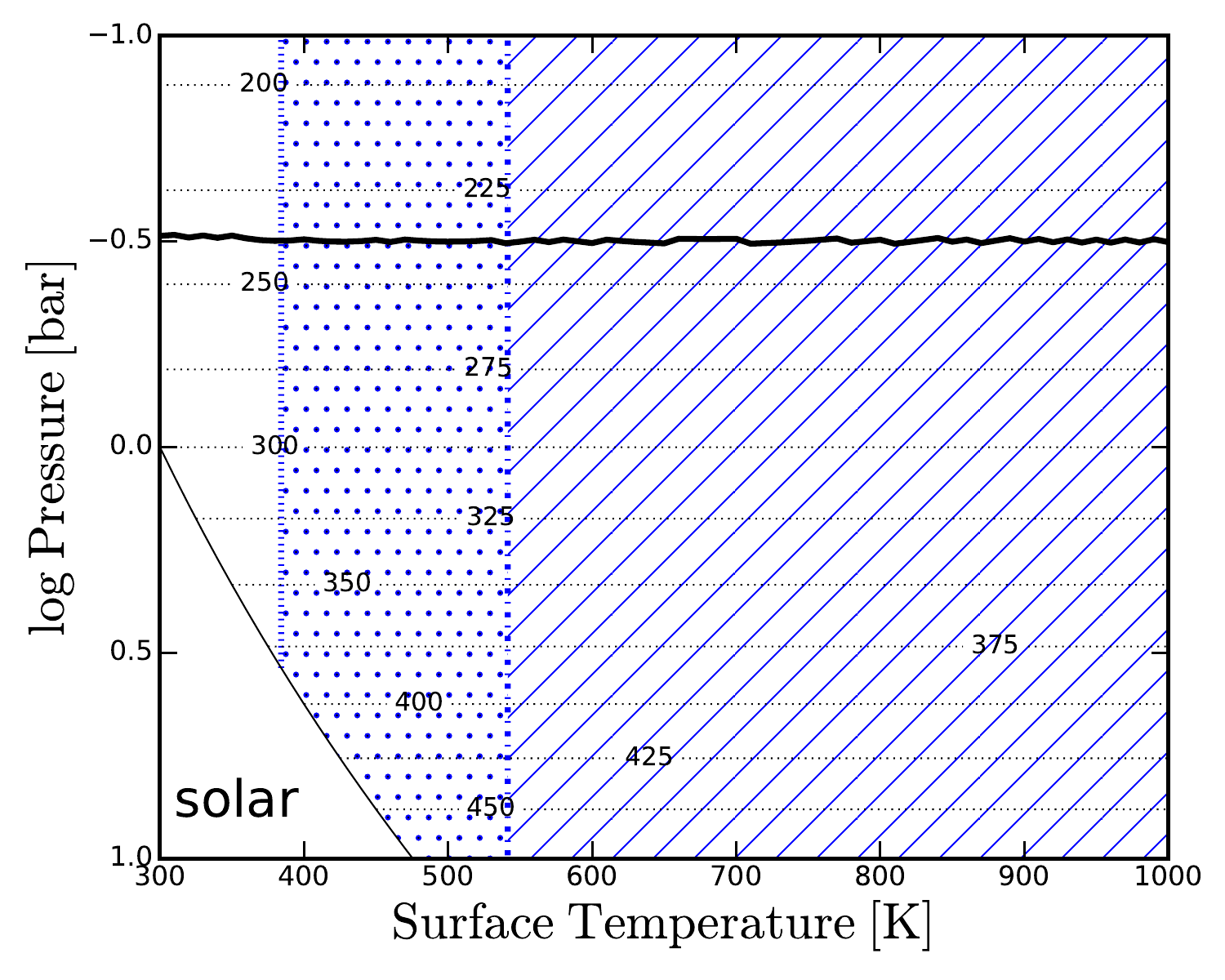}
\caption{As in Fig.~\ref{fig:NutriC}, but for all S bearing molecules. The colours refer to \ce{COS} (orange); \ce{SO2} (darkred); \ce{S2O} (red); \ce{H2S} (blue); \ce{Sx} (green); \ce{H2SO4} (salmon)}
\label{fig:NutriS}
\end{figure*}

\begin{figure*}
\centering
\includegraphics[width = .32\linewidth, page=7]{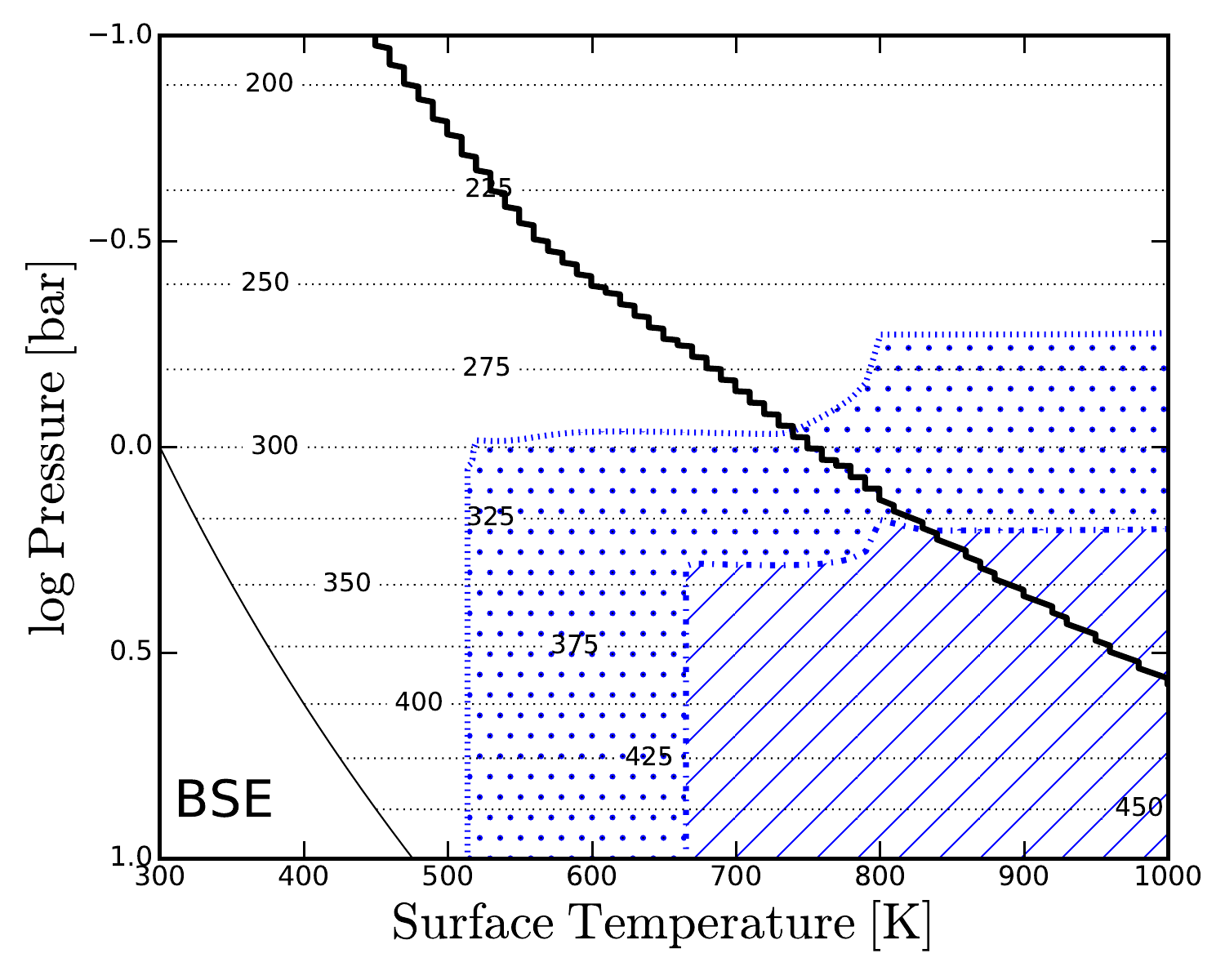}
\includegraphics[width = .32\linewidth, page=7]{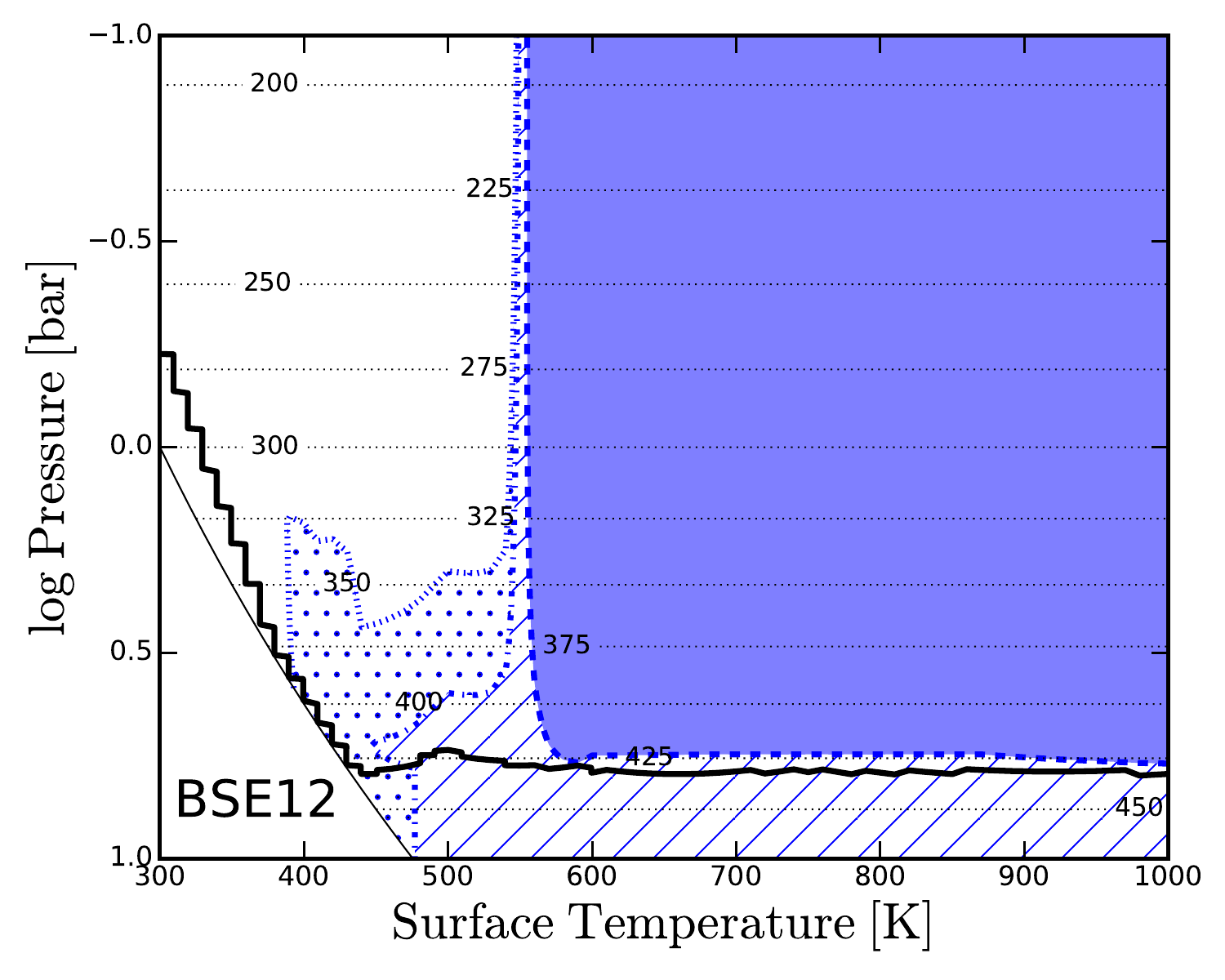}
\includegraphics[width = .32\linewidth, page=7]{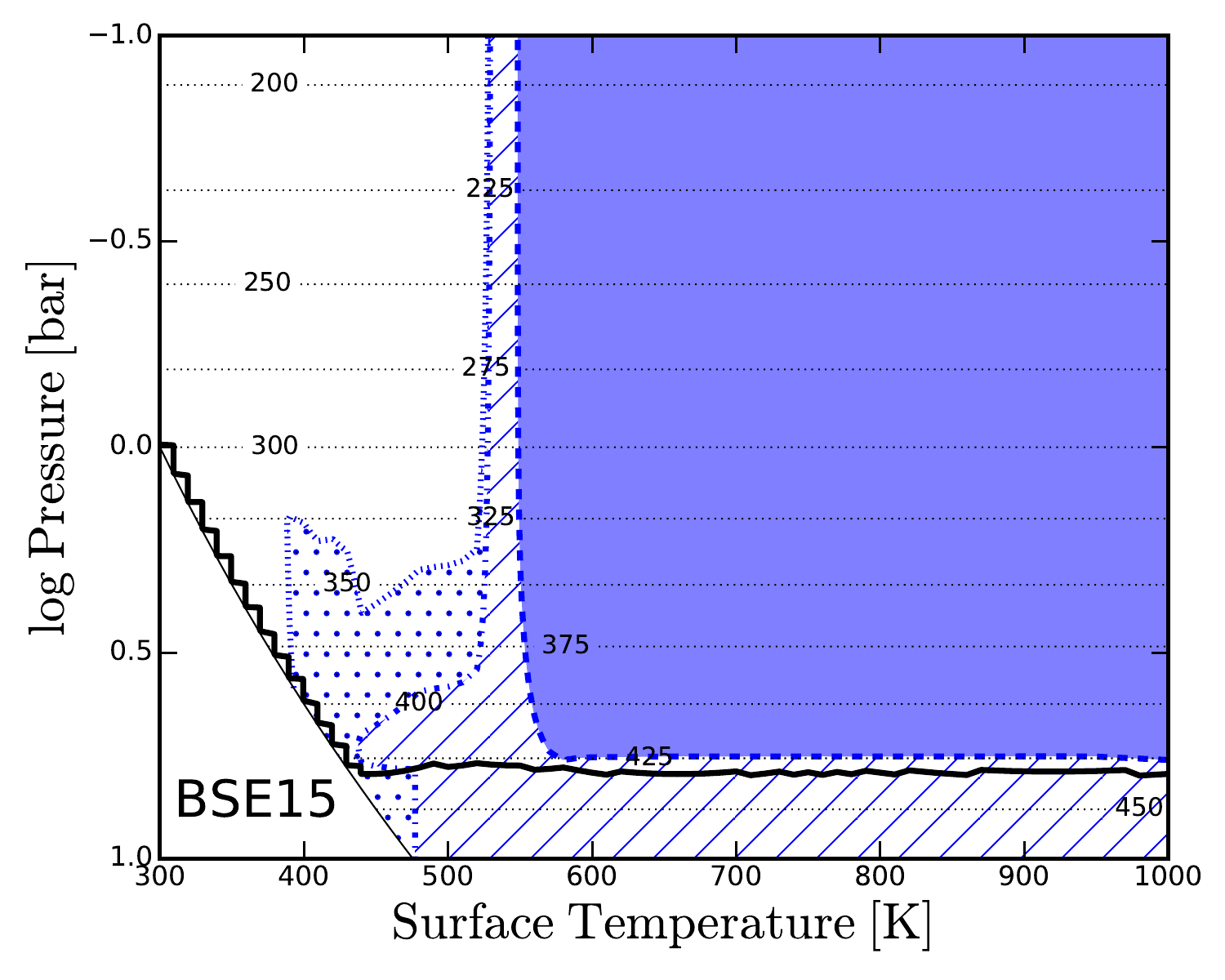}\\
\includegraphics[width = .32\linewidth, page=7]{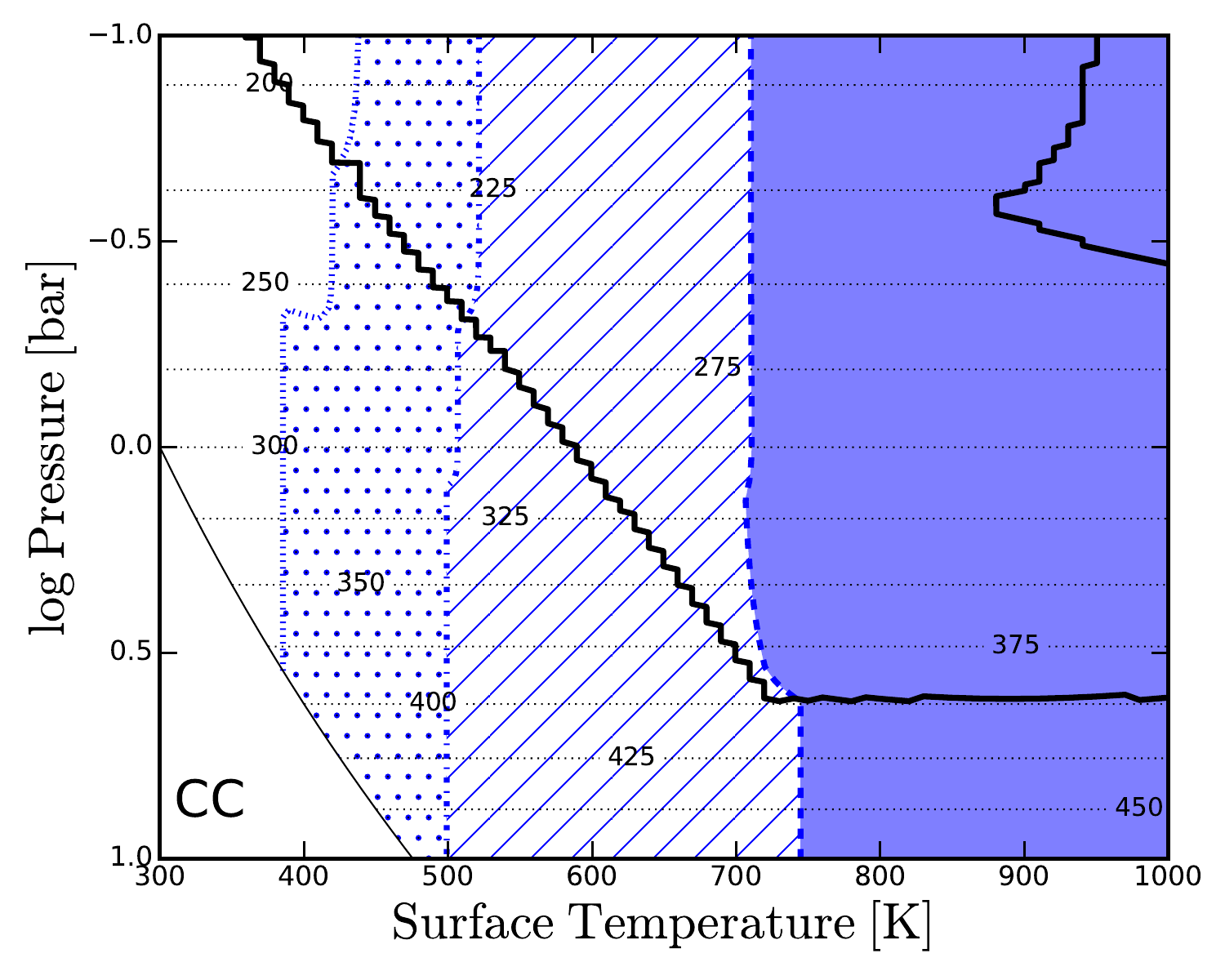}
\includegraphics[width = .32\linewidth, page=7]{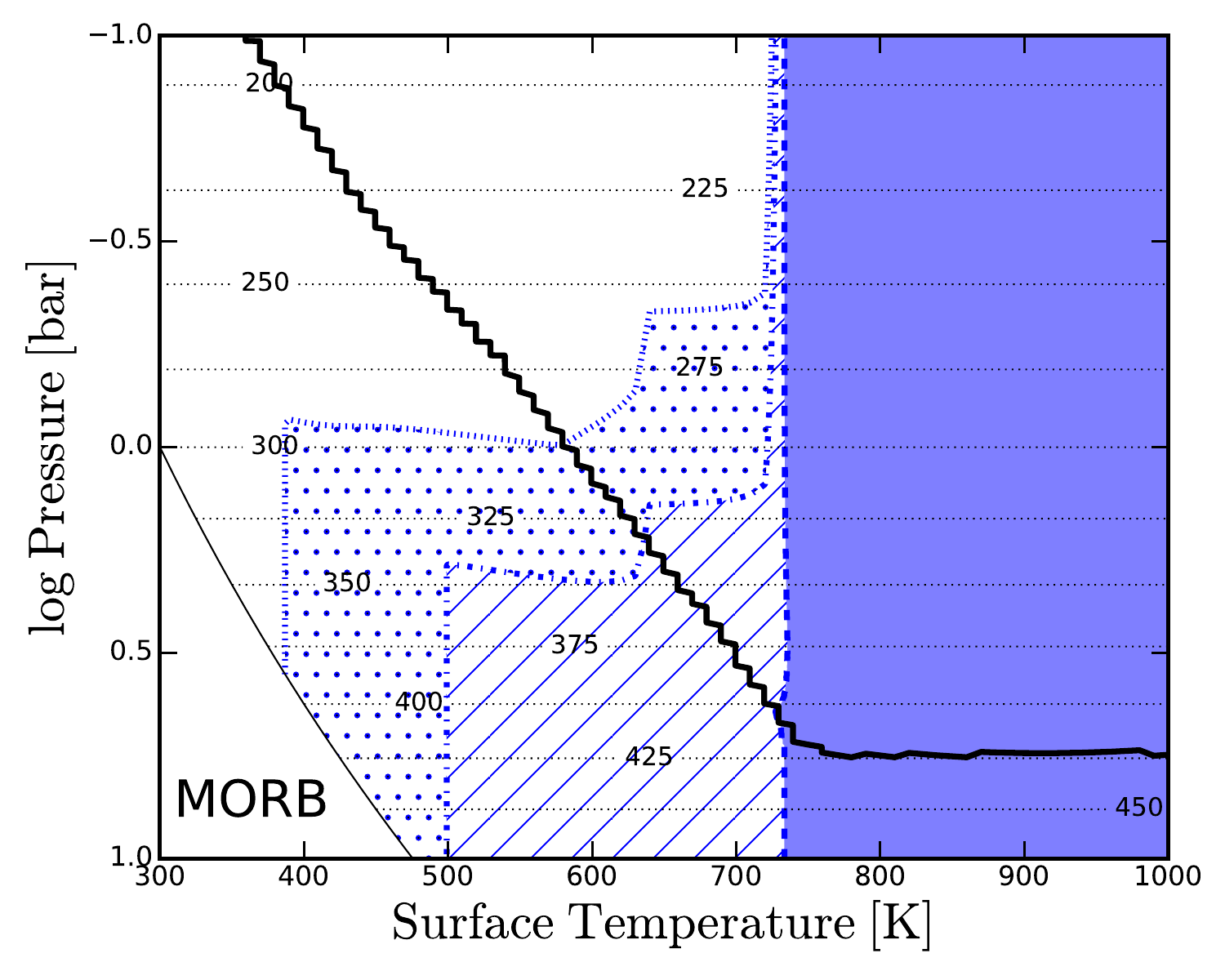}
\includegraphics[width = .32\linewidth, page=7]{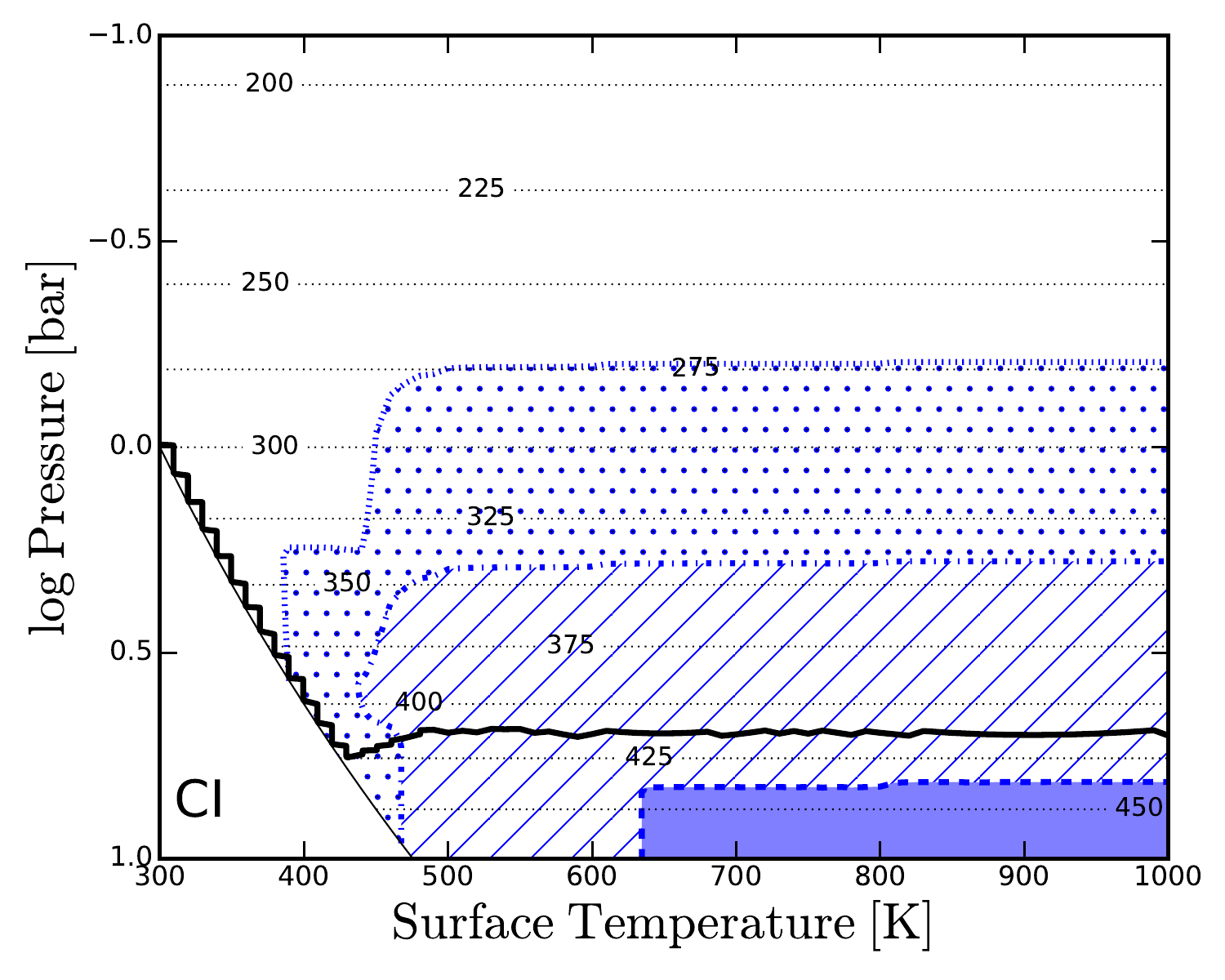}\\
\includegraphics[width = .32\linewidth, page=7]{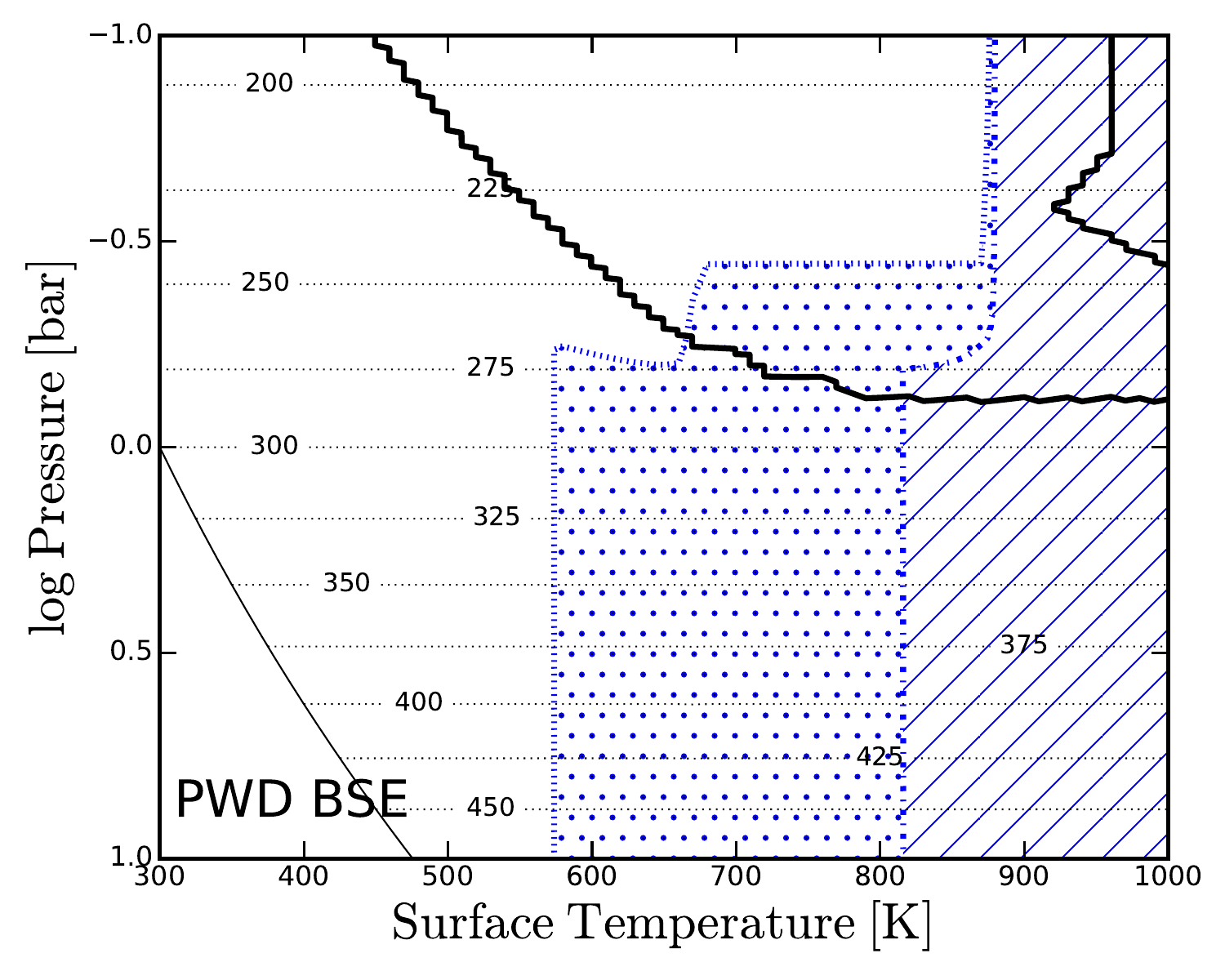}
\includegraphics[width = .32\linewidth, page=7]{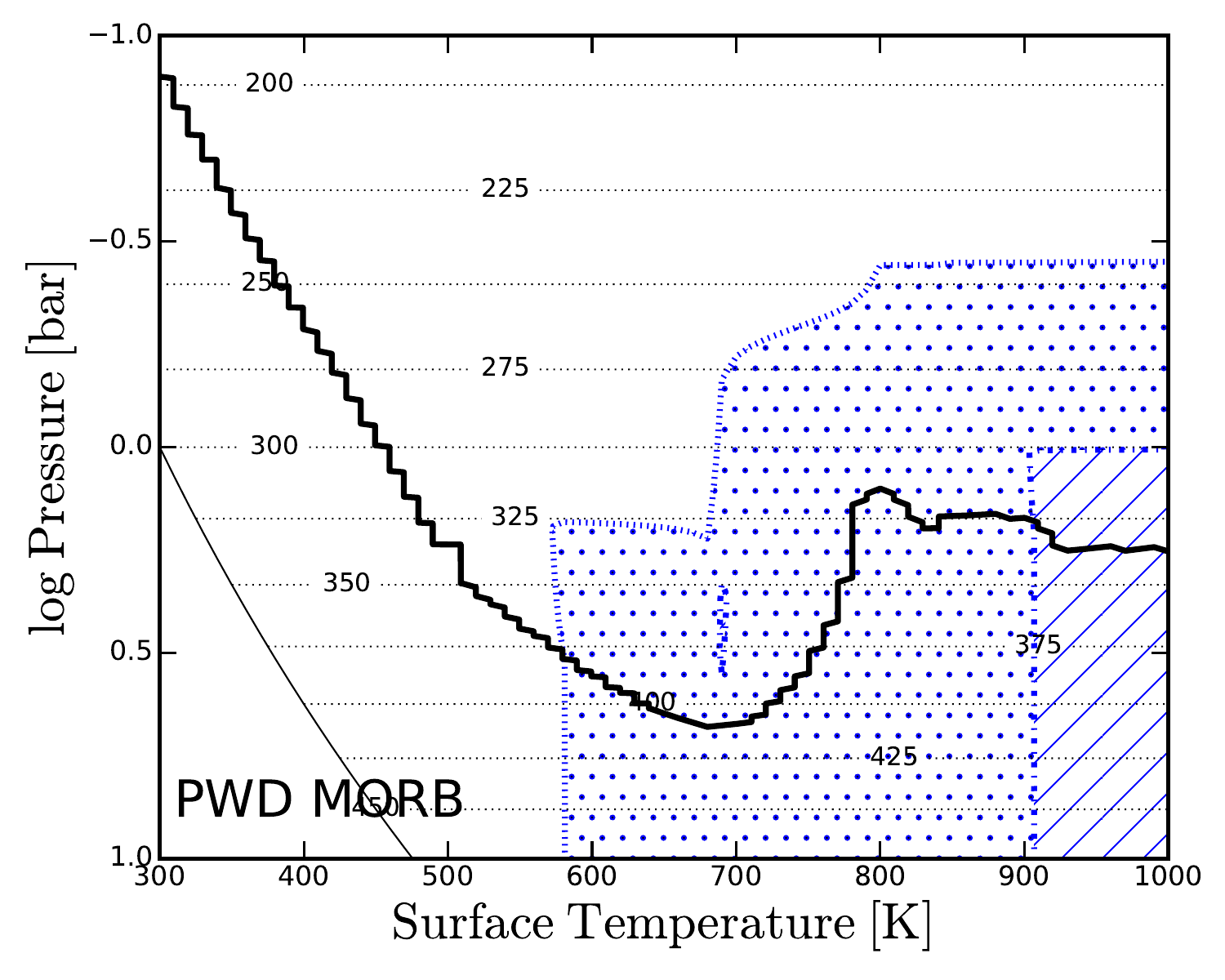}
\includegraphics[width = .32\linewidth, page=7]{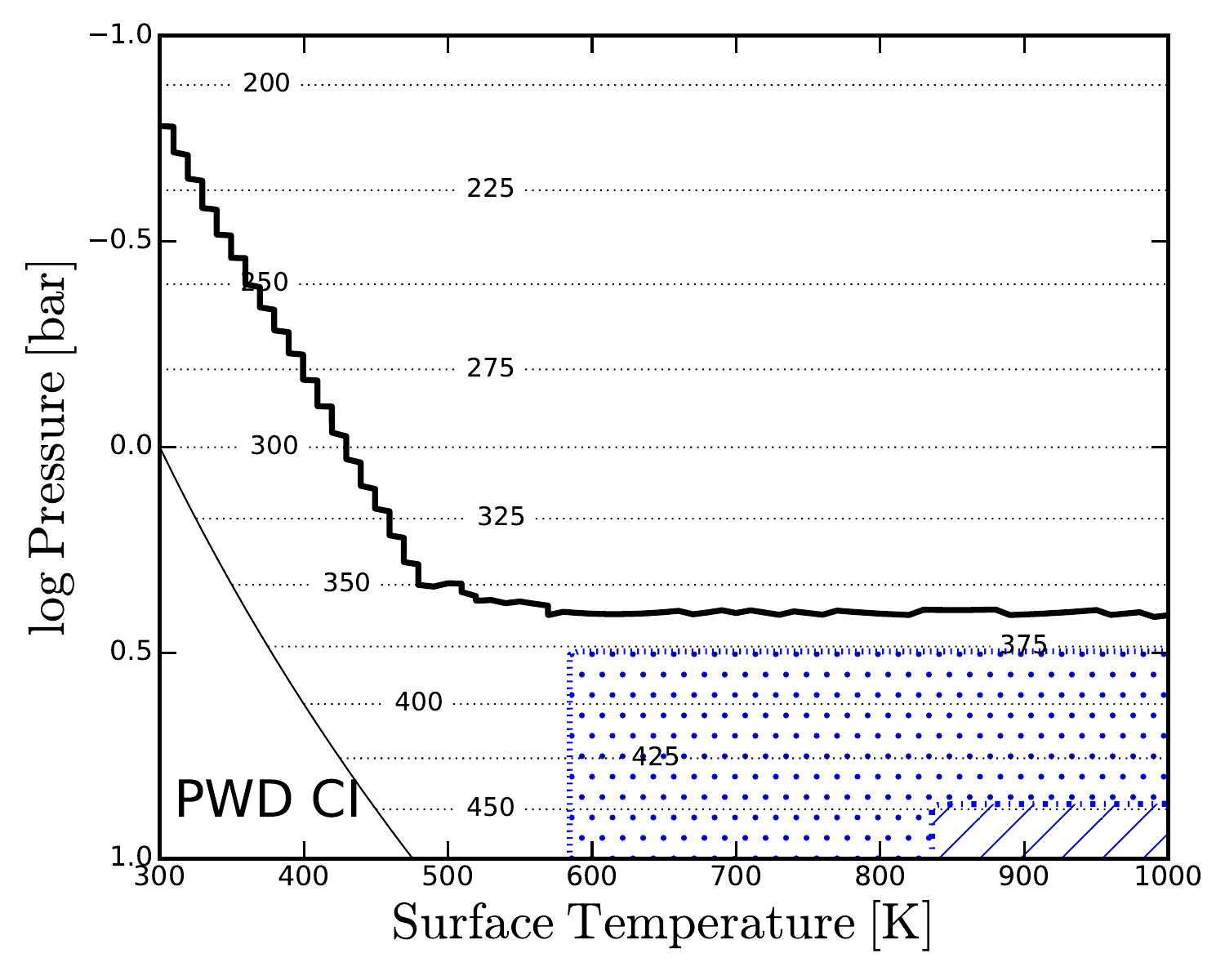}\\
\includegraphics[width = .32\linewidth, page=7]{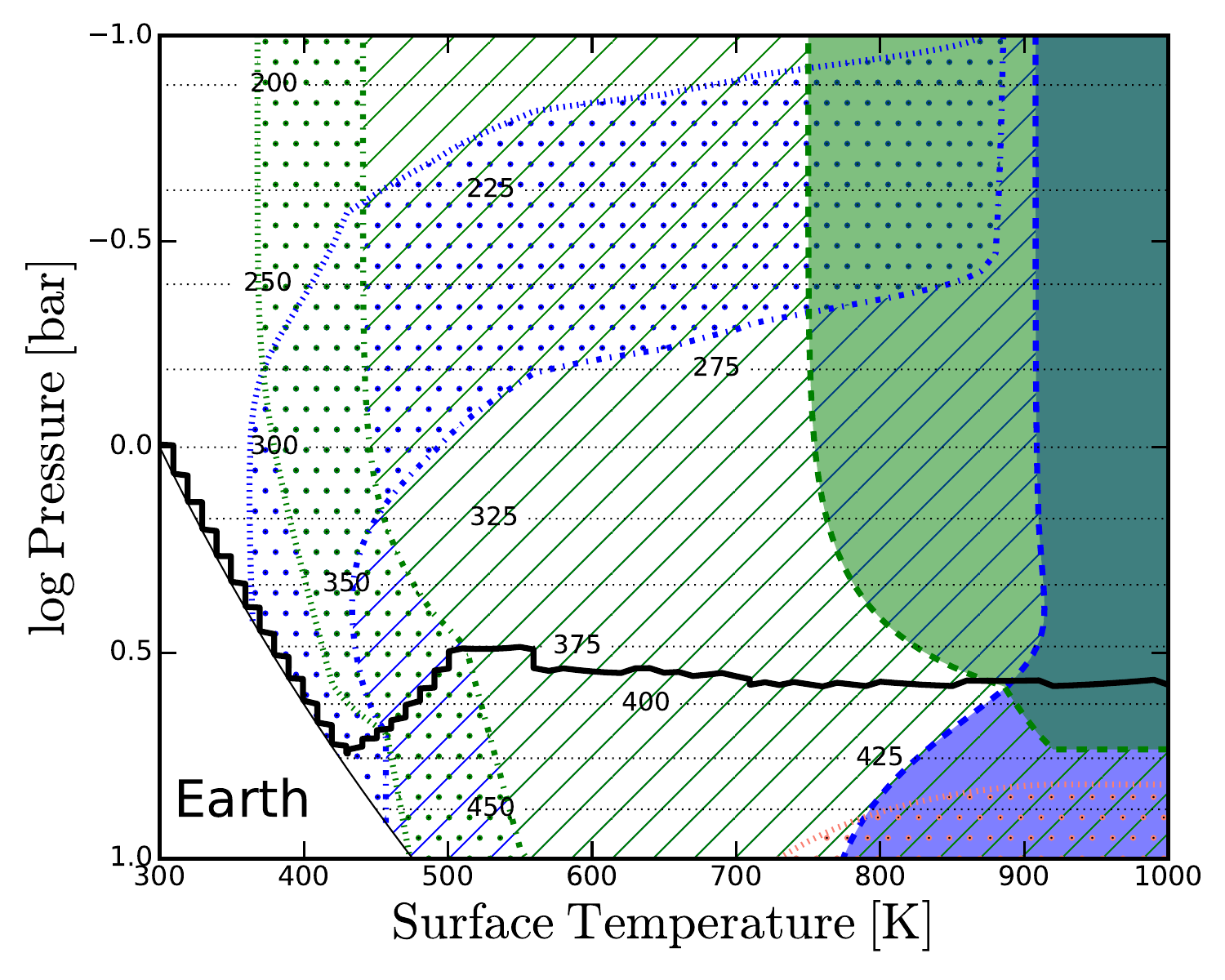}
\includegraphics[width = .32\linewidth, page=7]{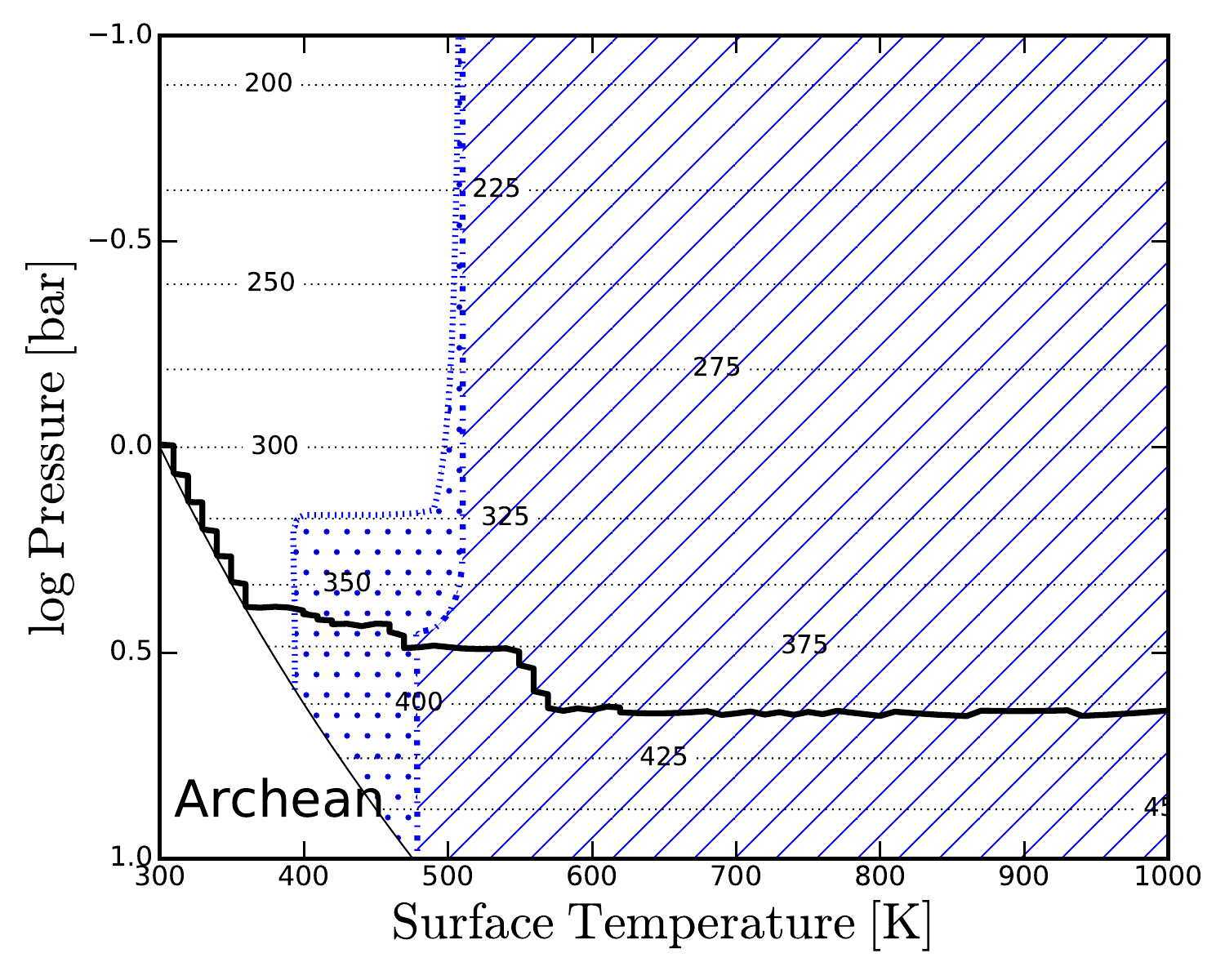}
\includegraphics[width = .32\linewidth, page=7]{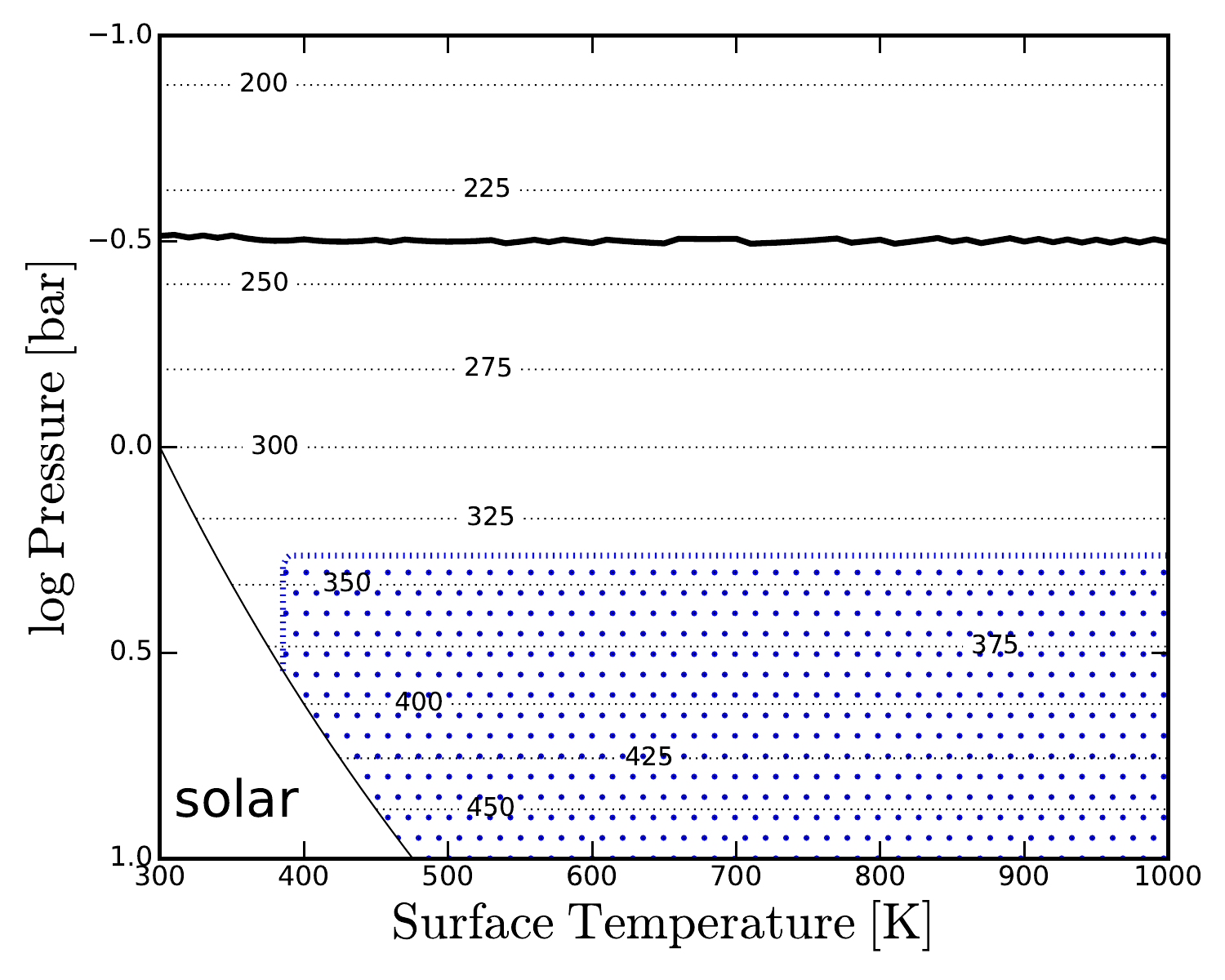}
\caption{As in Fig.~\ref{fig:NutriC}, but for all Cl bearing molecules. The colours refer to \ce{HCl} (blue), \ce{Cl2} (green), \ce{OHCl} (red).}
\label{fig:NutriCl}
\end{figure*}

\begin{figure*}
\centering
\includegraphics[width = .32\linewidth, page=8]{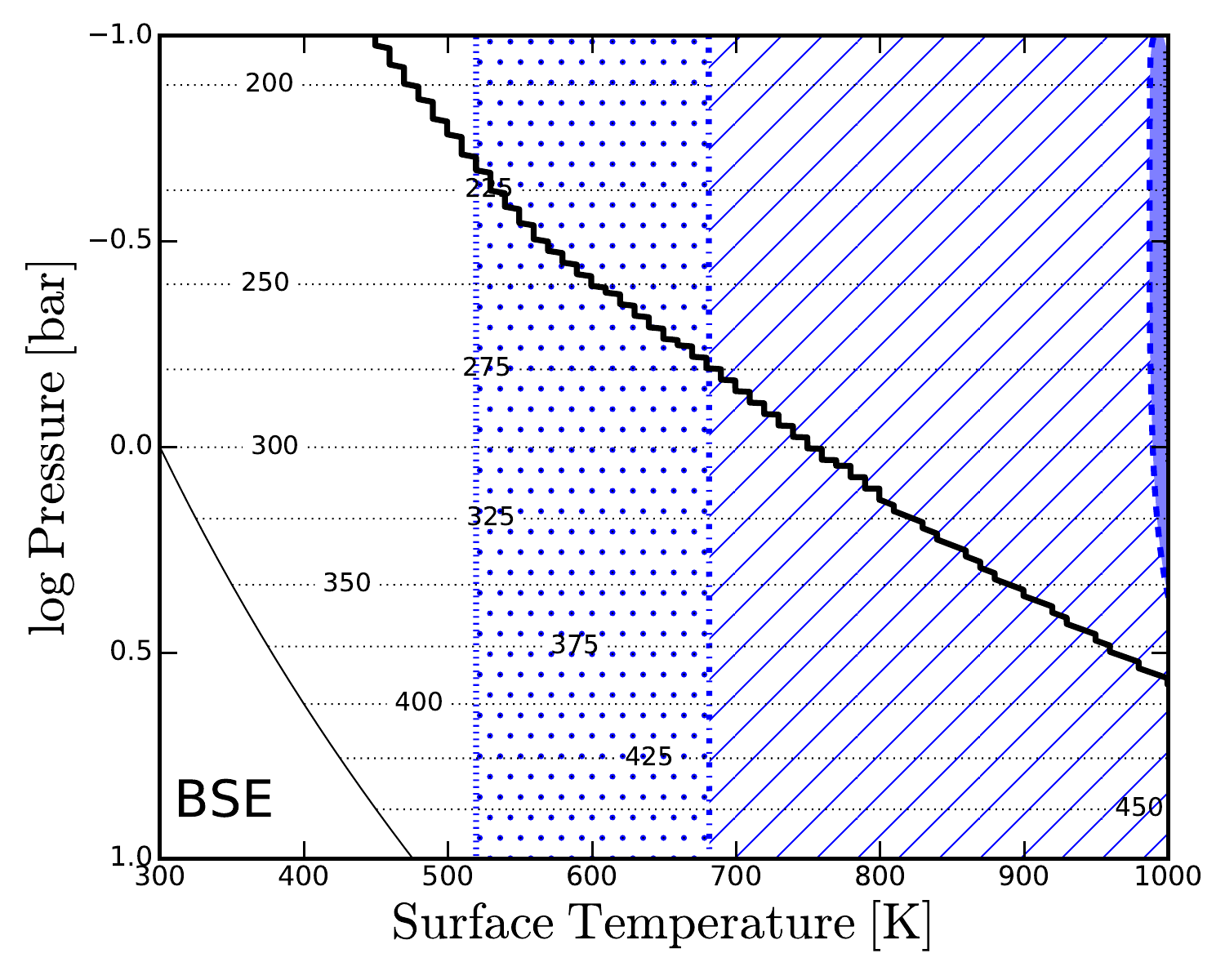}
\includegraphics[width = .32\linewidth, page=8]{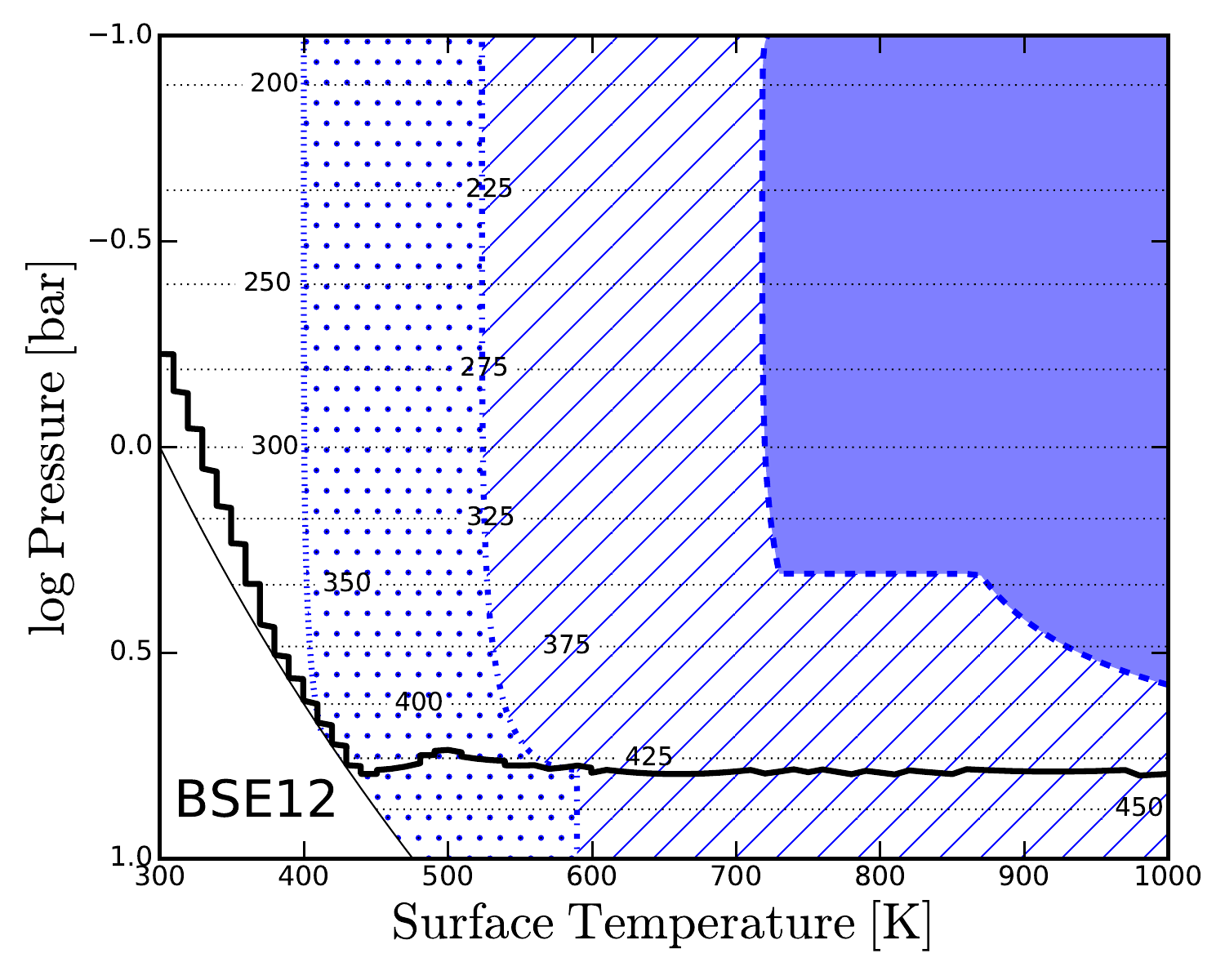}
\includegraphics[width = .32\linewidth, page=8]{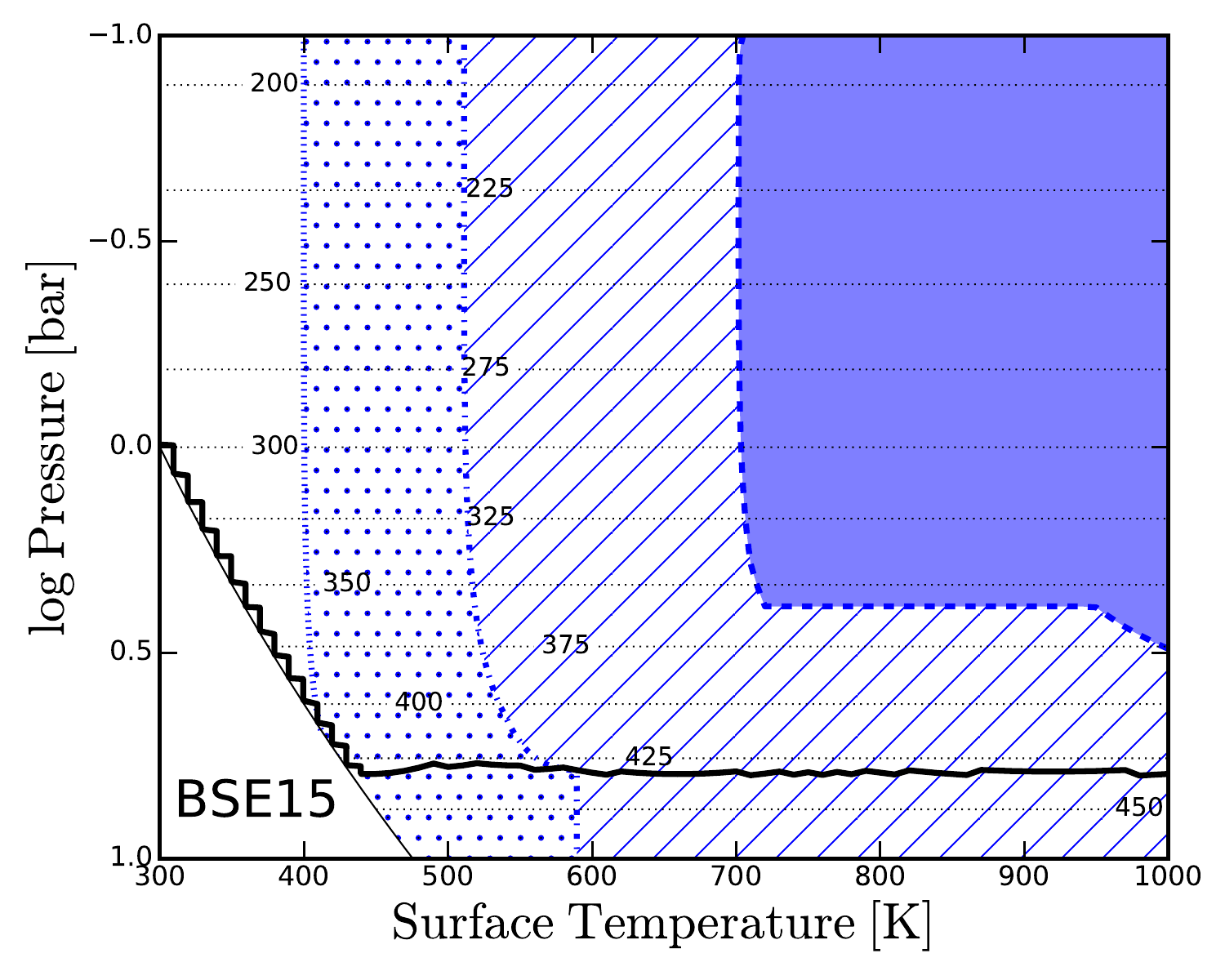}\\
\includegraphics[width = .32\linewidth, page=8]{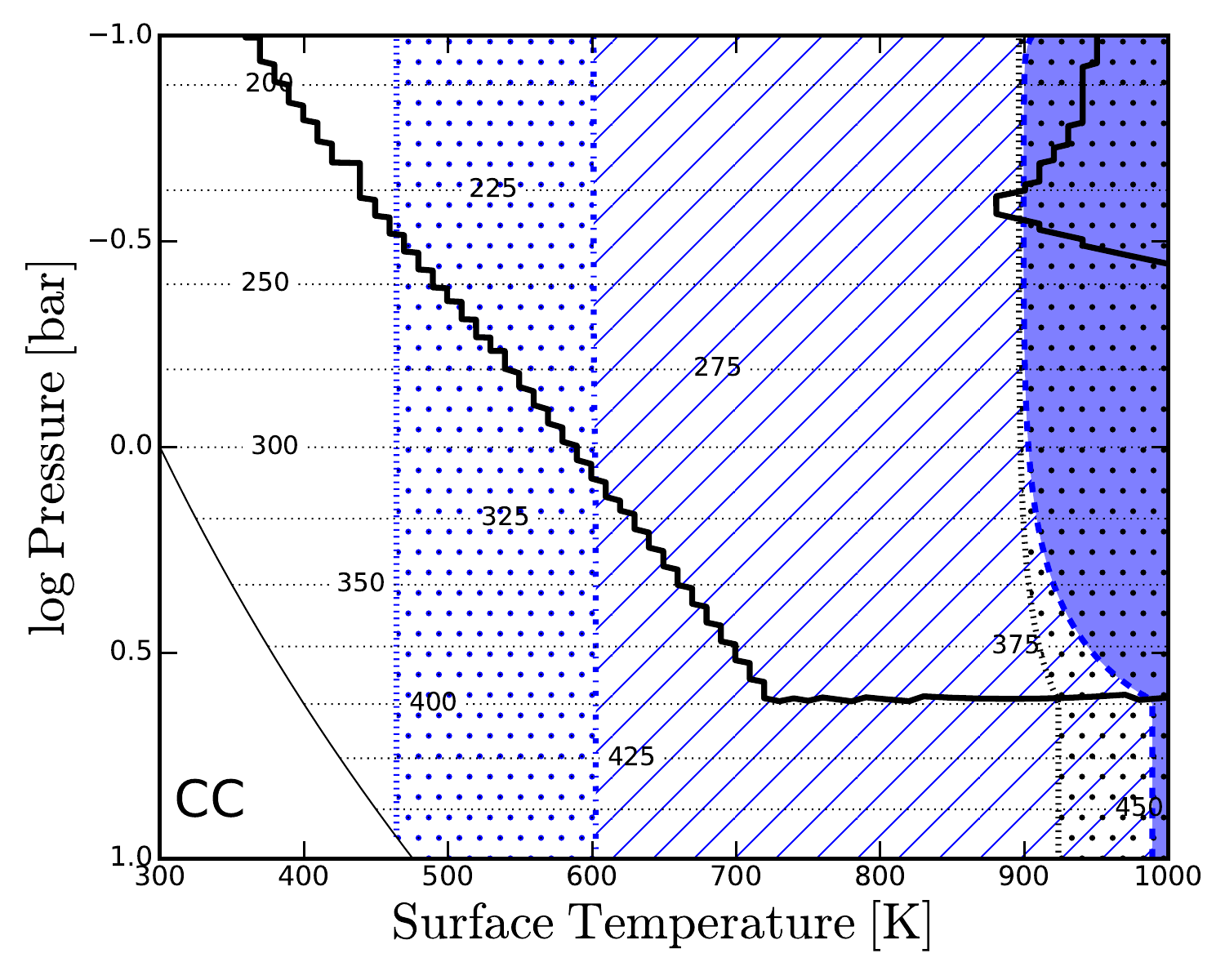}
\includegraphics[width = .32\linewidth, page=8]{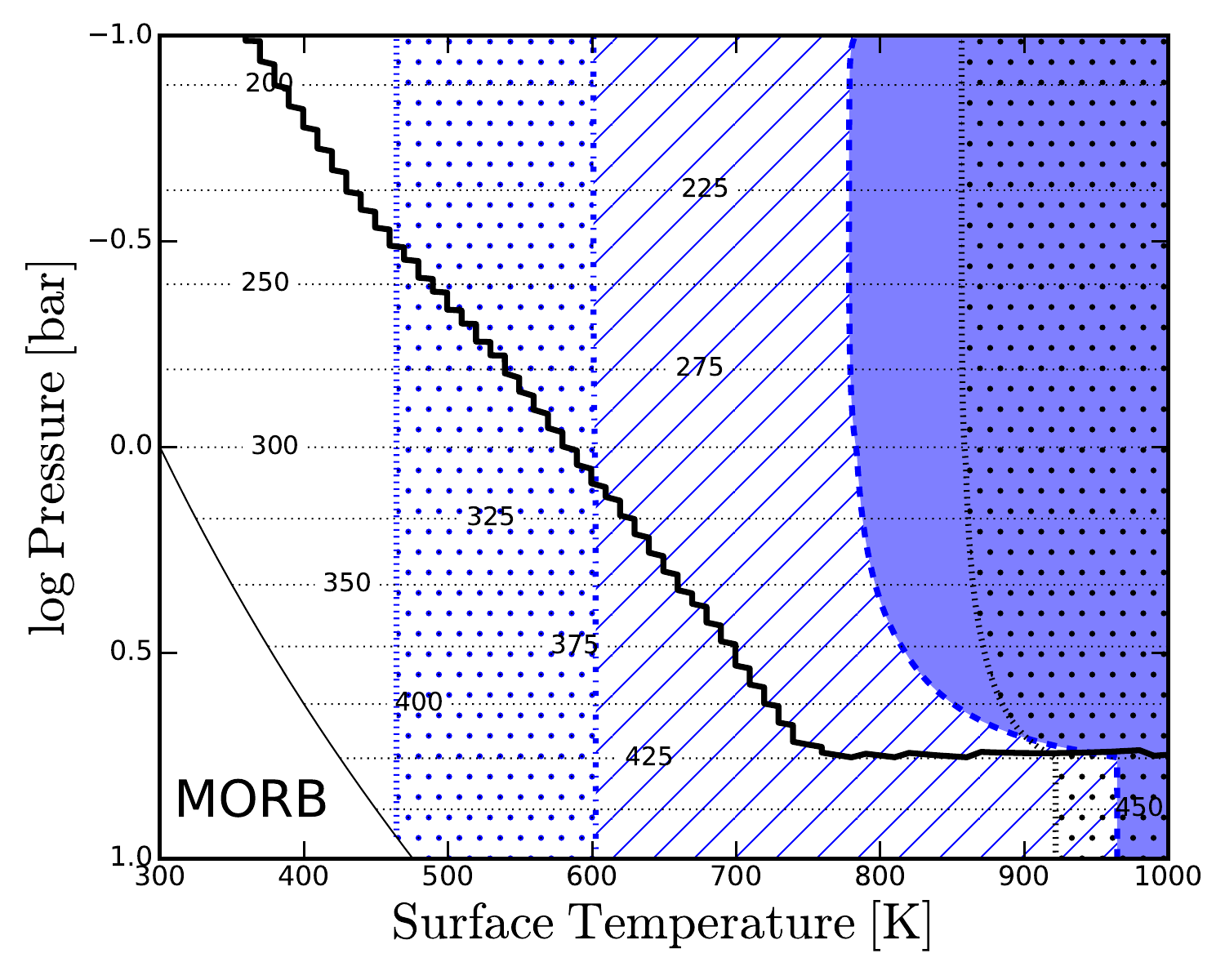}
\includegraphics[width = .32\linewidth, page=8]{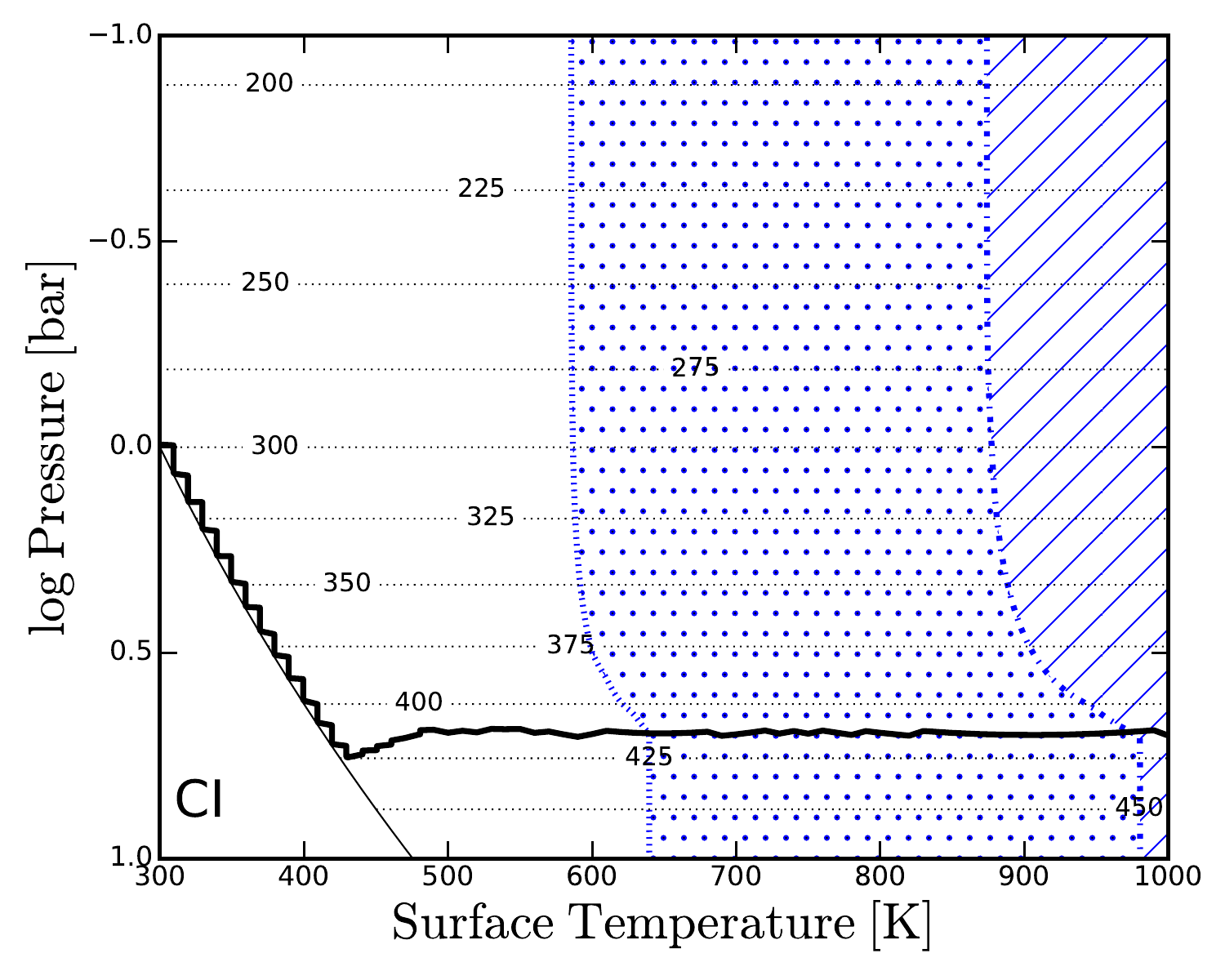}\\
\includegraphics[width = .32\linewidth, page=8]{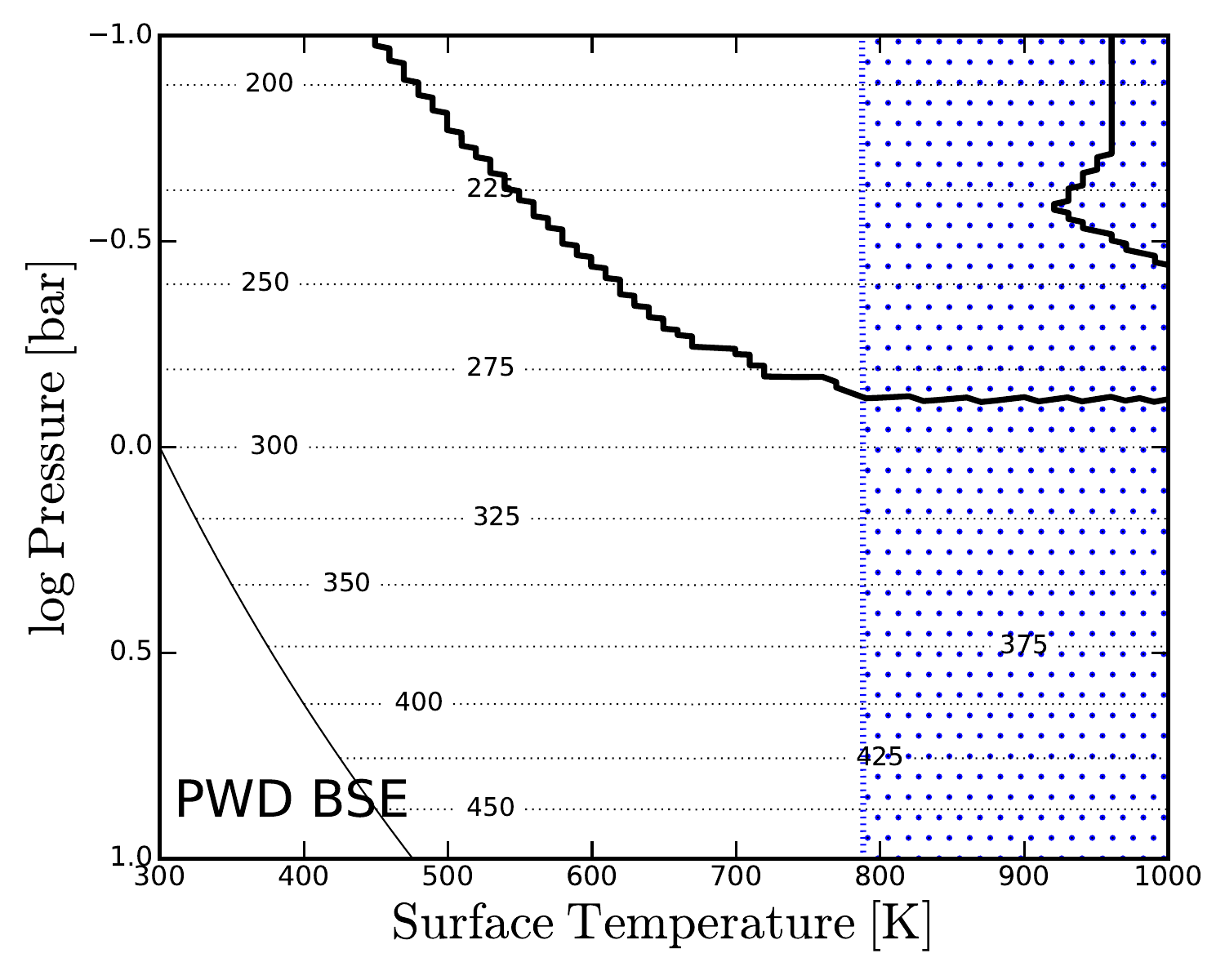}
\includegraphics[width = .32\linewidth, page=8]{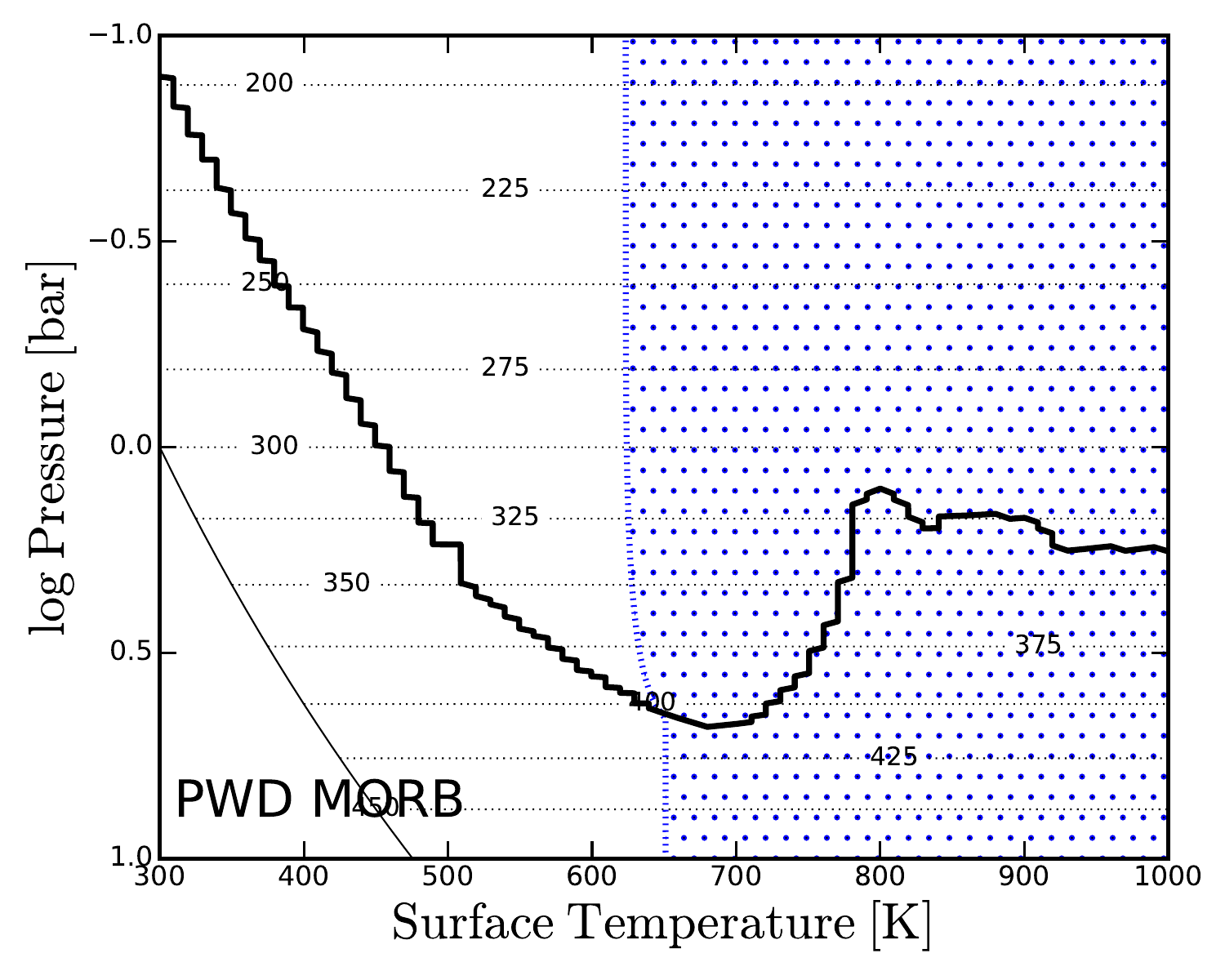}
\includegraphics[width = .32\linewidth, page=8]{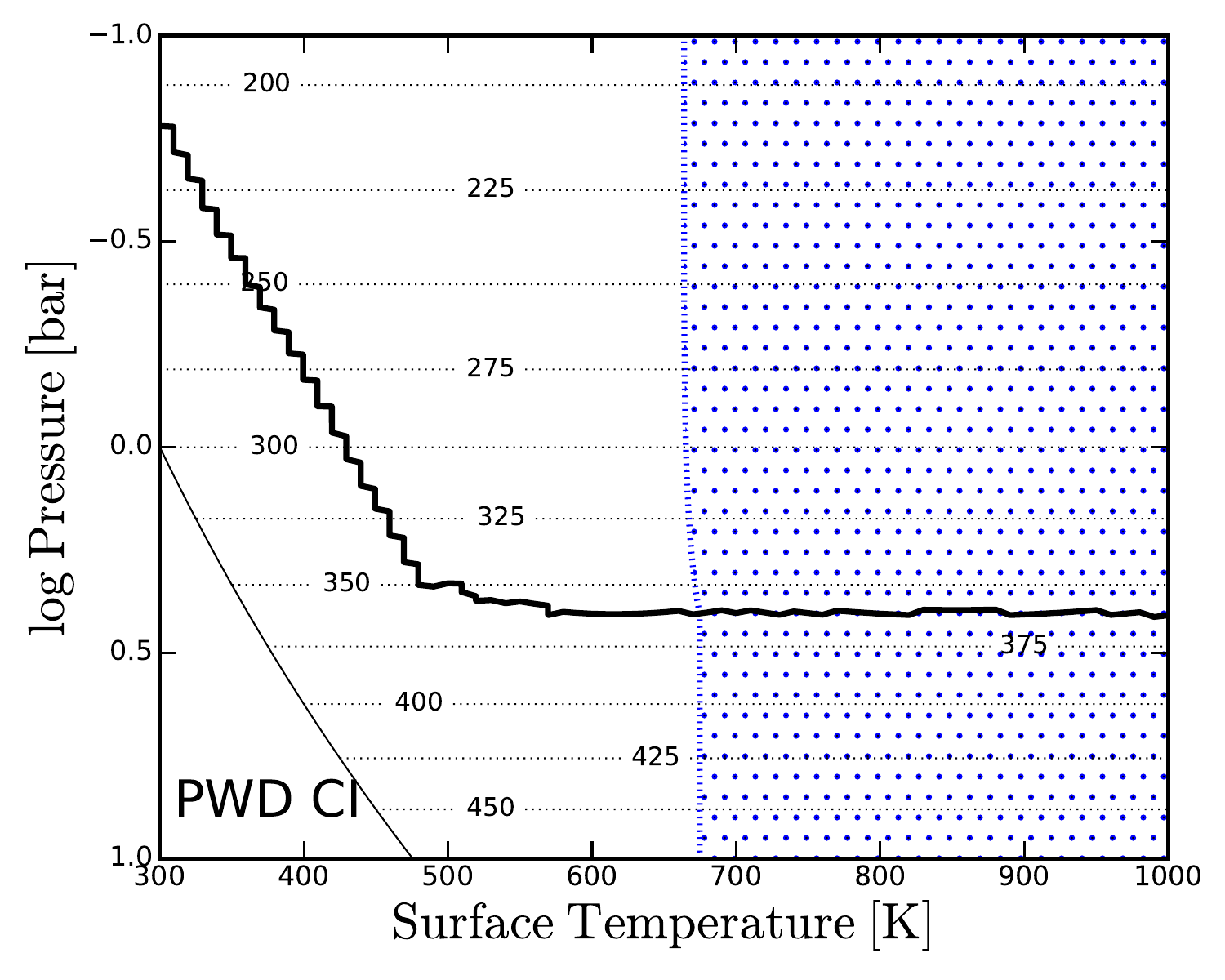}\\
\includegraphics[width = .32\linewidth, page=8]{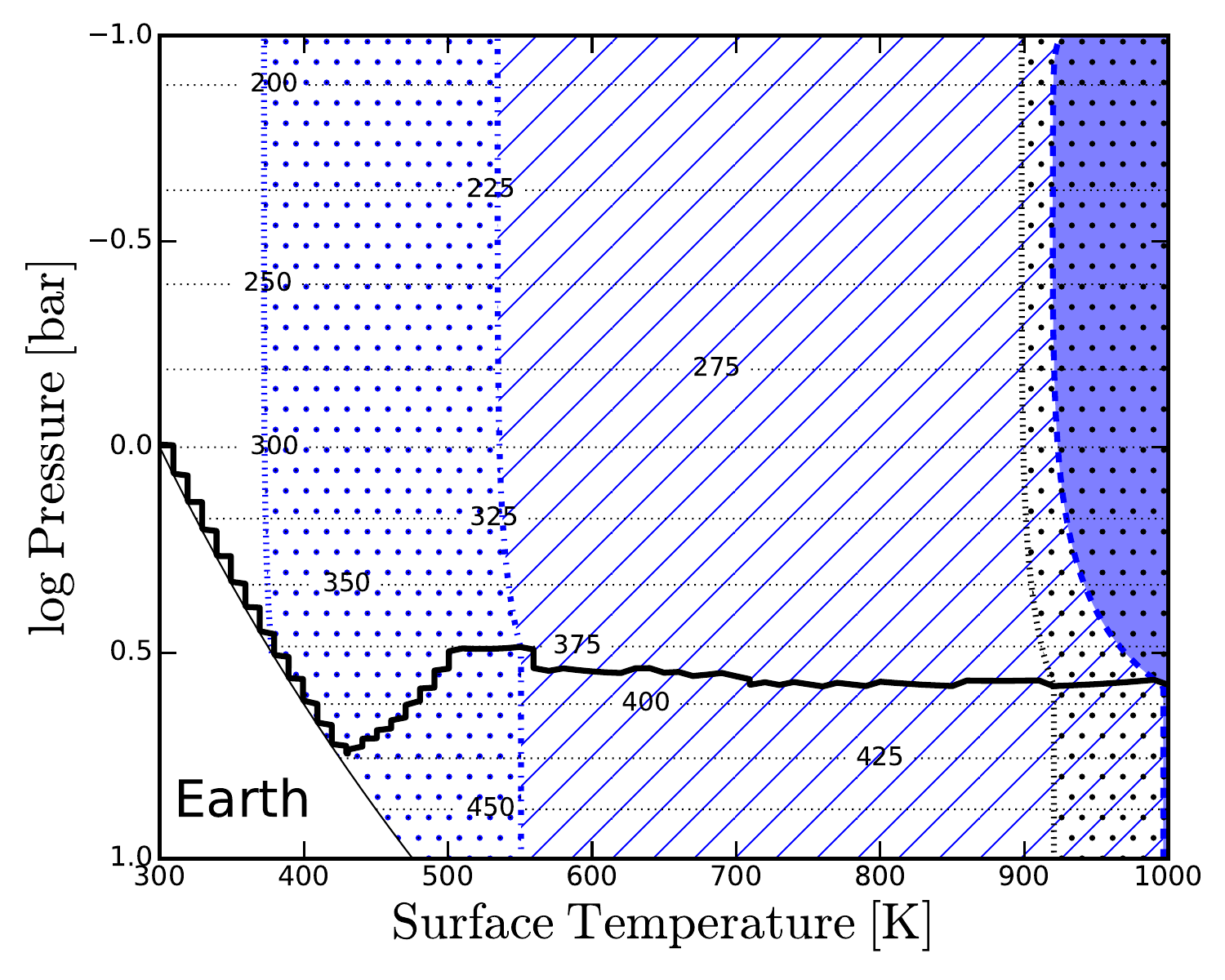}
\includegraphics[width = .32\linewidth, page=8]{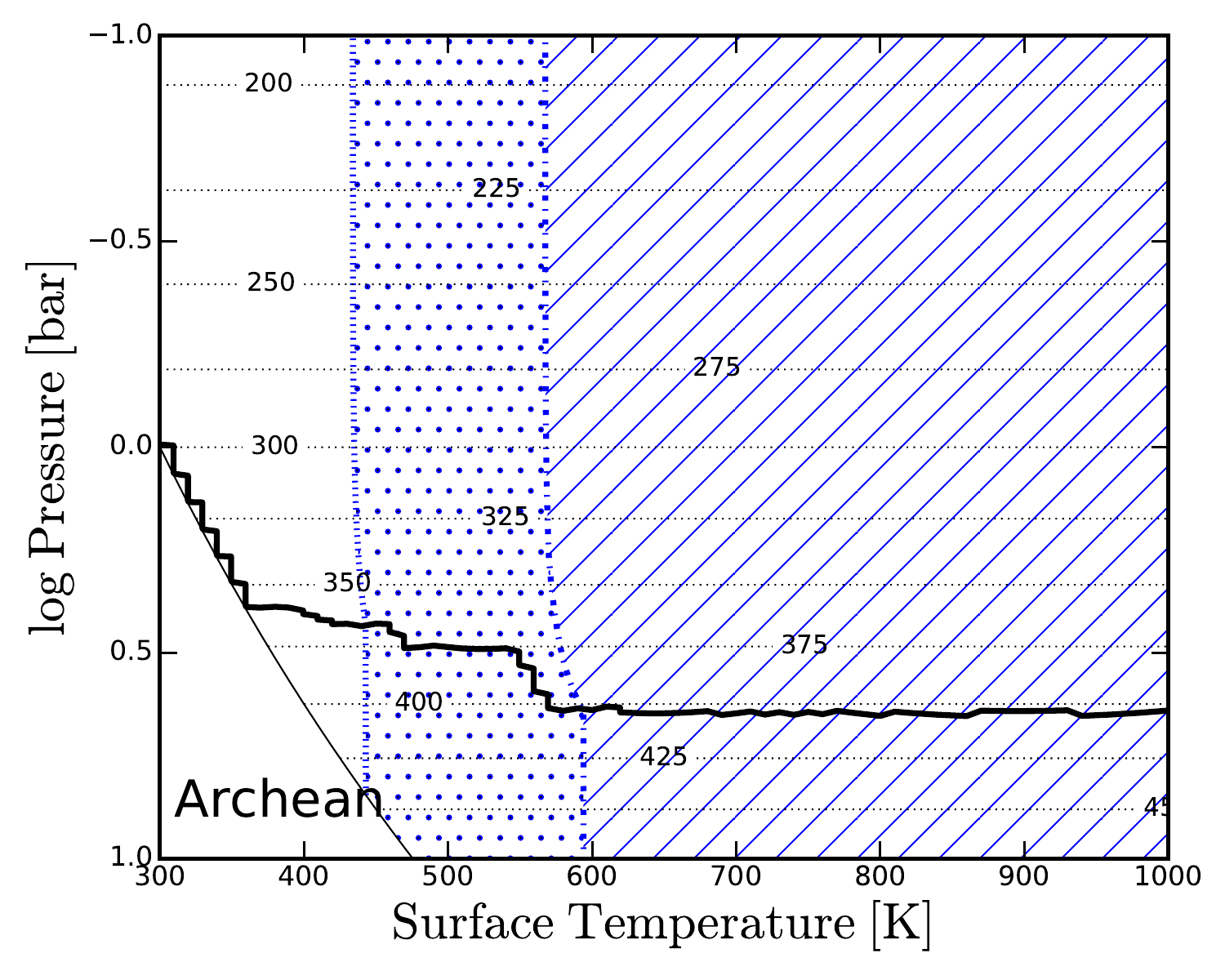}
\includegraphics[width = .32\linewidth, page=8]{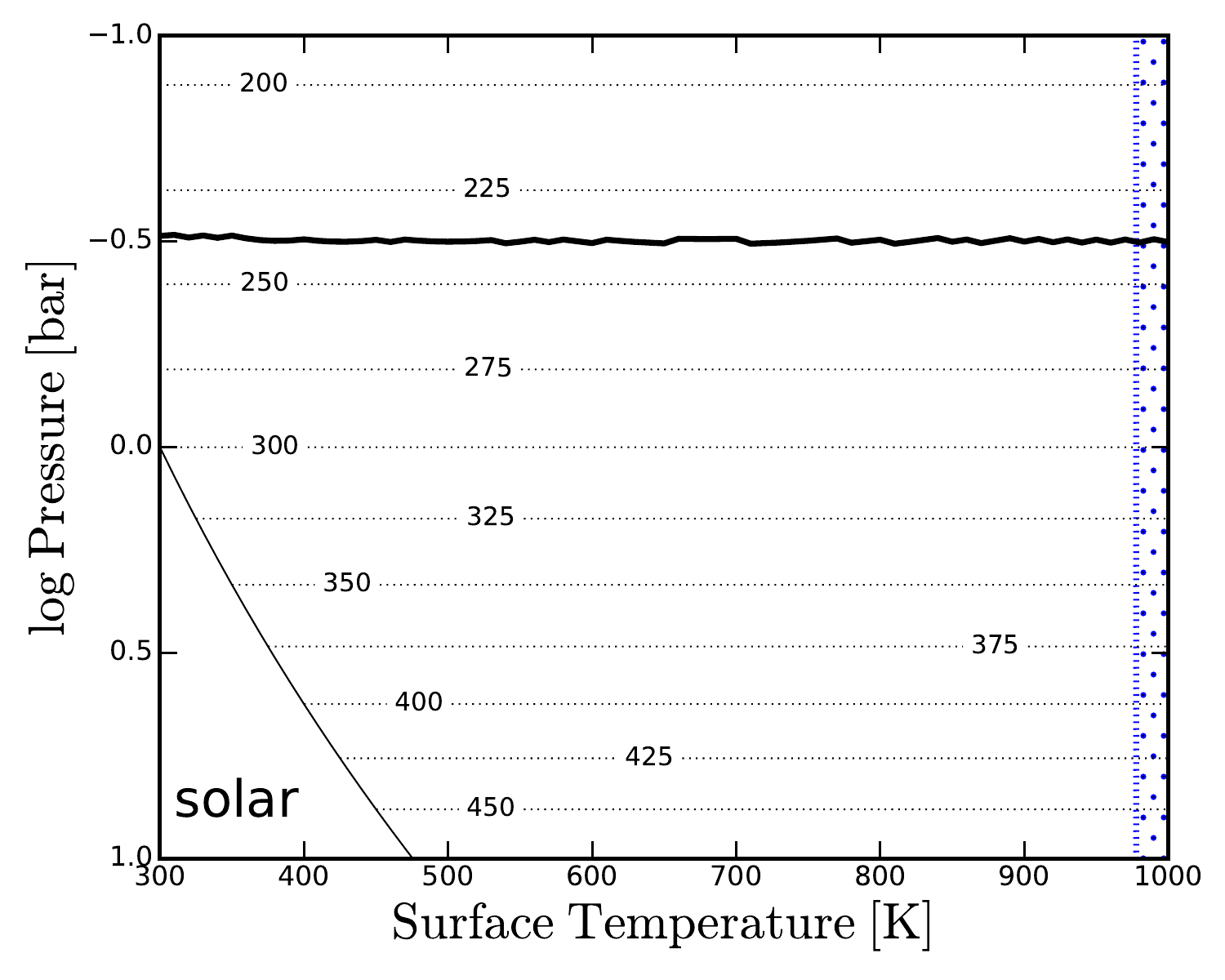}
\caption{As in Fig.~\ref{fig:NutriC}, but for all F bearing molecules. The colours refer to (\ce{HF})$_x$ (blue) and \ce{SiF4} (black)}
\label{fig:NutriF}
\end{figure*}

\begin{figure*}
\centering
\includegraphics[width = .32\linewidth, page=9]{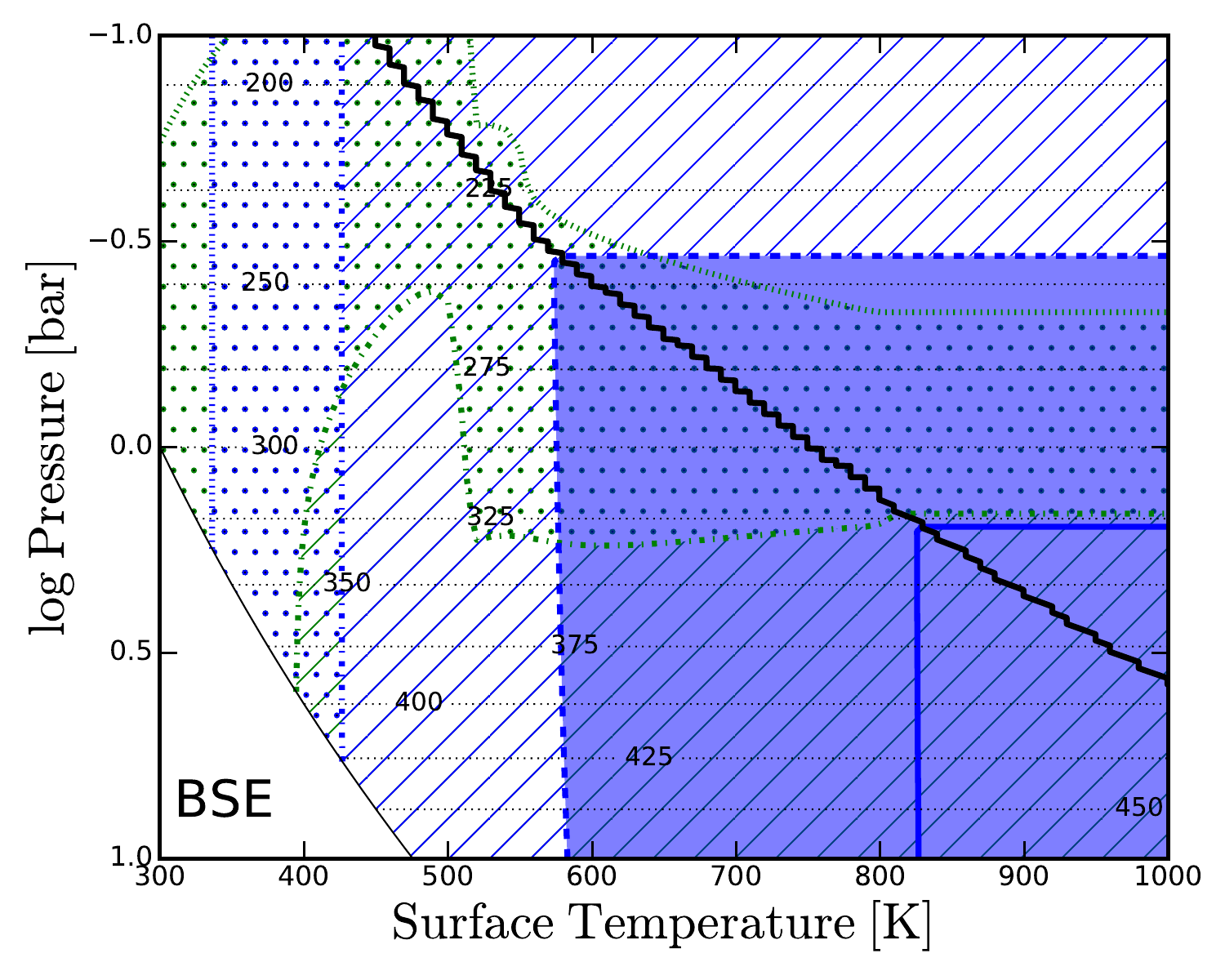}
\includegraphics[width = .32\linewidth, page=9]{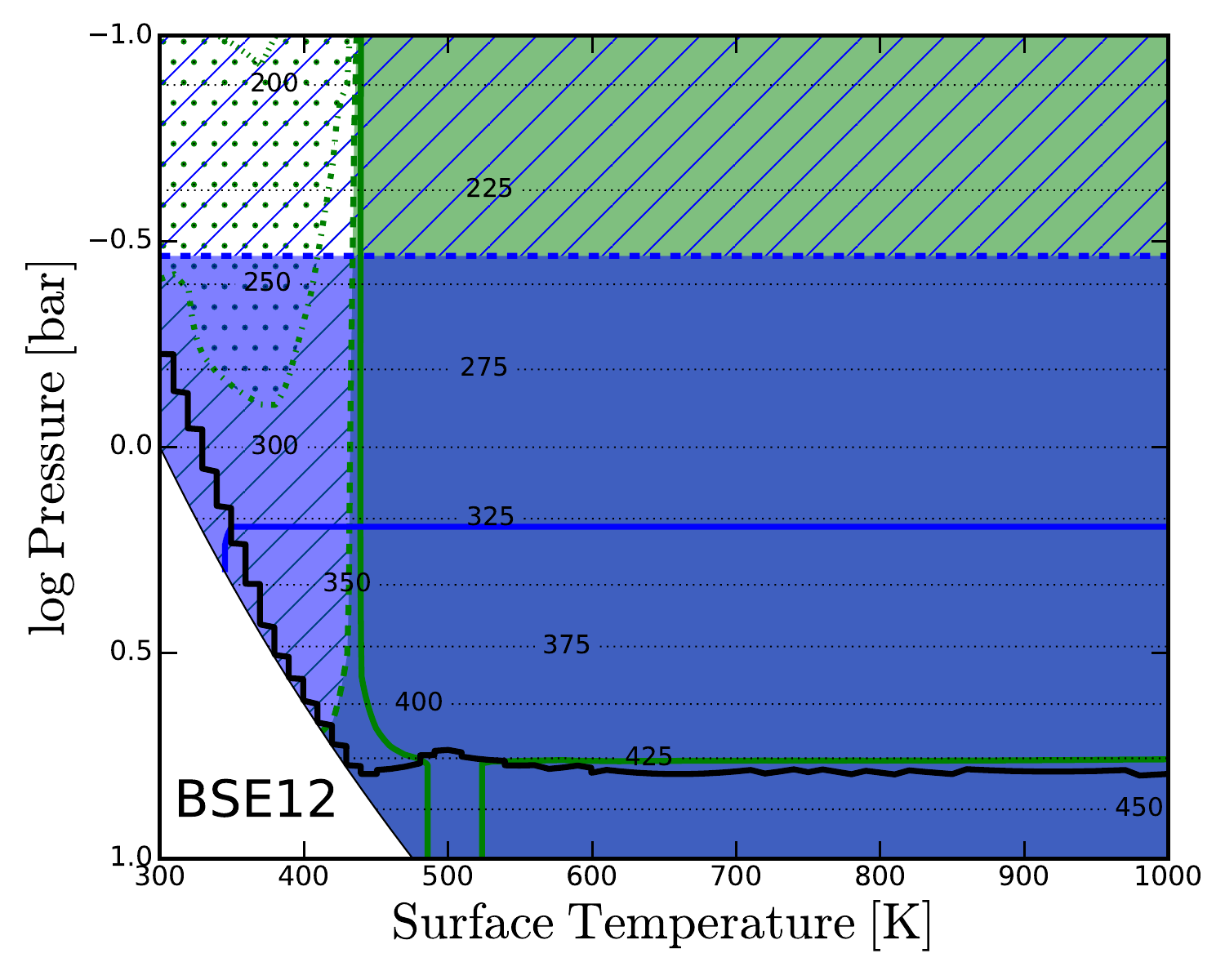}
\includegraphics[width = .32\linewidth, page=9]{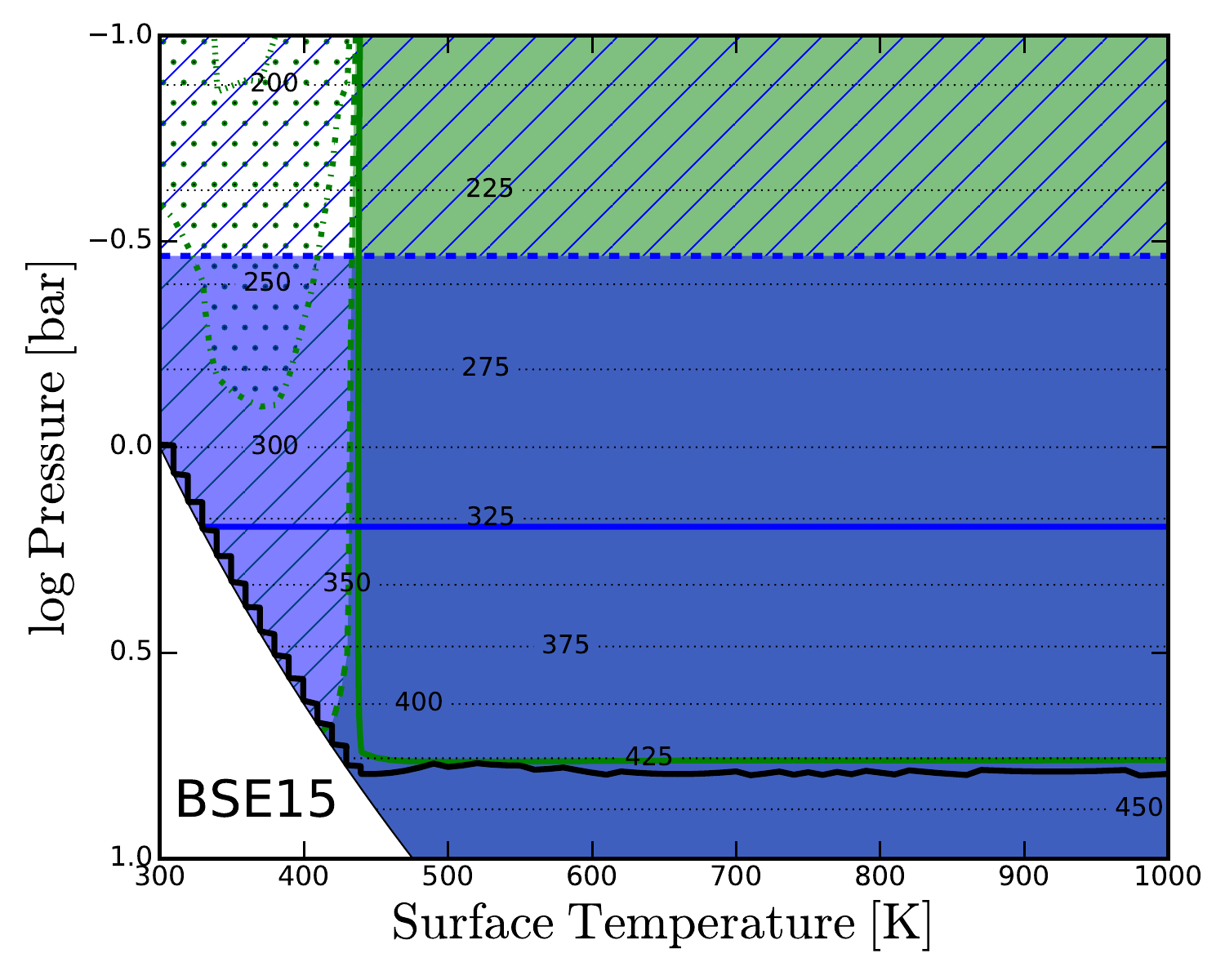}\\
\includegraphics[width = .32\linewidth, page=9]{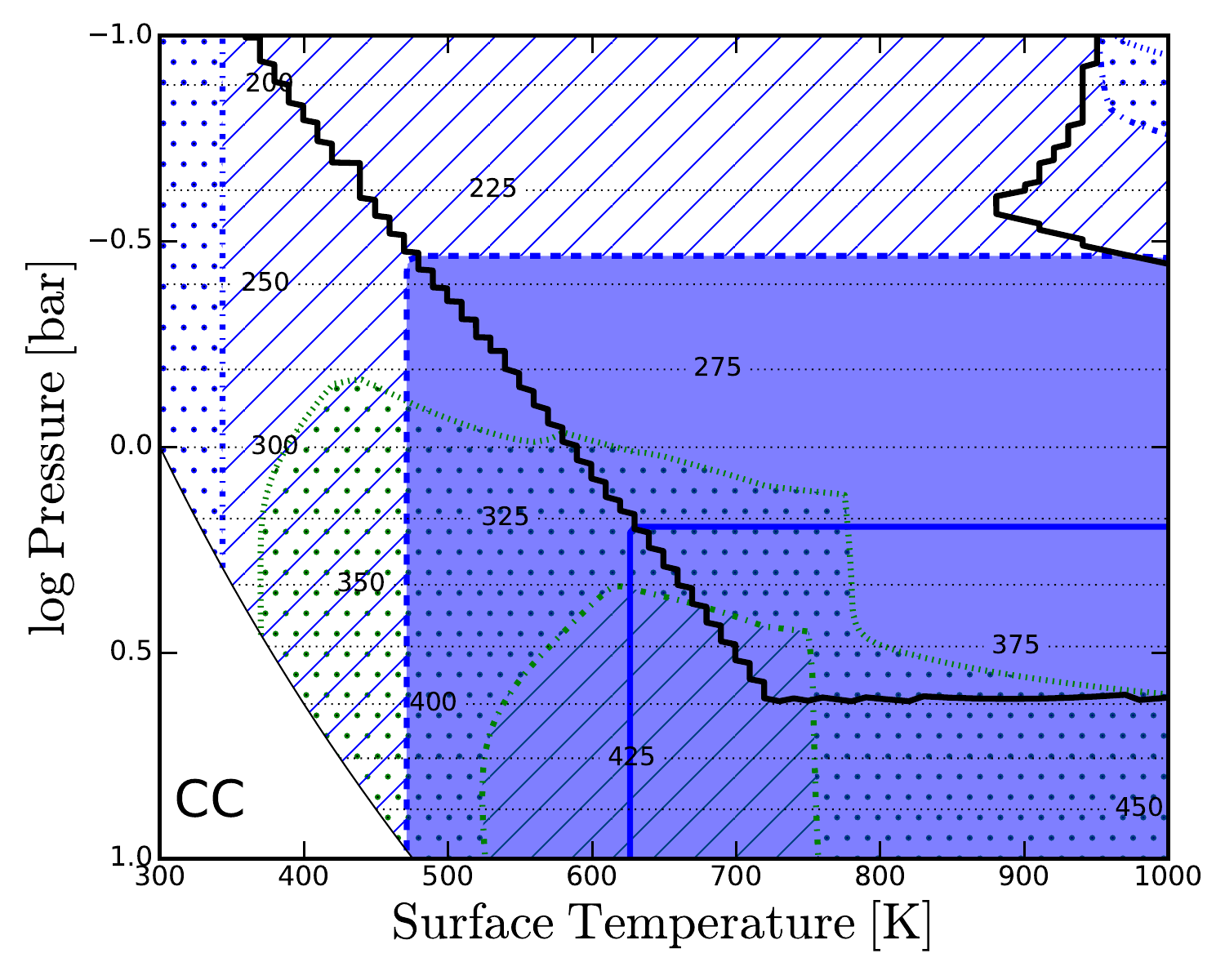}
\includegraphics[width = .32\linewidth, page=9]{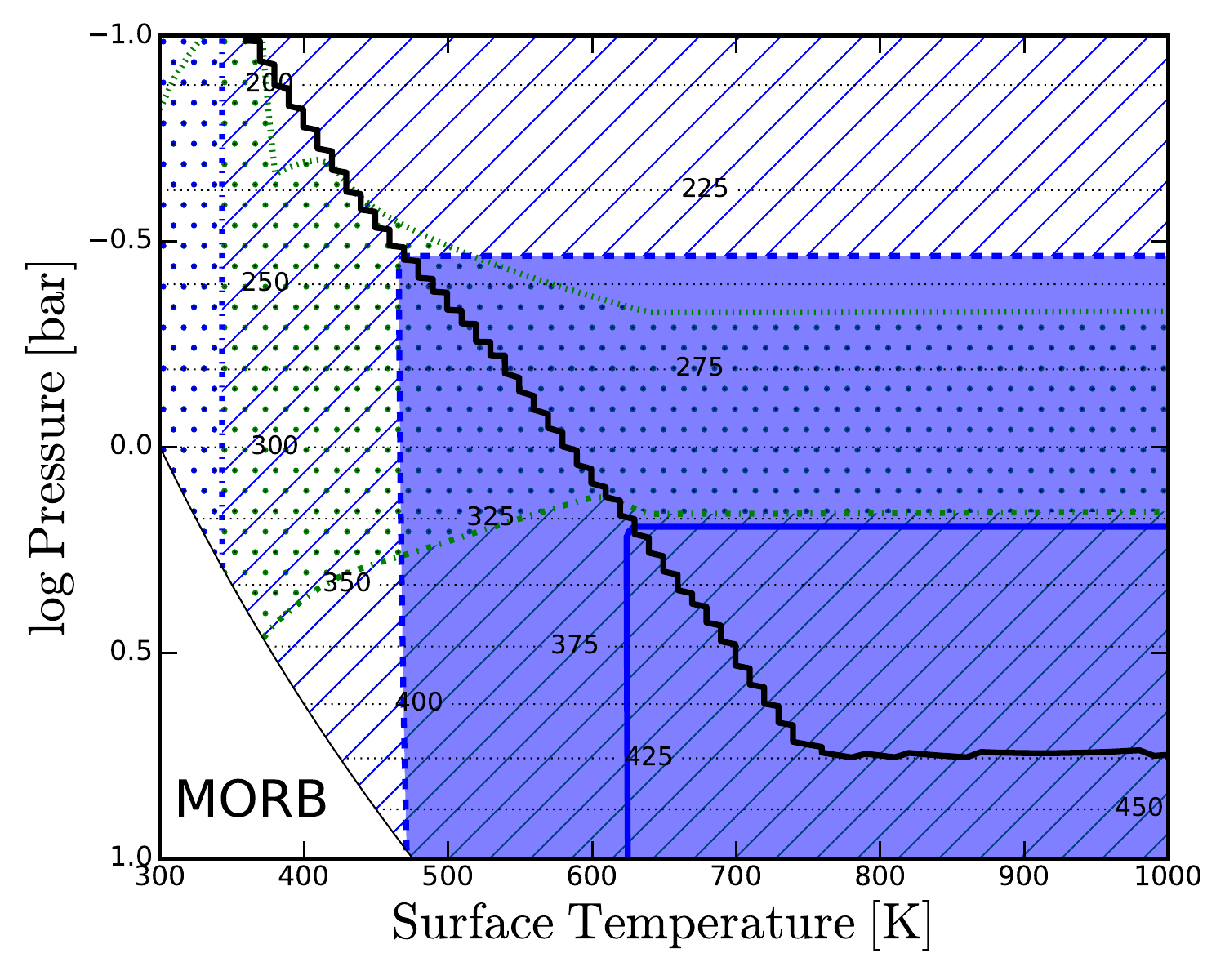}
\includegraphics[width = .32\linewidth, page=9]{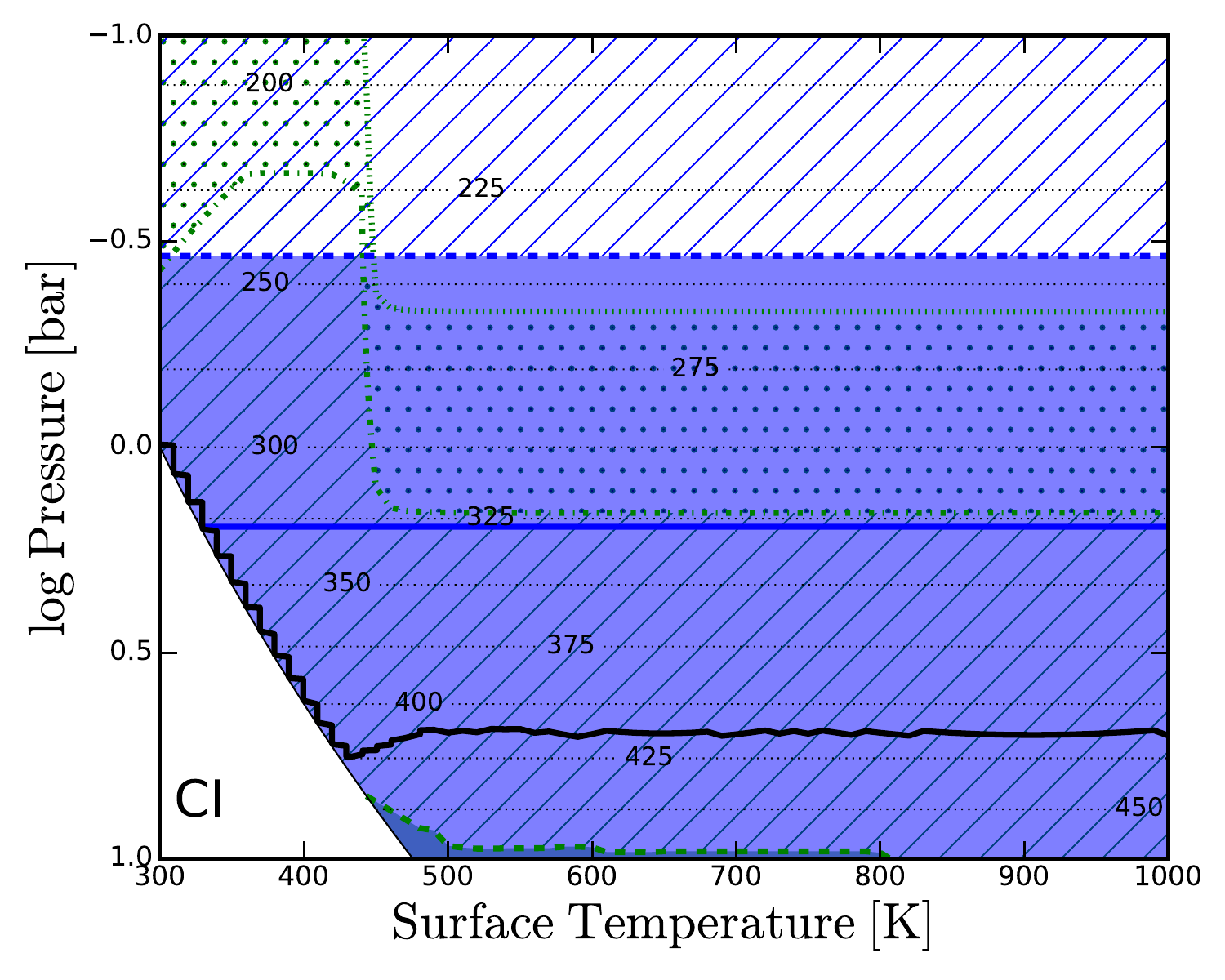}\\
\includegraphics[width = .32\linewidth, page=9]{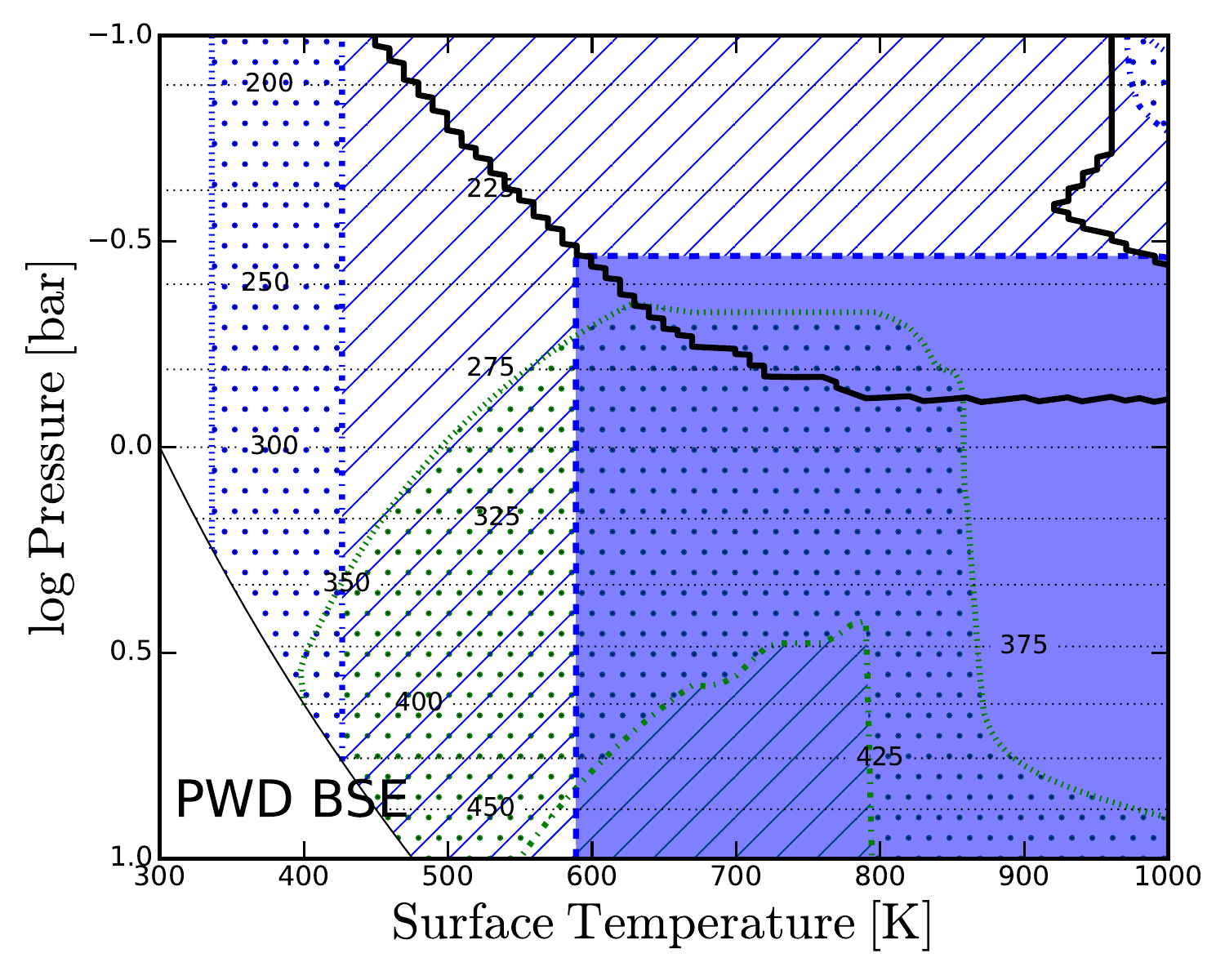}
\includegraphics[width = .32\linewidth, page=9]{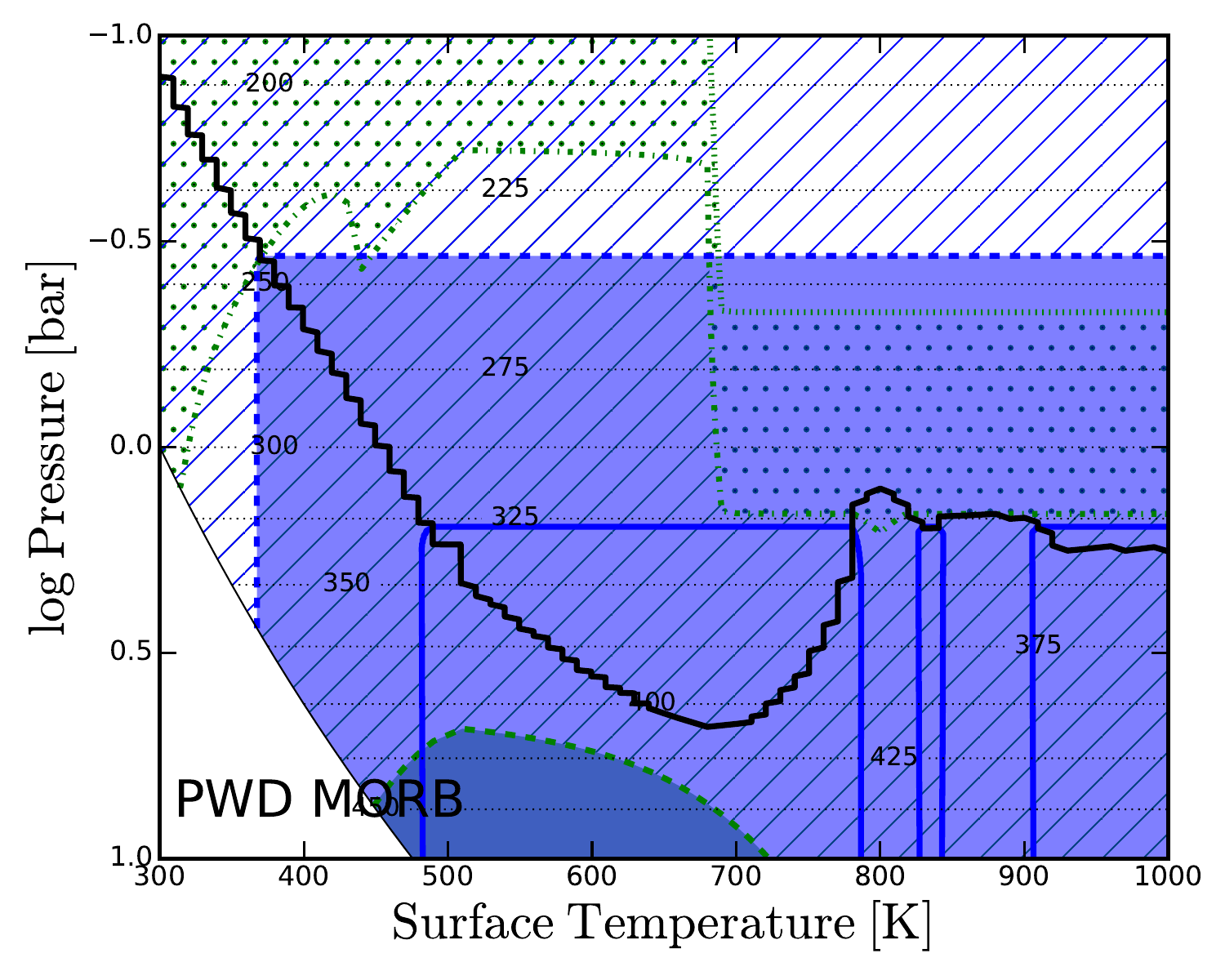}
\includegraphics[width = .32\linewidth, page=9]{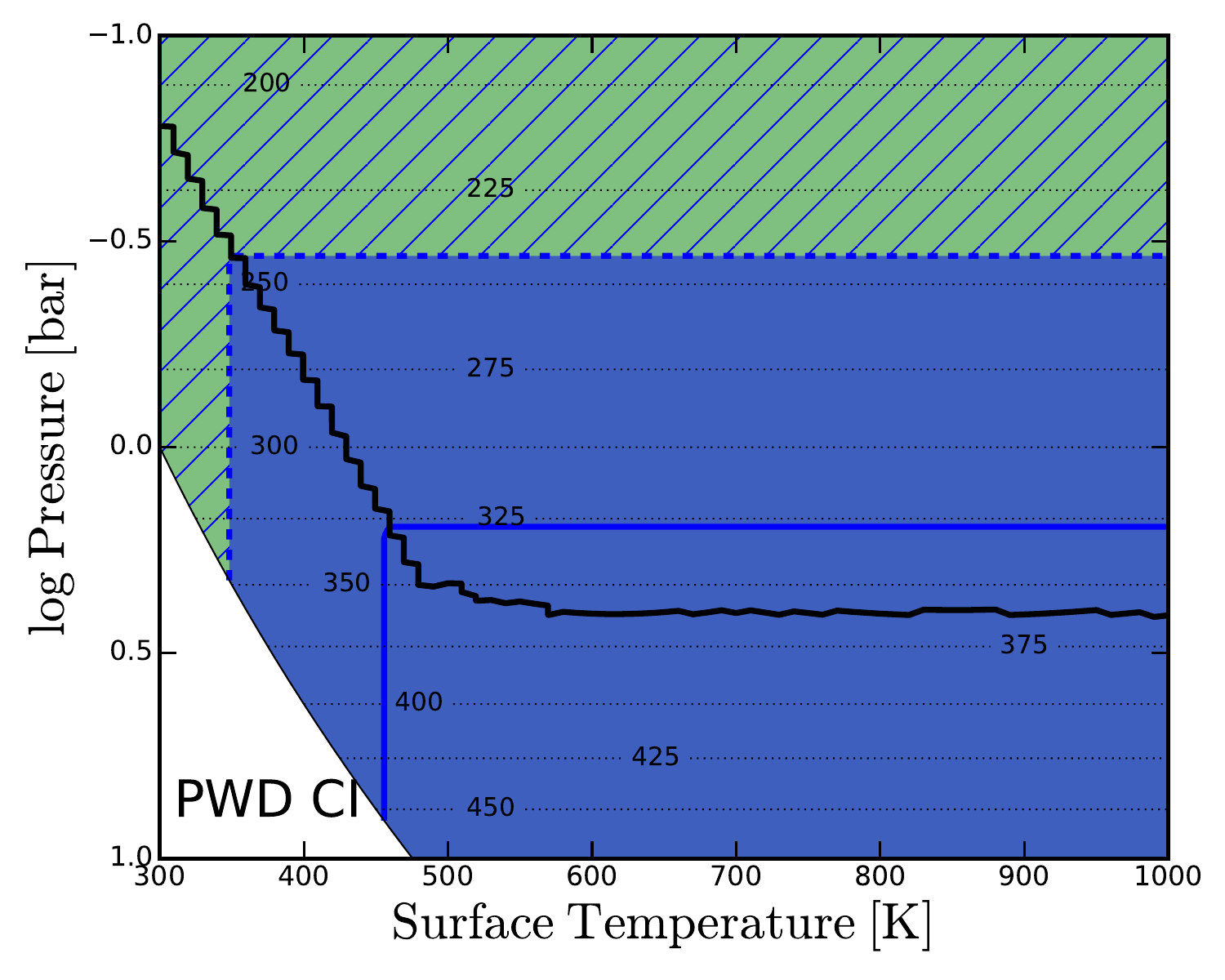}\\
\includegraphics[width = .32\linewidth, page=9]{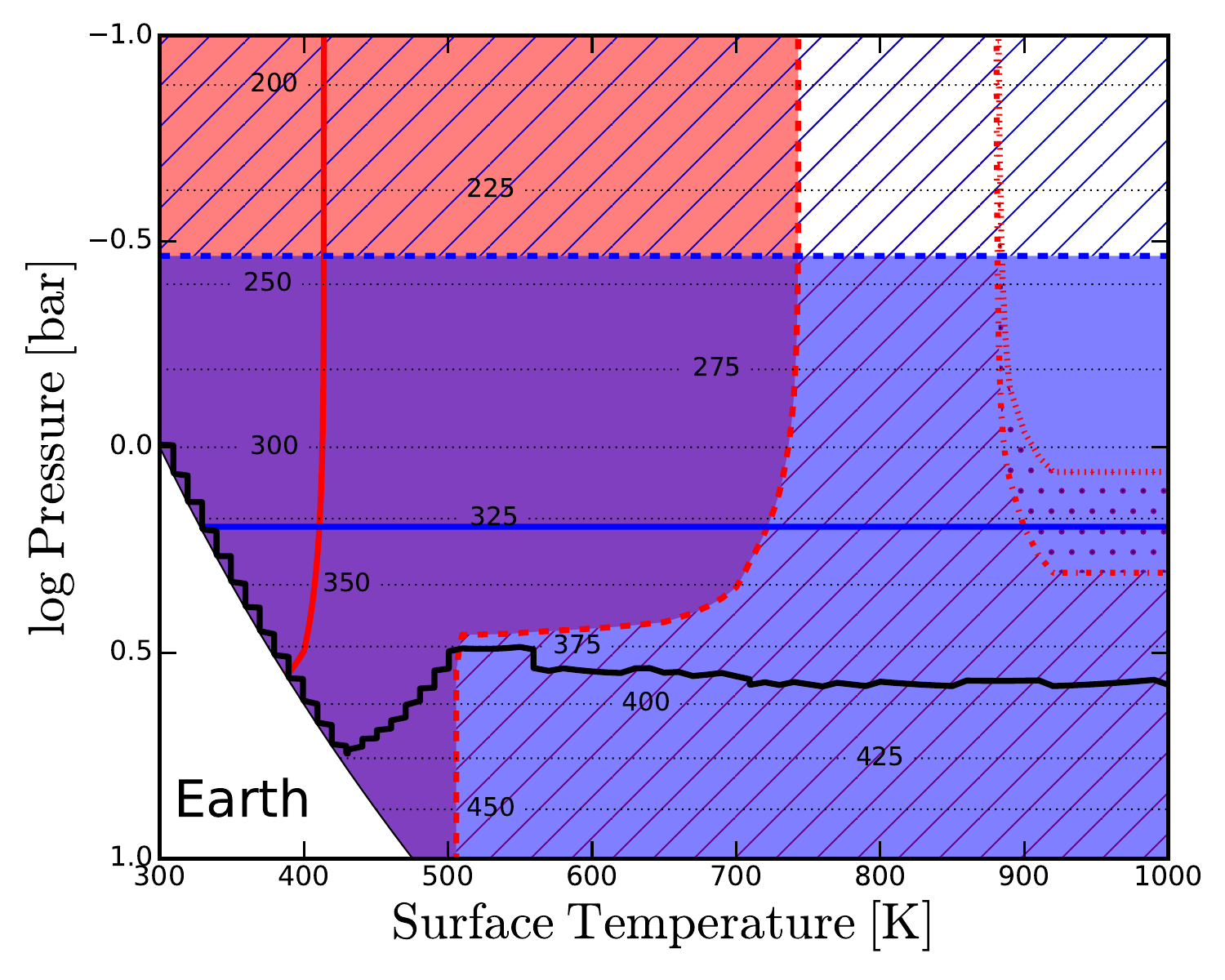}
\includegraphics[width = .32\linewidth, page=9]{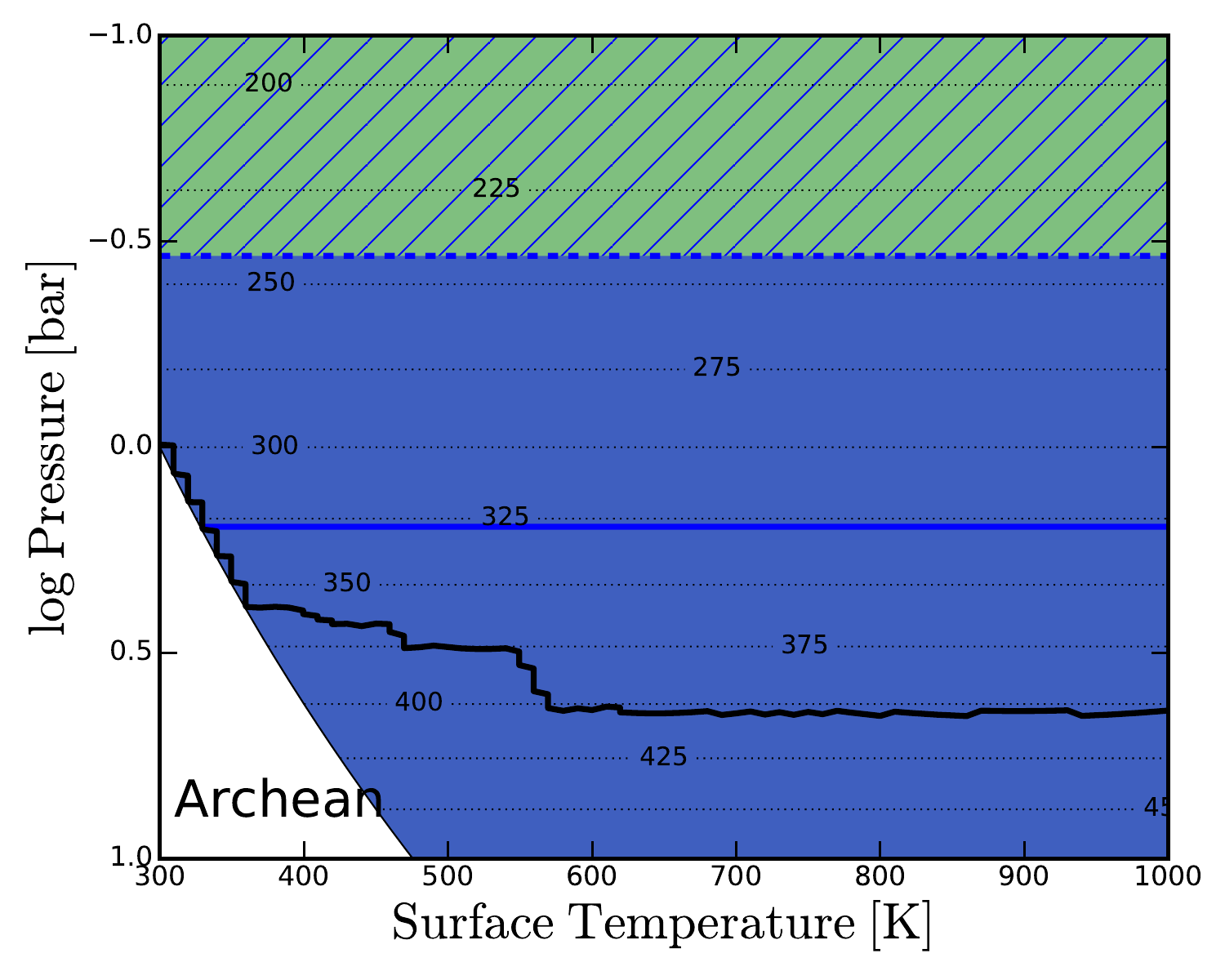}
\includegraphics[width = .32\linewidth, page=9]{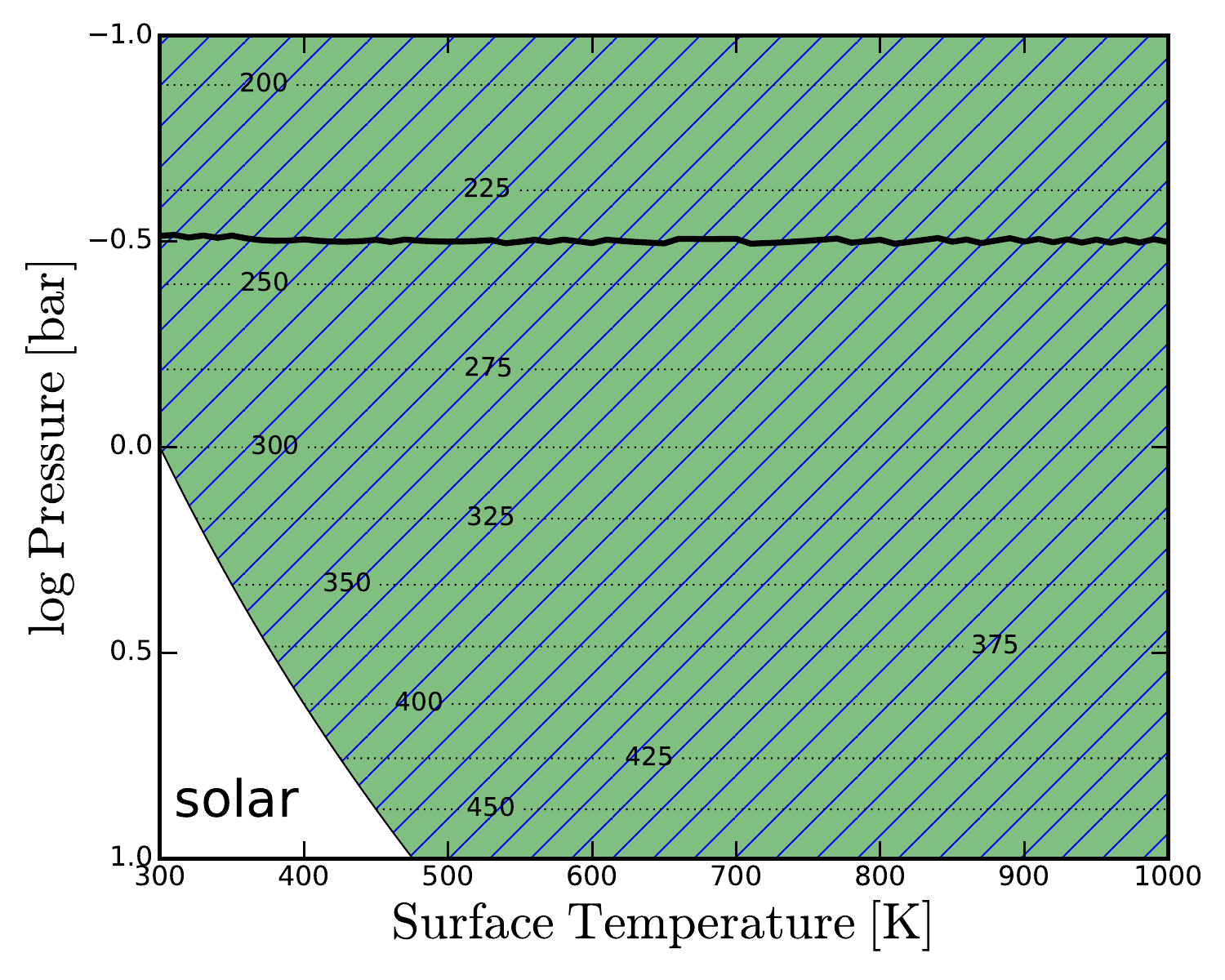}
\caption{As in Fig.~\ref{fig:NutriC}, but only for \ce{H2} (green), \ce{O2} (red), and \ce{H2O} (blue).}
\label{fig:NutriHO}
\end{figure*}

\end{appendix}
\end{document}